# Faculty of Physics and Astronomy

University of Heidelberg

Diploma thesis
in Physics

submitted by

**Alexander Westphal**

**July 2001**

[...] Πλατονος, ως φησι Σωσιγενης, προβλημα τουτο ποιησαμενου τοις περι ταυτα εσπουδακοσι, τινων υποτεθεισων ομαλων και τεταγμενων κινησεων διασωθη τα [...] φαινομενα.

(σιμπλικιος, commentarium in Aristotelis de caelo pg. 498a 46 – 498b 3)

[...] nachdem Platon, wie Sosigenes sagt, es als Problem für alle, die diese Dinge ernsthaft studieren, aufgestellt hatte, herauszufinden, welches die gleichförmigen und geordneten Bewegungen sind, durch deren Annahme die Erscheinungen [...] gerettet werden können.

# Quantum Mechanics
# and Gravitation

**This diploma thesis has been carried out by** *Alexander Westphal* **at the**

INSTITUTE OF PHYSICS

UNIVERSITY OF HEIDELBERG

**Under the supervision of**

**Priv. Doz. Dr. Hartmut Abele**

## Quantum Mechanics and Gravitation

In summer 1999 an experiment at ILL, Grenoble was conducted. So-called ultra-cold neutrons (UCN) were trapped in the vertical direction between the Fermi-potential of a smooth mirror below and the gravitational potential of the earth above [Ne00, Ru00]. If quantum mechanics turns out to be a sufficiently correct description of the phenomena in the regime of classical, weak gravitation, one should observe the forming of quantized bound states in the vertical direction above a mirror. Already in a simplified view, the data of the experiment provides strong evidence for the existence of such gravitationally bound quantized states.

A successful quantum-mechanical description would then provide a convincing argument, that the so-called first quantization can be used for gravitation as an interaction potential, as this is widely expected. Furthermore, looking at the characteristic length scales of about 10 µm of such bound states formed by UCN, one sees, that a complete quantum mechanical description of this experiment additionally would enable one to check for possible modifications of Newtonian gravitation on distance scales being one order of magnitude below currently available tests [Ad00]. The work presented here deals mainly with the development of a quantum mechanical description of the experiment.

## Quantenmechanik im Gravitationsfeld

Im Sommer 1999 wurde am ILL in Grenoble, Frankreich, ein Experiment durchgeführt, in dessen Verlauf sogenannte ultrakalte Neutronen (UCN) zwischen dem Schwerefeld der Erde und einem hochpolierten Planspiegel aus Glas eingefangen wurden. Wenn nun die Quantenmechanik eine hinreichend präzise Beschreibung der mikroskopischen Phänomene im Bereich der klassischen, nicht-relativistischen Gravitation darstellt, so erwartet man die Ausbildung quantisierter Zustände der vertikalen Bewegung von massiven Teilchen in einer solchen gravitativen Kavität. Bereits eine einfache, rein phänomenologische Analyse der experimentellen Daten erbrachte starke Hinweise darauf, daß hier zum ersten Mal gravitativ quantisierte Zustände von Teilchen „gesehen” worden sind.

Gelänge hier nun eine vollständig quantenmechanische Beschreibung, so würde dies ein entscheidendes Argument für die Anwendbarkeit der ersten Quantisierung auf die Gravitation als klassische, nicht-relativistische Kontinuumswechselwirkung darstellen. Ganz generell wird hier der Einfluß der Gravitation auf die Bewegung von Teilchen bei einer Abstandsskala von mehreren zehn µm gemessen, was erstmal den Test des Gravitationsgesetzes überhaupt auf einer Skala von unter einhundert µm erlaubt und damit gegenwärtige Resultate [Ad00] um knapp eine Größenordnung unterbietet. Die vorliegende Arbeit wird sich hauptsächlich mit der Suche nach einer geschlossenen quantenmechanischen Beschreibung dieses Experiments befassen.

# Introduction

Gravitation is the interaction, that is perhaps the most basically one in daily life. Besides the artificial phenomena and technologies, the facts, that we are bound to the earth, and that most of the wanted or unwanted motions of objects are caused more or less directly by gravitational forces represent the very first experiences of our lives.

Furthermore, the large-scale structure of the celestial motions as well as of the entire universe itself is controlled (nearly) entirely by gravitation. Einstein gravity dictates the dynamics of the world as a whole.

This situation arises from three facts. First there is the match of charge quantization of protons and electrons which are the constituents of ordinary matter (besides neutrons). This results in exact cancellation of the immense electromagnetic forces inside all-day's matter, leaving mechanical, chemical and electromagnetic properties of matter as the only macroscopic phenomena to be generated by electromagnetism. Next, one has to notice the short-ranged nature of the strong and weak interaction limiting their effective coupling distance to $10^{-15}$ m and $10^{-18}$ m, which reduces their effective macroscopic actions to generating the atomic structure of matter and its radioactive properties. And last but not least there is the cumulative nature of gravitation – there is only one type of "gravitational charge" known, usually called *heavy mass*. This situation is the only reason for gravitation dominating the macroscopic phenomena on earth despite the fact, that it is more than 38 orders of magnitude weaker at the elementary particle level than all of the other interactions known today.

Considering this predominance of gravitation at the macroscopic scales and its apparent ability to govern the development of planetary systems, galactic structures and even the whole universe, one has to realize a fundamental inconsistency in describing the interactions. Despite having one of the most beautiful classical field theories to describe gravitation on macroscopic scales at hand with "General Relativity", which is capable to totally reduce the gravitational force into geometry of space and time, we are not yet able to reformulate this theory within the framework of quantum field theory, which is tremendously successful in explaining the structure of the non-gravitational interactions.

Only within the past ten years we have developed the so-called string theories as a way, which perhaps points out a possibility to both geometrize and quantize all interactions in a consistent unification. This way consists of relaxing the hypotheses of point-like elementary objects and four-dimensionality of space-time at all scales. Recent developments on this field suggest, that the well-known non-relativistic law of gravitation first derived by Newton might only be an approximation sufficiently close to reality only at distances above 0.1 mm. Since experimental tracking of the law of gravitation has so far succeeded down to 0.2 mm distance of two test masses, there is a submillimeter regime where something completely new could be found out concerning the nature of gravitation [Ar98, Di96].

Looking closer at the state of experimental tractation of gravitation, one realizes, that in contrast to the other interactions there are attempts to quantize gravitation together with all the other interactions without having even proved completely the validity of quantum mechanics under influence of the classical non-relativistic gravitational field. Though the existence of interference patterns caused by the earth's gravitation has been shown using neutrons in a crystal interferometer [Co75], there the necessity to show the existence of quantum mechanical binding and discretization in a gravitational field remains.

In summer 1999 an experiment at ILL, Grenoble was conducted. So-called ultra-cold neutrons (UCN) were trapped in the vertical direction between the Fermi-potential of a smooth mirror below and the gravitational potential of the earth above [Ne00, Ru00]. If quantum mechanics turns out to be a sufficiently correct description of the phenomena in the regime of classical, weak gravitation, one should observe the forming of quantized bound states in the vertical direction above a mirror. Already in a simplified view, the data of the experiment provides strong evidence for the existence of such gravitationally bound quantized states.

A successful quantum-mechanical description would then provide a convincing argument, that the so-called first quantization can be used for gravitation as an interaction potential, as this is widely expected. Furthermore, looking at the characteristic length scales of about 10 μm of such bound states formed by UCN, one sees, that a complete quantum mechanical description of this experiment additionally would enable one to check for possible modifications of Newtonian gravitation on distance scales being one order of magnitude below currently available tests [Ad00]. The work presented here deals mainly with the development of a quantum mechanical description of the experiment.

From here one can see the outline of the diploma thesis presented here – to search for a quantum mechanical description of this experiment [Ne00, Ru00] and then eventually to use its results to track the law of gravitation in a distance range between 10 μm and 100 μm.

# Contents

# Chapter 1

## “Search for quantum states in the gravitational field”

As already mentioned in the introduction the main purpose of this thesis is the description of an experiment we did at the Institute Max von Laue – Paul Langevin for Neutron Physics (ILL) in Grenoble, France. This thesis demonstrates, that quantized bound states of so-called ultracold neutrons (UCN) in a gravitational cavity have been observed.

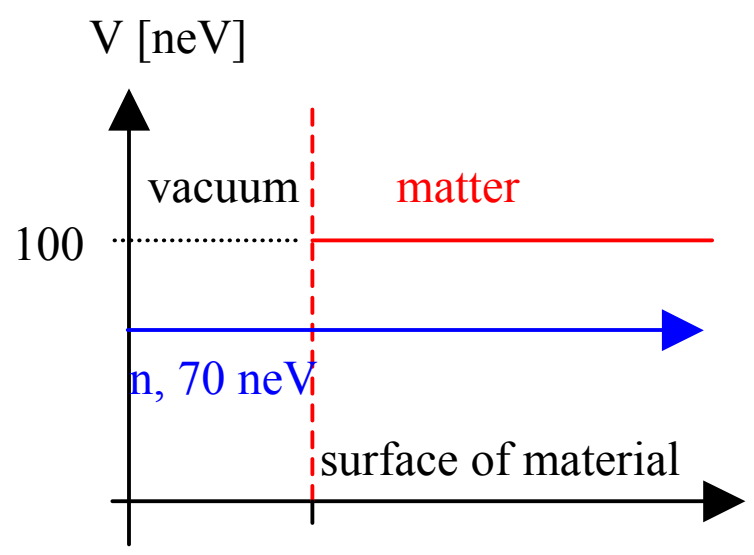

Fig. 1.1 : Fermi Pseudopotential

UCN are neutrons with kinetical energies below about 100 neV, which due to such small energies are reflected totally from smooth surfaces of solid state matter under all angles of incidence. The phenomenological origin of this behaviour is the effective and most often repulsive so-called Fermi pseudopotential of 100 neV of order of magnitude (Fig. 1.1), that neutrons are exposed to.

The basic properties of neutrons leading to this “ultra cold” behaviour will be explained later. Here it is important to see that for such UCN a smooth matter surface acts as a mirror. Together with the earth’s gravitational potential it forms a gravitational potential well (Fig. 1.2). The basic properties of the corresponding Schrödinger equation then require the formation of eigenstates of the Hamiltonian of such a system with discretized eigenvalues of energy. For the idealized situation shown in Fig. 1.2 the corresponding Schrödinger equation can be solved analytically by means of a special linear combination of Bessel function of order $^2/_3$ , so called Airy functions of first type *Ai(ξ)*. As one can easily derive by using the mass of the neutron and the earth’s gravitational acceleration constant, the height scale shown in Fig. 1.3 corresponds to energies on the scale of a few $10^{-3}$ neV. This shows clearly that even UCN energies are orders of magnitude beyond a situation, where only the few lowest bound state would exist. However, there are quantized energies only in the vertical direction. In the horizontal directions there are no boundaries or potentials. That provides free propagation of neutrons in those directions.

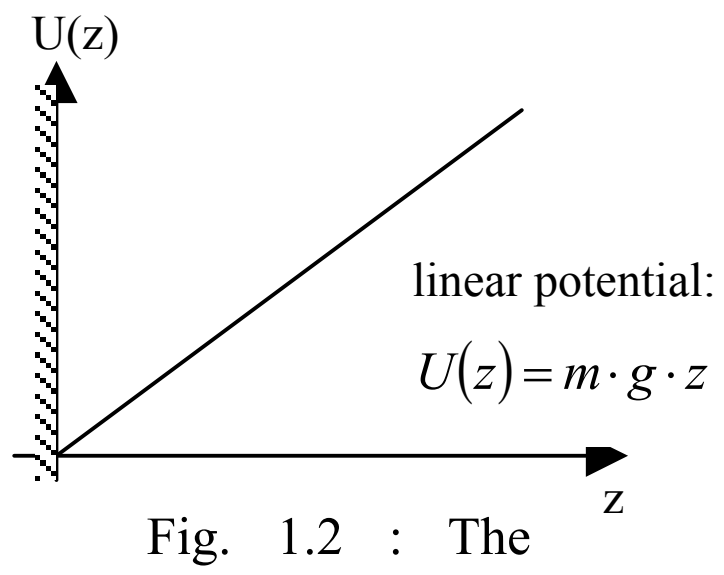

Fig. 1.2 : The gravitational well

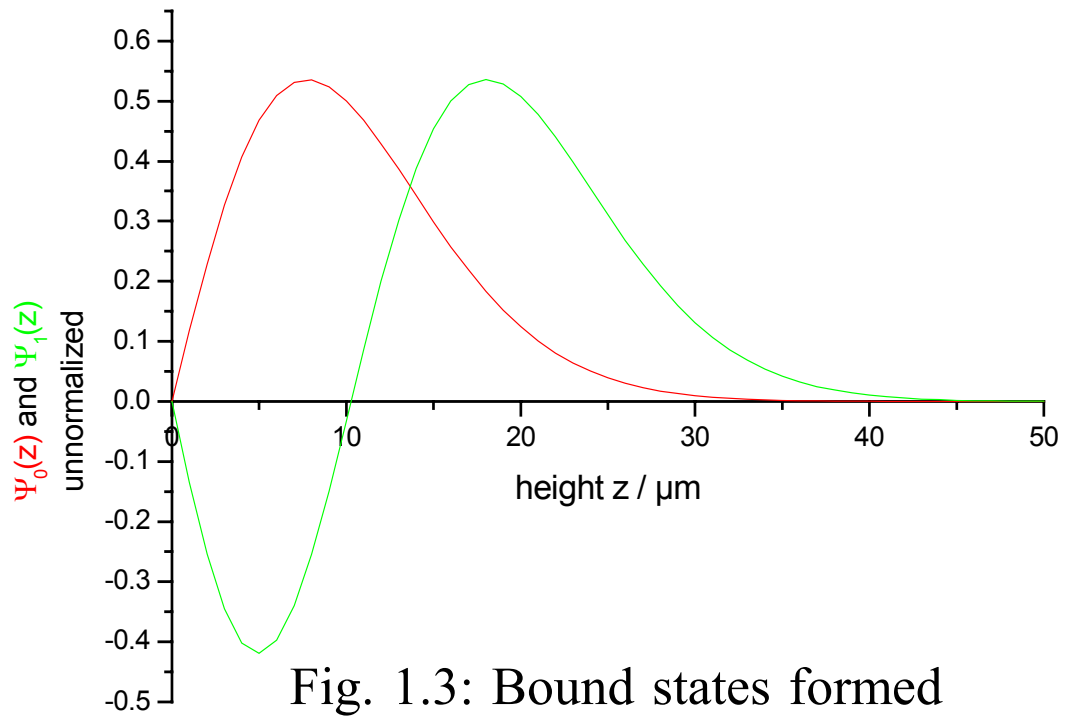

Fig. 1.3: Bound states formed by the potential Fig. 1.2

Suppose one is able to provide UCN entering the region above the mirror along classical parabolic trajectories, which have their turning points close to the front edge of the mirror. Such UCN would have inital vertical energies orders of magnitude below the mean energy of UCN of about 100 neV. The vertical velocity component of the UCN at the front edge of the mirror then is nearly zero. Such vertical velocities in the order of cm/s correspond to vertical energies comparable to the energy eigenvalues of the first few bound states expected to be formed above the mirror. Then only the few lowest states would be populated by UCN, which are prepared this way.

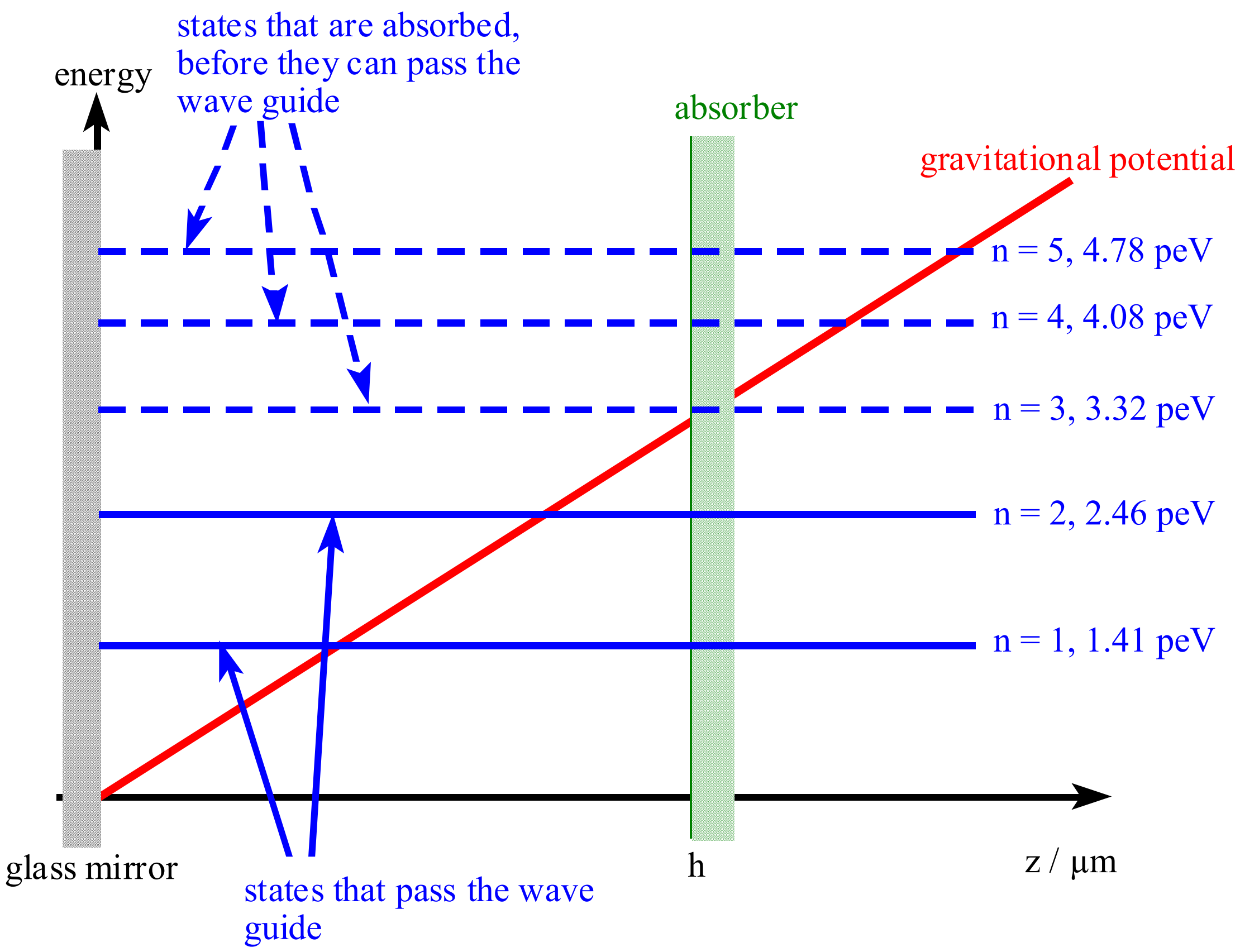

Fig. 1.4: Idea of the so-called “online” measurement.

Looking at Fig. 1.4 one could suggest that a neutron absorbing medium would start cutting away neutrons whose states have a sufficiently large probability of being inside the absorbing medium. The average spatial extension of the states increases with the state number (see Figs. 1.3 and 1.5). One would expect therefore the ability of the states to carry neutrons raising successively for each state one after another with the height of the absorbing medium, if this is lifted from the mirror. The result is, that the system's ability to carry neutrons in its bound states should show a more or less step-like dependence on the height of the absorbing medium.

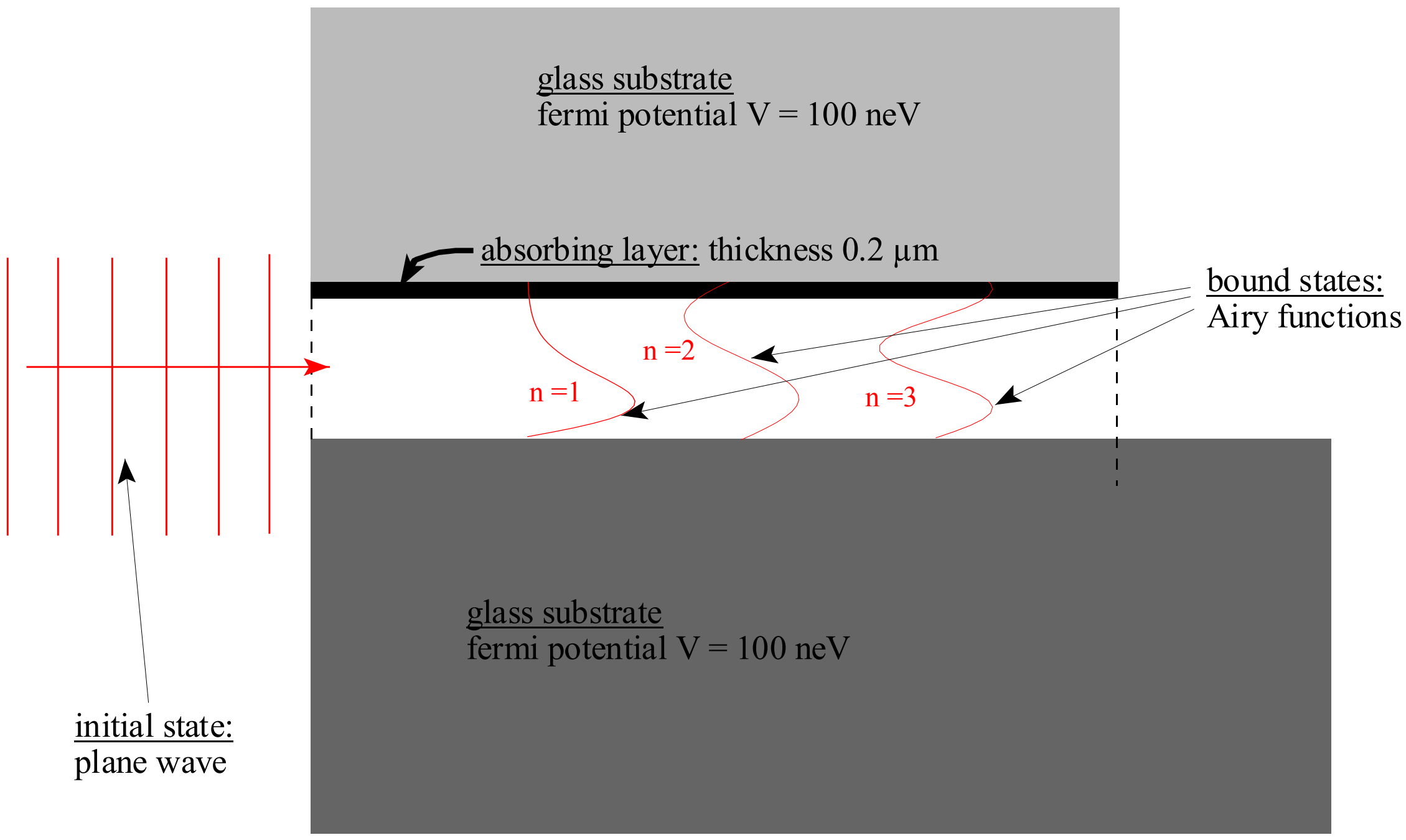

Fig. 1.5: Simplified view of the setup of the "online" measurement

The heights, where the distinct increase in the system's ability to carry neutrons occur, would be associated with the mean spatial extensions of the states and therefore with their energy eigenvalues. Measuring the transmitted neutron flux of horizontal motion above a mirror and below such an absorbing medium would show the ability of a given state to carry neutrons along a certain horizontal distance. A horizontal one-dimensional wave guide of variable width made up by a mirror at bottom and an absorbing medium at top and filled by UCN with very small vertical energies as mentioned above therefore could show the existence of gravitationally bound states by the more or less step-like structure of its transmittivity as a function of wave guide width.The measurement of the transmitted flux of such a wave guide indeed shows signs of such a behaviour:

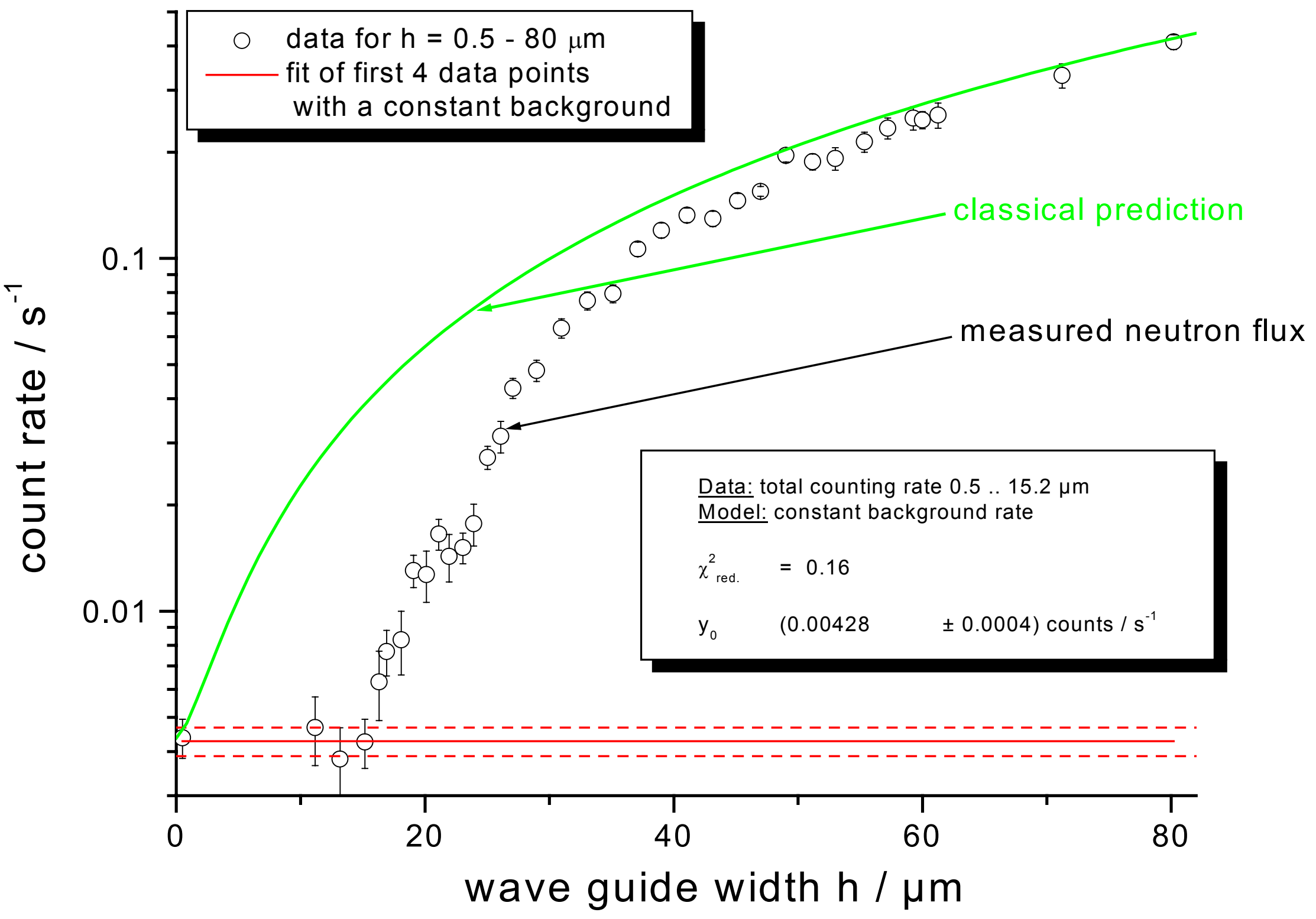

Fig. 1.6: Results of the measurement of transmitted neutron flux, classical prediction of the flux-width-dependency, and background determination.

The strongest evidence for such bound states would be expected from the situation where the width of the wave guide is smaller than the spatial extension of even the lowest bound state expected to exist. There should be no transmission at all then at a wave guide width of below about 15 µm as seen from Figs. 1.4 and 1.5. This effect would be the most significant sign of the existence of gravitationally bound states in such a wave guide.

Fig. 1.6 indeed shows an apparent region of non-penetration of the correct order of magnitude. The green curve shows the behaviour, that one would expect from classical particles moving in the gravitational field. The deviation of the data from this prediction is obvious.

Later in the thesis a full quantum mechanical description of this measurement will be developed, that is able to describe the data with a fit of only two free parameters.

The basic setup of the apparatus, that obtained the result displayed in Fig. 1.6, can be seen in Fig. 1.7:

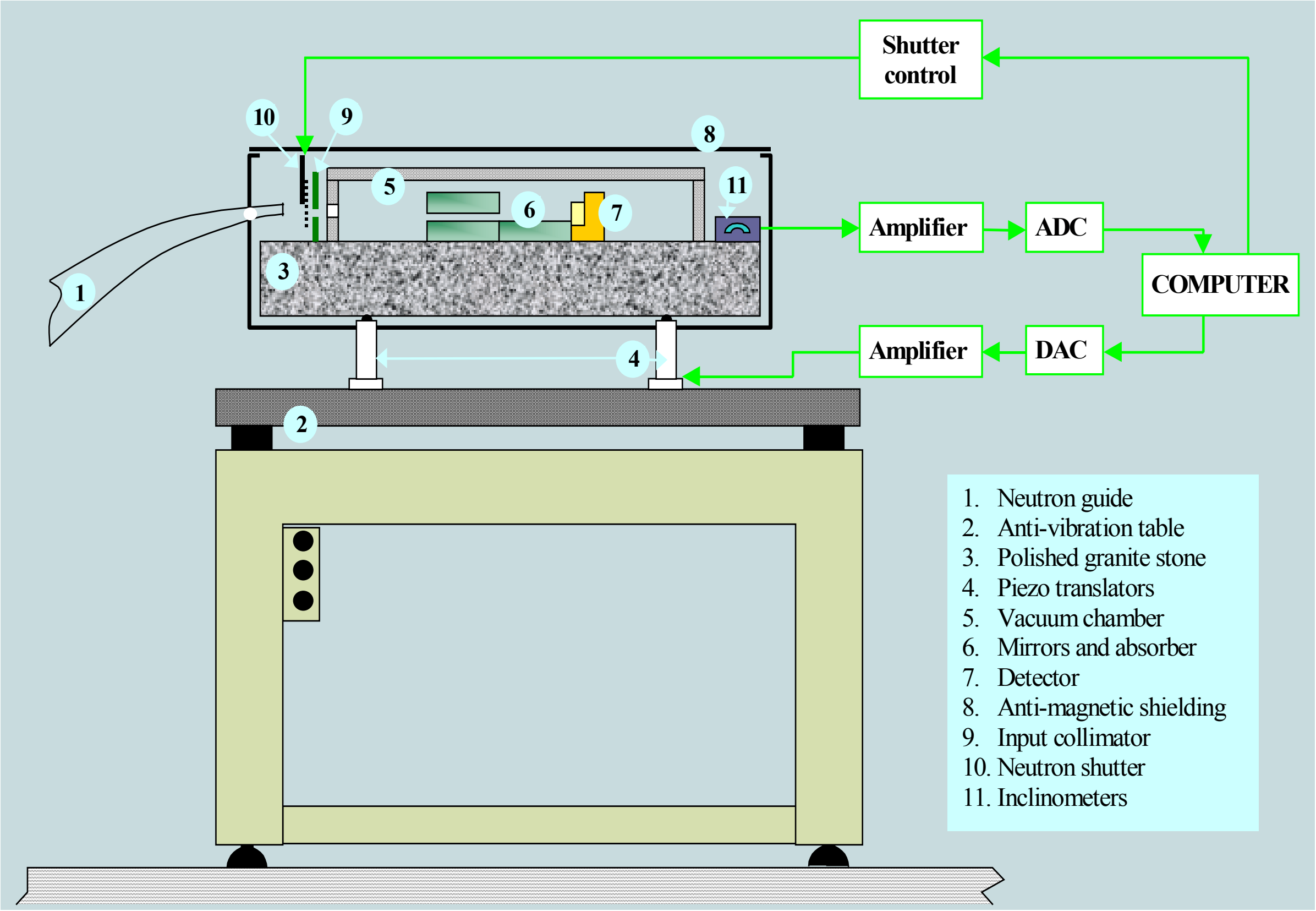

Fig. 1.7: The experimental setup – overview

In Fig. 1.7 one sees (Fig. 1.7 / 6) the setup of the horizontal one-dimensional wave guide described above. The condition of UCN entering the entrance of this wave guide classically spoken close to the turning points of their parabolic trajectories is to be fulfilled by a collimation system (Fig. 1.7 / 9). One has to use an anti-magnetic shielding (Fig. 1.7 / 8) to prevent neutrons from being immersed into potentials arising from the coupling of their spin to artificial magnetic or geomagnetic fields. Neutron detection (Fig. 1.7 / 7) is performed using a $^{3}He$ gaseous detector to catch the total flux of neutrons though the wave guide (Fig. 1.7 / 6). This detector provides a quantum efficiency of 50 % if used at the peak of the detection reaction's energy release:

$$n + {}^{3}He \rightarrow t + p\,, \quad \Delta Q = 0.764\ MeV$$

and of about 60 % if the full counted energy spectrum window is used. The other devices shown are used to provide horizontal leveling of the waveguide to an accuracy better then 10µrad and to provide decoupling from mechanical vibrations.

The bottom mirror (Fig. 1.7 / 6) is made from a glass plate, which has been polished to be smooth within a range of a few Å (see chapt. 2). The absorber (Fig. 1.7 / 6) uses the same type of polished glass plate as a substrate. However its surface then has been treated with an acid to yield a large microscopic roughness of the surface while maintaining its long scale flatness. Finally this rough surface has been coated by means of magnetron evaporation with a layer of 2000 Å made from an alloy consisting of 54 % *Ti*, 35 % *Gd* and 11 % *Zr*.

Gadolinium with a neutron absorption cross section of 48000 barn even for neutrons of thermal energy is the most pronounced candidate for a neutron absorbing material. As it will be shown later in chapter 3, its absorption cross section even increases by three orders of magnitude if one uses it in combination with UCN. The detailed influence of both the absorbing and the reflecting properties of such a coating and the roughness applied to it will be described later. This is one of the major tasks of chapters 2 and 3. Here it is sufficient to mention that the coating's alloy structure significantly reduces the repulsive character of the coating's Fermi pseudopotential, thus enabling the neutron's wave function to enter more easily the coating. Further the coating's roughness causes significant amounts of upscattering of neutrons carried by lowest bound states by means of strong non specular scattering at the rough surface of the coating. Both effects are enhancing the coating's effective absorbtivity. The second effect originates mainly because non-specularly upscattered neutrons would correspond to highly excited bound states. Due to their large spatial extension these neutrons would be excellently absorbed by the coating even if its position above the mirror, that is, the width of the wave guide was quite large. For simplification this coated glass substrate will be called "absorber" from now on.

The position of the absorber is set by means of three active piezo element leveling devices and controlled by three precise inclinometers. Zero position is obtained from touchdown onto the mirror surface. Then the absorber is shifted leg by leg to its targeted position obtaining its actual position by means of triangulation from inclinometer values. At each position of the absorber the total flux of the neutrons through the wave guide is then integrated over a certain time with the $^3$He-detector. The positioning system allows to place the absorber with an accuracy of about 1 µm [Ru00].

At the end of summer 1999 there were conducted three runs of described kind with identical setup conditions. Data acquisition points were distributed with 2 µm spacing in the range of 0 µm to 50 µm wave guide width, whereas 10 µm spacing was chosen for larger wave guide widths up to 160 µm. The bottom mirror of the setup was formed by two identical glass mirrors as described above of 6 cm length each and 10 cm width each. These two plates were shifted against each other in their vertical position by about $(5 \pm 1) \cdot \mu\mathrm{m}$. The absorber was a coated rough glass plate (see above) of 13 cm length and 10 cm width that covered the first 9 cm of the two bottom mirrors:

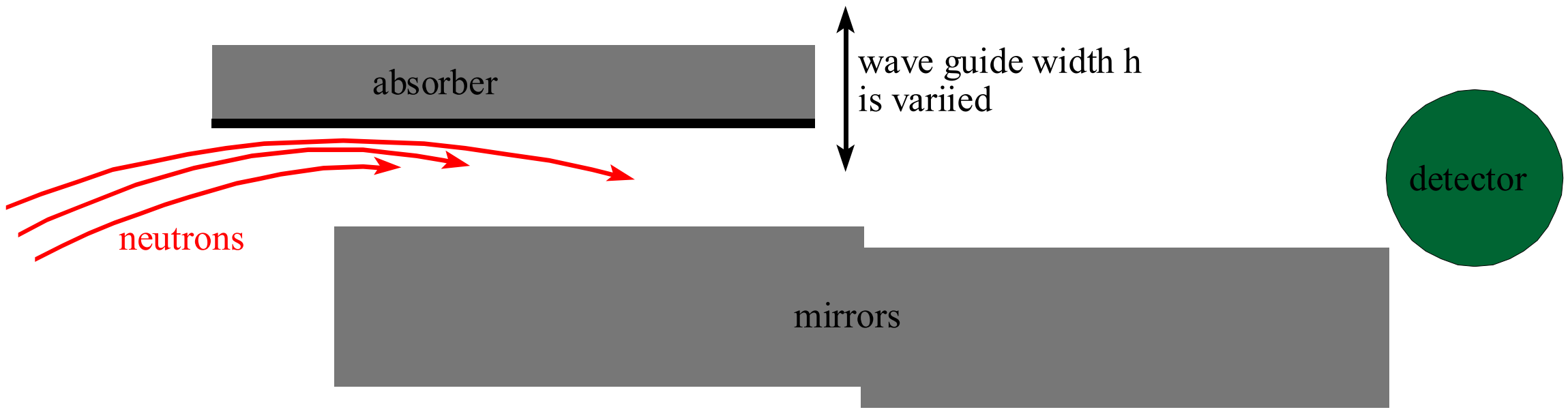

Because the three data acquisition runs were taken with identical setup and nearly identical acquisition steps, the data can be combined into one result describing the transmitted neutron flux of the wave guide shown in Fig. 1.6 as a function of its width. This result is given in Fig. 1.6 (see [Ru00]).

It is clearly visible that the neutron count rate vanishes to the background level for wave guides narrower than 15 µm. This regions size of about 15 µm has been checked independently of the piezo positioning system using foils placed between absorber and mirror, which provide constant distances. The background level of $4.3 \cdot 10^{-3} \cdot s^{-1}$ has been established several times by independent measurements.

Thus, as a very first result one can report a measurement of background directly from the plot shown in Fig. 1.4 averaging the first four data points, which is completely independent on the rest of the data:

$$background\ count\ rate = (4.28 \pm 0.4) \cdot 10^{-3} \cdot s^{-1} \ .$$

The existence of about 15 µm of non-transmission already indicates strongly – as shown in a mostly qualitative argument on pp. 2/3 – the formation of at least the gravitationally bound ground state of the system.

A second measurement was then performed, which aimed to record the neutron flux density distribution of the neutrons, which leave the wave guide and travel across the free far end of the bottom mirror. Therefore one needs an efficient position resolving neutron detector with d 2 µm spatial resolution in vertical direction to record the neutrons, which leave the far end of the bottom mirror. Such a device was placed at the far end of the bottom mirror. The density distribution of tracks in the detector caused by neutrons corresponds to the neutron flux density distribution (see [Ru00]).

The neutrons arriving at the end of the absorber-free part of the bottom mirrors should be either in a coherent or in an incoherent superposition of vertical eigenstates formed by the gravitational potential of the earth and the Fermi-pseudopotential of the bottom mirror. Thus the vertical distribution of probability density and therefore the neutron flux density should show a more or less oscillating structure.

The setup of the wave guide, the absorber-free far end of the bottom mirrors and the detector was chosen as follows:

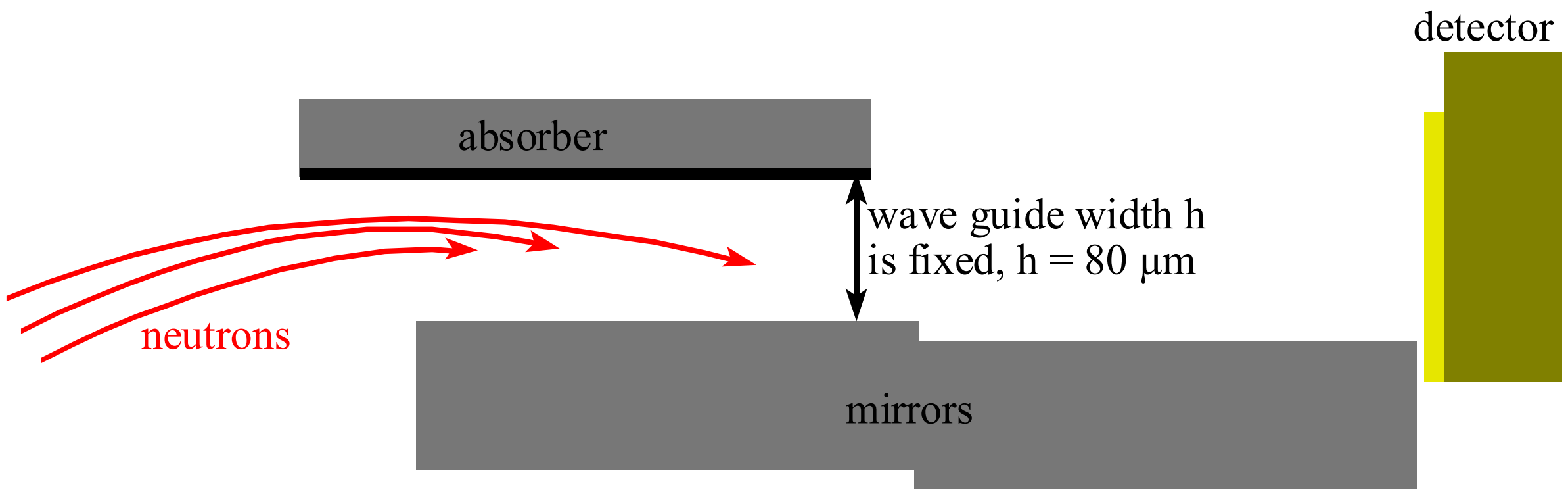

The position resolving detector was a so-called "CR 39 position sensitive track recording neutron detector". It consists of a plastic body coated with $^{235}$U. Neutrons with a surface-perpendicular velocity component above a critical value of about 4.5 $^{m}/_{s}$ are able to overcome uraniums Fermi-pseudopotential. They will enter the enter the coating, which is about 1 µm in thickness. There, neutrons will cause fission of $^{235}$U-nuclei with an overall quantum efficiency of about 10 %. Because there usually are emitted two high energetic fission fragments, one will leave the detector to the front while the other one causes a track inside the plastics. This plastics body will be chemically developed after irradiation of the detector. The tracks themselves are read out using automatic optical scanning microscopes in several layers of depth [Ru00]. This data has to be corrected,because eventually deformation of the plastics takes place during development. Finally one has to reconstruct the tracks in 3 dimensions from photos of sheets of different depths in order to localize the original point of neutron-caused fission. This position is within the thickness (1 µm) of the uranium coating identical with origi-nal position of a neutron entering the detector [Ru00]. Reconstruction process then itself eventually leaves a certain smearing of the initial position of neutrons entering the detector. Fig. 1.8 shows the result of a measurement after 6 days of detector irradiation at a wave guide width h = 80 µm. About 5000 tracks were recorded, to which the reconstruction process given in [Ru00] has been applied:

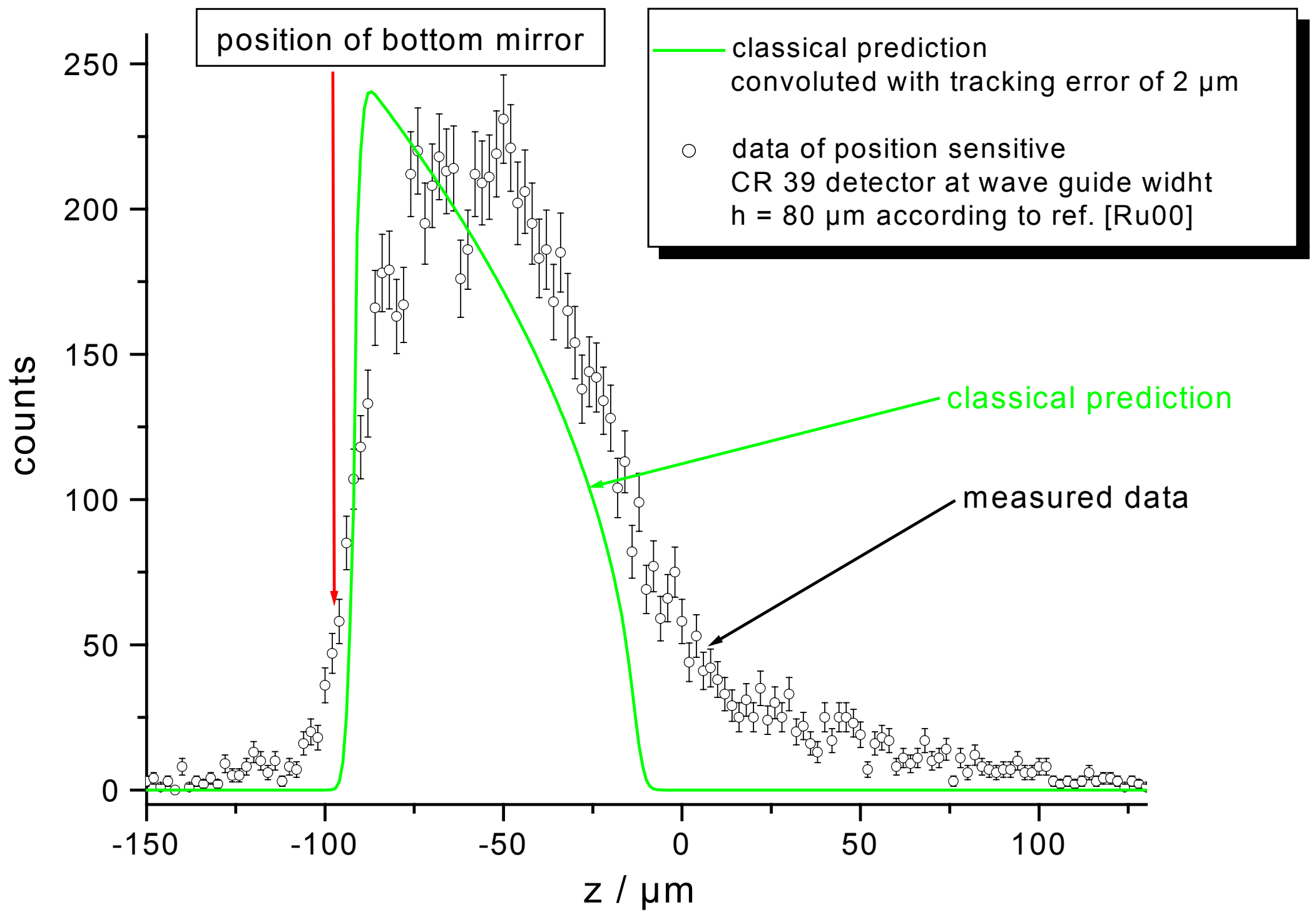

Fig. 1.8: Measurement of the vertical neutron flux density distribution after the wave guide, and classical prediction of this measurement.

Again one sees an obvious disagreement between the behaviour of classical particles moving in the gravitational field, that is given by the green curve, and the data.

For this measurement, too, a two-parametric quantum mechanical description will be presented later in the thesis.

# Chapter 2

# Properties of the setup

From chapter one we know now that two glass mirrors and an alloy-coated rough glass substrate, working as a neutron absorber, are essential parts of the wave guide system. As it will be shown in chapter three, section three, these surface properties will enter directly the calculation of the transmission of the wave guide. In particular, the roughness of the surfaces would be a free parameter, that one would have to adjust by fitting routines, if it was not measured. Therefore one should determine the surface properties of the wave guide system to avoid additional varying parameters.
Another critical part of the apparatus is the collimator system, that is placed in front of the wave guide entrance. This device allows significantly for a shaping of the arriving neutron velocity spectrum. This spectrum, however, has influence on the details of the coupling of the arriving neutron beam to the interior of the wave guide. Therefore, a discussion of the collimator is needed.

## 2.1) The surfaces of the mirrors and the absorber

From chapter one we know now that two glass mirrors and an alloy-coated rough glass substrate, working as a neutron absorber, are the essential parts of the wave guide system. Lets have a simple look at a formerly mirror-like reflecting metal surface, which has been grinded with a device, that generates an unevenness of the surface within one order of magnitude compared to the wavelength of visible light. We then observe that there is no longer any pronounced mirror-like or so-called “specular” reflection. We observe diffuse “non-specular” scattering of light from such a rough surface with a more or less pronounced peak of reflectivity in specular directions, depending on the details of the roughness.

It was already mentioned in chapter one that neutrons entering matter “feel it” in terms of a so-called Fermi pseudopotential, which is most often repulsive and more or less absorptive. The situation is a close analogon to the complex refractive index n with $\Re(n) = 1 - \varepsilon, \varepsilon << 1$ x-rays see inside matter compared to a purely real refractive index of unity for pure vacuum. The matter’s refractive index causes absorption of x-rays and for very small angles of incidence from the vacuum the phenomenon of total external reflection. Indeed the refractive index of matter seen by x-rays can be properly described by a so-called quasi-optical potential – a precise analogon of the Fermi pseudopotential seen by neutrons if they enter matter.
Now since the structure of scattering seems to depend on the roughness of the surface in comparison to the wavelength of scattered radiation, one could suggest that detailed information about the scattering of radiation with known properties from a given surface would enable one to extract characteristics of the roughness from data.
With this motivation in mind we will now start to develop a description of the scattering of plane waves with a given wavelength from a rough surface to the necessary degree of accuracy. We will use plane waves, since any arbitrary pulse structure can be generated from a superposition of plane waves due to Fourier’s theorem.
The first question that needs to be answered is how to describe the roughness of a surface in mathematical terms. Therefore let us first find out what the term roughness means. Generally, “roughness” is assigned to the phenomenon that the surface of any arbitrary piece of matter deviates somehow from the properties of the idealized mathematical term of a perfectly flat and smooth plane surface.
Looking at the interatomic distance scale this phenomenon has a fundamental cause, for the structure of matter is a discrete one. Since this is in contrast to the continuum description forming the basis of the description of a perfectly flat and smooth surface, on this scale we have roughness at the very beginning. Beyond interatomic scales roughness is the phenomenon of mechanical imperfections disturbing the smoothness of a surface viewed at a given distance scale. These imperfections give rise to the fact, that the “landscape” of a surface viewed at a certain length scale or spatial resolution is more or less hill-dominated.
Imperfections can be generated directly in a mechanical way by means of grinding, or they can originate from purely statistical processes like treatment of a surface with an acid. It is clear that the same surface might look either rough or smooth depending on the resolution of view. Thus roughness is a scaling phenomenon. One will not be able to determine the complete structure of roughness at all length scales from just looking at the surface at one given resolution. Yet, such a look will provide information about roughness approximately within the same order of magnitude as the length scale, which is used to examine the surface. Common characteristic of roughness at all length scales is a cut-off behaviour. Due to the finiteness of all real surfaces there is at least one hill being the highest of the whole surface viewed at each length scale.

Since the roughness applied to the surfaces in question (mirrors, absorber) is generated in one case by polishment with very fine-grained devices and in the other case by etching, we will restrict the discussion to the case of roughness generated by processes of "white" random nature possessing complete isotropy in space. The application of a thermalized acid consisting of molecules with Maxwell-Boltzmann and therefore in each direction independently gaussian distributed velocities is probably the closest case to such an idealization. Because both, acids and very fine-grained polishment devices, produce very smooth and homogenous interaction regions at scales being large compared with their own sub-micron structure, surfaces treated these ways will look very flat and smooth at scales comparable to their extension as a whole. Howeve, they will show a more or less developed roughness at scales between interatomic and macroscopic distances. Such a roughness is usually called "micro-roughness". Micro-roughness caused by such stochastic processes can be described – as the stochastic origin of such roughness itself – by means of a random variable.

If one has already confirmed the macroscopic flatness of such a surface, one can define an average plane of zero height by averaging over the heights of all points of the surface viewed at a given resolution window. Then the height of the surface points as a function of the average lateral coordinates defined by the average zero plane can be described by a random variable:

Fig. 2.1

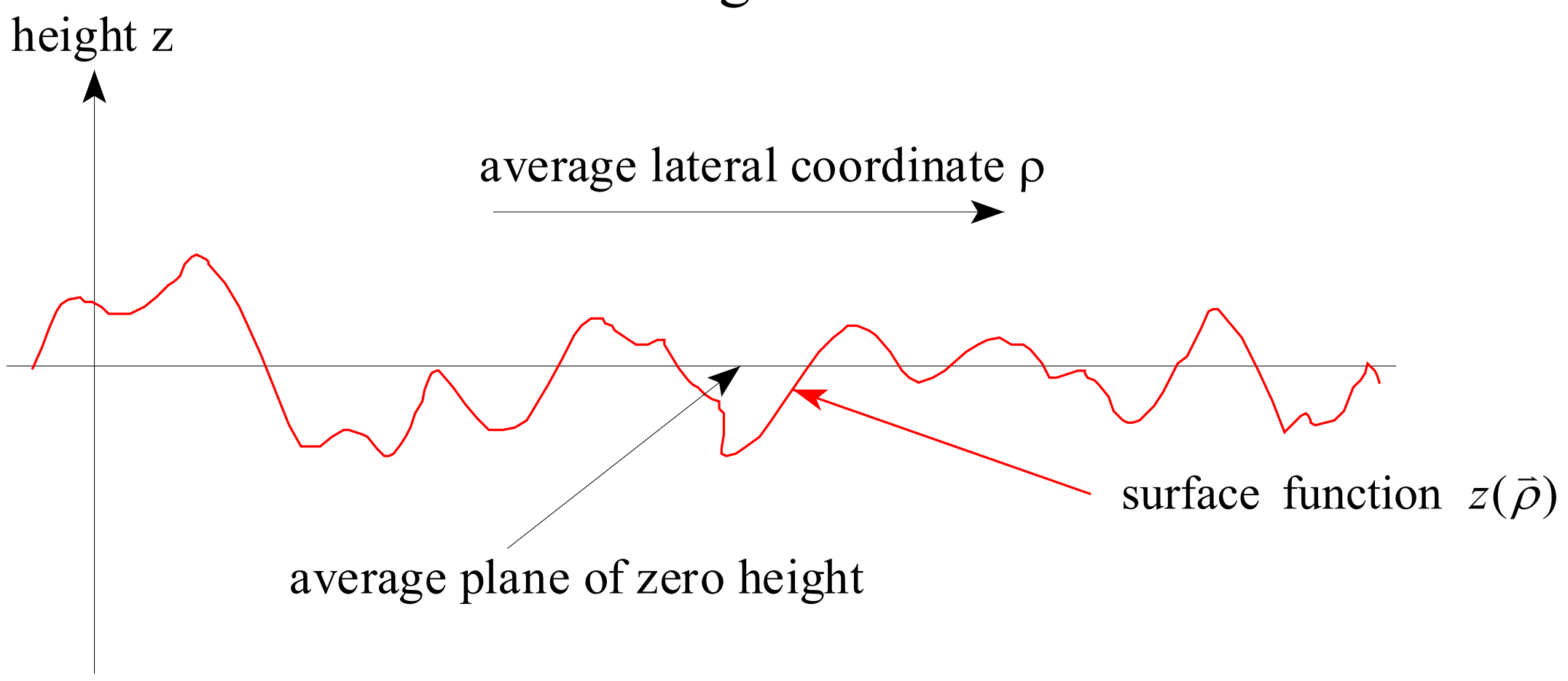

According to [Si88] it is possible for a wide range of rough surfaces to characterize the surface's height as a function of the lateral coordinates by means of a gaussian random variable. This is evident as the application of a thermalized acid of finite temperature to a surface leads to a stochastic origin of a surface's roughness with gaussian statistics. The main statistical information of a gaussian distributed surface is provided by the so-called "two-point correlation function" of the randomly distributed variable. The assumption, that the height of a rough surface is a gaussian random variable, then means, that it is governed by a height-height correlation function *G(x-x',y-y')*:

$$(2.1.1)\quad G(x-x',y-y')=\left\langle\left[z(x,y)-z(x',y')\right]^2\right\rangle=2\cdot\sigma^2\cdot\left(1-e^{-\frac{\rho}{\xi}}\right),\quad \rho=\left|(x-x',y-y')\right| .$$

Here $\sigma$ and $\xi$ denote respectively the root mean square roughness and the lateral correlation length of the roughness.

The height $z$ then is distributed gaussian with a probability density $w$:

$$(2.1.2)\quad w(z(x-x',y-y'))=w(\Delta z(\bar{\rho})=\frac{1}{\sqrt{2\pi\cdot G(\bar{\rho})}}\cdot e^{-\frac{\Delta z^2(\bar{\rho})}{2\cdot G(\bar{\rho})}}\ ,\quad \bar{\rho}=(x-x',y-y')\ .$$

From (2.1.2) it is visible, that points being very close together compared to $\xi$ have virtually no probability to be at different heights, while points with a separation much larger than $\xi$ are gaussian distributed in their probability to have different heights with a variance $\sigma^2$. Therefore, the correlation length $\xi$ describes a quantity best referred to as the "mean hill to hill next neighbour distance along the average zero plane" and $\sigma^2$ is a measure of the mean squared height variance, which is most often simply called the "roughness" of the surface.

Since we now have a formalism to describe a surface with its roughness, we will proceed to the next step, where we will have to describe the scattering of plane waves from a rough surface modelled as shown above.

Plane waves are infinitely extended in space. Therefore they average over all the surface height variations. If one, for instance, describes the interaction of the surface with a plane wave by means of a repulsive step potential with constant height at each surface point, the wave will see a smeared averaged potential which will have the structure of a gaussian error function as a function of the relative position vertical to the average zero plane of the surface. This can easily be derived from eq. (2.1.2) by integrating it for different semi-infinite z ranges. That leads to an interaction length of the surface on a plane wave of about $2\cdot\sigma$. If we now consider a weak potential, which is equivalent to wavelengths $\lambda_\perp$ far greater than its interaction length $2 \cdot \sigma$ or:

(2.1.3) $q_\perp \cdot \sigma << 1$

$q_\perp$ = wave vector transfer perpendicular to the average surface zero plane $= \frac{4\pi}{\lambda_\perp} \cdot \sin \vartheta$

there fortunately exists a quantum mechanical approach to plane wave scattering within the framework of perturbation theory. Within validity of eq. (2.1.3) scattering can be described by means of the first order Born approximation (BA). The detailed formalism is given in [Si88] and others and will show us the basic properties of scattering from rough surface as well as it will provide tools to extract the main roughness parameters $\sigma$ and $\xi$ from the data of a scattering experiment.

The part of this formalism, that is relevant for the description of scattering from a rough surface of a solid state material, is resumed in a summarized way in Appendix C. It is shown there, that the plane wave approach of BA has to be replaced by so-called "Fresnel eigenstates", which respect the fact, that solid state materials show the phenomenon of total external reflection of x-rays below a certain critical angle of incidence. The use of these Fresnel eigenstates modify the BA towards the so-called "distorted wave Born approximation" (DWBA), which to first order finally yields the expression (C.14) for the specular reflectivity of a rough surface

$$(C.14) \quad \frac{d\sigma}{d\Omega} \propto \left| \frac{q_z - q_z^+}{q_z + q_z^+} \right|^2 \cdot \left| e^{-q_z \cdot q_z^+ \cdot \sigma^2} \right|, \quad q_z^+ = q_z \sqrt{1-\delta} \, , \delta = 1 - n^2 \; .$$

The first factor describes the Fresnel reflectivity with respect to the average zero plane of the surface of a material with a refractive index *n*. The second one is a modified form of eq. (C.12) called "Nevot-Croce factor" after Nevot and Croce [Né76]. The emergence of $q_z^+$ in the Nevot-Croce factor compared to (C.12) leads to a constant reflectivity of unity for angles of incidence below a certain value given by $\delta$ if $\delta > 0 \wedge \delta \in$ Ñ. In the case of complex $\delta$, the Nevot-Croce factor shows a small suppression of the reflectivity below unity even inside the region of total external reflection. This is due to the fact, that a complex $\delta$ corresponds to absorptive media.

Such a small-angle-scattering measurement has been performed with one of the glass mirrors used in the wave guide experiment in summer 1999 at the small-angle-x-ray-scattering-laboratory (XSAS) of Ben K. Saidane, ILL Grenoble under his supervision [We99].
The setup of the specular scan is given in Fig. 2.3 below:

Fig. 2.3
specular scan

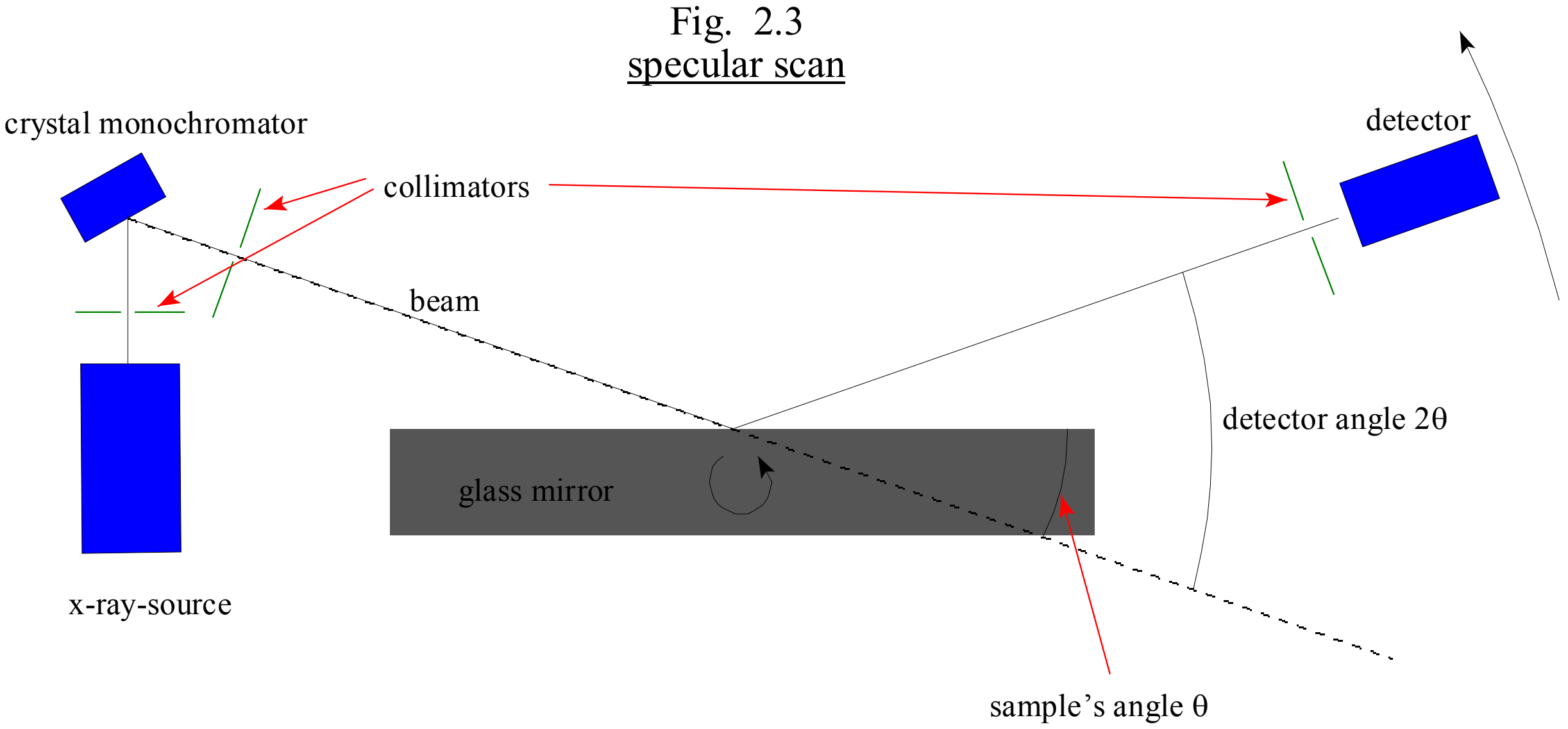

In this measurement both the sample and the detector were rotated counter-clock-wise in increments of $\Delta\theta$ and $2\Delta\theta$, respectively, because of the reflection condition $\theta_{counter} = 2 \cdot \theta_{sample}$ .
In the actual measurement $\Delta\theta = 0.005°$ was chosen and there 500 data points in the range of $\theta = 0° \ldots 2.14°$ were reorded with a counting time of 100 s for each data point. According to the long-term statistics of the XSAS-apparatus as a standard surface examination technique, systematical errors and drifts of the x-ray-source as well as the angle incrementing system are far below 1 %. The collimated x-ray-beam has a rectangular cross section of 100 µm by 10 mm in dimensions with a more or less gaussian intensity profile. The x-ray wavelength $\lambda$ was selected with a crystal monochromator and was $\lambda = (2.144 \pm 0.01)$ Å.
We will now rewrite eq. (C.14), which describes the specular scan (Fig. 2.3), in terms of the angles shown there. The solution of the Fresnel eigenstates obeyes Snellius' law. Using Snellius' law, which determines the angles of refraction for the part of x-rays entering the glass mirror:

Fig. 2.5

(2.1.4) $\cos(\theta) = n \cdot \cos(\theta_{trans})\ ,\ n = \sqrt{1-\delta}\ ,$

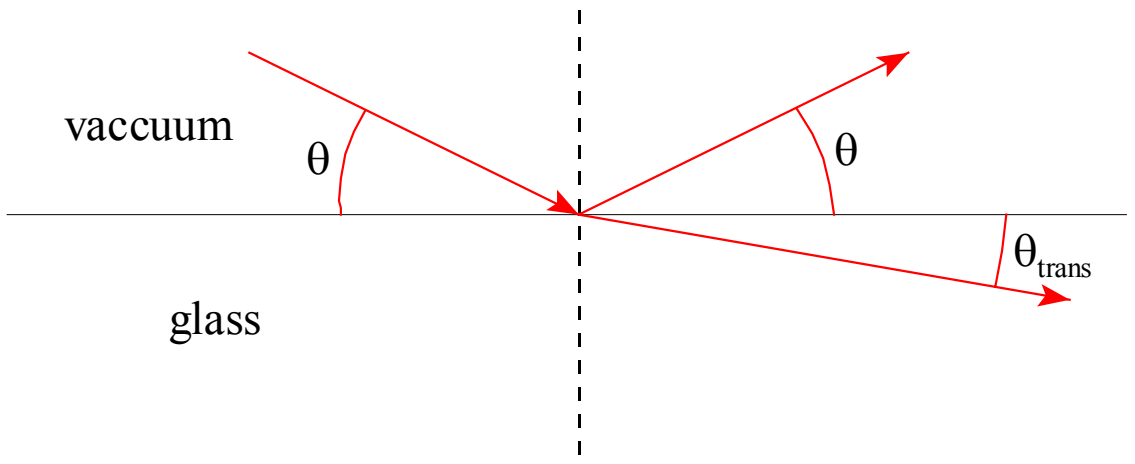

we can rewrite the wave vectors in eq. (C.14). Therefore, we can give now eq. (C.14) in terms of the measured quantities, where the scattering cross section is replaced by the x-ray-photon flux measured in counts, which is proportional to the cross section:

$$(2.1.5)\quad \Phi_{reflected} = background + scaling \cdot \left| \frac{\sin(\theta) - \sqrt{\sin^2(\theta) - \delta}}{\sin(\theta) + \sqrt{\sin^2(\theta) - \delta}} \right|^2 \cdot \left| e^{-16\cdot\pi^2 \cdot \frac{\sigma^2}{\lambda^2} \cdot \sin(\theta) \cdot \sqrt{\sin^2(\theta) - \delta}} \right| .$$

$$\text{where}: \theta[\text{rad}], \text{background}[\text{counts}], \text{scaling}[\text{counts}], \Phi_{\text{reflected}}[\text{counts}], \lambda[\text{Å}], \sigma[\text{Å}]$$

The proportionality constant is mainly a result of integration over solid angles. This is the fit function, which depends on four free parameters: *background*, *scaling*, the deviation of the glass' refractive index from unity, $\delta$, and the mean height roughness $\sigma$. The result of this measurement is shown in Fig. 2.6, where the data was normalized to unity for $\theta = 0$:

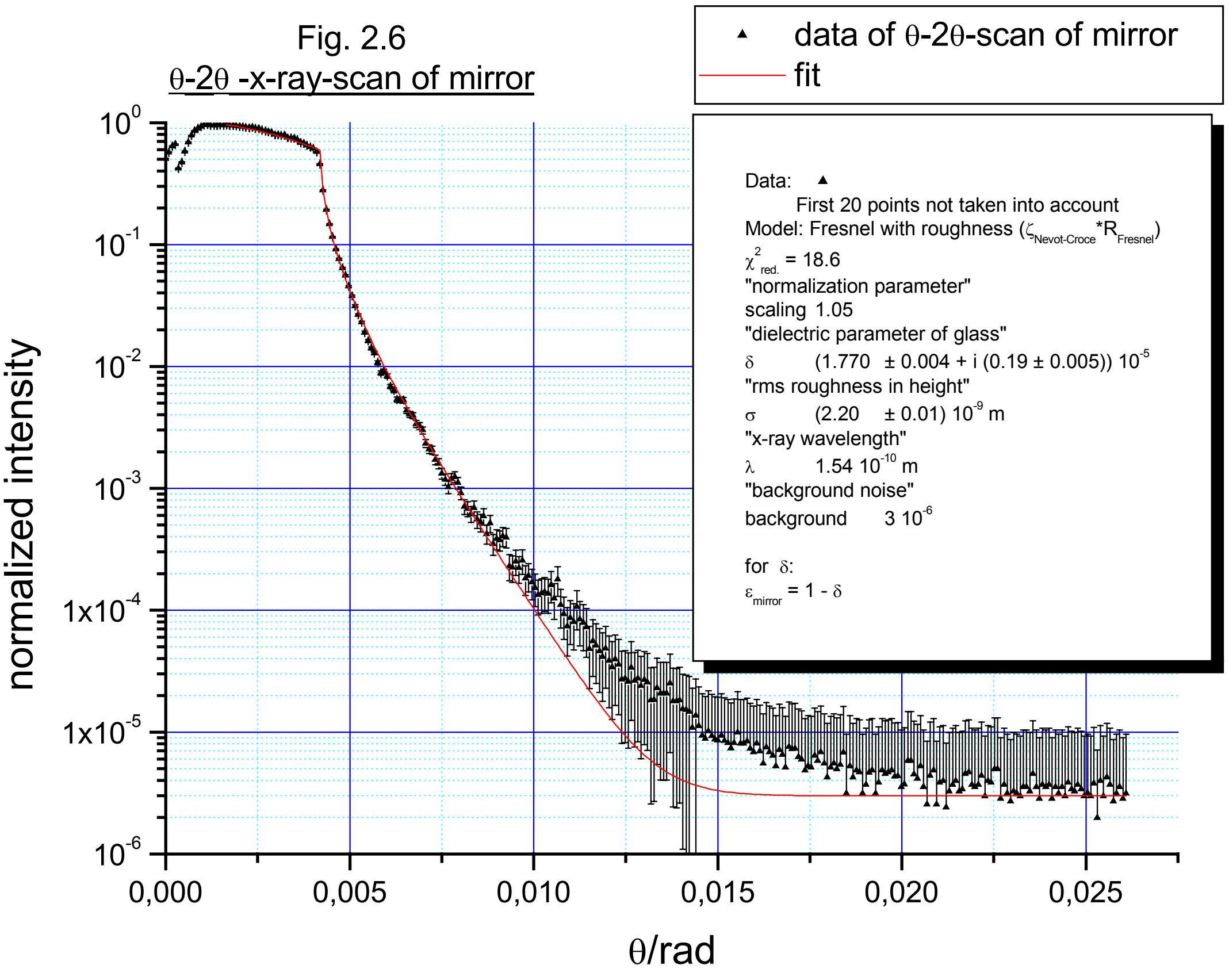

From this analysis we find the mean height roughness of the glass mirrors to be $\sigma = (22.0 \pm 0.1)$ Å.

Next a diffuse scan according to Fig. 2.4 was performed. The fixed angle between detector and the line of incidence was set to $\theta + \theta' = 0.52°$, which is equivalent to $\theta = 0.26° = 0.0045\ rad$ in Fig. 2.6.

Fig. 2.4
diffuse scan

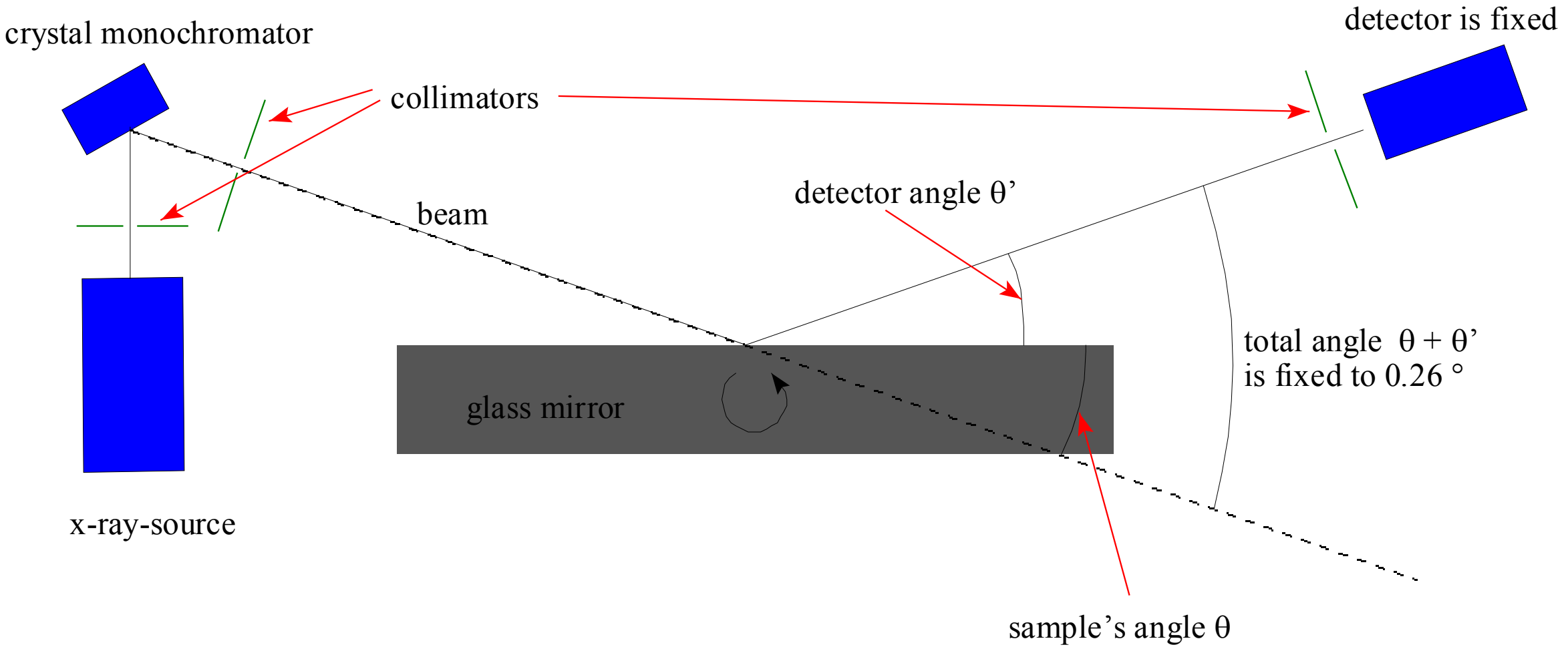

This angle was chosen to avoid the region of total reflection. This choice also guarantees enough intensity in the specular reflex to gain enough statistics in the non-specular combinations of $\theta$ and $\theta'$ (see Fig. 2.6 at θ = 0.0045 rad). The sample's angle was then incremented in steps of $\Delta\theta$ = 0.005° from $\theta$ = 0° to $\theta$ = 0.52°. The sampling time was chosen to be 100 s for positions, where reasonable fluxes where detected. However, for positions far away from the specular point the data accumulation time adapted in order to get at least a minimum of 20 counts for each data point. After the measurement data was renormalized to the normalized flux at $\theta = 0.26° = 0.0045\ rad$ in Fig. 2.6. The fit function which is to be used here one gets from (C.12) by rewriting it with the measured angles and adding to it a narrow gaussian describing the intensity profile of the beam which smears out the flux as function of angle deviation from specular condition close to the specular point. Thus we get here from (C.13):

$$\Phi_{diffuse} = N_0 + N \cdot \left[ \frac{2\pi \cdot \frac{\sigma^2}{\xi} \cdot \frac{\theta + \theta'}{\lambda}}{\left[ (\theta - \theta')^2 + \left( 2\pi \cdot \frac{\sigma^2}{\xi} \cdot \frac{\theta + \theta'}{\lambda} \right)^2 \right]^{3/2}} + A \cdot e^{-\frac{(\theta - \theta' - \theta_{offset})^2}{2 \cdot w^2}} \right]$$

$$\wedge \quad \theta + \theta' = 0.52° = const. \Rightarrow FWHH := 2\pi \cdot \frac{\sigma^2}{\xi} \cdot \frac{\theta + \theta'}{\lambda} = const.$$

$$\Rightarrow \quad (2.1.6) \quad \Phi_{diffuse} = N_0 + N \cdot \left[ \frac{FWHH}{\left[ (\theta - \theta')^2 + FWHH^2 \right]^{3/2}} + A \cdot e^{-\frac{(\theta - \theta' - \theta_{offset})^2}{2 \cdot w^2}} \right] \quad .$$

It has been used, that the cos can be expanded due to θ and θ' being small:

$$(2.1.7)\quad q_\rho^2 = k^2\left(\cos(\theta) - \cos(\theta')\right)^2 \approx \frac{k^2}{2}\cdot(\theta+\theta')^2\cdot(\theta-\theta')^2 \ .$$

Thus the fit function for a measurement of type Fig. 2.4 contains 5 free parameters: the background $N_0$ [counts], the scaling $N$ [counts], the full width half height of the Lorentzian describing the diffuse scattering according to eq. (2.1.6), *FWHH* [rad], a small offset of the position of the point of specular scattering, $\theta_{offset}$ [rad], and the width of the gaussian beam intensity profile $w$ [rad]. *FWHH* depends on both the mean height roughness $\sigma$ as well as the lateral height-height correlation length $\xi$. Because $\sigma$ is known from the specular scan (Fig. 2.3), one can determine $\xi$ according to eq. (2.1.6) if *FWHH* is known from a diffuse (Fig. 2.4). Now results of such a diffuse scan are given in Fig. 2.7:

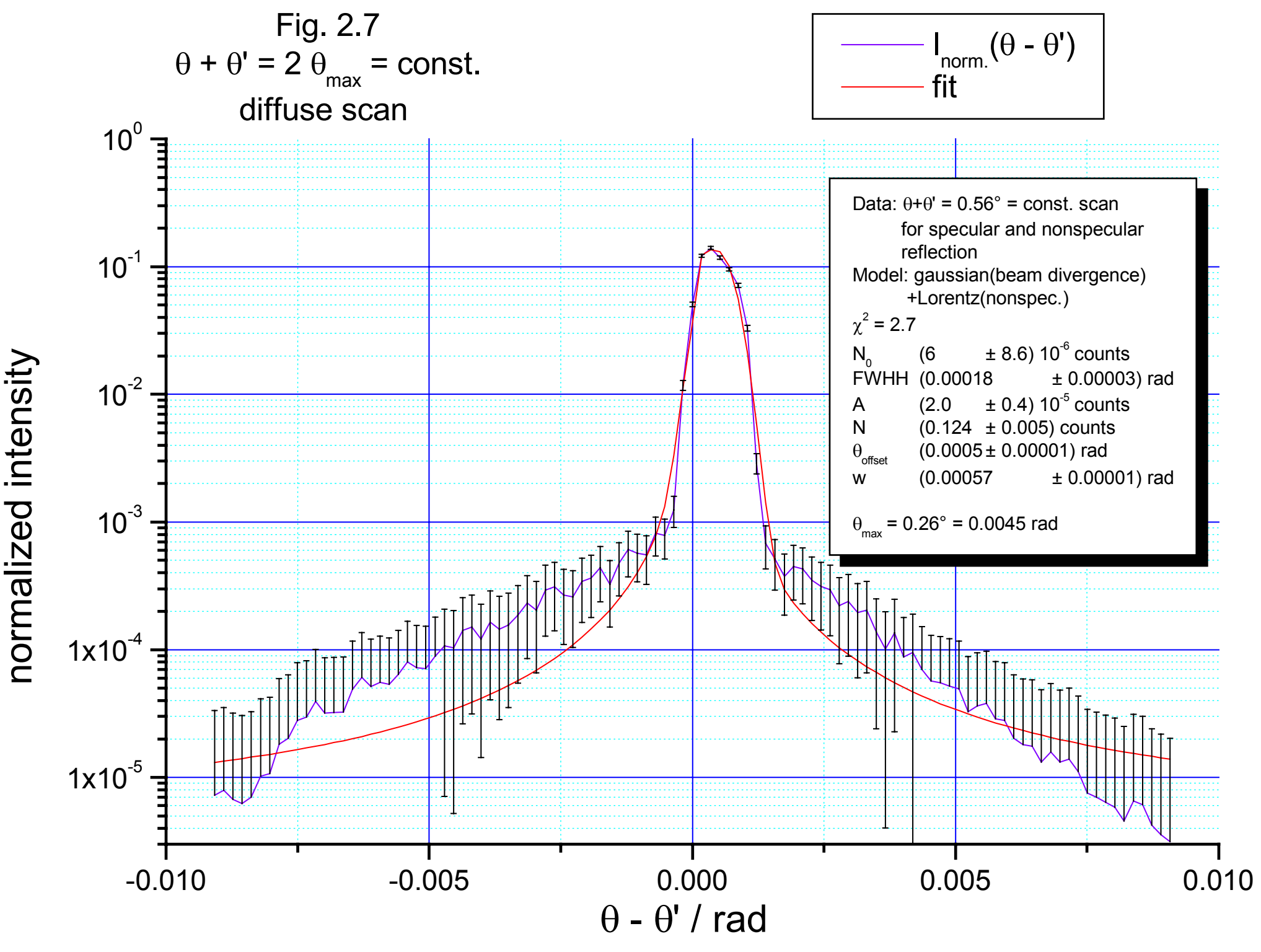

For convenience, the data is plotted as a function of an angle variable, which places the point of specular scattering at zero position. An analysis of the data with eq. (2.1.6) gives the fit shown above, and from *FWHH* one can, according to eq. (2.1.6), with $\sigma$ already known, calculate the correlation length to $\xi = (10 \pm 2)$ μm.

Thus, a detailed analysis of the data collected in the two basic types of scattering schemes conducted with XSAS, which is based on the DWBA approach [Si88, Bo94] presented above, provides information about the characteristical quantities describing the micro-roughness of the glass mirrors with an accuracy of about 10 %, as it was initially desired. It should be emphasized here, that a much more detailed and thorough analysis of scattering theory would have to be developed to determine these quantities to a higher degree of accuracy.

Exactly the same kind of measurement and subsequent analysis can be applied to the absorber, consisting of a rough glass plate coated with an *Gd-Ti-Zr* alloy,which is 2000 Å thick. However, without any detailed analysis there are some things one already is able to estimate concerning the results of such measurements.

First, as a reminder one should think about the scaling properties of roughness. With x-rays of a few Å wavelength it is not possible to determine parameters of roughness if its characteristical scale differs by more than one to two orders of magnitude from the wavelength used. The mean height roughness as well as the correlation length of the absorber is expected to be around a few μm due to process of generating roughness by means of etching the glass surface. From these facts it can already be concluded, that XSAS will not be able to determine even the correct order of magnitude of the absorber's roughness parameters.
On the contrary, one may deduce from the magnitude of angles of incidence suitable for reasonable statistics – being at maximum one order of magnitude larger than the critical angle of total reflection, which is usually between 0.005 rad and 0.01 rad – and the magnitude of the roughness, which is around a μm, that due to pure geometrical shadowing of the surface, x-rays of a few Å wavelength will see an effective mean height roughness of around 100 Å in a specular scan.

Second one has to take into account the presence of the coating, which can be represented by two rough surfaces separated by 2000 Å. This should lead to the existence of some phase generated interference patterns, i.e. oscillations in the cross section. Yet, the separation of 2000 Å expressed in terms of x-ray wavelengths of a few Å is so large, that the interference would only be visible at angles of incidence, which are so large, that the suppression of the scattered intensity due to the roughness makes it impossible to measure such effects.

Nevertheless there have been to short runs of the types Fig.s 2.3 and 2.4 with the absorber as the sample to establish at least the qualitative estimations made above in their orders of magnitude. The results of a specular and a diffuse scan are shown in Figs. 2.8 and 2.9:

Fig. 2.8
specular scan

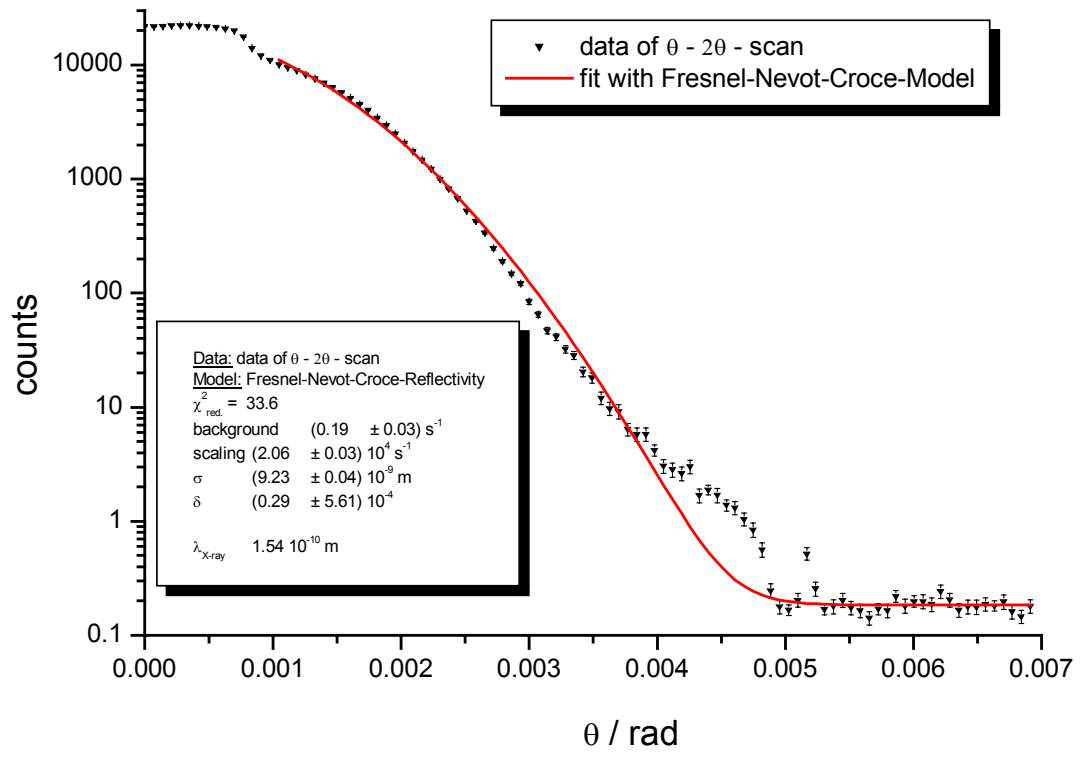

Fig. 2.9
diffuse scan

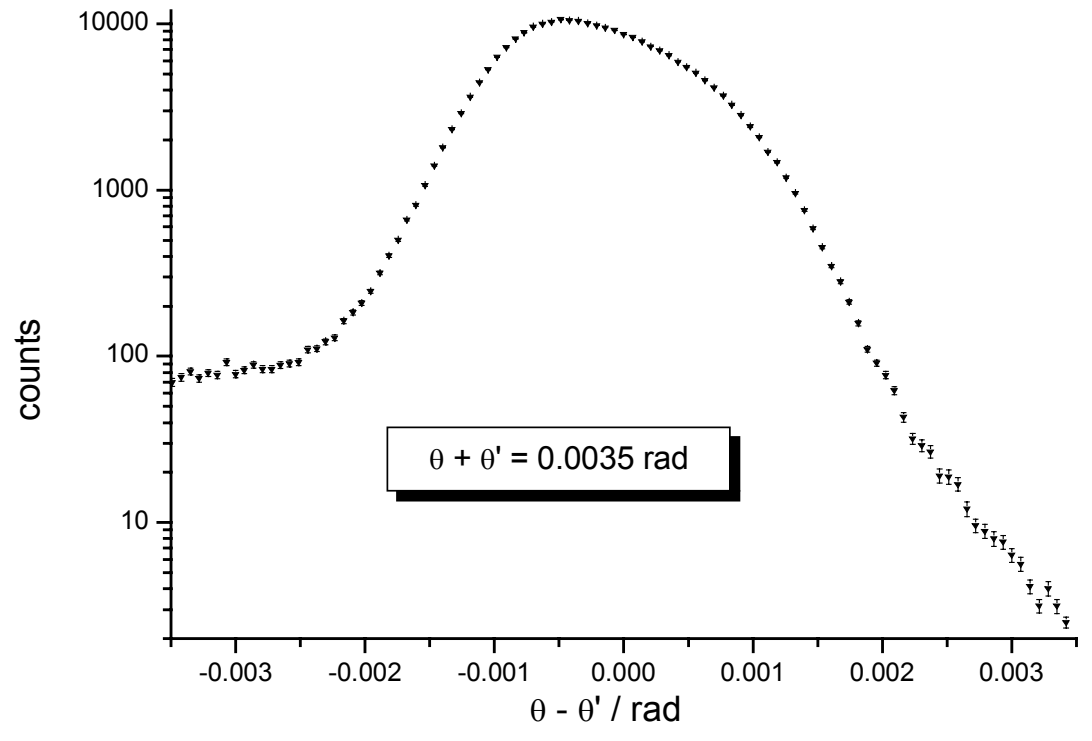

The specular scan indeed provides an effective roughness of 92 Å, as it was expected from the qualitative argumentation given above.

Therefore we have to search for an alternative to determine the roughness of the absorber. In principle, there are two different ways:

First, it is possible to use the scaling properties of the roughness and to repeat the scans of the types shown in Fig.s 2.3 and 2.4 with light of suitable wavelength, i.e. to perform XSAS with wavelength in the 100 nm regime of UV light.

Second ,one should think of a way of directly mapping the roughness.

Fortunately, there emerged an opportunity to solve the problems of small-angle-scattering at highly rough surfaces by directly imaging the surface with a resolution of around 100 nm by means of an atomic force microscope (AFM) at the ESRF in Grenoble, France [Dc01], in February of 2001. Its micro-imaging laboratory can perform surface images of given samples up to atomic resolution, routineously. The striking property of the AFM there is, that it is able to perform scans over quadratic fields with edges up to 90 μm in length.

This feature enables one to map certain parts of the surface in dimensions of around ten times the correlation length. If several sample fields are mapped this way, one gets the surface function *z(x,y)* directly for an area which is large enough, compared to the expected range of the roughness parameters, so that these parameters can be determined by a statistical analysis of the surface function.

Thus, five scans were performed with the AFM, mapping the surface of the absorber in five squares of 80 µm by 80 µm each. The positioning of the mapping squares on the surface is given in Fig. 2.10:

Fig. 2.10
AFM-scan-geometry

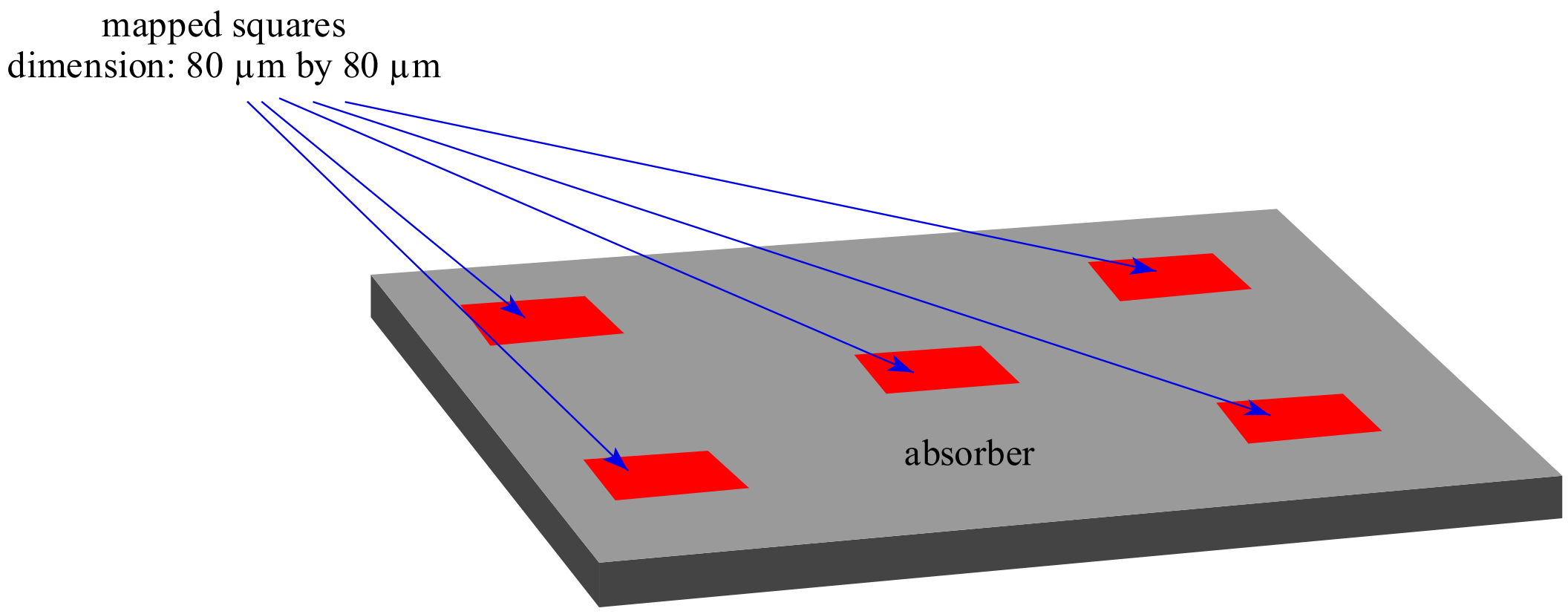

Each scan was performed by determining the vertical z-position of the AFM-needle at the points of a 256 by 256 grid, which covered the square to be mapped, thus producing a 256 by 256 matrix with integers as the result. The integer values of these matrices then have to be symmetrized with regard to the position of the average zero plane of the square, and renormalized with a scale factor to give the heights of the measured grid points relative to the average zero plane. The data acquisition software of the apparatus [Dc01] allows one to perform an addition of the absolute values of all properly scaled heights. That gives the two dimensional integral of the surface function *z(x,y)*, given by the elements of the matrices for each square independently, which we will call *RMS*:

$$(2.1.8)\quad RMS = \iint_{square\ area} dxdy \cdot |z(x,y)| \ .$$

This can directly be linked to the mean height roughness $\sigma$ of a gaussian rough surface. The probability distribution of the surface function *z(x,y)* is then gaussian, so eq. (2.1.8) becomes:

$$(2.1.9)\quad RMS = \iint_{square\ area} dxdy \cdot |z(x,y)| = \int dz \cdot |z| \cdot \frac{1}{\sigma \cdot \sqrt{2\pi}} \cdot e^{-\frac{z^2}{2 \cdot \sigma^2}}$$

$$= 2 \cdot \int_0^\infty dz \cdot z \cdot \frac{1}{\sigma \cdot \sqrt{2\pi}} \cdot e^{-\frac{z^2}{2 \cdot \sigma^2}} = \sigma \cdot \sqrt{\frac{2}{\pi}} \quad .$$

The *RMS*-value, directly calculated with the microscope's software, thus provides us with the mean height roughness $\sigma$. Fig.s 2.11 and 2.12 show two screen-shots of the results of the scanning of one square:

Fig. 2.11

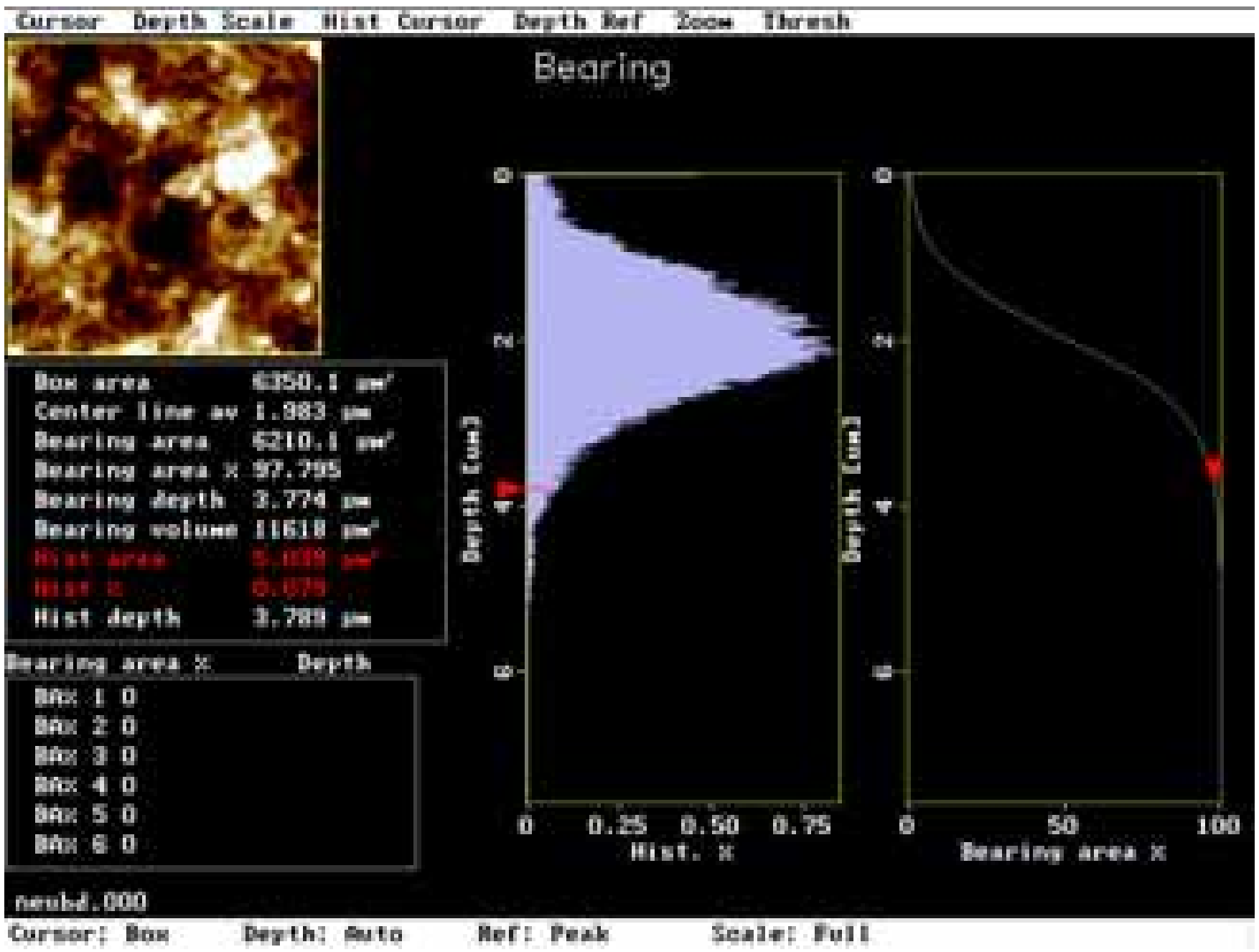

Fig. 2.11: In the upper left there is an image of the surface. The grey-scaling is proportional to the relative heights of the sampled points. The middle of this shot displays a histogram of the points' heights of the sampled square, showing directly and strikingly the gaussian nature of the surface's roughness, while the right of the shot shows, why a rough surface has an "interaction length" perpendicular to the average zero plane. The part of the surface seen by something approaching from above is increasing with a gaussian error function as seen in the right part of this screen-shot.

Fig. 2.12

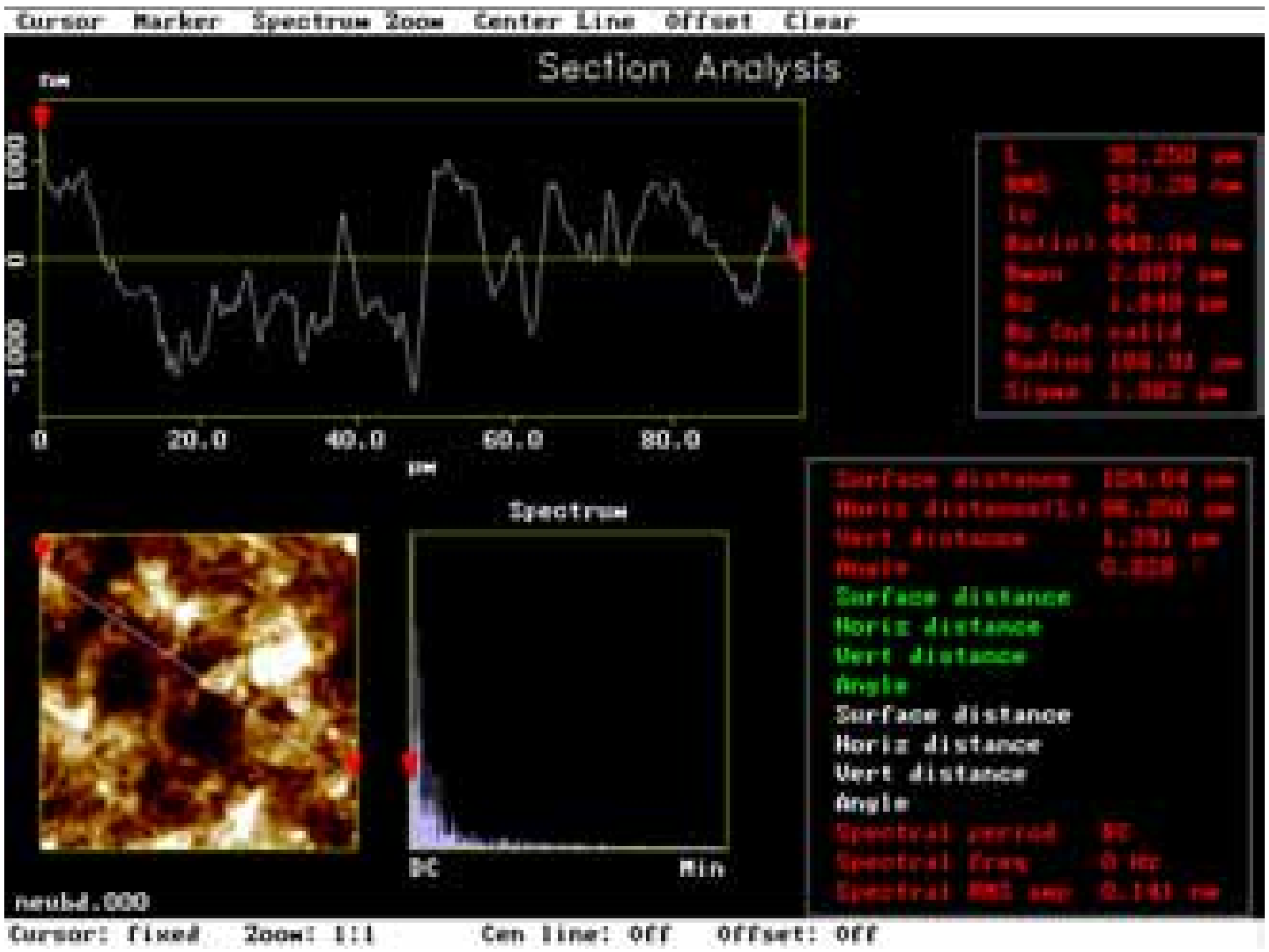

Fig. 2.12: The same scan is shown together with a section of the surface, which is shown graphically above the grey-scale surface picture. The stochastically fluctuating structure of the surface is clearly visible as well as a roughness of about 1.5 µm in vertical direction and several µm along the surface.

One can calculate the mean height roughness range $2\sigma$, using eq. (2.1.9) and the *RMS*-values calculated from the scanned squares:

| | square 1 | square 2 | square 3 | square 4 | square 5 |
|---|---|---|---|---|---|
| $2\cdot\sigma\,[\mu\text{m}]$ | 1.5683 | 1.5659 | 1.5455 | 1.4367 | 1.4139 |

This leads to: $2\cdot\sigma = (1.51 \pm 0.03)\cdot \mu m$.

Next, one has to look for a method to extract the lateral height-height correlation length from the surface function *z(x,y)* given in the matrices. The simplest way is to determine the height autocorrelation function. If this function is peaked around zero and has fast vanishing values for all nonzero arguments, then the full width half height *FWHH* is two times the correlation length $\xi$.

Now the determination of the autocorrelation function from the surface function *z(x,y)* is possible due to an already known formalism based on the so-called "Wiener theorem", which links the Fourier transform of *z(x,y)* and its autocorrelation function.

Let us first consider the one-dimensional case. The autocorrelation function $C(\Delta x)$ of a more or less random valued function *z(x)* is formally defined as a convolution of *z(x)* with itself but with shifted argument, where the shift $\Delta x$ gives the correlation interval:

$$(2.1.10)\quad C(\Delta x) = \lim_{R\to\infty} \frac{1}{2\cdot R}\cdot \int_{-R}^{R} dx \cdot z(x)\cdot z(x+\Delta x) \quad .$$

Next, the spectral intensity of the Fourier transform of *z(x)* is to be found. The average power of the statistically fluctuating function *z(x)* is defined as:

$$(2.1.11)\quad \overline{P} = \lim_{R\to\infty} \frac{1}{2\cdot R}\cdot \int_{-R}^{R} dx \cdot |z(x)|^2 = \lim_{R\to\infty} \frac{1}{2\cdot R}\cdot \int \frac{dk}{2\pi}\cdot |\tilde{z}_R(k)|^2 \,,$$

where:

$$(2.1.12)\quad \tilde{z}_R(k) = \int_{-R}^{R} dx \cdot z(x)\cdot e^{i\cdot k\cdot x}$$

is the bounded Fourier transform of *z(x)*. It is then natural to define the spectral intensity (or power density) $\rho_P(k)$ of *z(x)* as:

$$(2.1.13)\quad \rho_{\overline{P}}(k) = \frac{d\overline{P}}{dk} = \lim_{R\to\infty} \frac{1}{2\cdot R}\cdot |\tilde{z}_R(k)|^2 \ .$$

Now consider the Fourier transform of (2.1.13):

$$\tilde{\rho}_{\overline{P}}(\Delta x) = \frac{1}{2\pi}\cdot\int dk\cdot\rho_{\overline{P}}(k)\cdot e^{i\cdot k\cdot\Delta x} = \frac{1}{2\pi}\cdot\lim_{R\to\infty}\frac{1}{2\cdot R}\cdot\int dk\cdot e^{i\cdot k\cdot\Delta x}\cdot\tilde{z}_R^*(k)\cdot\tilde{z}_R(k)$$

$$= \lim_{R\to\infty}\frac{1}{2\cdot R}\cdot\int_{-R}^{R} dx\cdot z(x)\cdot\int dk\cdot e^{i\cdot k\cdot(x+\Delta x)}\cdot\tilde{z}_R^*(k) = \lim_{R\to\infty}\frac{1}{2\cdot R}\cdot\int_{-R}^{R} dx\cdot z(x)\cdot z(x+\Delta x)$$

$$\Rightarrow \quad (2.1.14) \quad C(\Delta x) = \frac{1}{2\pi}\cdot\int dk\cdot\rho_{\overline{P}}(k)\cdot e^{i\cdot k\cdot\Delta x} \quad .$$

This last equation is called the Wiener theorem. Its meaning is that by generating a Fourier transform (by FFT for instance) from the surface function *z(x,y)* we get its autocorrelation function by using eq.s (2.1.13) and (2.1.14). This procedure can be generalized to two dimensions respecting the two-dimensional nature of the scanned area.

However the process of roughening made by etching the surface leads to a thermal phenomenon, which is controlled by the Maxwell-Boltzmann statistics. It provides very high degrees of uniformity and isotropy of the surface's roughnes. Therefore the two-dimensional analysis can be reduced to a one-dimensional one by applying eq.s (2.1.13) and (2.1.14) e.g. to each row of a surface scan matrix. Each row autocorrelation function found this way gives a correlation length $\xi_j = {}^1/_2\ FWHH_j$ . These values will then be averaged. Fig. 2.13 shows one such row autocorrelation function:

Fig. 2.13

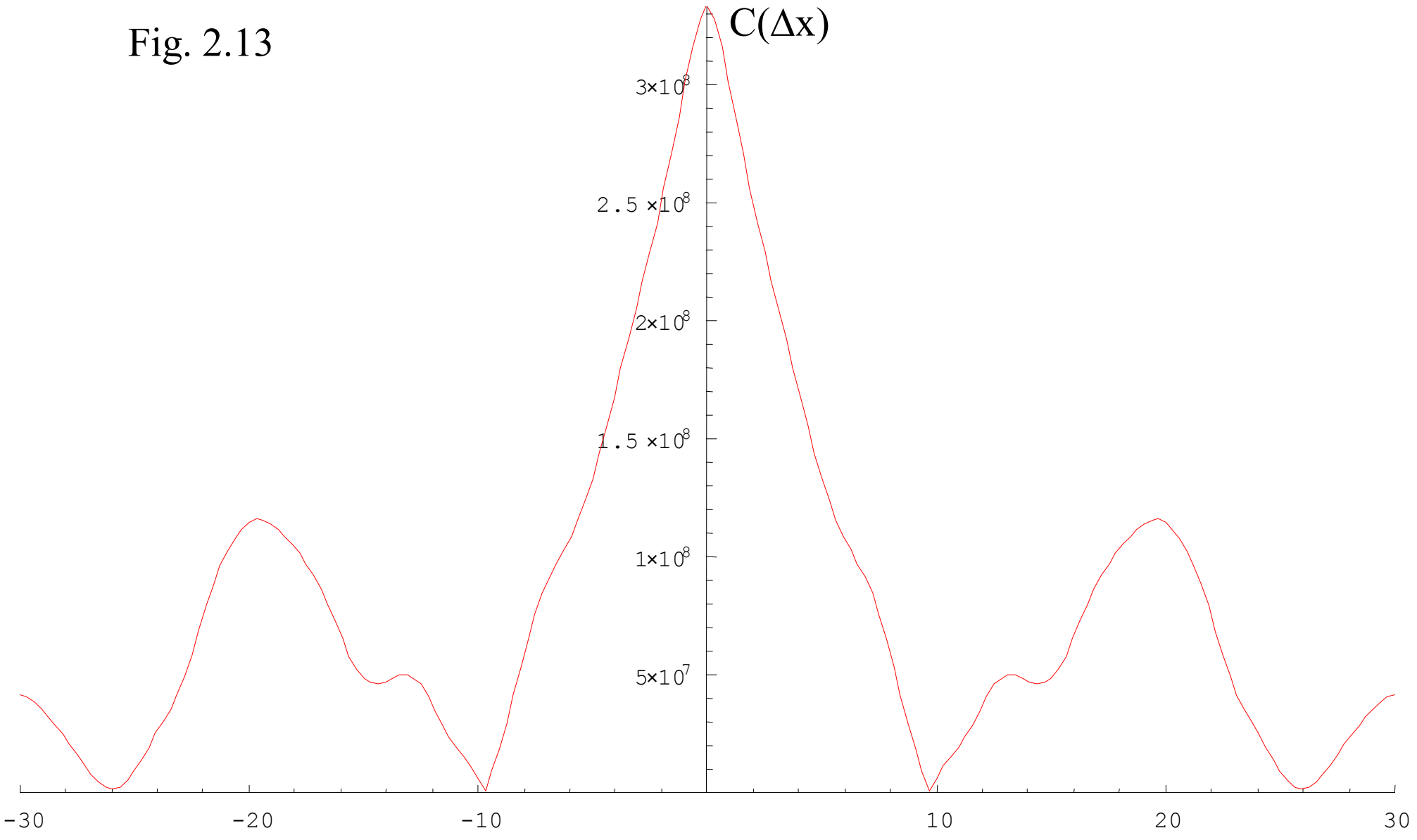

There seems to be some indication for short range order at scales of around 20 µm. However this might be smeared out by summing up all rows of a given matrix. Determining all $FWHH_j$ as mentioned above and averaging them, one gets the value for the lateral height-height correlation length of the absorber surface $\xi = (4.7 \pm 0.2)\cdot\mu m$ .

### 2.2) The collimator system

This section deals with the details of the collimation system, that is placed in front of the wave guide's entrance.

Fig. 2.14 gives a more detailed view of the collimation system (Fig. 1.4 / 9). Together with the information, that characteristical UCN energies of about 100 neV correspond to mean velocities of 5 – 10 $^{m}/_{s}$ , one can derive from Fig. 2.14 the ability of the collimation system to provide a very small vertical velocity component and therefore a very small vertical energy of UCN entering the wave guide at the right part of Fig. 2.14. Furthermore it selects certain parts of the spectrum of the horizontal velocity component.

Fig. 2.14: The collimation system and the wave guide

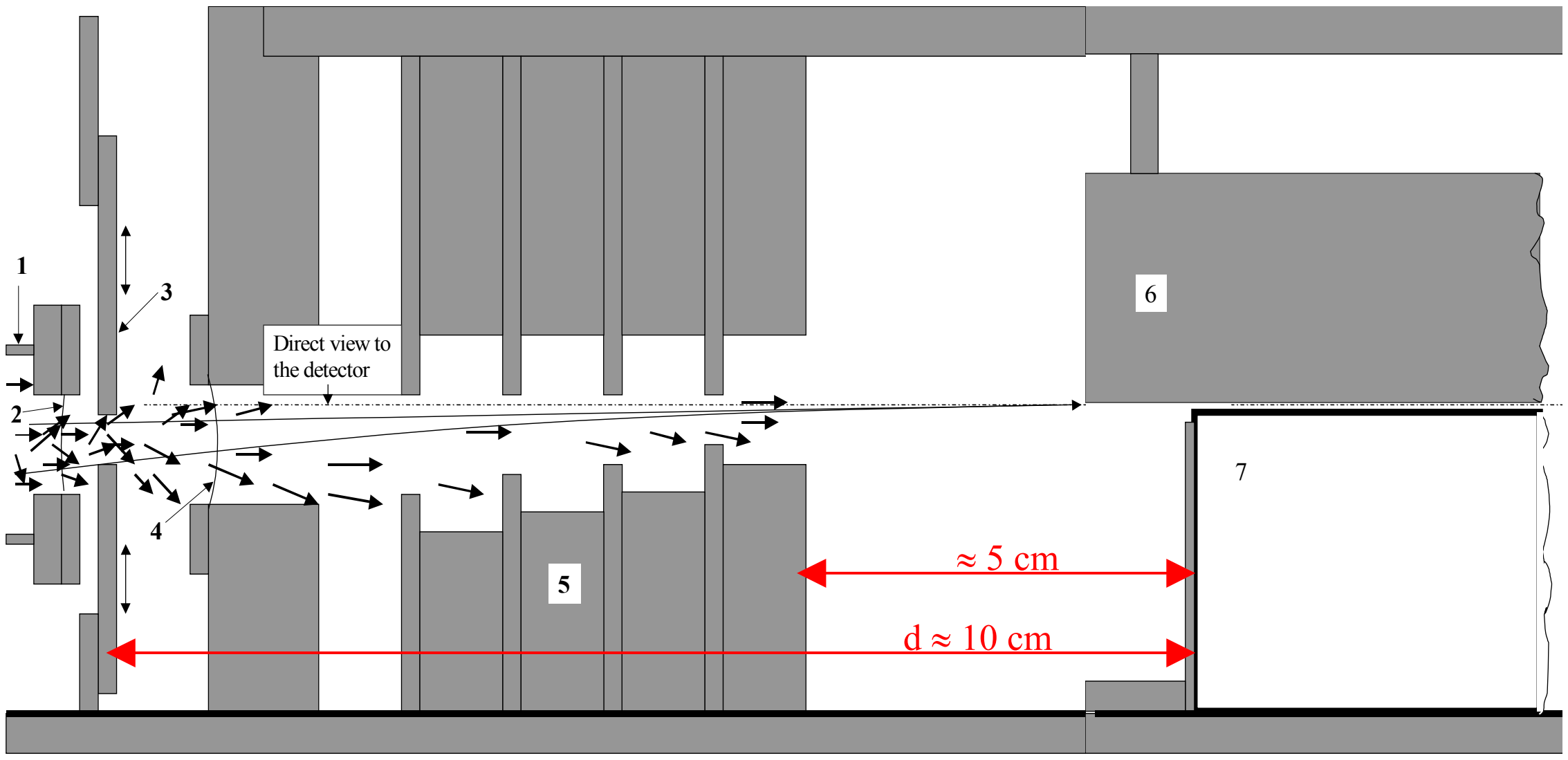

As one can see in Fig. 2.14, a combination of two titanium plates (Fig. 2.14 / 3) and several collimating slits (Fig. 2.14 / 5) block all neutrons, which are not approaching the wave guide formed by the absorber (Fig. 2.14 / 6) and the bottom mirror (Fig. 2.14 / 7) along classical parabolical trajectories. These trajectories are defined by the ends of the titanium plates (Fig. 2.14 / 3) and the entrance opening of the wave guide. It is clearly visible, too, that a direct view from the wave guide entrance towards the window (Fig. 2.14 / 1), which couples in the UCN from the feeding UCN guide, is blocked by the upper titanium blend (Fig. 2.14 / 3). The distances shown are not to be taken as properly scaled. The distance *d* between the titanium slit (Fig. 2.14 / 3) and the wave guide entrance is about 10 cm, the last collimating slit of (Fig. 2.14 / 5) at the right is about 5 cm before the wave guide entrance.

The collimating properties of the system shown in Fig. 2.14 can be derived as follows. Suppose, one places the lower end of the upper titanium plate (Fig. 2.14 / 3) at $\Delta z$ = 220 µm below the line of direct view and the upper end of the lower titanium plate (Fig. 2.14 / 3) $\Delta z$ = 2000 µm below the line of direct view. Only neutrons having parabolic trajectories with turning points close to the wave guide's (Fig. 2.14 / 6 + 7) entrance will enter it. The limiting trajectories are those corresponding to the slit formed by the titanium plates (Fig. 2.14 / 3). Because the wave guide's bottom mirror is aligned with the direct line of view, neutrons moving along these two limiting trajectories do have initial vertical velocities of:

$$E^z_{kin\,,\,initial} = \frac{m}{2} \cdot \left(\mathrm{v}^z_{initial}\right)^2 = m \cdot g \cdot \Delta z = \Delta E_{grav}$$

(2.2.1)
$$\Rightarrow \begin{cases} \mathrm{v}^z_{initial}\Big|_{upper\ plate} = \sqrt{2 \cdot g \cdot \Delta z_{upper\ plate}} \approx 0.06 \cdot \dfrac{m}{s}, & \Delta z_{upper\ plate} \approx 220 \cdot \mu m \\ \mathrm{v}^z_{initial}\Big|_{lower\ plate} = \sqrt{2 \cdot g \cdot \Delta z_{lower\ plate}} \approx 0.2 \cdot \dfrac{m}{s}, & \Delta z_{lower\ plate} \approx 2000 \cdot \mu m \end{cases}$$

From the distance $d$ of about 10 cm and the time $\Delta t_z$ the neutrons on those limiting trajectories need to reach their turning points at the wave guide's, which one derives from:

(2.2.2) $$\mathrm{v}^z_{initial} = g \cdot \Delta t_z \quad ,$$

one get the limiting values of the selected part of the spectrum of horizontal velocity component as follows:

(2.2.3) $$\mathrm{v}_{hor.} = \frac{d}{\Delta t_z} \quad \Rightarrow \quad \begin{cases} \mathrm{v}^{upper\ plate}_{hor.} \approx 15 \cdot \dfrac{m}{s} \\ \mathrm{v}^{lower\ plate}_{hor.} \approx 5 \cdot \dfrac{m}{s} \end{cases} .$$

From (2.2.1) and (2.2.3) as well as from Fig. 2.14 one easily sees, that by varying the positions of the titanium plates (Fig. 2.14 / 3) one can control the selected part of the spectrum of the horizontal and vertical velocity components simultaneously. Furthermore one sees that the collimation system provides a spectrum of vertical velocity component which range is at least one order of magnitude below the range of horizontal velocity component, which is comparable to UCN mean velocities.

# Chapter 3

# Theoretical treatment of the experiment

It is now time to start a discussion of how an experimental setup presented in chapter one with the properties presented and analyzed in the first two chapters should behave from theoretical points of view. In chapter 1 it has been argued qualitatively, that effects of the quantization of the canonical observables of motion should show up if neutrons are trapped between a mirror at the bottom and a linearly increasing gravitational potential above – a system called “gravitational wave guide”.

Before entering a quantum theoretical treatment of the setup, one should consider the expectations deriving from a purely classical description of the measurements described in chapter one. Yet, whatever picture is used, there is always one irreducible part at the basics of the whole experiment’s idea – besides the presence of the earth’s gravitational field. And that is the existence of UCN reflecting mirrors. The surface properties of such mirrors and their scattering properties have been analyzed thoroughly in the last chapter. However, the argument given in the first chapter, why UCN are reflected from ordinary solid state matter, was only a qualitative one. Therefore, we should start with establishing the effects of ordinary matter of a homogenous solid state type on neutrons in general – that is, where the Fermi pseudopotential comes from.

These considerations then aim to derive the magnitude of the Fermi potential of a metallic alloy of Gadolinium. Since Gd is highly efficient neutron absorber, one expects its Fermi potential to have a high imaginary part. However, calculation of this Fermi potential for a highly neutron absorbing material is non-trivial, since its imaginary part seems to decrease proportional to the wave number $k$ of neutrons for very small neutron energies. This phenomenon now corresponds to a saturation of the absorption cross section of Gd at very small $k$ .

The saturation can be understood from an intuitive argument, which is nevertheless only a plausibility consideration:

The effective absorption cross section ceases to rise proportional to $k^{-1}$ if the diameter of the circular disk, that corresponds to the cross section, equals the interatomic distance of the medium. For then the disks would start to overlap, and the cross section must remain constant, if multiple scattering is omitted.

The section finishes therefore with a dicussion of the quantum mechanical derivation of the Fermi pseudopotential of highly absorptive materials. These theoretical considerations are yet not completely fortified. Furthermore, the best measurements available [Ra99] demonstrate absorption cross sections of Gd only above neutron velocities of about 3 m/s, which is two orders of magnitude above the vertical velocity components present in the wave guide.

### 3.1) The Fermi pseudopotential

The neutron is a massive particle of about 938 MeV, that is formed by two d-quarks and one u-quark. Their electrical charge add up to a total zero charge of a neutron, which has been proved to a limit of $q_{neutron} \leq (-0.4 \pm 1.1) \cdot 10^{-21} e^{-}$ according to [Gae89], where $e^{-}$ denotes the electron's charge.

The half-integer spins of the three constituent quarks of a neutron cause it to be a fermion of spin ½ . Thus, the neutron, while not being able to interact directly with the electrical charge, can couple to magnetic fields. Therefore, efficient magnetic shielding is necessary for experiments at extremely low energies.

However, the main interaction neutrons will experience if they encounter matter, is the strong force, which couples to the nuclei of matter, if neutrons collide more or less directly with them. A "collision" is needed due to the short-ranged nature of the strong force and the very small extension of nuclei (of about 1 fm compared to atomic diameters of about 1 Å). The result is, that neutrons feel the presence of matter either via strong interactions if they hit the nuclei, or eventually via magnetic coupling to the nuclei's magnetic moment.

Here we deal with so called "ultracold neutrons (UCN)" of energies in the range of 100 neV, which correspond to de Broglie wave lengths of about 500 Å. That is several hundred times of an averaged interatomic distance and several million times of a nucleus' diameter. Thus the internal structure of the matter's nuclei can be omitted without any loss of accuracy in the description. Having the numbers given above in mind, we can easily describe the nucleus by means of an attractive spherical box potential with a depth of about –50 MeV:

$$(3.1.1) \quad U_{nucleus}(\vec{r}) = U_{nucleus}(r) \approx \begin{cases} -50 \cdot MeV \,, r < R \\ 0 \,, r \geq R \end{cases} \quad , \quad R \approx 10^{-14} \cdot m \quad .$$

*R* here is the averaged radius of a nucleus.

What one has to do now is to calculate the scattering of a neutron, described by an incident wave packet, at a single nucleus and to derive its scattering amplitude. The incident wave packet can be written quite generally:

$$(3.1.2) \quad \psi_0(\vec{r},t) = \int \frac{d^3k}{(2\pi)^3} \cdot a_{\vec{k}} \cdot e^{i \cdot \vec{k} \cdot \vec{r}} \quad .$$

One then rewrites eq. (3.1.2) in terms of the stationary eigenfunctions $\psi_{\vec{k}}$ of the Hamiltonian constructed from the potential (3.1.1) which posesses energy eigenvalues $E_{\vec{k}}$:

$$(3.1.3)\quad \begin{cases} E_{\vec{k}} = \dfrac{\hbar^2 \cdot \vec{k}^2}{2 \cdot m} \geq 0 \\ \hat{H}_{nucleus} \cdot \psi_{\vec{k}}(\vec{r}) = \left[ -\dfrac{\hbar^2}{2 \cdot m} \cdot \vec{\nabla}^2 + U_{nucleus}(r) \right] \cdot \psi_{\vec{k}}(\vec{r}) = E_{\vec{k}} \cdot \psi_{\vec{k}}(\vec{r}) \end{cases} .$$

The $\psi_{\vec{k}}$ can be found in their general form using the Green's function of the free Schrödinger equation to be:

$$(3.1.4)\quad \psi_{\vec{k}}(\vec{r}) = e^{i \cdot \vec{k} \cdot \vec{r}} - \frac{m}{2\pi \cdot \hbar^2} \cdot \int d^3 r' \cdot \frac{e^{i \cdot k \cdot |\vec{r} - \vec{r}'|}}{|\vec{r} - \vec{r}'|} \cdot U_{nucleus}(r') \cdot \psi_{\vec{k}}(\vec{r}') \quad .$$

One can now treat eq. (3.1.4) with the usual formalism of a partial wave expansion combined with the limites of small neutron energy and distances, which are large compared to the scattering region, as this is shown e.g. in [Schw]. Then one arrives at the s-wave scattering limit, that connects the scattering cross section with the s-wave scattering phase $\delta_0$ .

Because for small $k$ – low neutron energies – only s-wave scattering is dominant, the relations between the cross section and scattering phase read as (see [Schw]):

$$(3.1.5)\quad \begin{cases} \sigma_{el.} = \dfrac{4\pi}{k^2} \cdot |\delta_0|^2 + O\left(\delta_0{}^3\right) \\ \sigma_{inel.} = \dfrac{4\pi}{k^2} \cdot \delta_0^{(i)} \cdot (1 - 2 \cdot \delta_0^{(i)}) + O\left(\left|\delta_0^{(i)}\right|^3\right) \end{cases} .$$

In the regime of UCN energies even $\delta_0 \propto k$ will be small. In this case one can use the so-called (first order) Born approximation, which provides the scattering phase $\delta_l$ to become:

$$(3.1.6)\quad \delta_l \approx -\frac{2m \cdot k}{\hbar^2} \int dr \cdot r^2 \cdot U_{nucleus}(r) \cdot \left[ j_l(k \cdot r) \right]^2 \quad .$$

Additionally, the fact, that for UCN s-wave scattering is dominant and even $\delta_0$ is small due to $\delta_0 \propto k$, allows one to replace the spherical Bessel function $j_l(kr)$ by simple plane waves, thus rewriting eq. (3.1.6) as:

$$(3.1.7)\quad \delta_l \approx -\frac{1}{4\pi} \cdot \frac{2m \cdot k}{\hbar^2} \int d^3 r \cdot U_{nucleus}(r) \cdot e^{i \cdot (\vec{k} - \vec{k}') \cdot \vec{r}} \quad .$$

For our case of very small $k$, the potential eq. (3.1.1) can be approximated with a delta function with complex norm factor, containing a negative real part. This describes the potential of the nucleus, which is complex in general:

$$(3.1.8)\quad U_{nucleus}(r) = A \cdot \delta(\vec{r}),\ A\ complex \quad .$$

This approximated nuclear potential is called the Fermi pseudopotential. Using this in eq. (3.1.7), one gets in the limit of vanishing energies:

$$(3.1.9)\quad \delta_0 \approx -\frac{2m}{\hbar^2}\cdot\frac{k\cdot A}{4\pi}\ .$$

The coupling A has to be chosen to represent the coupling strength of the potential eq. (3.1.1), and it must be complex to account for inelastic phenomena such as neutron capture by the nucleus. In general, A will depend on the neutron energy. In the case, however, where neutron energy is negligible compared to the 50 MeV magnitude of the nuclei's potential, it is a constant. With this information and eq. (3.1.9) one derives from $\delta_0$ small, i.e.

$$\delta_0 << 1 \Rightarrow \delta_0^{(i)} << 1 \Rightarrow \left[\delta_0^{(i)}\right]^2 << \delta_0^{(i)},$$

by using eq. (3.1.5), that:

$$(3.1.10)\quad \begin{cases} \sigma_{inel.} \propto \dfrac{1}{k} \\ k\cdot\sigma_{el.} \propto k \end{cases} ,$$

which are the known velocity dependencies of elastic and inelastic neutron scattering cross sections observed in matter. Furthermore, one can combine eq.s (3.1.5) and (3.1.9) to yield direct relations between the cross section and the amplitude of the one-nucleus Fermi pseudo-potential eq. (3.1.8):

$$(3.1.11)\quad \begin{cases} \sigma_{inel.} = \sigma_{inel.}(k) = -\dfrac{4\pi}{k}\cdot\dfrac{2m}{\hbar^2}\cdot\dfrac{\Im(A)}{4\pi} \\ \sigma_{el.} = -4\pi\cdot\left(\dfrac{2m}{\hbar^2}\right)^2\cdot\left[\left(\dfrac{\Re(A)}{4\pi}\right)^2+\left(\dfrac{\Im(A)}{4\pi}\right)^2\right] \end{cases} .$$

Since the potentials of several nuclei are additive, all nuclei together form a potential, which is the average of eq. (3.1.8) over a volume M, thus giving:

$$(3.1.12)\quad U_{nuclei}(\vec{r}) = V(\vec{r}) + i\cdot W(\vec{r}) = \frac{1}{M}\cdot\sum_{\{j|\vec{r}_j\in M\}}\int_M d^3r\cdot A\cdot\delta(\vec{r}-\vec{r}_j) = n(\vec{r})\cdot A\ ,$$

where $n(\vec{r})$ is the volume density of the nuclei as a function of the location inside the considered material. For materials, which are homogeneously distributed, heterogenic mixtures of pure elements, the nuclei density can be written as:

$$(3.1.13)\quad n(\vec{r}) = n_0^{(j)}\cdot\partial M(\vec{r})\ ,$$

where $\partial M(\vec{r})$ is the boundary of the piece of volume M filled by matter, and $n_0^{(j)}$ is the homogenous nuclei density of the element *j* present in the material. $\partial M(\vec{r})$ is defined as:

$$(3.1.14)\quad \partial M(\vec{r}) = \begin{cases} 1\,,\ \vec{r} \in M \\ 0\ \ else \end{cases} \Leftrightarrow \int d^3 r \cdot \partial M(\vec{r}) = M$$

One can now rewrite the Fermi pseudopotential, eq. (3.1.12), for such materials as:

$$(3.1.15)\quad U_{nuclei}(\vec{r}) = U_{Fermi} \cdot \partial M(\vec{r}) = (V + i \cdot W) \cdot \partial M(\vec{r})\ ,$$

where:

$$(3.1.16)\quad \begin{cases} V = \sum_j V^{(j)} \\ W = \sum_j W^{(j)} \end{cases}$$

and:

$$(3.1.17)\quad \begin{cases} V^{(j)} = n_0^{(j)} \cdot \Re\left(A^{(j)}\right) = \pm \dfrac{\hbar^2}{2m} \cdot n_0^{(j)} \cdot \sqrt{4\pi \cdot \sigma_{el.}^{(j)} - \left(k_0 \cdot \sigma_{inel.}^{(j)}\right)^2} \\ W^{(j)} = n_0^{(j)} \cdot \Im\left(A^{(j)}\right) = -\dfrac{\hbar^2}{2m} \cdot k_0 \cdot n_0^{(j)} \cdot \sigma_{inel.}^{(j)} \end{cases} .$$

In general, $k_0$ must be replaced by the component, which is locally normal to the surface, since the scattering phases of the nuclei add up to destructive, and therefore repulsive interference in the direction normal to the surface.

This now establishes the existence of an effective potential, to which the neutron is exposed to, which allows one to develop the formalism of an effective single-particle problem inside this potential of the medium.

It can now be shown, that even for vanishing $k_0$ of the incident plane wave the absorption cross section of the medium stays to be finite. This can be seen from the following argument:

A plane wave with a wave vector $k_0$ entering a medium from outside perpendicularly to the surface will have complex wave number $k_{med} = k_{med}^{(r)} + i \cdot k_{med}^{(i)}$ inside the medium. Thus, the plane wave inside can be written as:

$$\psi_{med} = \sqrt{\frac{2k_{med}^{(i)}}{k_{med}^{(r)}}} \cdot e^{i \cdot k_{med} \cdot z} = \sqrt{\frac{2k_{med}^{(i)}}{k_{med}^{(r)}}} \cdot e^{i \cdot k_{med}^{(r)} \cdot z} \cdot e^{-k_{med}^{(i)} \cdot z}$$

$$= \int \frac{dp}{2\pi} \cdot c(p) \cdot e^{i \cdot p \cdot z}\ ,\ c(p) = \sqrt{\frac{2k_{med}^{(i)}}{k_{med}^{(r)}}} \cdot \frac{1}{k_{med}^{(i)} - i \cdot \left(p - k_{med}^{(r)}\right)} .$$

The wave vector $k$ inside the medium is given by:

$$k_{med}^2 = k_0^2 - \frac{2m}{\hbar^2} U_{Fermi} \quad .$$

Comparison of this expression with eq.s (3.1.15), (3.1.16), and (3.1.17), as well as with $k_{med} = k_{med}^{(r)} + i \cdot k_{med}^{(i)}$ shows, that

$$k_{med}^{(i)} = \frac{1}{2} \cdot n_0 \cdot \frac{k_0}{k_{med}^{(r)}} \cdot \sigma_{inel.}$$

$$\left(k_{med}^{(r)}\right)^2 = k_0^2 \mp \alpha + \frac{1}{4}\left(\frac{k_0}{k_{med}^{(r)}} \cdot n_0 \cdot \sigma_{inel.}\right)^2 \quad ,$$

where

$$\alpha = n_0 \cdot \sqrt{4\pi \cdot \sigma_{el.} - \left(k_0 \cdot \sigma_{inel.}\right)^2} \quad .$$

For $\sigma_{inel.}$ the relation

$$(3.1.18a) \quad \sigma_{inel.} = \frac{a}{k_0}$$

holds according to eq. (3.1.11). The proportionality constant $a$ can be extracted from tabulated inelastic cross section values at thermal neutron energies. Thus, it is:

$$(3.1.18b) \quad a = \left(k_0 \cdot \sigma_{inel.}\right)\Big|_{v(k_0)=2200 m/s} \quad .$$

One sees, that $\alpha$ depends not on $k_0$ , and can be calculated the same way as the quantity $a$. Eq. (3.1.18a) can now be inserted into the expression for $k_{med}^{(r)}$ , which gives for the case, where $k_0 << \alpha \quad \wedge \quad k_0 << n_0 \cdot \sigma_{inel.}$ holds:

$$\left(k_{med}^{(r)}\right)^2 = k_0^2 \mp \alpha + \frac{1}{4} \frac{\left(n_0 \cdot a\right)^2}{\left(k_{med}^{(r)}\right)^2} \quad .$$

Since $\alpha$ depends not on $k_0$ , this expression denotes the existence of an absolute minimum of the possible values of $k_{med}^{(r)}$

$$\left(k_{med}^{(r)\ \min}\right)^2 = \mp\frac{\alpha}{2} + \frac{1}{2} \cdot \sqrt{(n_0 a)^2 + \alpha^2}$$

– even for the case, where $k_0 = 0$ is approached! It is clear from eq. (3.1.18a), that a global minimum of $k_{med}^{(r)}$ corresponds to the fact, that the absorption cross section $\sigma_{inel.}$ will saturate at a finite value for a vanishing incident wave vector $k_0$ , since then the incident wave vector $k_0$ has to be replaced with the effective wave vector $k_{med}^{(r)}$ inside the medium to calculate the effective absorption cross section, that is formed by both the potential of one individual scattering nucleus and the effective potential of the surrounding nuclei.

As argued in [Gu62], this finite saturation value of the absorption cross section evaluates to:

$$(3.1.19)\quad \sigma_{inel.}^{\max} = \frac{a}{\sqrt{1/2\cdot\sqrt{n_0^2 a^2+\alpha^2}-1/2\cdot\alpha}} \approx \sqrt{\frac{2a}{n_0}} \quad , \quad \alpha << n_0 a \quad .$$

The analysis presented since eq. (3.1.12) is largely a modification of the calculations displayed in [Gu62]. They arrive at the same result eq. (3.1.19). However, their expression for the averaged Fermi potential eq. (3.1.17) uses $k_{med}^{(r)}$ instead of $k_0$ . I question this, since the averaged potential is derived from scattering at single nuclei surrounded by empty space, and $k_{med}^{(r)}$ can only be stated after the averaged potential has been established.

Therefore, it should be emphasized here, that the saturation effect takes solely place for the absorption cross section, since the effective medial wave number enters here. The expression eq. (3.1.17) does not explicitly depend on $k_{med}^{(r)}$ but on $k_0$ instead, because this effective Fermi potential is derived from the spatial average of single nuclei scattering events solely! Yet, [Gu62] introduce $k_{med}^{(r)}$ already in eq. (3.1.17), although they derive it the same way from single nucleus scattering. However, this overestimates the effects of the bulge of matter surrounding a nucleus such, that the effective medial wave number $k_{med}^{(r)}$ enters twice into eq. (3.1.17) for the potential and (3.1.19) for the saturation cross section. The potential, however, must not use $k_{med}^{(r)}$ , since $k_{med}^{(r)}$ is derived only after the potential has been formulated before.

The values of the imaginary part of the Fermi pseudopotential by [Gu62] and the modified method shown here disagree by about one order of magnitude, if they are calculated for the complete alloy that is used as the absorber. It should be mentioned here again, however, that today no direct measurement of the absorption cross section of Gd at neutron velocities of about 1 cm/s exists [Ra99]. Furthermore, if one uses the imaginary part of the alloy's Fermi pseudopotential as a varying parameter, that has to be fitted to the experimental data with the models presented in the sections 3.3.1 and 3.3.2, then these fits result in values for the imaginary part of the alloy's Fermi pseudopotential, that are close to the one that is obtained by the method given above.

Since we know from chapter one, that the energies corresponding to the wave vector component perpendicular to the mirror and absorber surface are around $10^{-11}$ eV, in the experiment the Gadolinium absorption cross section in particular will saturate. Natural Gd then has a constant absorption cross section for slow neutrons with velocities in the order of cm per s of:

$$(3.1.20) \quad \sigma_{Gd}^{abs}\Big|_{UCN} \approx 55 \cdot Mbarn \ ,$$

where eq.s (3.1.19) and (3.1.18b) have been used.
One has therefore the result, that the imaginary part of the bound coherent scattering length, $\Im(b)$, which is given by the optical theorem as

$$\Im(b) = \frac{k_0 \cdot \sigma_{inel.}}{4\pi} ,$$

decreases for very small $k_0$. This loweres the imaginary part of the Fermi pseudopotential eq. (3.1.17) compared to the value it takes for thermal neutrons.

As mentioned at the begin of this chapter, the saturation of the absorption cross section also can be understood from a very intuitive argument – the cross section ceases to rise proportional to $k_0^{-1}$ if the diameter of the circular disk, that corresponds to the cross section, equals the interatomic distance of the medium.

Eq. (3.1.16) has to be calculated from eq. (3.1.17) for the alloy used in the experiment. For Gd the value of eq. (3.1.20) is used in eq. (3.1.17). $k_0$ corresponds to a vertical velocity of about 5 cm/s for energies, that correspond to heights of several ten µm.Eq. (3.1.16) then is calculated by adding up the potentials of the three components given by eq. (3.1.17).
In eq. (3.1.17) one encounters the problem of calculating the fractional nuclei densities of the components in the alloy. According to [Ne01] the alloy is described by the fractions of mass each component carries in one mass unit of alloy. Since the mass fractions of the components are known, one has to know the mass density of the alloy to obtain the fractional nuclei densities $n_0^{(j)}$. Unfortunately, there is no method known, according to which one can calculate the mass density of an alloy, if the mass fractions of the alloy's components are known. Since the atomic radii and weights, as well as the mass densities of the three components in question are sufficiently close together in this case, the alloy's density is approximated by the weighted arithmetic mean of the components' densities. The weights are given by the mass fraction percentages of the components.

Then the potential of eq. (3.1.16) for an alloy consisting of 54 % *Ti*, 35 % *Gd*, and 11 % *Zr* in mass proportions is:

$$U_{Fermi}^{alloy} = (7.7 - i \cdot 0.44) \cdot 10^3 \cdot peV \quad .$$

It is clear, that the imaginary part of this potential only depends stronly on the mass fraction of *Gd*. The absorption cross sections of *Ti* and *Zr* are negligible compared to *Gd*. However, the real part of eq. (3.1.21) is a mixture of the real parts of the Fermi pseudopotential of all three components. Its value is already one order of magnitude below values calculated for usual solid state matter, because *Ti* has a negative real part in its Fermi pseudopotential. Thus, relatively small changes in the proportion values can cause the real part of eq. (3.1.21) to vary by one order of magnitude. Now it should be remembered, that the absorbing layer was attached to the surface of the glass substrate by means of magnetron evaporation (see [Ne00]). Furthermore, one finds, that *Zr* has a boiling point temperature of about 4200 K, which is several hundred Kelvin beyond the boiling point temperature of the other components of the alloy. These facts together point out, that the proportion of *Zr* inside the actual alloy on the glass surface should probably somewhat smaller than the value given for the bulge composition in [Ne00]. The precision of bulge composition fractions allow for deviations of about 1 % from the values given in [Ne00]. Therefore a composition of about 55 % *Ti*, 34.5 % *Gd* and 10.5 % *Zr*, for example, should be considered as a better approximation to the true composition of the actual alloy, which is coated onto the glass. With these numbers the Fermi potential eq. (3.1.16) of the absorbing alloy is:

$$(3.1.21) \quad U_{Fermi}^{alloy} = (6.0 - i \cdot 0.44) \cdot 10^3 \cdot peV$$

Next, the “form factor” of the absorbing alloy layer $\partial M(\vec{r})$ in eq. (3.1.16) has to be determined. On a microscopic scale this factor is simply given by the rough surface profile function determined with AFM scans as shown in chapter two. Yet, only classical particles or neutrons with wave lengths small enough would see the full rough surface profile. From chapter one we know, that the vertical velocity spectrum has a width of about 10 – 20 $^{cm}/_{s}$ . This corresponds to de Broglie wave lengths of several μm, which is large compared to the mean height roughness of 0.75 μm. Therefore, it is suitable to view neutrons as plane waves, having uniform probability density in space. This uniformity leads to the fact, that the neutron wave function will more or less average over the absorber’s surface with respect to the directions parallel to this plane. The probability of locally passing the surface is then distributed as a simple gaussian with $\sigma$ = 0.75 µm, as shown in chapter two. $\partial M(\bar{r})$ of a single rough surface will thus be a gaussian error function with limiting values of zero and one, respectively. Since the absorber is made of an alloy with 0.2 µm in thickness, that is coated onto a rough glass surface, there will be two error functions separated by 0.2 µm, describing the front and the back surface of the alloy.

$\partial M(\bar{r})$ of the alloy coating is therefore:

$$(3.1.22) \quad \partial M_{alloy}(z) = Erf\left(\frac{z-(h-d)}{\sigma\sqrt{2}}\right) - Erf\left(\frac{z-h}{\sigma\sqrt{2}}\right) .$$

Here, *h* is the position of the average zero plane of the back of the alloy coating, which is identical with the average zero plane of the glass substrate. *Erf(z)* denotes the standard gaussian error function. *d* = 0.2 µm is the thickness of the coating and $\sigma$ = 0.75 µm the mean height roughness.
The Fermi pseudopotential of the absorber is therefore:

$$(3.1.23) \quad U_{Fermi}^{absorber}(z) = U_{Fermi}^{alloy} \cdot \partial M(z)_{alloy} ,$$

where $\partial M(z)_{alloy}$ and $U_{Fermi}^{alloy}$ are given by eq.s (3.1.21) and (3.1.22).

### 3.2) Classical description of the wave guide

This section will deal with the experimental results, which would be obtained if the setup behaved completely classically. The analysis of the collimator system presented in chapter one will therefore be taken as a beginning.

From this analysis we know, that the vertical velocity component of all neutrons entering the wave guide is below about 10 $^{cm}/_{s}$ , thus giving entirely total reflection from the mirrors for all those neutrons. Compare the horizontal motion, corresponding to de Broglie wave lengths of about 100 nm, and the vertical motion corresponding to wave lengths of about 1 ... 10 µm, with the parameters of the mirror's roughness from chapter two – mean height roughness $\sigma$ = 22 Å and correlation length $\xi$ = 10 µm. This leads to the conclusion, that the mirrors reflect all those neutrons, which enter the wave guide, in a fairly specular way. That suppresses any mixing of wave vector components by several orders of magnitude compared to the specular reflection probability.

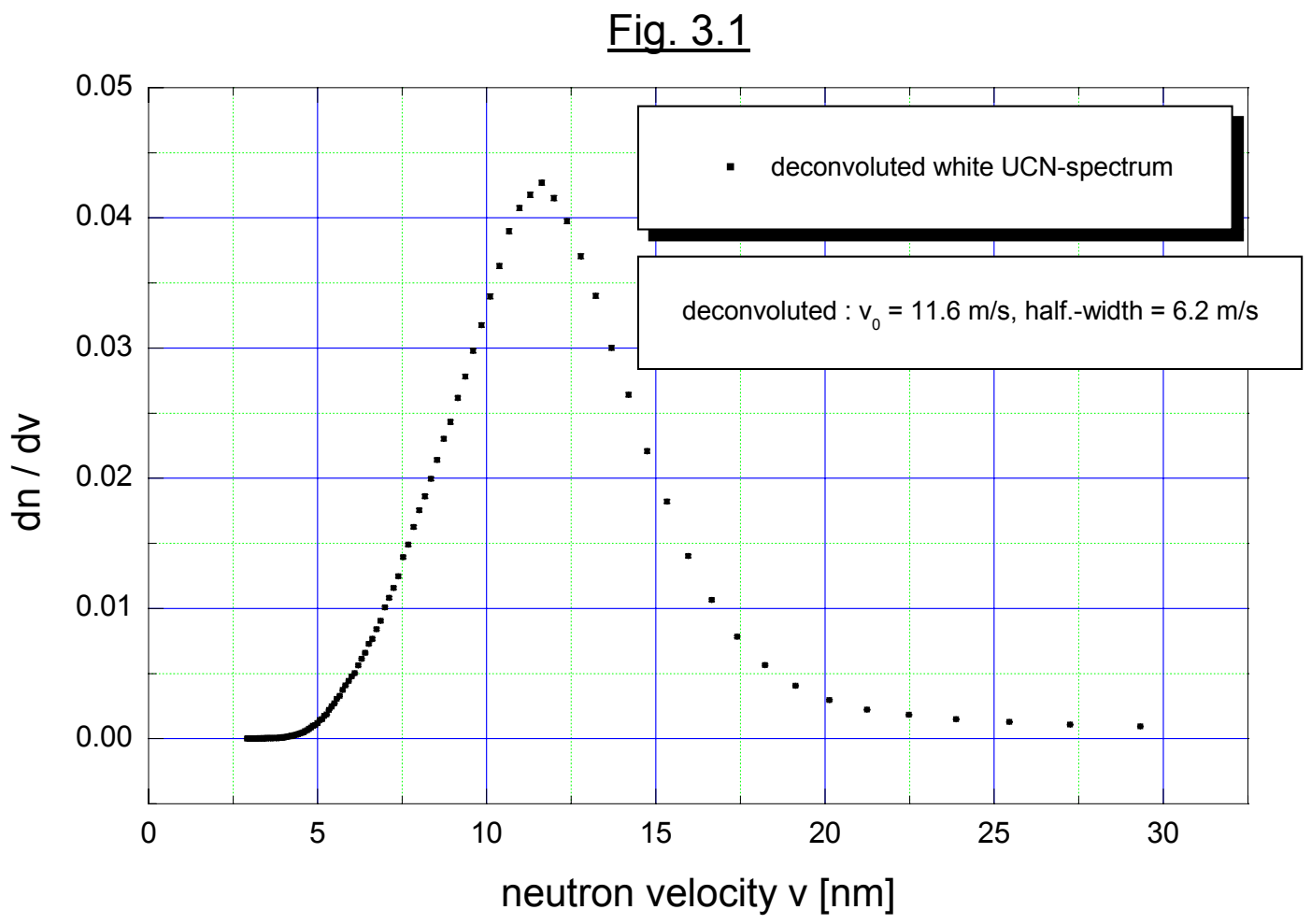

Fig. 3.1: UCN velocity spectrum *dn / dv = f(v)*

With this information, one is now able to derive a classical picture of the wave guide's transmission. Fig. 3.1, [Kl00], contains to major regions of neutron behaviour. Neutrons below about 10 $^{m}/_{s}$ are governed by a $T$ = 40 K Maxwell-Boltzmann distribution. Since $kT$ corresponds to about 500 $^{m}/_{s}$ and $v << 500 \cdot m/s$ everywhere, neutrons should be described by the low-energetic part of this distribution. The three dimensional Maxwell-Boltzmann distribution is quadratic in velocity in its low-energetic part, and this behaviour one sees for $v < 10 \cdot m/s$ in Fig. 3.1. Neutrons above 13 $^{m}/_{s}$ cut off exponentially, since these are lost in the neutron guides, and the losses increase exponentially with the neutron velocity. Since the spectrum is measured for the beam, that leaves the neutron guide, it is weighted once more with the velocity v . Therefore, the rising part of the spectrum has (approximately) a $v^3$-dependency.

Yet, the vertical velocities of the neutrons entering the wave guide are far below 1 $^{m}/_{s}$ . Therefore, Maxwellian distribution for this velocity component for $T = 40$ K predicts, that all vertical velocities are equally populated with neutrons.

This originates from the fact, that the one dimensional Maxwellian is a gaussian:

$$(3.2.1) \quad dN(\mathrm{v}_j) \propto e^{-\frac{m\cdot \mathrm{v}_j^2}{2\cdot k_B\cdot T}} \cdot d\mathrm{v}_j \quad ,$$

and that $m \cdot \mathrm{v}_j^2 << 2 \cdot k_B \cdot T$ for $\mathrm{v}_j << 500 \cdot \frac{m}{s}$, which is true for the whole UCN spectrum ($k_B$ is the Boltzmann constant). Therefore it can be concluded, that neutrons with a uniform vertical velocity distribution form the initial conditions in the classical picture.

The last notion needed for the classical approach is the assumption, that the absorber is a very good one, which is formulated in the statement, that every neutron touching the absorber at a certain time is lost. This means that neutrons, which are moving on parabolic trajectories inside the wave guide with repeatedly bouncing off the mirrors, will only pass it if their parabola's turning point lies below the absorber. With this idealization, the transmission of the wave guide as a function of its width – i.e. the height of the absorber – can be calculated analytically in the classical view:
The rate of neutrons passing the absorber-mirror-system is proportional to the phase space volume allowed by the absorber, as it is known from classical mechanics. Consider all the neutrons, which enter the wave guide at a given height $h' < h = h_{absorber}$ within an infinitesimal range $dh'$, and with vertical velocities small enough not to touch the absorber:

$$(3.2.2) \quad |\mathrm{v}_z(h')| \leq \mathrm{v}_z^{\max}(h') \; .$$

Their number is:

$$dn(h') = dh' \cdot \int_{-\mathrm{v}_z^{\max}(h')}^{\mathrm{v}_z^{\max}(h')} e^{-\frac{m\cdot \mathrm{v}_z^2}{2\cdot k\cdot T}} d\mathrm{v}_z \quad .$$

This expression controls the phase space volume available for the neutrons. The total counting rate in the classical picture is therefore the integral of eq. (3.2.2) over all possible heights $h' < h$:

$$(3.2.2') \quad \dot{N}_{tot} \propto \int_{-\infty}^{\infty} e^{-\frac{m \cdot v_x^2}{2 \cdot k \cdot T}} dv_x \cdot \int_{-\infty}^{\infty} e^{-\frac{m \cdot v_y^2}{2 \cdot k \cdot T}} dv_y \cdot \int_0^h dh' \cdot \int_{-v_z^{max}(h')}^{v_z^{max}(h')} e^{-\frac{m \cdot v_z^2}{2 \cdot k \cdot T}} dv_z \quad .$$

Eq. (3.2.2') now is essentially the relevant classical phase space volume of the wave guide system.

Eq. (3.2.2) can be expressed in terms of the potential, which governs motion of the particle inside the wave guide:

$$(3.2.2'') \quad v_z^{max}(h') = \sqrt{2 \cdot (\phi(h) - \phi(h'))} = \sqrt{2g \cdot (h - h')} \quad .$$

This gives:

$$\dot{N}_{tot} \approx const. \cdot \int_o^h dh' \cdot \sqrt{2g \cdot (h - h')} = const.' \cdot \frac{2}{3} \sqrt{2g} \cdot h^{3/2}$$

$$\frac{m}{2} \cdot \left[ v_z^{max}(h') \right]^2 << k \cdot T \quad \forall h'$$

$$\Rightarrow \quad (3.2.3) \quad \dot{N}_{tot} \propto h^{3/2} \quad .$$

This expectation can be checked by means of a Monte Carlo simulation, which will consist of classical mass-points moving inside a one-dimensional hollow wave guide with a perfect bottom mirror and a perfect absorber above in a homogenous vertical gravitational field. "Perfect" is assigned to the fact, that the absorber is considered to be an ideal one, absorbing every neutron that touches it. The simulation is performed two dimensionally, taking into account the vertical direction and one horizontal dimension in direction of optical axis of the wave guide. The horizontal transverse direction is omitted, since it is not spatially limited and will only contribute with a constant factor to the total transmitted neutron flux. Total count rate of neutrons passing the wave guide is then "measured" as a function of the wave guide width $h$.

The results of such a Monte Carlo simulation are:

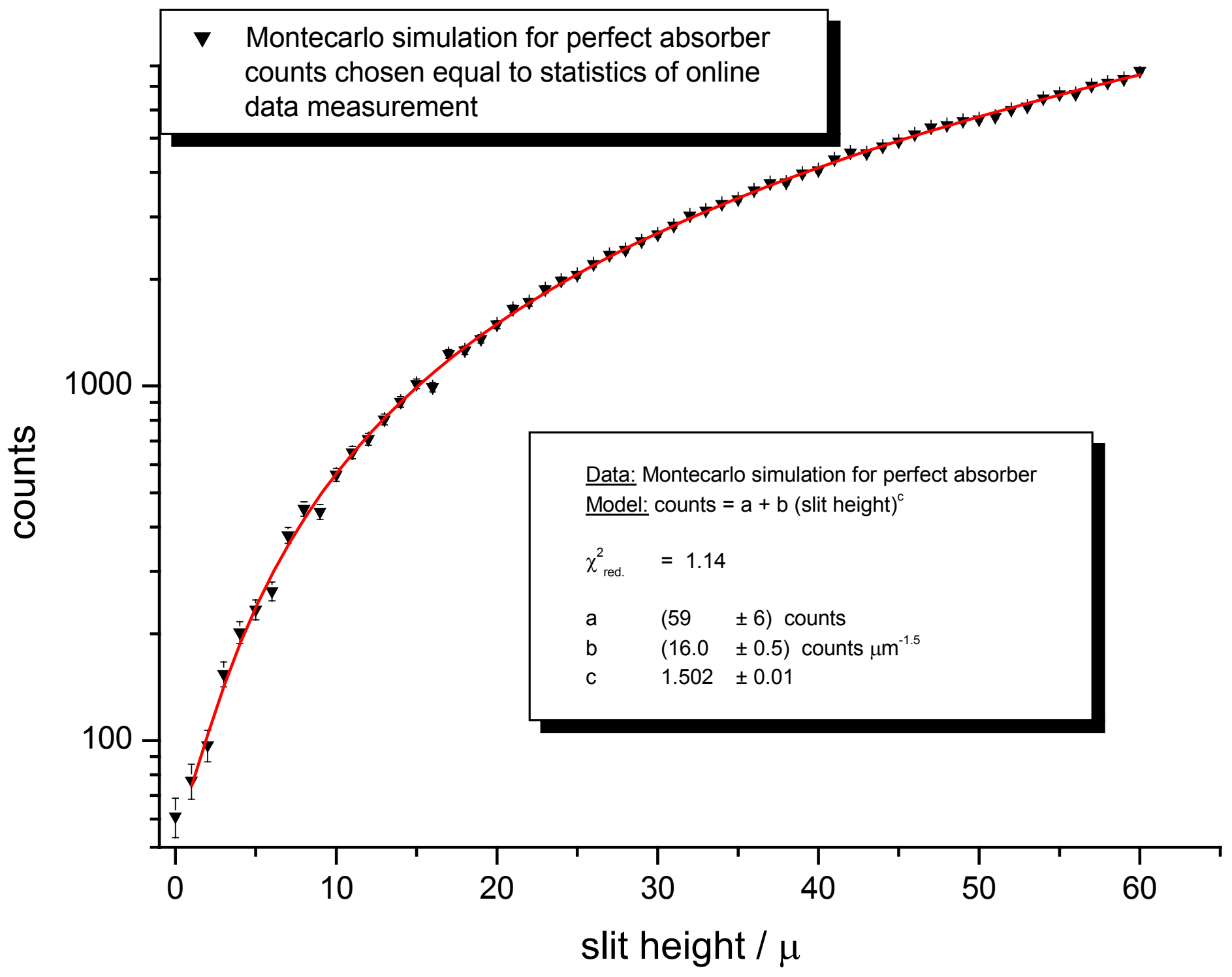

If one turns of gravitation inside the wave guide then Monte Carlo simulation will yield the following behaviour:

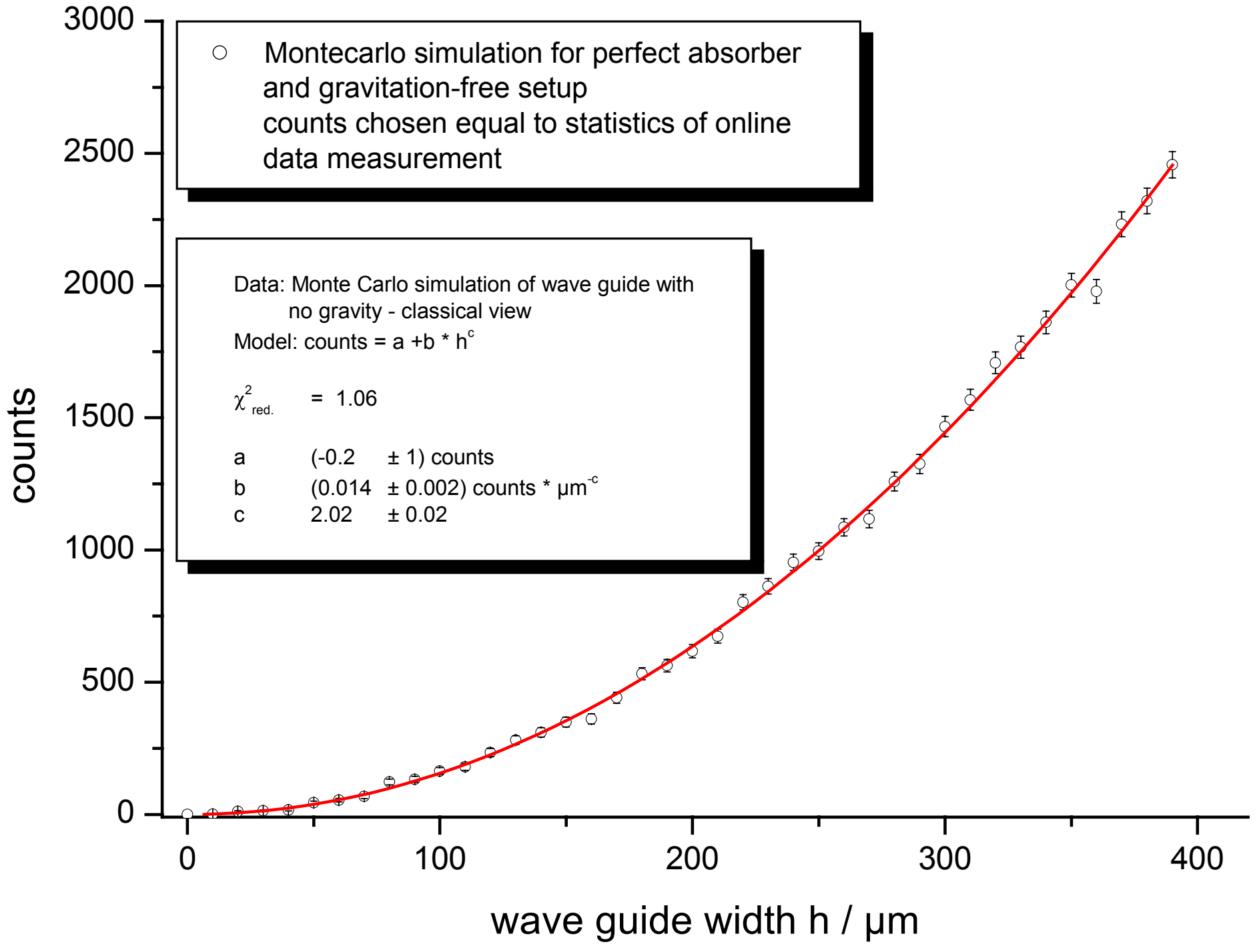

Comparing the results clearly shows, that one has to expect a behaviour given by eq. (3.2.3) for the quasi-classical regime.

In contrast, the case without gravitation shows an unfamiliar situation. Simple intuition would expect a linear dependency on the wave guide width *h*. Yet, the transmission of a one dimensional wave guide consisting of one mirror and one absorber shows a quadratical dependency on *h*, if gravitation is absent.

Here a look at the phase space volume of the system is useful. First, one realizes, that the range of the vertical velocity, that a neutron of a given horizontal velocity, that enters the wave guide at the position *z*, must not exceed to avoid touching the absorber directly or via reflecting once at the bottom mirror, is given by:

$$-\frac{h+z}{T_{flight}} = -\frac{l}{\mathrm{v}_{hor.}} \cdot (h+z) < \mathrm{v}_{vert.} < \frac{l}{\mathrm{v}_{hor.}} \cdot (h-z) \ .$$

Now one can apply eq. (3.2.2'):

$$(3.2.2') \quad \dot{N}_{tot} \propto \int_{-\infty}^{\infty} e^{-\frac{m\cdot \mathrm{v}_x^2}{2\cdot k\cdot T}} d\mathrm{v}_x \cdot \int_{-\infty}^{\infty} e^{-\frac{m\cdot \mathrm{v}_y^2}{2\cdot k\cdot T}} d\mathrm{v}_y \cdot \int_0^h dz \cdot \int_{\mathrm{v}_z^{\min}(z)}^{\mathrm{v}_z^{\max}(z)} e^{-\frac{m\cdot \mathrm{v}_z^2}{2\cdot k\cdot T}} d\mathrm{v}_z$$

$$\Rightarrow \quad \dot{N}_{tot} = const. \cdot \int_0^h dz \cdot \frac{2\cdot l}{\mathrm{v}_{hor.}} \cdot z = const. \cdot \frac{l}{\mathrm{v}_{hor.}} \cdot h^2 \quad .$$

The Monte Carlo simulations shows the correct behaviour! However, one should note, that the exponent of the power law depends on the fact, that at all heights *h* in question the available range of the vertical velocity is large enough to ensure the validity of the cutoff condition given above. If the wave guide becomes so wide, that nearly all of the vertical velocities, that are present in the beam, can pass the wave guide, the power law will become linear, since the wave guide width alone decides about the number the neutrons, that enter the guide. This is satisfied the better, the larger the distance of the neutrons to the absorber is, where they enter the guide.

The next plot shows a Monte Carlo simulation, that elucidates this behaviour:

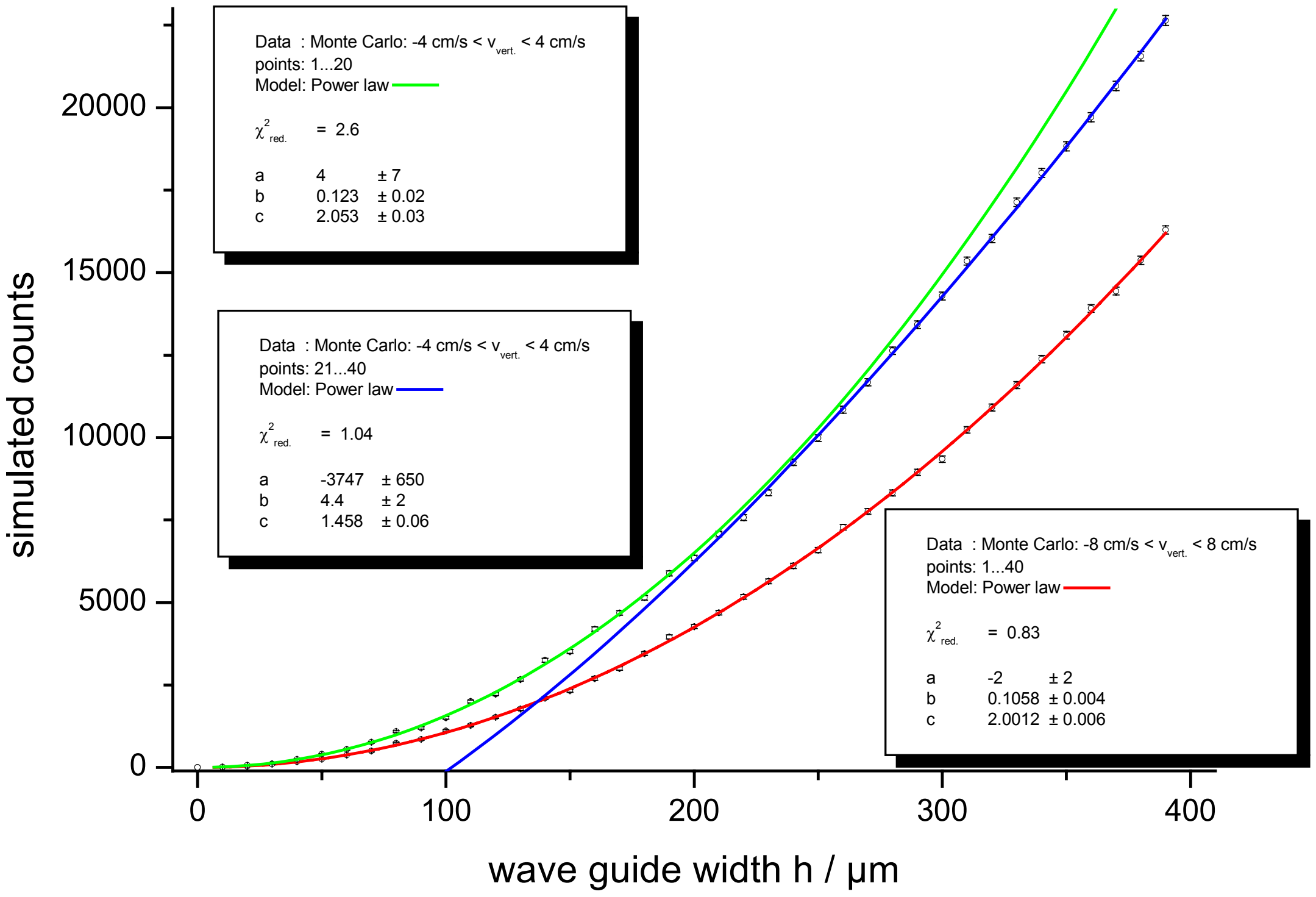

Figure: Monte Carlo simulation of classical neutrons transmitted by the wave guide, if there is no gravitational field. Different regions of the simulated data have been fitted by a power law $counts = a + b \cdot h^c$. The range of the vertical velocity was $\pm 4\, cm/s$ and $\pm 8\, cm/s$, respectively.

One clearly sees, that the power law, that is obeyed by the classical transmission of the wave guide without a gravitational field, depends on the range of the vertical velocity as well as on the range of the wave guide widths $h$.

## 3.3) Quantum mechanical approach

### 3.3.1) The general formalism

Now that we have gathered enough information about the setup of the system, we can try to develop a quantum mechanical description of the measurements. The setup can be described within quantum mechanics, if it is possible to reduce the problem of a neutron moving inside the setup to an effective one-particle problem. In that case, a Schrödinger equation can be formulated in general, which then has to be solved.

The neutron interacts with the setup in two different ways. First there is the weak, but long-range interaction with the earth as a whole due to the earth's gravitational field. Second, we have interaction with matter due to the strong force coupling to the nucleons.

The gravitational interaction causes the neutron to feel a resulting gravitational potential generated from the earth, the mirrors at the bottom and the absorber above. The vertical positions of the neutron are less than 200 µm inside the wave guide. Since the glass substrates of the mirrors and the absorbers have thicknesses of about 1 cm, the neutrons will be about $\bar{z} = 0.5$ cm from the center of mass of the glass mirror. Thus, we can determine the relative strength of the gravitational contributions, which arise from the mirror and the earth itself, respectivley.

The gravitational potential of the earth is essentially:

$$(3.3.1.1) \quad \phi_{grav}^{earth} = -G \cdot M \cdot \left( \frac{1}{r} - \frac{1}{r'} \right) = -G \cdot M \cdot \left( \frac{1}{R+z} - \frac{1}{R} \right) = -\frac{G \cdot M}{R} \cdot \left( \frac{1}{1+\frac{z}{R}} - 1 \right)$$

$$\approx \frac{G \cdot M}{R^2} \cdot z = g \cdot z \quad .$$

The order of magnitude of the gravitational potential caused by the mirrors and the absorber is:

$$(3.3.1.2) \quad \phi_{grav}^{plates} \approx -\frac{G \cdot m_{glass\ mirror}}{z} = -\frac{G \cdot M}{R^2} \cdot \frac{m_{glass\ mirror}}{M} \cdot \frac{R^2}{z} \approx -g \cdot z \cdot \frac{m_{glass\ mirror}}{M} \cdot \frac{R^2}{\bar{z}^2} \ ,$$

where

$$(3.3.1.3) \quad \frac{m_{glass\ mirror}}{M} \cdot \frac{R^2}{\bar{z}^2} \approx 10^{-7} \ .$$

This number gives the ratio of the strengths of the earth's and the mirror's gravitational potential felt by a neutron inside the wave guide.
Thus neutrons inside the wave guide move in a gravitational potential given by:

$$(3.3.1.4) \quad \phi_{grav}(z) = g \cdot z \,, \quad \text{where:} \quad g \approx 9.81 \cdot \frac{m}{s^2} \,,$$

since the contribution of the gravitational potential, that arises from the mirror (as well as from the absorber due to the same argument), is neglibible.

It was shown in section 3.1), that the interaction of low energetic neutrons with matter can be reduced to an effective one-particle interaction potential, the Fermi pseudopotential. Thus, the absorber will be represented by a potential given by eqs. (3.1.21), (3.1.22) and (3.1.23). The same way the glass substrate of the absorber and the mirrors are then represented by Fermi pseudopotentials. They are essentially real, corresponding to the very small absorption of neutrons by glass. The Magnitude of the glass' Fermi potential is about 100 neV. The form factor, eq. (3.1.22), for the glass mirrors is a gaussian error function with the mirrors' roughness amplitude of 22 Å as its width. From chapter two we know, that reflection of neutrons with wave lengths of about 500 Å from such mirrors will be fairly specular. Therefore, the roughness of the mirrors can be omitted without any loss of information and the mirrors' Fermi pseudopotential is:

$$(3.3.1.5) \quad U_{Fermi}^{mirrors}(z) = \theta(-z) \cdot U_{Fermi}^{glass} \quad , U_{Fermi}^{glass} \approx 100 \cdot neV \;.$$

The Fermi potential of the absorber's glass substrate has the same magnitude. Yet, it would be smeared out by an error function of 0.75 µm width due to the roughness, as it is the case for the absorber's Fermi potential itself. However, the absorbing layer has a real part of the Fermi potential of about 6 neV. Furthermore, the gravitational potential energy of 10 µm height corresponds to ony about 1 peV, and the vertical velocity spectrum of 10 $^{cm}/_{s}$ range corresponds to about 20 peV in energy. The neutrons will therefore virtually not feel the roughness of the glass substrate. The tiny fraction of a neutron's wave function, which arrives at the backward average plane of the coating, will just feel a steep potential rising towards 100 neV. This will suppress the wave function's tail inside the glass by some further orders of magnitude.

The argument shows, that it makes sense to approximate the absorber's glass substrate by a Fermi potential like eq. (3.3.1.5):

$$(3.3.1.6) \quad U_{Fermi}^{glass\ substrate\ absorber}(z) = \theta(z-h) \cdot U_{Fermi}^{glass} \quad , U_{Fermi}^{glass} \approx 100 \cdot neV \,.$$

Putting all pieces together, we can formulate an interaction potential seen by a neutron propagating inside the wave guide:

$$(3.3.1.7)\quad U(z)=\begin{cases} U_{Fermi}^{glass}\ ,\ z\leq 0\wedge x>0 \\ m\cdot g\cdot z+\partial M_{alloy}(z)\cdot U_{Fermi}^{alloy}\ ,\ 0<z<h\wedge 0\leq x\leq l_{abs} \\ U_{Fermi}^{glass}\ ,\ z\geq h\wedge 0\leq x\leq l_{abs} \\ m\cdot g\cdot z\ ,\ z>0\wedge x>l_{abs} \end{cases}$$

$$where:\quad \begin{cases} \partial M(z)_{alloy}=Erf\left(\frac{z-(h-d)}{\sigma\sqrt{2}}\right)-Erf\left(\frac{z-h}{\sigma\sqrt{2}}\right) \\ \sigma\approx 0.75\cdot\mu m \\ d\approx 0.2\cdot\mu m \\ U_{Fermi}^{glass}\approx 10^{5}\cdot peV \\ U_{Fermi}^{alloy}\approx(6.0-i\cdot 0.44)\cdot 10^{3}\cdot peV \end{cases}\quad .$$

The smearing of the absorber's Fermi pseudopotential with an error function form factor $\partial M_{alloy}(z)$, that averages over the local scattering events in the absorbing layer, is an appropriate way to account for the roughness of the absorber, since only specular scattering is viewed. Specular scattering does only depend on the wave vector component orthogonal to the average zero plane, as it is shown in chapter two and Appendix C. In the limit of:

$$q_{\perp}\cdot\sigma<<1$$

$$q_{\perp}=\text{wave vector transfer perpendicular to the average surface zero plane}=\frac{4\pi}{\lambda_{\perp}}\cdot\sin\vartheta$$

the Born approximation describes the scattering amplitude as the Fourier transform of the horizontally averaged surface potential, which is essentially an error function for a gaussian roughness. Since $\lambda_{\perp}\approx 15\,\mu\text{m}>>\sigma$ for the ground state and $\sin\vartheta=\text{v}_{\perp}/\text{v}_{hor.}<<1$, it is justified here to incorporate the roughness of the absorber simply by using a form factor $\partial M_{alloy}(z)$.

This is the potential that essentially describes the setup of Fig. 3.2:

Fig. 3.2
Setup of Measurement

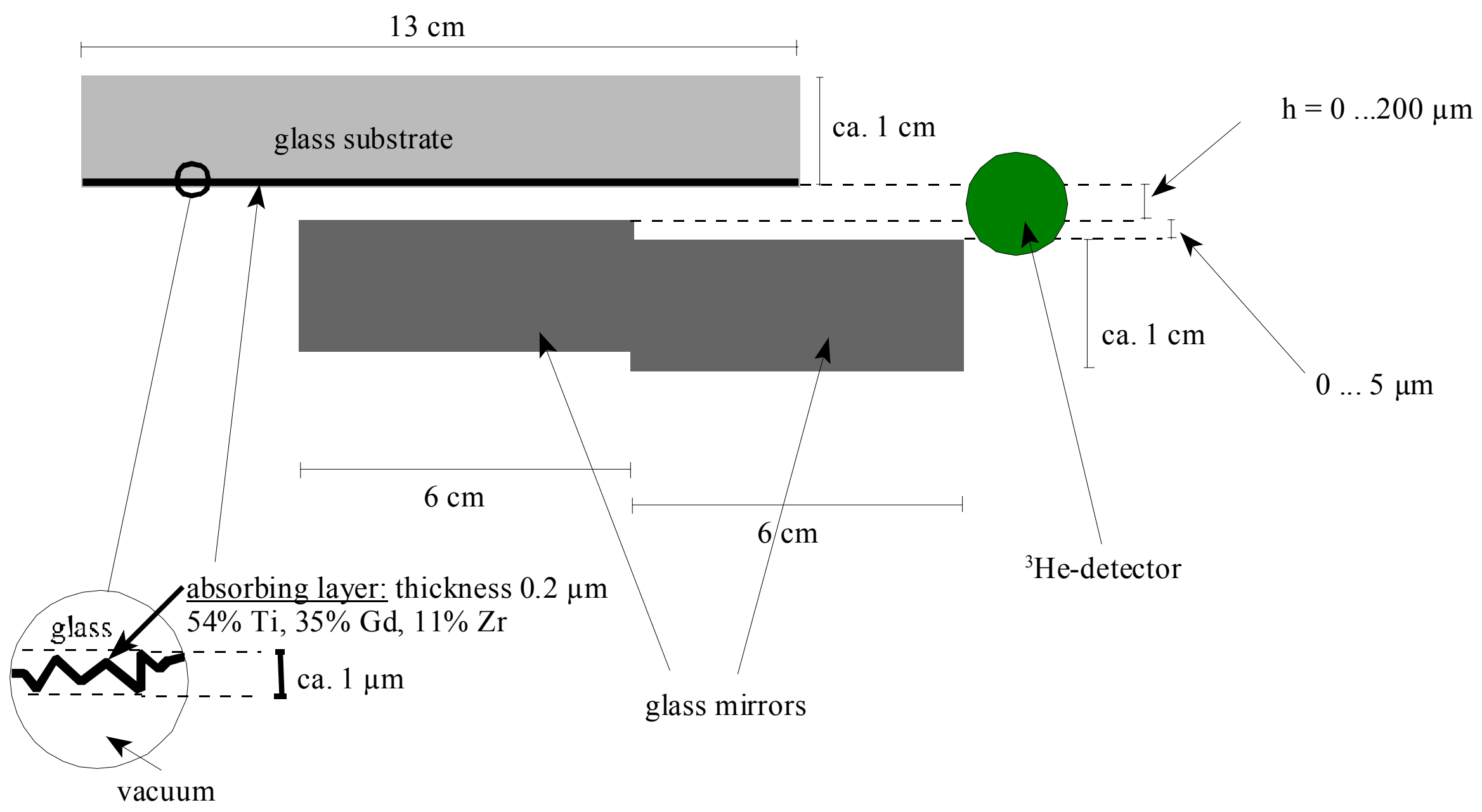

The large scale structure of the potential eq. (3.3.1.7) for the region between the absorber and the bottom mirror is seen in Fig. 3.3:

Fig. 3.3
distribution of potentials in experimental setup

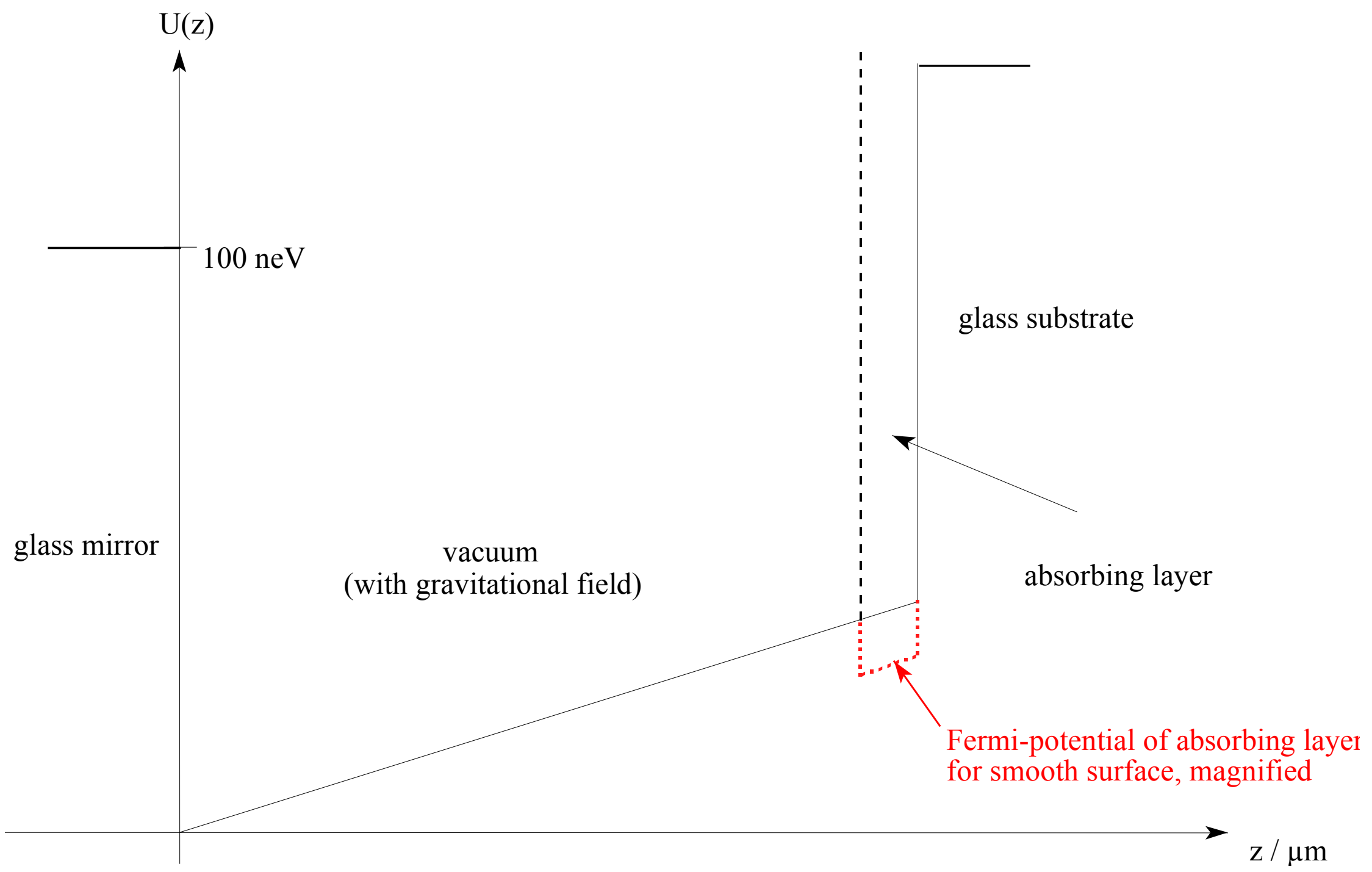

The form factor of the absorber's Fermi potential is shown in more detail in Fig. 3.4:

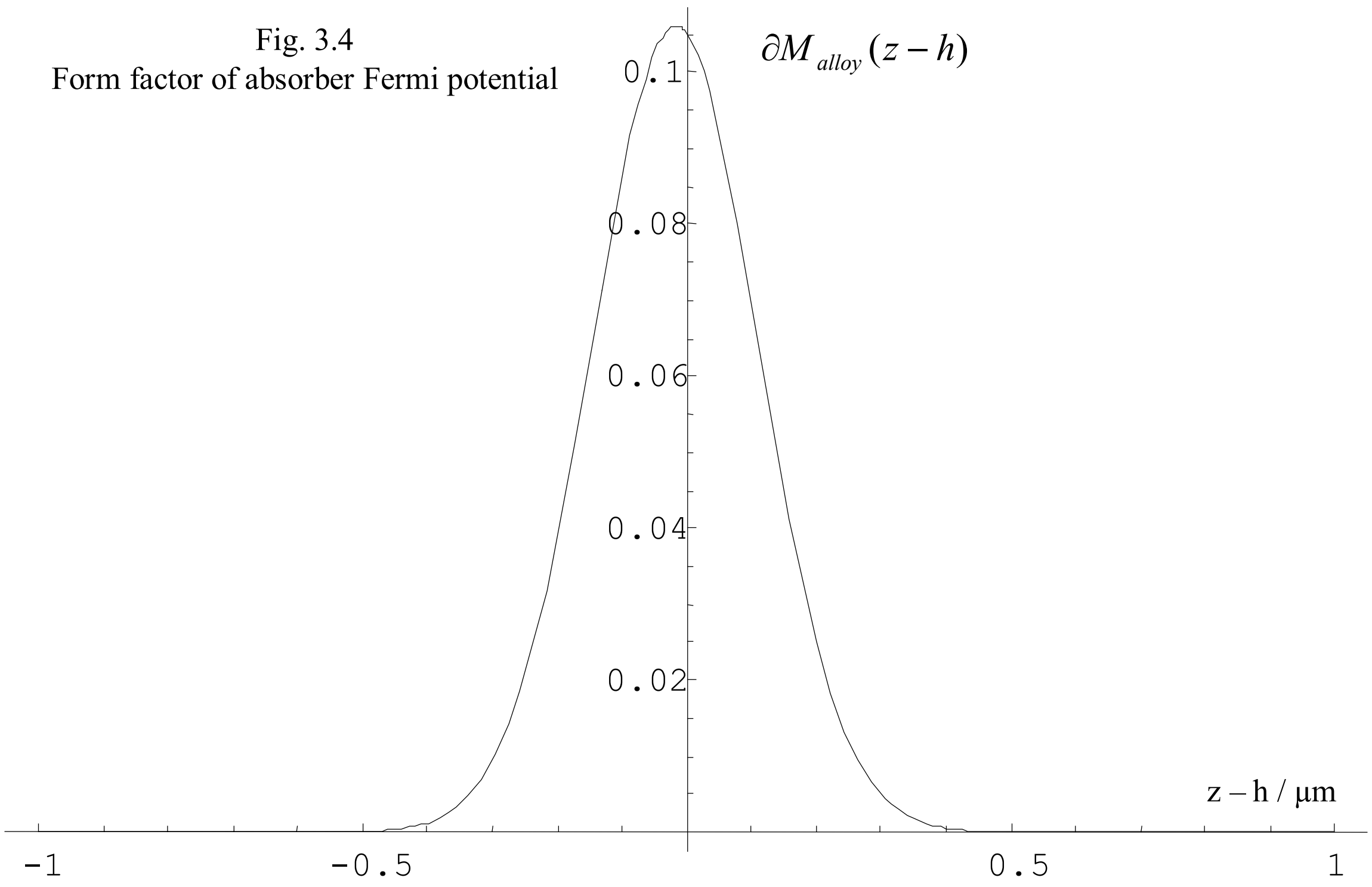

Fig. 3.4
Form factor of absorber Fermi potential

The potential of eq. (3.3.1.7) describes an effective single-particle problem. The Schrödinger equation is:

$$(3.3.1.8)\quad i\hbar\cdot\frac{\partial}{\partial t}\psi(\vec{r},t)=\left[-\frac{\hbar^2}{2\cdot m}\cdot\Delta+U(z)\right]\cdot\psi(\vec{r},t).$$

The wave guide has a geometry which suggests the use of rectangular coordinates. The wave function can be written quite generally as

$$(3.3.1.9)\quad \psi(\vec{r},t)=\varphi(z)\cdot\int d^2k\cdot e^{i\cdot k_x\cdot x}\cdot e^{i\cdot k_y\cdot y}\cdot e^{-i\cdot\frac{E(k_x,k_y)}{\hbar}\cdot t}\ .$$

Consider one mode of the composition eq. (3.3.1.9) with given total energy *E*:

$$(3.3.1.9b)\quad \psi_{\vec{k}}(\vec{r},t)=\varphi(z)\cdot X(x)\cdot Y(y)\cdot e^{-i\cdot\frac{E}{\hbar}\cdot t}=\varphi(z)\cdot e^{i\cdot k_x\cdot x}\cdot e^{i\cdot k_y\cdot y}\cdot e^{-i\cdot\frac{E}{\hbar}\cdot t}\ ,\ E=E_x+E_y+E_z\ .$$

Inserting this into (3.3.1.8) will lead to separation of the Schrödinger equation into:

$$(3.3.1.10)\quad \begin{cases}(a)\quad E_x\cdot X(x)=\dfrac{\hbar^2\cdot k_x^2}{2\cdot m}\cdot X(x)\\ (b)\quad E_y\cdot Y(y)=\dfrac{\hbar^2\cdot k_y^2}{2\cdot m}\cdot Y(y)\\ (c)\quad E_z\cdot\varphi(z)=\left[-\dfrac{\hbar^2}{2m}\cdot\dfrac{\partial^2}{\partial z^2}+U(z)\right]\cdot\varphi(z)\end{cases}\quad\Rightarrow\quad E=k_x^2+k_y^2+E_z\quad .$$

Since we have chosen $x$ to be oriented along the optical axis of the wave guide, and $z$ to be the vertical direction, $y$ represents one free laterally transverse dimension, which we will omit in the further calculations. Eq.s (3.3.1.8) and (3.3.1.9b) are then written in terms of $x$ and $z$ thus splitting (3.3.1.8) into (3.3.1.10a) and (3.3.1.10c).

Now eigenfunctions of eq. (3.3.1.10c) have to be found. A look at the potential of eq. (3.3.1.7) shows, that the eigenfunctions of eq. (3.3.1.10c) will decline exponentially inside the glass substrate of the absorber and the mirrors, if the energy eigenvalues $E_z$ are smaller than the magnitude of the Fermi potentials. This means, that

$$(3.3.1.11) \quad E_z < U_{Fermi}^{glass} \Rightarrow \begin{cases} \lim\limits_{z\to-\infty} \varphi(z) = 0 \\ \lim\limits_{z\to\infty} \varphi(z) = 0 \end{cases} .$$

are proper boundary conditions for the solutions of eq. (3.3.1.10c) in the $E_z$-range given in eq. (3.3.1.11). The eigenfunctions will therefore be restricted at both sides, if their eigenvalues are small enough. As the problem of eq. (3.3.1.10c) is additionally one-dimensional, eq. (3.3.1.10c) will have a finite set of eigenfunctions with discrete eigenvalues $E_n^{(z)}$. Discreteness of the eigenvalues arises from the boundary conditions eq. (3.3.1.11). Furthermore, the set of eigenfunctions and eigenvalues $E_n^{(z)}$ fulfilling eq. (3.3.1.11) will be finite, because the potential eq. (3.3.1.7) does not show a behaviour like $z^{-a}$ with a ≥ 1, while it has a constant magnitude of about 100 neV. Therefore, eq. (3.3.1.10) will change to:

$$(3.3.1.10') \quad \begin{cases} (a) \quad E_x \cdot X(x) = \dfrac{\hbar^2 \cdot k_x^2}{2 \cdot m} \cdot X(x) \\ (c) \quad E_n^{(z)} \cdot \varphi_n(z) = \left[ -\dfrac{\hbar^2}{2m} \cdot \dfrac{\partial^2}{\partial z^2} + U(z) \right] \cdot \varphi_n(z) \end{cases}$$

$$\Rightarrow \quad E = k_x^2 + E_n^{(z)} \quad , \quad E_n^{(z)} \le U_{Fermi}^{glass} \quad .$$

Lets now have one more look at the Fermi potential of the glass. If the total energy E of a solution eq. (3.3.1.10') is below 100 neV, then the wave function inside the glass will essentially be

$$\varphi_n(z) \propto e^{-\kappa \cdot z} \quad , \quad \kappa = \sqrt{\frac{2m}{\hbar^2} \cdot \left( U_{Fermi}^{glass} - E_n^{(z)} \right)} \, .$$

If furthermore $E_n^{(z)} \ll U_{Fermi}(glass)$ = 100 neV, then $\kappa$ will corresponds to an attenuation length of:

$$(3.3.1.12) \quad attenuation\ length = \kappa^{-1} \approx 140 \cdot \text{Å} \, .$$

This means, that the neutron enters the glass substrates with only a very tiny fraction of its wave function over a very small range. This justifies to define the glass' Fermi potential as being infinitely high, since the boundary conditions at the glass' surface then simplify to:

$$(3.3.1.13)\quad \begin{cases} \varphi_n(z \le 0) = 0 \\ \varphi_n(z \ge h) = 0 \end{cases} .$$

As argued above, the potential eq. (3.3.1.7) is here modified to:

$$(3.3.1.13')\quad U(z) = \begin{cases} \infty\,,\, z \le 0 \wedge x > 0 \\ m \cdot g \cdot z + \partial M_{alloy}(z) \cdot U_{Fermi}^{alloy}\,,\, 0 < z < h \wedge 0 \le x \le l_{abs} \\ \infty\,,\, z \ge h \wedge 0 \le x \le l_{abs} \\ m \cdot g \cdot z\,,\, z > 0 \wedge x > l_{abs} \end{cases} .$$

Now the potential of eq. (3.3.1.7) as well as the setup of Fig. 3.2 can be split into several regions, where the potential does not have discontinuities. This is most clearly seen in Fig. 3.5:

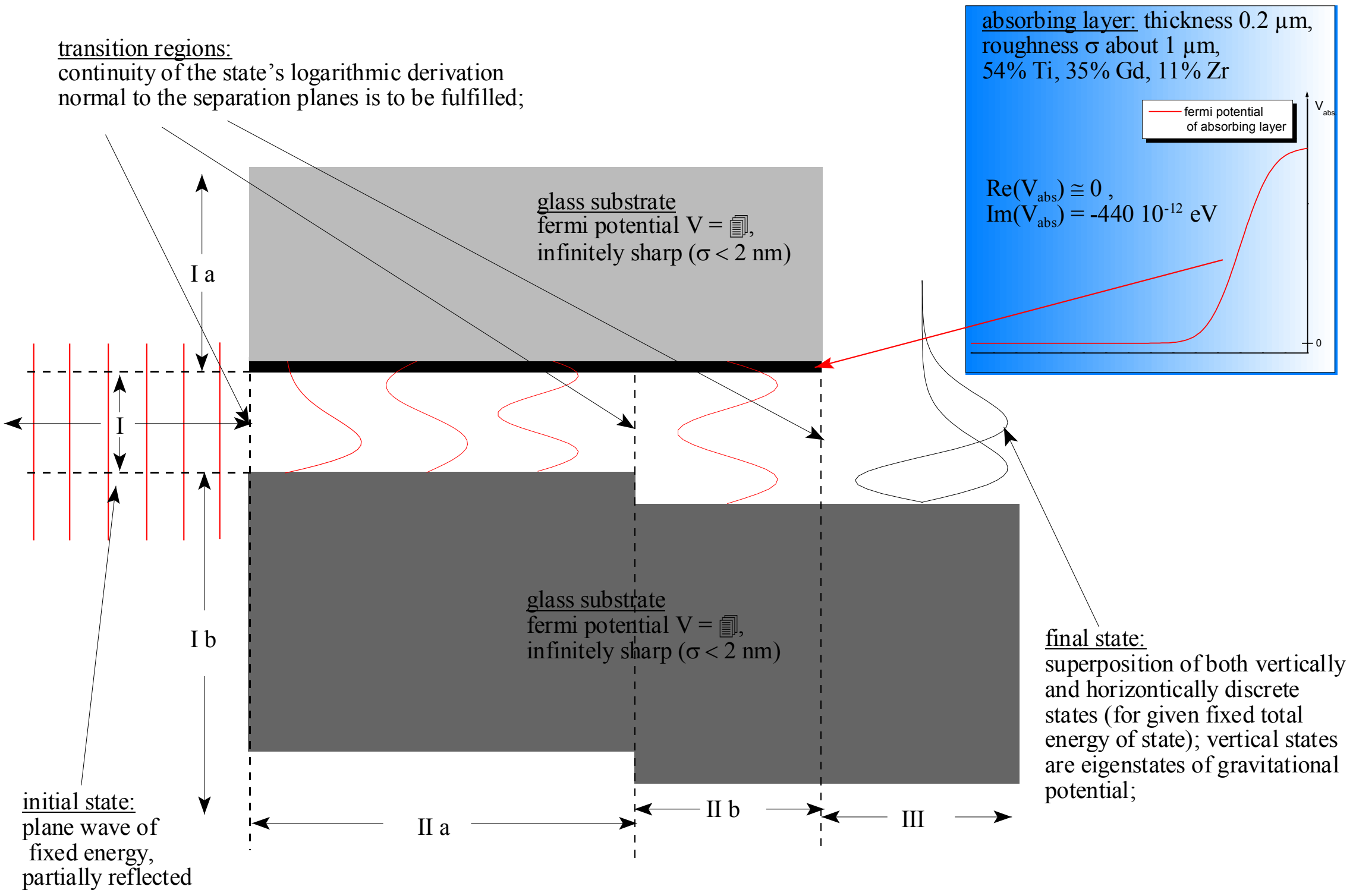

Fig. 3.5: The boundary conditions of the wave guide setup used for the "online" measurement

From this plot, we immediately obtain, that in the regions *I*, *Ia* and *Ib* the Schrödinger equation simplifies to:

$$(3.3.1.14)\quad i\hbar\cdot\frac{\partial}{\partial t}\psi_{\vec{k}}(\vec{r},t)=\left[-\frac{\hbar^2}{2\cdot m}\cdot\Delta+m\cdot g\cdot z\right]\cdot\psi_{\vec{k}}(\vec{r},t)\ .$$

The same equation holds for region *III*. Additionally, there is a boundary condition in this region:

$$(3.3.1.14b)\quad \varphi_n(z\le 0)=0\ .$$

The solution of (3.3.1.14) can be obtained analytically:
After separation we have

$$(3.3.1.14')\quad \begin{cases}(a)\quad E_x\cdot X(x)=\dfrac{\hbar^2\cdot k_x^2}{2\cdot m}\cdot X(x)\\ (c)\quad E_n^{(z)}\cdot\varphi_n(z)=\left[-\dfrac{\hbar^2}{2m}\cdot\dfrac{\partial^2}{\partial z^2}+m\cdot g\cdot z\right]\cdot\varphi_n(z)\end{cases}\quad\Rightarrow\quad E=k_x^2+E_n^{(z)}\ .$$

One can define $E_R$ and $R$ as characteristical energy and length scale, respectively, of (3.3.1.14'):

$$(3.3.1.15)\quad E_R=\left(\frac{\hbar^2}{2}\cdot m\cdot g^2\right)^{\frac{1}{3}}\approx 0.602\cdot 10^{-12}\cdot eV\quad and\quad R=\left(\frac{\hbar^2}{2\cdot m^2\cdot g}\right)^{\frac{1}{3}}\approx 5.87\cdot\mu m\ .$$

Then eq. (3.3.14'c) can be written as:

$$(3.3.1.16)\quad \varphi''(\eta)+[\varepsilon-\eta]\cdot\varphi(\eta)=0$$

$$where:\ z=\eta\cdot R\quad and\quad E^{(z)}=\varepsilon^{(z)}\cdot E_R\ .$$

This can now be solved analytically. Plane waves are the solutions of eq. (3.3.14'a). The solution of eq. (3.3.1.14) and thus of eq. (3.3.14'c) is given by:

$$(3.3.1.17)\quad \varphi(\eta)=C_A\cdot\phi_A\left(\eta-\varepsilon^{(z)}\right)+C_B\cdot\phi_B\left(\eta-\varepsilon^{(z)}\right)\ .$$

$\phi_A$ and $\phi_B$ are the so-called Airy functions of type AiryAi and AiryBi. They are defined as a linear combination of Bessel functions (see Appendix A (A.2)):

$$(3.3.1.18)\quad \begin{cases} \phi_A(\eta) = \frac{1}{3}\sqrt{\eta}\cdot\left[I_{-1/3}\left(\frac{2}{3}\eta^{3/2}\right) - I_{+1/3}\left(\frac{2}{3}\eta^{3/2}\right)\right] \\ \phi_B(\eta) = \frac{\sqrt{3}}{3}\sqrt{\eta}\cdot\left[I_{-1/3}\left(\frac{2}{3}\eta^{3/2}\right) + I_{+1/3}\left(\frac{2}{3}\eta^{3/2}\right)\right] \end{cases} .$$

For the behaviour of the two types of Airy functions with real arguments see Appendix A. In region *I*, *Ia* and *Ib* there is either no or at most one boundary condition for the movement in the vertical direction. This can be seen from Fig. 3.2. Therefore, the energy eigenvalues $E^{(z)}$ of eq. (3.3.1.15) will have a continuous spectrum in these regions.

The situation in region *III* is different. The solution eq. (3.3.1.17) is restricted from below by eq. (3.3.1.14b), and it must vanish for $z \to \infty$, since the potential eq. (3.3.1.13') rises to infinity in this limit. Thus, only the first type Airy function $\phi_A$ can be used to form the eigenstates eq. (3.3.1.17), i.e. $C_B = 0$. The boundary condition eq. (3.3.1.14b) then forces the energy eigenvalues $E^{(z)}$ to take the values of the zeros of $\phi_A$ , that is: $E_{III}^{(z)} = E_n^{(z)III}$ , $n \in$ Í – the spectrum of $E_n^{(z)}$ is discrete! The energy eigenvalues of the vertical motion in region III are therefore given by:

$$(3.3.1.19)\quad E_n^{(z)} = \varepsilon_n^{(z)} \cdot E_R \quad ,where\ \phi_A\left(-\varepsilon_n^{(z)}\right) = 0\ \forall n \quad .$$

Here is a list of the first four energy eigenvalues $E_n^{(z)\,III}$ of the vertical motion in region *III*:

| n | $E_n^{(z)\,III}$ / $10^{-12}$ eV | classical turning point / µm | expectation value of $z$ / µm |
|---|---|---|---|
| 1 | 1.407 | 13.722 | 9.150 |
| 2 | 2.460 | 23.992 | 16.00 |
| 3 | 3.322 | 32.400 | 21.60 |
| 4 | 4.083 | 39.831 | 26.56 |

The two columns at the right give the turning point height, that a classical particle with initial kinetical energy $E_n^{(z)\ III}$ would have, and the expectation value $z_{expect.}$ of the corresponding eigenstate $\phi_n(\eta)$, which is defined as:

$$(3.3.1.20)\quad z_{expect.} = R \cdot \int d\eta \cdot \varphi_n^*(\eta) \cdot \eta \cdot \varphi_n(\eta) .$$

Determination of the eigenfunctions and the energy eigenvalues of the vertical motion in region *II* (*IIa* and *IIb*) is more complicated. Yet, we know something general from the fact, that only the vertical motion is bound and quantized in energy:

The energy eigenvalues of a one-dimensional motion form a non-degenerate set. The non-degeneracy corresponds to the fact, that the eigenstates of motion form a so-called "orthonormal system of functions". That means, that the normalized bound states of the vertical motion in the regions *II* and *III* generally fulfill the following relation:

$$(3.3.1.21a)\quad \langle \varphi_m(\eta) | \varphi_n(\eta) \rangle = \delta_{mn} \quad \textit{in region II and III} .$$

For the continuous spectrum of the region *I* a similar relation holds:

$$(3.3.1.21b)\quad \left\langle \varphi_{\varepsilon'^{(z)}}(\eta) \middle| \varphi_{\varepsilon^{(z)}}(\eta) \right\rangle = \delta\left(\varepsilon'^{(z)} - \varepsilon^{(z)}\right) \quad \textit{region I} .$$

Using these relations will enable one to directly work out the conclusions of applying the boundary conditions. This will be done next, since until now we only used the "quantizing" boundary conditions, formed by the wave guide.

First, the situation in region *I* can be simplified once more. Since the energy spectrum of all directions of motion is continuous in this region, the error will be small if we also use plane waves for the vertical motion. Thus, we will use a three dimensional plane wave of a given total energy *E(k)* as the initial state of a neutron sufficiently far away from the wave guide entrance:

$$(3.3.1.22)\quad |\psi\rangle_I = \left(e^{i \cdot k_x^I x} + R \cdot e^{-i \cdot k_x^I x}\right) \cdot e^{i \cdot k_y y} \cdot e^{i \cdot k_z^I z}$$

$$\textit{where}:\ E(k) = E_0 = \frac{\hbar^2}{2m} \cdot \left(\left(k_x^I\right)^2 + k_y^2 + \left(k_x^I\right)^2\right) .$$

The coefficient $R$ has been introduced to account for possible reflections from the wave guide system. The $y$-direction will again be omitted (see above). We already know from chapter one, that $k_z$ is limited to < 10 $^{cm}/_{s}$ , whereas $k_x$ is in the range of 10 $^{m}/_{s}$ . Thus $E(k) = E(k_x)$ .
The glass substrates of the mirrors and the absorber form infinitely high potential discontinuities. Therefore, the initial state eq. (3.3.1.22) must be continued into the wave guide such, that the vertically bound states inside maintain the conditions of continuous differentiability. Furthermore, there is direct total reflection from the glass substrates in regions *Ia* and *Ib*.
Let's first have a look at these regions. The wave field, that is reflected from two planes separated by a gap, can be deduced from Babinet's principle. This theorem states, that the reflected wave field is equivalent to the wave field, which passes a small reflecting piece of matter with the same shape as the gap between our glass surfaces:

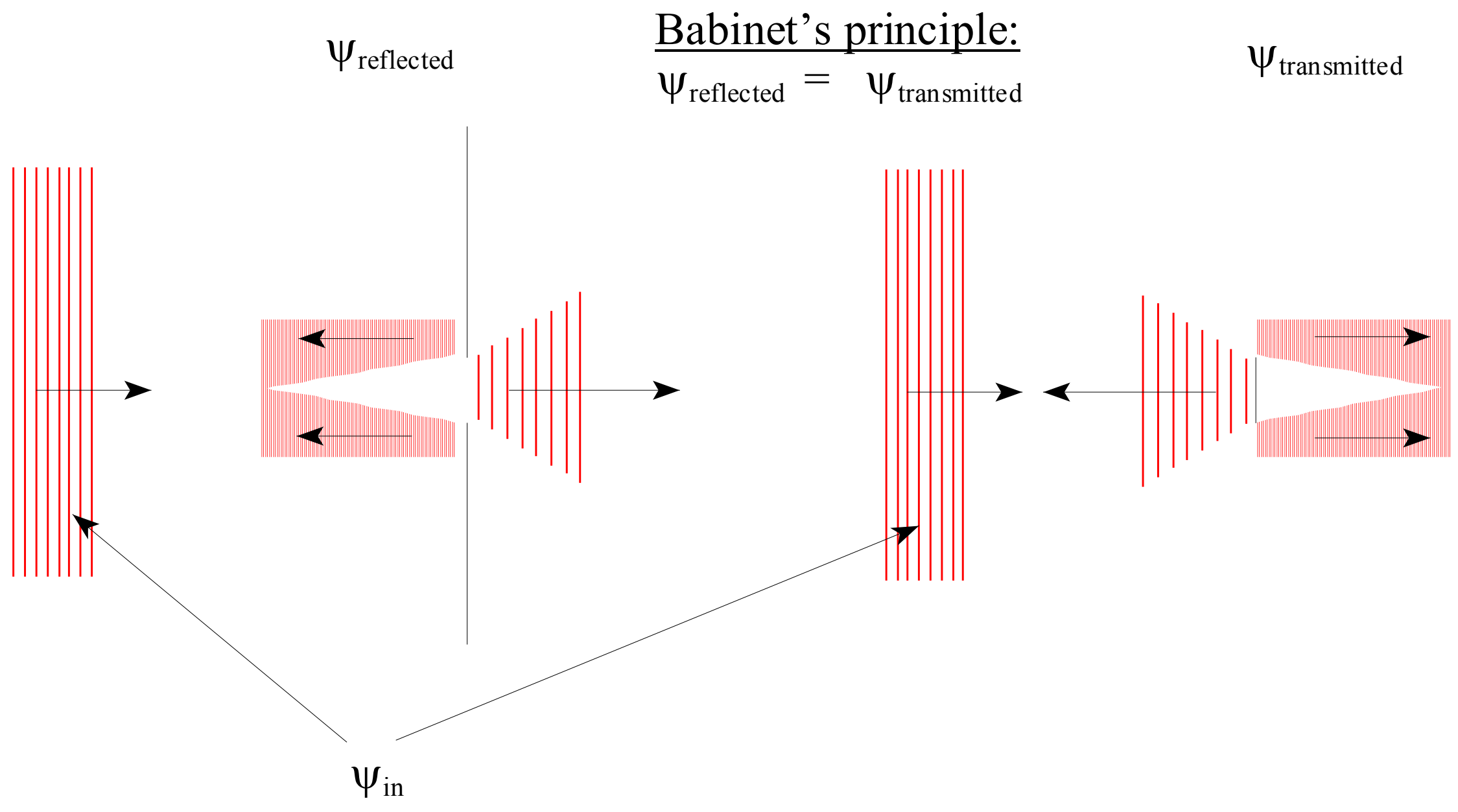

If we look at the complementary situation of a small reflecting rectangle, we can tell from the comparison of the neutron's wave length of about 50 Å and the wave guide's width of 10 to 200 μm, that the relative contribution of the diffracted wave field will be small even at about 1 mm behind the rectangle. According to [Somm], the diffracted wave field behind a semi-infinite plane can be written as the difference between two line integrals, $F(w)$ and $F(\infty)$ along the so-called Cornu spiral in the complex plane:

$$(Fresnel) \quad I_{diffr}(x) = \frac{I_0}{\sqrt{2}} \cdot \left[ F\left( -x \cdot \sqrt{\frac{2}{\lambda \cdot \Delta\rho}} \right) - F(\infty) \right]$$

$$\frac{I_{diffr}}{I_0} \approx 0.1 \Leftrightarrow -x \cdot \sqrt{\frac{2}{\lambda \cdot \Delta\rho}} \approx 1 \quad ,$$

where $\Delta\rho$ is the distance perpendicular to the semi-infinite plane, $x$ the position parallel to the plane and perpendicular to its edge ($x = 0$ denotes the boundary of the geometrical shadow), $\lambda$ the wave length, and $F(w)$ a so-called Fresnel integral:

$$F(w) = \int_0^w d\tau \cdot e^{i\pi\cdot\tau^2/2} .$$

If $\Delta\rho \approx 10 \cdot h_{absorber}$ is chosen, then one finds less than 10 % of the incident intensity being diffracted forth about 4 μm into the region of geometrical shadow even about 0.8 mm behind the rectangle for an absorber height $h_{absorber}$ = 80 μm. 4 μm, however, is small compared to most values of the absorber height.

For the experimental situation, this means, that there will only be small perturbations from the wave field, reflected from the glass substrates in region *Ia* and *Ib*, at distances closer than about 1 mm before the wave guide entrance. Using Babinet's theorem, it therefore is possible to omit these regions and to consider only the pure initial state eq. (3.3.1.22) in region *I*. The error made by this simplification will be small, as argued above.

For the moment we will omit the step between the two bottom mirrors. That leaves us with the problem of continuing eq. (3.3.1.22) from region *I* into region *II* and further into region *III*.

Since we have chosen the total energy of the states with (3.3.1.22), we have to write a general wave function in the regions *II* and *III* as a superposition of the vertical motion's eigenstates. Every eigenstate of the vertical motion is multiplied by a plane wave mode in $x$-direction in the regions *II* and *III* such, that the resulting $x$-$z$-state has the same total energy $E$ as the initial state eq. (3.3.1.22). The superposition of these $x$-$z$-states mentioned above is then one of modes of the same total energy $E$. Thus, the energy conservation of the neutrons, which fly through the apparatus, is preserved.

Following this argument, we write the general wave function in region *II* and *III* as follows:

$$\text{(3.3.1.23)} \quad \begin{cases} |\psi\rangle_{II} = \sum_j \left( A_j \cdot e^{i\cdot k_j^{(x)II}\cdot x} + B_j \cdot e^{-i\cdot k_j^{(x)II}\cdot x} \right) \cdot |\varphi_j^{II}(\eta)\rangle \quad , k_j^{(x)II} = \sqrt{\frac{2m}{\hbar^2}\cdot\left(E_0 - E_j^{(z)II}\right)} \\ |\psi\rangle_{III} = \sum_n T_n \cdot e^{i\cdot k_n^{(x)III}\cdot x} \cdot |\varphi_n^{III}(\eta)\rangle \quad , k_n^{(x)III} = \sqrt{\frac{2m}{\hbar^2}\cdot\left(E_0 - E_n^{(z)III}\right)} \end{cases} .$$

Continuing the initial state eq. (3.3.1.22) from *I* into *II* and *III* then requires:

$$(3.3.1.24)\quad \begin{cases} (a)\ \ |\psi\rangle_I\big|_{x=0} = |\psi\rangle_{II}\big|_{x=0} \ \wedge\ \dfrac{\partial}{\partial x}|\psi\rangle_I\big|_{x=0} = \dfrac{\partial}{\partial x}|\psi\rangle_{II}\big|_{x=0} & \forall z \in [0,h] \\ (b)\ \ |\psi\rangle_{II}\big|_{x=l} = |\psi\rangle_{III}\big|_{x=l} \ \wedge\ \dfrac{\partial}{\partial x}|\psi\rangle_{II}\big|_{x=l} = \dfrac{\partial}{\partial x}|\psi\rangle_{III}\big|_{x=l} & \forall z \in [0,h] \end{cases} .$$

This is a system of four equations for the four unknowns $R$, $A_j$ , $B_j$ and $T_n$ . The situation, that $A$, $B$ and $T$ are sets with indices running over the modes of region *II* and *III*, can be reduced to the problem of determining the quadruple ($R$, $A_j$ ,$B_j$ ,$T_n$) for each continuation $|\psi\rangle_{in} \to |j\rangle_{II} \to |n\rangle_{III}$ along the single modes $|j\rangle_{II}$ in region *II* and $|n\rangle_{III}$ in region *III*. This is possible, using the orthonormality relations of eq (3.3.1.21a) to break up the sums in eq. (3.3.1.23). Then one can use (3.3.1.21a) to express $A_j$ and $B_j$ in terms of R as well as $T_n$ in terms of $A_j$ and $B_j$ . Using these results – two equations in $R$ and $T_n$ – together it is possible to determine $T_n$ . The result is:

$$(3.3.1.25)\quad T_n = 4\cdot k_x^I \cdot \sum_j \frac{c_j^{II}(k_z^I)\cdot c_{nj}^{III}\cdot k_j^{(x)II}\cdot e^{i\cdot\left(k_j^{(x)II}-k_n^{(x)III}\right)\cdot l}}{e^{2i\cdot k_j^{(x)II}\cdot l}\cdot\left(k_x^I - k_j^{(x)II}\right)\cdot\left(k_j^{(x)II}-k_n^{(x)III}\right)+\left(k_x^I + k_j^{(x)II}\right)\cdot\left(k_j^{(x)II}+k_n^{(x)III}\right)}$$

$$where: \begin{cases} c_j^{II}(k_z^I) = \left\langle \varphi_j^{II}\left(\dfrac{z}{R}\right)\middle| k_z^I, z\right\rangle = \displaystyle\int dz\cdot\left[\varphi_j^{II}\left(\dfrac{z}{R}\right)\right]^*\cdot e^{i\cdot k_z^I\cdot z} \\ c_{nj}^{III} = \left\langle \varphi_n^{III}(\eta)\middle|\varphi_j^{II}(\eta)\right\rangle \end{cases} .$$

Now we shall look at the detector:

The $^3$He-proportional-counter counts neutrons with an overall efficiency of 60%. Detection of a total transmission of the wave guide structure with a beam operating in the continuous mode means, that the counter will measure the neutron flux, as long as the detection is independent of the energy. We have already seen in section 3.2), that for thermal neutrons the neutron absorption cross sections are generally velocity-dependent with a $v^{-1}$-characteristics. However, it was shown there, too, that in the regime of UCN energies this dependency ceases to hold. There is a saturation of the absorption cross sections at these energies. Therefore, the $^3$He-detector will directly measure the area-integral of the neutron flux density, which is the flux.

Therefore, the total probability current density of all modes in region *III* populated according to eq. (3.3.1.25) has to be calculated. Since the neutrons are moving essentially in the $x$-direction, and the detector averages over the $y$-direction, we will determine the component $j_x$ of current density, which is:

$$(3.3.1.26)\quad j_x^{III} = \frac{\hbar}{2i \cdot m} \cdot \left( \left|\psi\right\rangle_{III}^{*} \frac{\partial}{\partial x} \left|\psi\right\rangle_{III} - \left|\psi\right\rangle_{III} \frac{\partial}{\partial x} \left|\psi\right\rangle_{III}^{*} \right) = \frac{\hbar}{m} \cdot \Re\left( \left|\psi\right\rangle_{III}^{*} \frac{\partial}{\partial x} \left|\psi\right\rangle_{III} \right) .$$

Inserting eq. (3.3.1.23), we get for the total current density of region *III* in $x$-direction:

$$(3.3.1.27)\quad j_x^{III} = \frac{\hbar}{m} \cdot \left[ \sum_n \Re\left(k_n^{(x)III}\right) \cdot \left|\varphi_n^{III}\right|^2 \cdot \left|T_n\right|^2 + 2 \cdot \Re\left( \sum_{n<n'} k_n^{(x)III} \cdot \left(\varphi_{n'}^{III}\right)^{*} \varphi_n^{III} \cdot e^{i\cdot\left(k_n^{(x)III} - k_{n'}^{(x)III}\right)\cdot x} \cdot T_{n'}^{*} T_n \right) \right]$$

The total flux through the $y$-$z$-plane is then given by:

$$(3.3.1.28)\quad \Phi_x^{III}(k_x^I, k_z^I) = \int_0^b dy \int_0^{z_{window}} dz \cdot j_x^{III}(z) = b \cdot \frac{\hbar}{m} \cdot \sum_n k_n^{(x)III} \cdot \left|T_n\left(k_x^I, k_z^I\right)\right|^2 .$$

Here, $z_{window}$ gives the vertical width of the detector's entrance window and $b$ is the horizontally transverse width of the wave guide as well as of the mirrors in the $y$-direction. The dimension of (3.3.1.28) is [$s^{-1}$]. Yet, this flux is only correct, if neutrons described by a monochromatic initial state eq. (3.3.1.22) enter the apparatus. But in reality, a neutron spectrum enters the guide. Therefore, the question arises, whether the total flux over the whole spectrum must be calculated by coherently superposing all initial states eq. (3.3.1.22) or whether an incoherent average of eq. (3.3.1.28) over the $k_x$-$k_z$-spectrum is adequate:

Let's look at the entrance collimator of the setup again. It is fuelled by a UCN "gas" from the neutron guide. Therefore, it is a vastly incoherent neutron source, concerning the horizontal velocity component. The vertical direction is governed by the collimator width of about 2 mm (see chapter one). A broadening of the neutron beam in the vertical direction due to diffraction at the collimator, therefore gives a measure for the vertical coherence length, $L_z$ , of the neutron "radiation":

$$(3.3.1.29)\quad L_z \approx D_{coll.-wave\ guide} \cdot \Delta\vartheta_{diffr.} = \frac{D_{coll.-wave\ guide}}{k_z^I \cdot coll.width} \approx 0.5 \cdot \mu m$$

$$for \begin{cases} k_z^I \mathrel{\hat{=}} 10 \cdot \frac{cm}{s} \\ D_{coll.-wave\ guide} = 5 \cdot cm \\ coll.width = 2 \cdot mm \end{cases} .$$

The longitudinal coherence length $L_x$ is more difficult to determine. Yet, the neutron spectrum is very wide in $k_x$. Therefore, we will derive an estimate for $L_x$ the same way, as the coherence length of a vastly incoherent "white" light source is determined. The spectral width will provide us with the order of magnitude of $L_x$ . Thus we have:

$$(3.3.1.30)\quad L_x \approx \frac{1}{\Delta k_x^I} = \frac{\lambda_x^2}{2\pi \cdot \Delta\lambda_x} \approx 15 \cdot nm$$

$$.$$

$$for \quad \lambda_x \approx 50 \cdot \text{Å} \quad and \quad \Delta\lambda_x \approx 25 \cdot \text{Å}$$

Both the eq.s (3.3.1.29) and (3.3.1.30) show, that it is well justified to describe the entering UCN as an incoherent mixture of plane wave states eq. (3.3.1.22). Therefore, the total flux of neutrons averaged over the arriving spectrum via eq.(3.3.1.28) is given by:

$$(3.3.1.31)\quad \Phi_x^{III} = \int dk_z^I \cdot f_1(k_z^I) \cdot \int dk_x^I \cdot f_2(k_x^I) \cdot \Phi_x^{III}(k_x^I, k_z^I)$$

$$= b \cdot \frac{\hbar}{m} \cdot \sum_n k_n^{(x)III} \cdot \int dk_z^I \cdot f_1(k_z^I) \cdot \int dk_x^I \cdot f_2(k_x^I) \cdot \left| T_n\left(k_x^I, k_z^I\right)\right|^2 ,$$

where $f_1(k_z^I)$ and $f_2(k_z^I)$ are the form factors of the spectrum, which arrives at the wave guide's entrance.

The form factors: According to [Kl00] the deconvoluted velocity spectrum, that arrives at the collimator, can be described by the following empirical function:

$$\frac{dN(v)}{dv} = const. \cdot v^{n-2} \cdot \left( e^{\left. -k\frac{s}{v} - t_1 \right/ a_1} + 1 \right)^{-1} \cdot \left( e^{\left. -k\frac{s}{v} - t_2 \right/ a_2} + 1 \right)^{-2} ,$$

where $v$ denotes the neutron velocity, $s$ the length of the time-of-flight line, $t_1$, $t_2$, $a_1$, $a_2$, and $n$ are fitted parameters, and $k$ the time scale of the multi channel analyzer. It looks like:

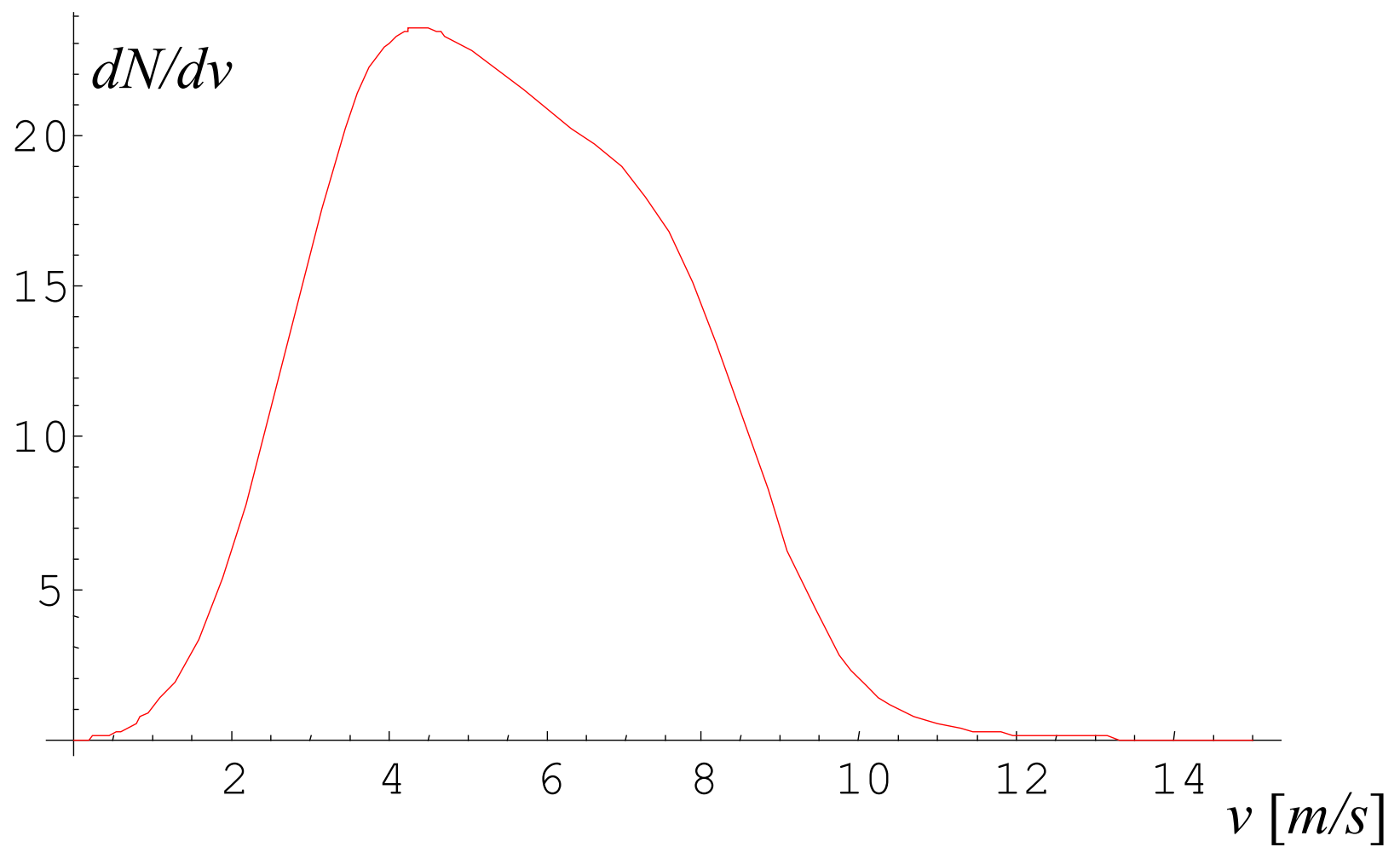

From chapter two, section two, one knows, that the collimator cuts velocities above 15 m/s in order to be able to sufficiently block the direct line of view. According to [Ne01] the measurements of the neutron transmission of the wave guide have been performed with an average horizontal velocity of about 8 m/s. Integrating the expression given above, one therefore finds, that the spectrum of the horizontal velocity provided by the collimator ranges from 7 m/s to 15 m/s. The spectral expression given above together with this velocity range given by the collimator thus provides the spectral form factor of the neutron beam, that arrives at the wave guide's entrance.

One should now consider the 5 μm vertical shift between the two bottom mirrors. What effect does this have? A look at eq. (3.3.1.24) and Fig. 3.5 shows, that the step introduces a third boundary. Two additional equations of the type of eq. (3.3.1.24) would emerge, introducing two new sets of constants, $C_l$ and $D_l$. Thus the expression replacing eq. (3.3.1.25) contains a double summation over $j$ and $l$. Its exponentials split into two partial lengths of $l_1$ = 6 cm and $l_2$ = 3 cm, whereas a third matrix of overlap integrals between the *IIa*- and the *IIb*-states multiplies the expression.
The overlap integrals usually have to be performed numerically. The double summation over $(20 \times 20)$ – matrices depends on the $k$-vector component of the arriving initial state eq. (3.3.1.22). Thus, the spectral average eq. (3.3.1.31) of the flux will be performed on sums, containing several ten thousand terms, which depend on the $k$-vector. Numerical integration of eq. (3.3.1.31) will therefore take impracticable large times. However, the numerical task re-mains managable, if we work with eq. (3.3.1.25) and try to incorporate the step without performing the whole formal apparatus shown above.

As mentioned above, the main effect of the step will be an additional summation over overlap integrals between states of region *IIa* and *IIb*. Therefore we should calculate the overlap integral matrix:

$$(3.3.1.32) \quad c_{lj}^{step} = \left\langle \varphi_l^{IIb} \middle| \varphi_j^{IIa} \right\rangle .$$

The states $l$ in region *IIb* will get populations, which are approximately given by a sum over $j$, which corresponds to the state $j$ arriving from the region IIa, of the element $lj$ of this transition matrix, which has first to be multiplied by the damping factor of this state $j$. The damping factor is an exponential, arising from the first 6 cm of the absorber length. Then the population of the state $l$ in the region *IIb* just after the step will be approximately:

$$(3.3.1.33) \quad P_l^{step} \approx \left| \sum_j c_{lj}^{step} \cdot e^{-\Im\left(k_j^{(x)IIa}\right)\cdot l_1} \right|^2 .$$

Once these quantities are calculated for all relevant wave guide widths, they can be inserted into eq. (3.3.1.25) modifying it to:

$$(3.3.1.25') \quad T_n = 4 \cdot k_x^I \cdot \sum_j \frac{\sqrt{P_j^{step}(h)} \cdot c_j^{II}(k_z^I) \cdot c_{nj}^{III} \cdot k_j^{(x)II} \cdot e^{i\cdot\left(k_j^{(x)II} - k_n^{(x)III}\right)\cdot l}}{e^{2i\cdot k_j^{(x)II}\cdot l} \cdot \left(k_x^I - k_j^{(x)II}\right)\cdot\left(k_j^{(x)II} - k_n^{(x)III}\right) + \left(k_x^I + k_j^{(x)II}\right)\cdot\left(k_j^{(x)II} + k_n^{(x)III}\right)} .$$

Thus, the presence of the step can be included in an approximate way with a reasonable amount of calculation.

The last thing, that can be done now without specific knowledge of the vertically bound states in region *II*, is an approximate determination of the coefficients $c_j^{II}(k_z^I)$ . These overlap integrals describe the coupling of the arriving initial states to the vertically bound states inside the wave guide. Since the initial population of the wave guide modes will have strong effects on the total transmitted flux, the discussion of these coefficients is somehow crucial. Yet, we see from eq. (3.3.1.31), that the measurement depends on the incoherent spectral average over the squared absolute values of $T_n$ . $|T_n|^2$ represents a double sum. In $T_n$ the $c_j^{II}(k_z^I)$, fortunately, depend only on $k_z^I$ . Therefore, the spectral average over $k_z^I$ of eq. (3.3.1.31) reduces to:

$$(3.3.1.34) \quad \overline{c}_{jj'}^{II} = \int dk_z^I \cdot f_1(k_z^I) \cdot \left(c_{j'}^{II}(k_z^I)\right)^* \cdot c_j^{II}(k_z^I) .$$

A measurement of the total transmitted flux does not directly depend on the $c_j^{II}(k_z^I)$ but on the averaged quantities eq. (3.3.1.34). These quantities depend on the specific form $f_1(k_z^I)$ of the vertical velocity spectrum. However, we know from chapter one, that this spectrum, with a range of 20 $^{cm}/_{s}$ corresponding to about 50 peV in energy units, is far wider than the energy eigenvalues of the lowest bound states of region *III*, for example. Thus, it makes sense to look at eq. (3.3.1.34) in the limit of an infinite wide and uniform spectrum of vertical velocities. Then, eq. (3.3.1.34) becomes:

$$(3.3.1.35)\quad \bar{c}^{II}_{jj'}(\infty) = \int dk^I_z \cdot \left(c^{II}_{j'}(k^I_z)\right)^* \cdot c^{II}_j(k^I_z) = \int dk^I_z \int dz \int dz' \cdot \left[\varphi^{II}_j\left(\frac{z}{R}\right)\right]^* \cdot \left[\varphi^{II}_{j'}\left(\frac{z'}{R}\right)\right] \cdot e^{i \cdot k^I_z \cdot (z-z')}$$

$$= 2\pi \int dz \int dz' \cdot \left[\varphi^{II}_j\left(\frac{z}{R}\right)\right]^* \cdot \left[\varphi^{II}_{j'}\left(\frac{z'}{R}\right)\right] \cdot \delta(z-z') = 2\pi \left\langle \varphi^{II}_j \middle| \varphi^{II}_{j'} \right\rangle = 2\pi \cdot \delta_{jj'} \quad .$$

In the case of such an infinitely wide uniform vertical velocity spectrum, the coupling of the initial plane waves to the wave guide modes lead therefore to a completely uniform initial population of the wave guide modes, regardless of their specific form.

As argued above, the real spectrum of the experiment is not infinitely. Yet, it is considerably wide compared to the bound state energy eigenvalues. Therefore, we should expect a nearly uniform initial population of the wave guide modes, which may additionally show some dependence on the wave guide width *h*.

Some overlap integrals in eq.s (3.3.1.25’) and (3.3.1.33) remain to be calculated. Then, with eq. (3.3.1.31), a prediction for the measurement of the total transmitted neutron flux as a function of the wave guide width *h* can be made.

Thus, it is now necessary to determine the precise form of the eigenstates and the energy eigenvalues of the vertical motion in region *II*. We see from the potential eq. (3.3.1.13’), that in this region the potential between the mirror and the glass substrate of the absorber is a sum of a linear term and two complex-valued gaussian error functions . Unfortunately, there is no analytic solution to the corresponding ordinary differential equation. Therefore the eigenstates can only be determined to some approximation.

One possibility to deal with this problem, is the basic behaviour of the solutions in region *III*. They vanish exponentially for large *z*. If the absorber is higher than the main spatial extension of a certain state of region *III*, this state will only feel a tiny fraction of the absorber. Therefore, for sufficiently large absorber heights, the lowest states in region *II* should be virtually unchanged compared to region *III*. In this case, the effects of the absorber’s Fermi potential are exponentially suppressed due to the general shape of the states themselves.

This suppression can be used to argue as follows:
Calculate the states in region *II*, which are valid for an absent absorber, but still two glass mirrors below and above. The eigenstates of such a gravitational Fabry-Pérot interferometer are given analytically as linear combinations of the two types of Airy functions. The energy eigenvalues arise from the boundary conditions at the mirror surfaces. Determine for each such state the height of the upper mirror, from where the state's wave function is decreasing exponentially towards the upper mirror. Then, the exponentially suppressed Fermi potential of an absorptive coating of the upper glass allows the calculation of its effects on the eigenvalues, using stationary perturbation theory.

The first step of this procedure determines so-called "two-mirror eigenstates" (one mirror above, one below). The boundary conditions of such a structure are:

$$\text{(3.3.1.36)} \quad \begin{cases} \varphi_{II}(0) = 0 \\ \varphi_{II}\left(\dfrac{h}{R}\right) = 0 \end{cases} .$$

Inserting the general solution eq. (3.3.1.17) of region *II* into eq. (3.3.1.36) gives two equations. The combination of these determines the multiplicative constants of eq. (3.3.1.17) and provides an equation to calculate the energy eigenvalues by numerically searching its zeros:

$$\text{(3.3.1.37)} \quad \phi_A(-\varepsilon)\cdot\phi_B\left(\frac{h}{R}-\varepsilon\right)-\phi_A\left(\frac{h}{R}-\varepsilon\right)\cdot\phi_B(-\varepsilon)=0 .$$

The solutions give the energy eigenvalues $\varepsilon_n^{(z)\,free,II}$ of the wave guide problem with no absorber. For each width $h$ there exists an infinite but countable set of positive, real energy eigenvalues, thus, the $n^{\text{th}}$ eigenvalue $\varepsilon_n$ is a function of the width $h$, $\varepsilon_n^{(z)}(h)$. This clearly shows, that solutions are valid only as long as $\varepsilon_n$ << 100 neV, because the glass' Fermi-potential is finite in reality. Therefore, the number of bound states is large, but finite. The combination of eq.s (3.3.1.37) and (3.3.1.36) allows one to determine the solutions eq. (3.3.1.17) for the two-mirror wave guide once the energy eigenvalues are known:

$$\text{(3.3.1.38)} \quad \phi_n(\eta)=\phi_A\left(\eta-\varepsilon_n^{(z)\,free,II}\right)-\frac{\phi_A(-\varepsilon)}{\phi_B(-\varepsilon)}\cdot\phi_B\left(\eta-\varepsilon_n^{(z)\,free,II}\right).$$

The following plots show the variation of the free energy eigenvalues with the height *h*.

<u>First 7 energy eigenvalues:</u>

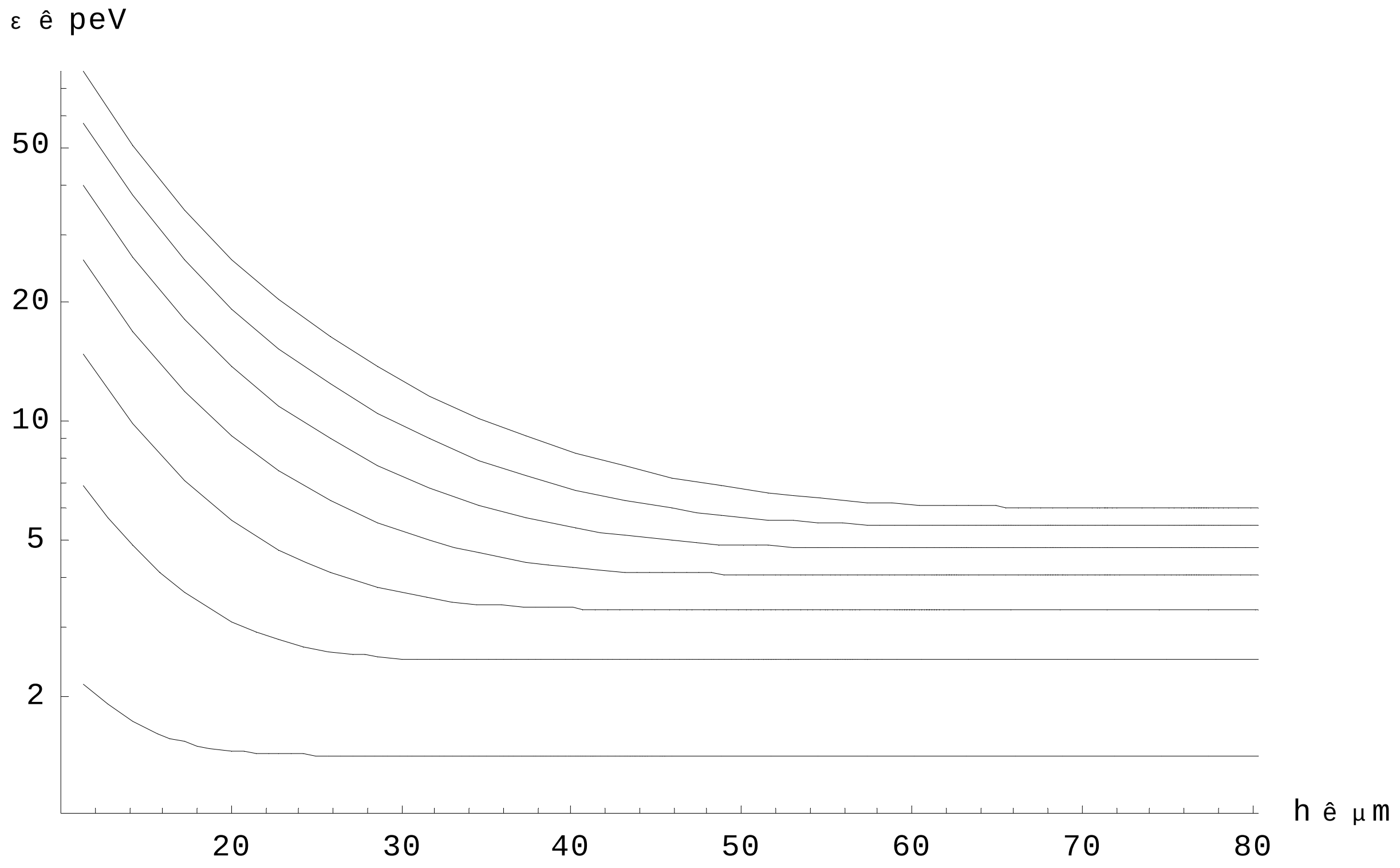

The term “free energy eigenvalues” we will assign to the set of eigenvalues, $\varepsilon_n^{(z)\,free,II}$ , of vertical motion between two mirrors in absence of any absorber.

It is clearly visible, that for small widths, the energies are increasing nearly quadratically with the state number – which is the characteristic of a field free box – and that for large widths, the gravitational potential dominates, generating fixed energy eigenvalues independent of the width *h*. This behaviour would be expected as an overall systematics of the bound states.

The second step deals with the absorbing layer through perturbation theory. The absorber’s Fermi potential corresponds to the gaussian error functions in eq. (3.3.1.13’). Fig. 3.4 clearly shows, that the layer’s potential is essentially small compared to the glass’ potential. Furthermore, the bound states are predominated by the linearly increasing gravitational potential for increasing wave guide width *h*. For sufficiently large widths, the states do not feel the upper substrate any longer. This happens for the ground state above h ~ 15 μm first, and then subsequently for all higher states at greater widths *h*. Because the absorbing layer’s potential has a spatial extension, that is small compared to 15 µm, the former statement made for the even stronger glass substrate's potential is also true for the layer. Thus, the layer's potential is small compared to the gravitational potential, if the wave guide width $h$ is sufficiently large (which means larger than ~ 15 µm for the ground state and higher values for higher states). The usage of perturbation theory is therefore justified for wave guide widths, which are not too small.
The perturbation Hamiltonian $H_I$ is given by:

$$(3.3.1.39)\quad H_I = \begin{cases} 0\,,\ z \le 0 \wedge z \ge h \\ \partial M_{alloy}(z)\cdot U_{Fermi}^{alloy}\ ,\ 0 < z < h \end{cases} .$$

If the normalized unperturbed bound states are given according to eq. (3.3.1.38)

$$(3.3.1.40)\quad \left|n^{(0)}\right\rangle = \varphi_n^{II}(\eta) = C_n \cdot \Phi_n(\eta) \ ,$$

then the first order correction on the energy eigenvalues is:

$$(3.3.1.41)\quad \left(\varepsilon_n^{(z)II}\right)^{(1)} = \frac{1}{E_R}\cdot\left\langle n^{(0)}\right|H_I\left|n^{(0)}\right\rangle = \frac{1}{E_R}\cdot\int_0^{h/R}\left|\varphi_n^{II}(\eta)\right|^2 H_I\cdot d\eta \ ,$$

the second order correction is:

$$(3.3.1.42)\quad \left(\varepsilon_n^{(z)II}\right)^{(2)} = \frac{1}{E_R^2}\cdot\sum_{m\neq n}\frac{\left|\left\langle m^{(0)}\right|H_I\left|n^{(0)}\right\rangle\right|^2}{\varepsilon_n^{(z)free,II} - \varepsilon_m^{(z)free,II}} \ ,$$

and the third order correction is:

$$(3.3.1.43)\quad \left(\varepsilon_n^{(z)II}\right)^{(3)} = \frac{1}{E_R^3}\cdot\sum_{\substack{m\neq n\\ j\neq n}}\frac{\left\langle j^{(0)}\right|H_I\left|n^{(0)}\right\rangle\left\langle m^{(0)}\right|H_I\left|j^{(0)}\right\rangle\left\langle n^{(0)}\right|H_I\left|m^{(0)}\right\rangle}{\left(\varepsilon_n^{(z)free,II} - \varepsilon_m^{(z)free,II}\right)\cdot\left(\varepsilon_n^{(z)free,II} - \varepsilon_j^{(z)free,II}\right)}$$
$$-\frac{1}{E_R^3}\cdot\varepsilon_n^{(1)}\cdot\sum_{m\neq n}\frac{\left|\left\langle m^{(0)}\right|H_I\left|n^{(0)}\right\rangle\right|^2}{\left(\varepsilon_n^{(z)free,II} - \varepsilon_m^{(z)free,II}\right)^2} .$$

As long as the condition of perturbation theory are fulfilled – i.e. the absorber is high enough for a given state – the first order contribution of eq. (3.3.1.41) will be dominate. It is proportional to $H_I$ , which itself is of the order of $U_{Fermi}{}^{alloy}$. Therefore, it holds for the energy eigenvalues in the presence of the absorber, that obey:

$$(3.3.1.44) \quad \begin{cases} \Re\left(\varepsilon_n^{(z)^{II}}\right) > \varepsilon_n^{(z)^{free,II}} \\ \Im\left(\varepsilon_n^{(z)^{II}}\right) < 0 \end{cases} .$$

This result of the perturbation theory is quite general, because it connects the deviation of the energy eigenvalues from the absorber-free case with the complex amplitude of the absorber's Fermi pseudopotential. It can be used to check other methods of determining the eigenstates in the presence of the absorber qualitatively.

However, the numerical results from the perturbation theory in first order overestimates the corrections on the eigenvalues. Furthermore, the convergence of the perturbation series becomes slow, if the transmission of a certain state is considerably reduced, since the absorber's influence on the state is then no longer weak, which invalidates the use of perturbation theory.

Yet, with the eigenstates eq. (3.3.1.38) it is now possible to determine the quantities of eq. (3.3.1.33) as functions of the step height, which enables one to calculate eq. (3.3.1.25') and therefore finally, the total transmitted flux eq. (3.3.1.31) as a function of the wave guide width *h*. The reason for that is, that for wave guide widths h larger than 10 µm the true eigenstates will only deviate slightly from the states eq. (3.3.1.38), since then the absorber regions covers less than 10 % of the whole wave guide width. To illuminate the behaviour of eq. (3.3.1.33), the quantities eq. (3.3.1.33) were evaluated for the second and the first bound state in region *II* as function of the wave guide width *h*. The step was chosen to be 5 µm [Ru00, p. 50].

The next plot shows the ratio of these quantities evaluated for the first state to the same quantities evaluated for the second state:

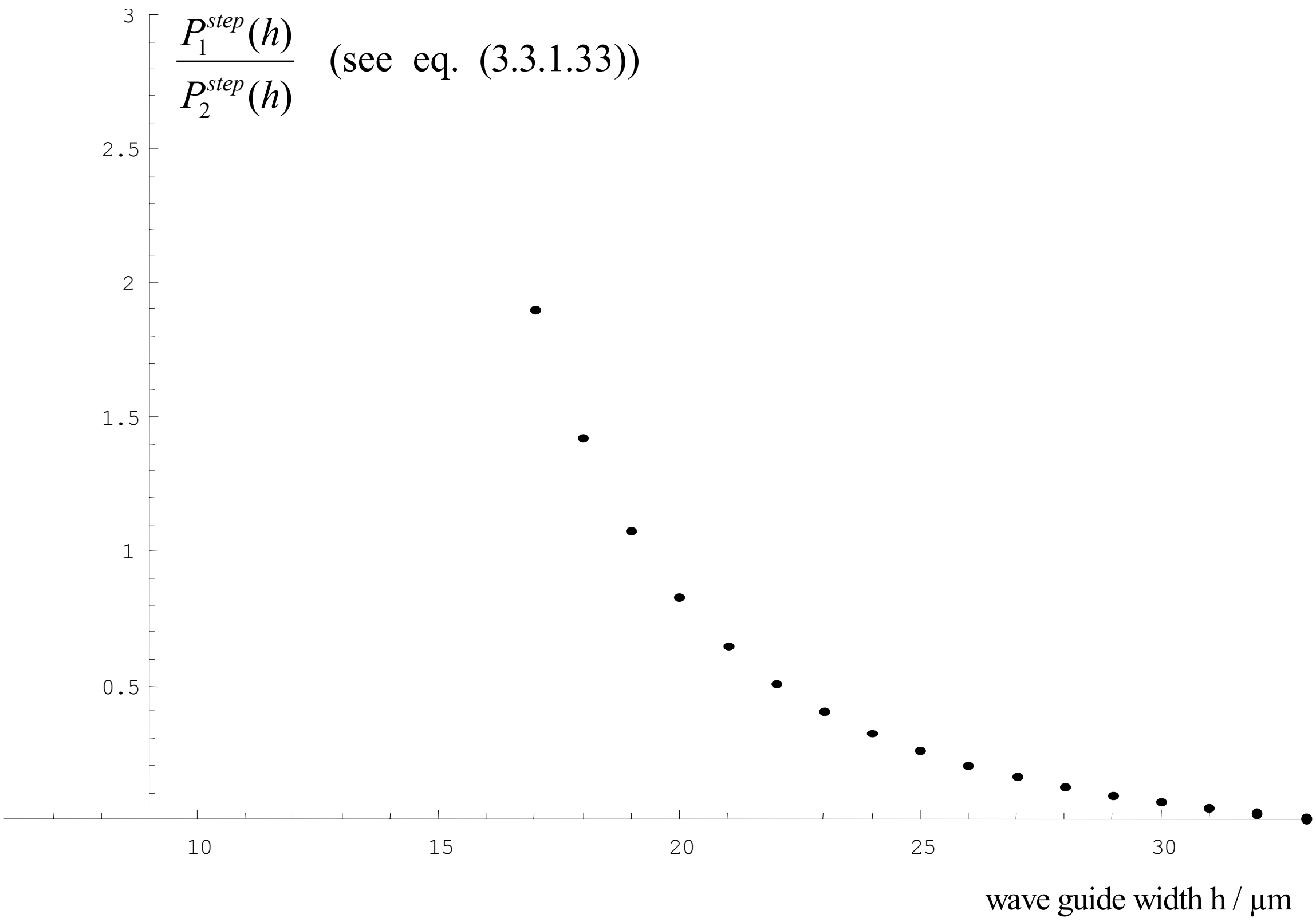

We see, that above a wave guide width of about 20 µm, where both the first and the second state arrive virtually undamped by the absorber at the step, the population of the ground state is suppressed compared to the second state, since the ratio, which is shown in the plot, is below one. The mean value of the ground state's population for $h$ > 25 µm is around 15% to 25% of the population of the second state. This can be understood, since the overlap integrals between the first two states before and after the step have significantly large zero parts due to the step. The effect is particulary strong for wave guide widths $h$ of below 25 µm, which is comparable to the step height. If the wave guide width is large compared to the step, then it will work more like a small perturbation. Therefore, it will be necessary to include eq. (3.3.1.33) in eq. (3.3.1.25'), at least by assigning a population $P_1^{step} = 0.2\ P_j^{step}$ for all $j > 1$ to the ground state – the step leads to an effective suppression of the ground state!

### 3.3.2) Eigenvalues

It is now the time to search for a method, that is capable to calculate the energy eigenvalues of the vertical motion inside the wave guide with a certain reliability. I will develop here two approaches, that can be used to derive approximations to the true eigenvalues of the vertical motion.

The first one is the well-known WKB method. WKB quantization of the motion leads to the Bohr-Sommerfeld quantization rules for the energy eigenvalues. It is known, that, in general, WKB quantization usually gives the energy eigenvalues of the lowest bound states only to an accuracy of about 20 %, whereas the quality increases for higher levels. Therefore, most often WKB quantization is applied to the semi-classical behaviour of system.
However, one should first take a look at WKB quantization of the eigenvalue problem given by eq.s (3.3.1.14b) and (3.3.1.16), which form the basic structure of the problem of the gravitationally bound states. There, however, is a problem. The WKB method demands, that the wave function can penetrate into all regions of evanescent wave propagation. Yet, the bottom mirror forms an infinitely high and sharp potential step. The WKB method cannot deal with such potential steps. Fortunately, one can avoid this problem by modifying the potential

$$U(z) = \begin{cases} m \cdot g \cdot z\,, & z \geq 0 \\ \infty\,, & z < 0 \end{cases}$$

towards

$$U(z) = m \cdot g \cdot |z| \quad .$$

Then there are no potential steps, and the classical motion of a particle in this potential has symmetrical turning points. However, the condition eq. (3.3.1.14b), which must be satisfied from the WKB solution as well, tells one, that only solutions with odd parity, that have a zero at z = 0, can be used to construct the eigenvalues, since these solutions are automatically solutions to the boundary value problem eq. (3.3.1.14b).

The WKB quantization rule for a one-dimensional motion is derived from the condition, that the phase shift $\Delta\phi$ along a closed integration contour $a \to b \to a$ must be equal to $2\pi \cdot (n + 1/2)$. Therefore one has:

$$\begin{aligned}
\Delta\phi &= \oint k(z) \cdot dz = 2 \cdot \int_a^b dz \cdot k(z) \\
&= 4 \cdot \int_0^{E/m \cdot g} dz \cdot \sqrt{\frac{2m}{\hbar^2}(E - m \cdot g \cdot z)} \\
&= 4 \cdot \sqrt{\frac{2m^2 g}{\hbar^2}} \cdot \int_0^{E/m \cdot g} dz \cdot \sqrt{\frac{E}{m \cdot g} - z} \\
&= \frac{8}{3} \cdot \sqrt{\frac{2}{\hbar^2 m \cdot g^2}} \cdot E^{3/2} \overset{!}{=} 2\pi \cdot (n + 1/2)
\end{aligned}$$

$$\Rightarrow \quad E_n = E_R \cdot \left[ \frac{3\pi}{4} \cdot \left( n + \frac{1}{2} \right) \right]^{2/3} \quad .$$

The WKB states are given by:

$$\psi_n^{WKB}(z) \propto \cos\left( \int_a^z dz' \cdot k(z') - \frac{\pi}{4} \right) .$$

Thus, one has to satisfy eq. (3.3.1.14b) at z = 0:

$$\psi_n^{WKB}(0) \propto \cos\left( \int_a^0 dz' \cdot k(z') - \frac{\pi}{4} \right) = \cos\left[ \frac{\pi}{2} \cdot (n + 1/2) - \frac{\pi}{4} \right] = \cos\left[ \frac{\pi}{2} \cdot n \right] \overset{!}{=} 0$$

$$\Rightarrow \quad n = 2 \cdot s + 1 , \quad s \in \text{Í} \quad .$$

This gives the energy eigenvalues of the boundary value problem eq.s (3.3.1.16) and (3.3.1.14b) in WKB quantization:

$$E_n = E_R \cdot \left[ \frac{3\pi}{2} \cdot \left( n + \frac{3}{4} \right) \right]^{2/3} \quad , \quad n \in \text{Í} \quad ,$$

where *s* has been renamed to *n*.

Now one can calculate these eigenvalues and compare them with the exact values found from the exact solution of eq.s (3.3.1.16) and (3.3.1.14b), which is given by Airy functions.

| n | $E_n^{(z)\ exact}$ / $10^{-12}$ eV | $E_n^{(z)\ WKB}$ / $10^{-12}$ eV |
|---|---|---|
| 1 | 1.407 | 1.396 |
| 2 | 2.460 | 2.456 |
| 3 | 3.322 | 3.319 |
| 4 | 4.083 | 4.082 |

The agreement is better than 1 % even for the ground state! Therefore, this is the reason, why WKB quantization of the vertical motion inside the waveguide is an appropiate way to calculate the energy eigenvalues of the motion to an acceptable accuracy:

For all wave guide widths h larger than 10 μm the absorber covers less than 10 % of the whole wave guide width. WKB quantization of the motion inside an error function potential, that describes the absorber, certainly does not yield the lowest states of the motion there with an accuracy of 1 % or better.

Yet, if the linear gravitational potential covers more than 90 % of the whole wave guide width, where the motion is given by the WKB method to an accuracy better than 1 %, the presence of the absorber will cause an relative error of the WKB eigenvalues compared to the true eigenvalues, that is smaller than 10 %.

Thus, it makes sense to approximate the eigenvalues of the vertical motion inside the wave guide in presence of the absorber with WKB quantitzation of the motion, if one bears in mind, that these approximation resembles the true eigenvalues with a relative accuracy of about 10 % only for wave guide widths larger than 10 μm.

For this purpose one has to take the potential of region *II* inside the wave guide from eq. (3.3.1.7), and to symmetrize it as this was done in the simple case shown above.

$$(3.3.1.7') \quad U(z)=\begin{cases} m\cdot g\cdot |z| + \partial M_{alloy}(|z|)\cdot U_{Fermi}^{alloy}, & 0<z<h \wedge 0\le x\le l_{abs} \\ U_{Fermi}^{glass}, & |z|\ge h \wedge 0\le x\le l_{abs} \end{cases}.$$

Then one has to apply to the WKB quantization rule and to select out all states of odd parity to ensure a wave function, that vanishes (nearly) at z = 0, where the bottom mirror is. However, one has to obey, that only the real part of the line integral $\Delta\phi = \Re\left(\oint k(z)\cdot dz\right) = 2\cdot\Re\left(\int_a^b dz\cdot k(z)\right) \stackrel{!}{=} 2\pi\cdot(n+1/2)$ has to be matched in the phase condition, since the imaginary part describes the decline of the state, and therefore must not contribute to the phase matching.

Using these eigenvalues, it is now possible to calculate the total flux eq. (3.3.1.31), using eq. (3.3.1.25'). The result indeed shows the existence of a range of wave guide widths, where no neutron can pass the guide. This "range of non-penetration" should depend on both the magnitude of the absorber's Fermi potential, particularly its real part, and the roughness $\sigma$ of the absorber.

Therefore, the next four figures show the systematic dependency of the range of non-penetration on these quantities:

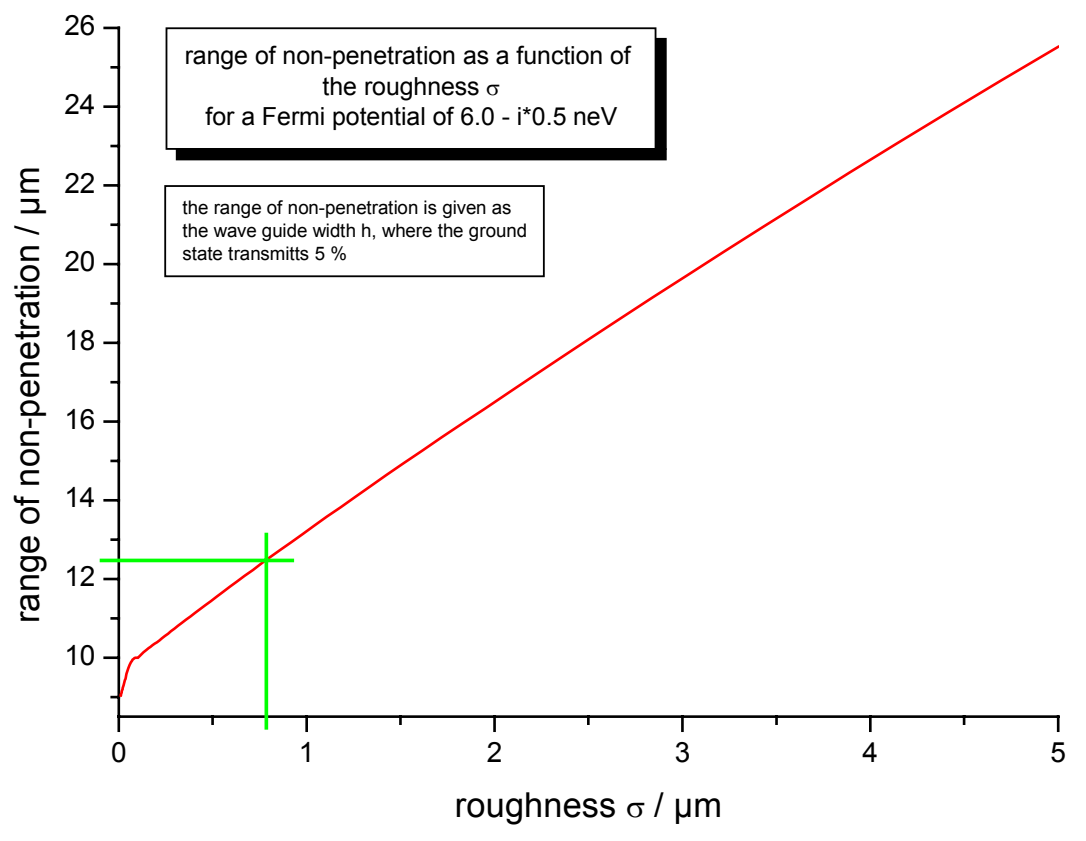

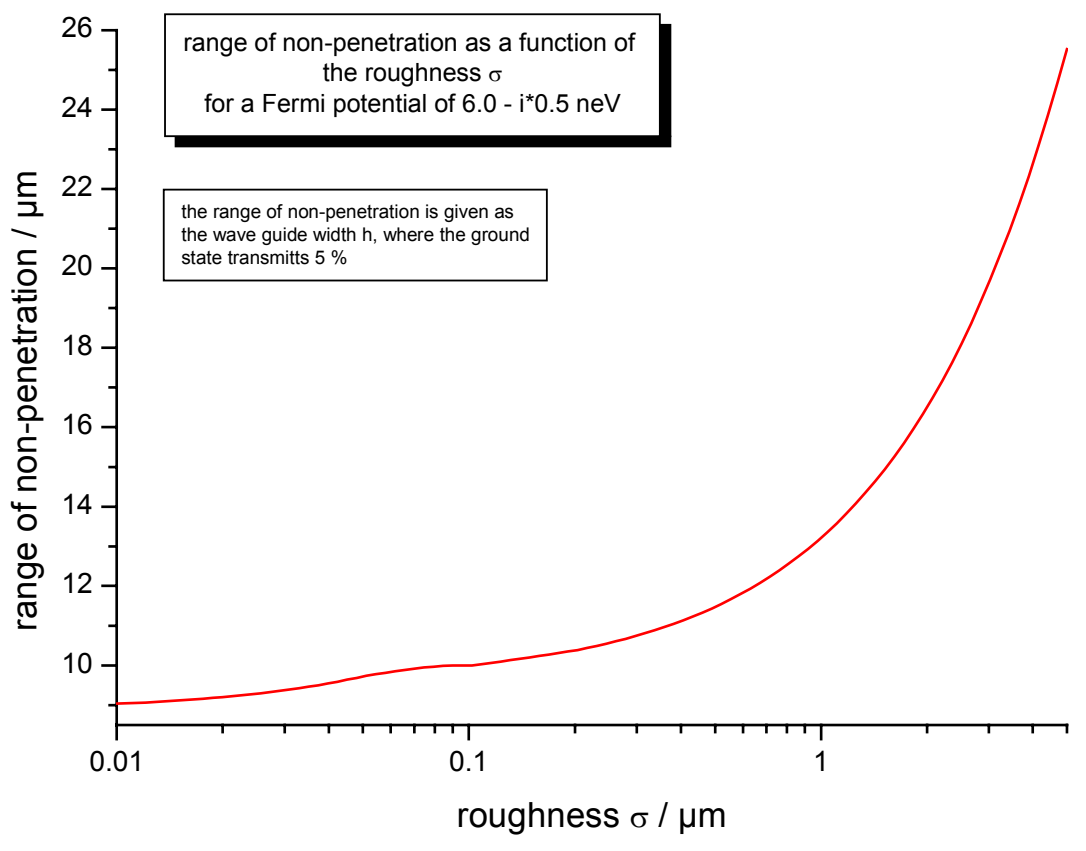

Figures: It is shown the dependency of the "range of non-penetration" in µm on the magnitude of the roughness of the absorber, $\sigma$, in µm.
The left figure shows this dependency in a linear plot, whereas the right one displays the same in a linear-log view. The green lines indicate the prediction for the measured roughness.

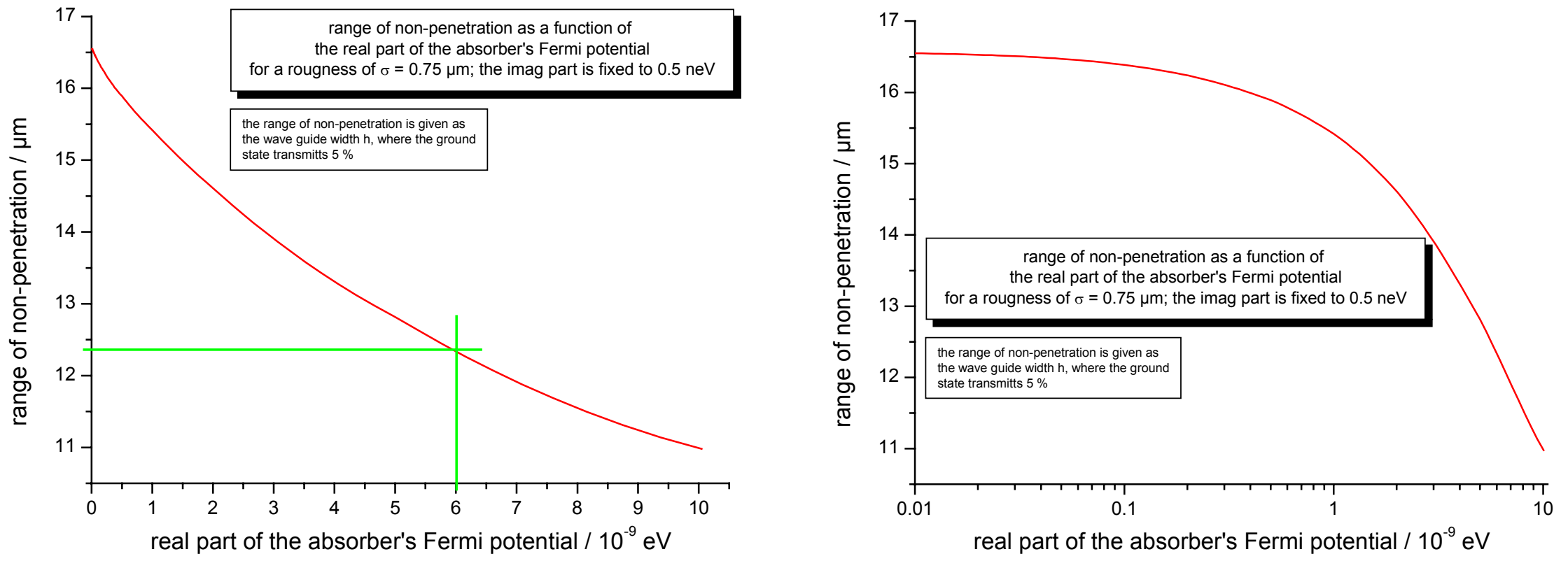

Figures: It is shown the dependency of the "range of non-penetration" in μm on the magnitude of the real part of the absorber's Fermi pseudopoential in $10^{-9}$ eV.
The left figure shows this dependency in a linear plot, whereas the right one displays the same in a linear-log view. The green lines indicate the prediction from the potential of eq. (3.1.21).

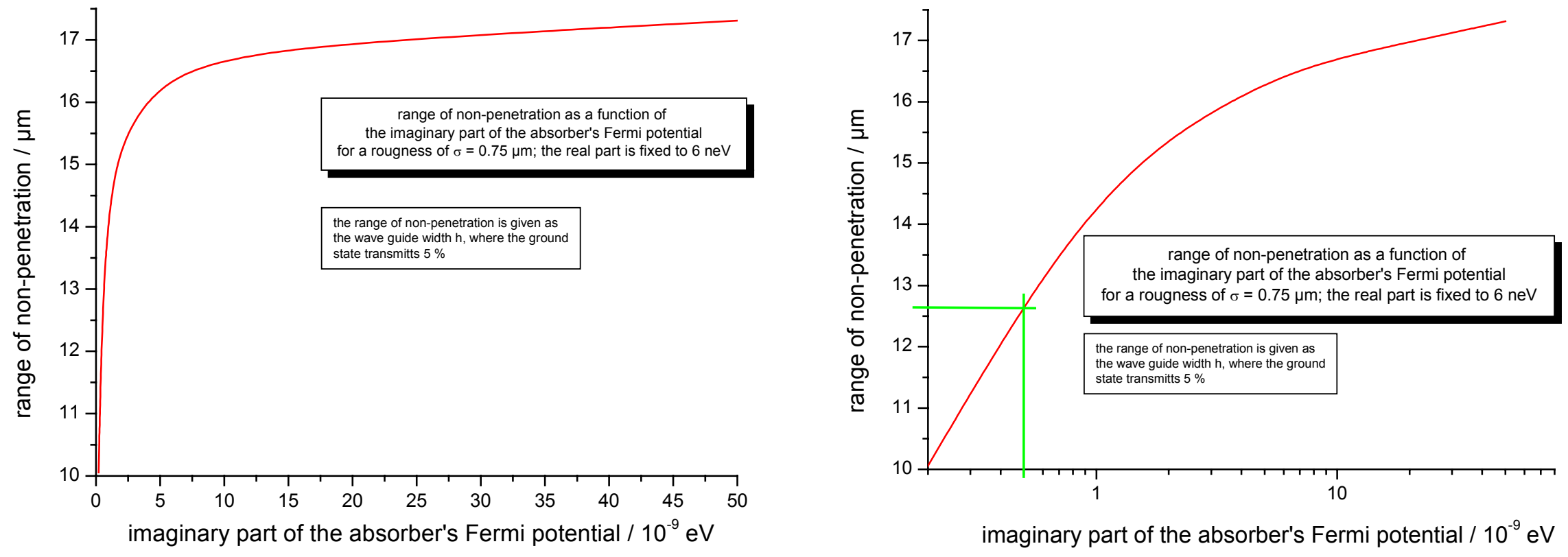

Figures: It is shown the dependency of the "range of non-penetration" in μm on the magnitude of the imaginary part of the absorber's Fermi pseudopoential in $10^{-9}$ eV.
The left figure shows this dependency in a linear plot, whereas the right one displays the same in a linear-log view. The green lines indicate the prediction from the potential of eq. (3.1.21).

One can now use the actual values of all quantities in the model:

$$
\begin{aligned}
&absorber: && U_{abs.}^{Fermi} = (6.0 - i \cdot 0.44) \cdot 10^{-9} \, eV \\
& && \sigma = 0.75 \ \mu m \quad (roughness) \\
&populations: && P_n = \begin{cases} 0.2 \, , \, n = 1 \\ 1 \, , \, n = 2..13 \end{cases} \qquad (shifted\ mirrors: 5 \mu m) \\
&wave\ guide\ length: && l = 0.09 \ m
\end{aligned}
$$

and try to derive a prediction of the total neutron transmission of the actual wave guide as a function of its width $h$ from eq.s (3.3.1.31) and (3.3.1.25'). Such a prediction looks like:

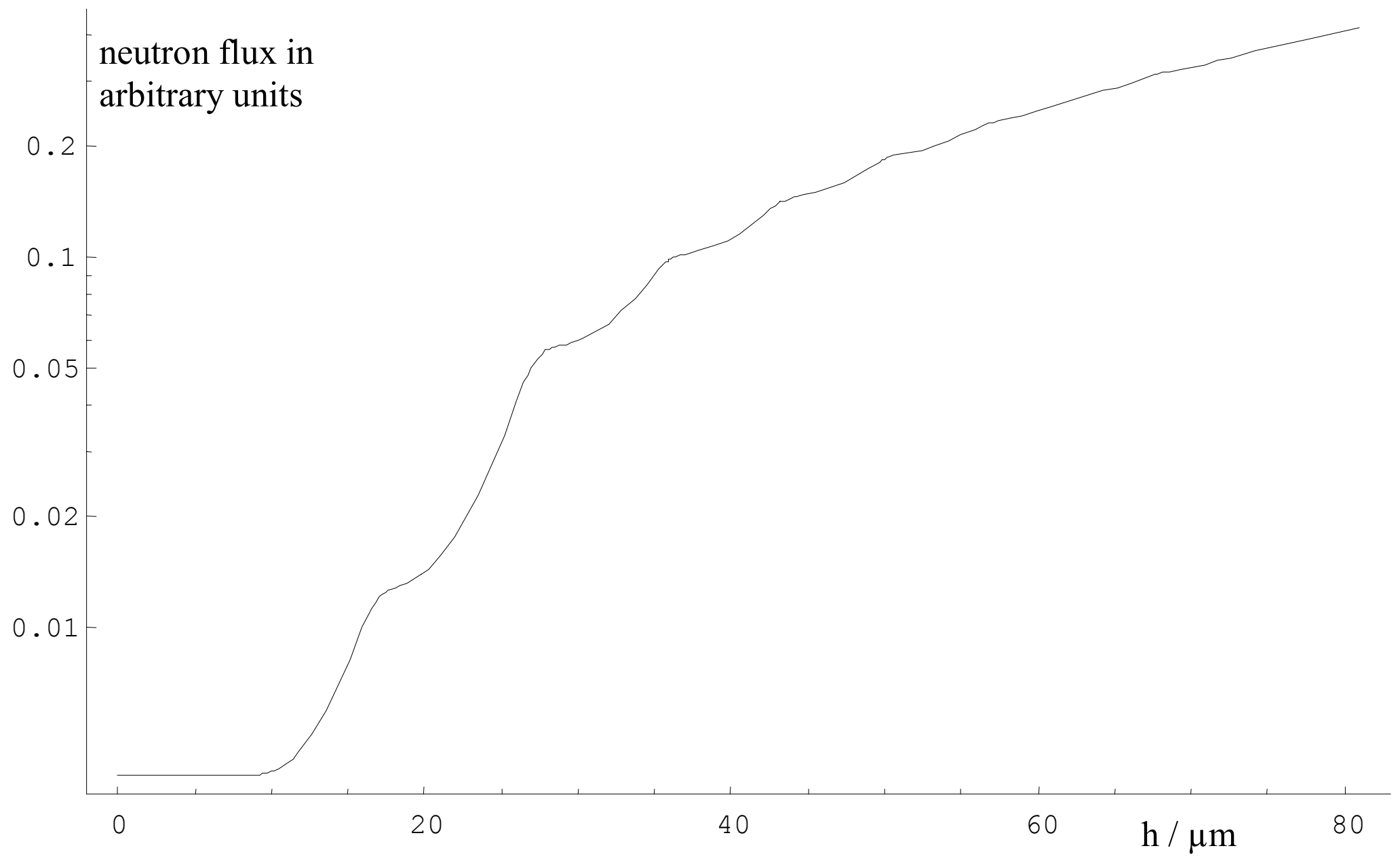

This theoretical prediction has to be treated for the finite spatial resolution of the actual measurement, which is reported to be 1 μm in [Ru00]. One can account for this fact by a convolution of the theoretical prediction with an area-normalized gaussian of $\sigma$ = 1 μm. The final prediction then, after convolution, looks like:

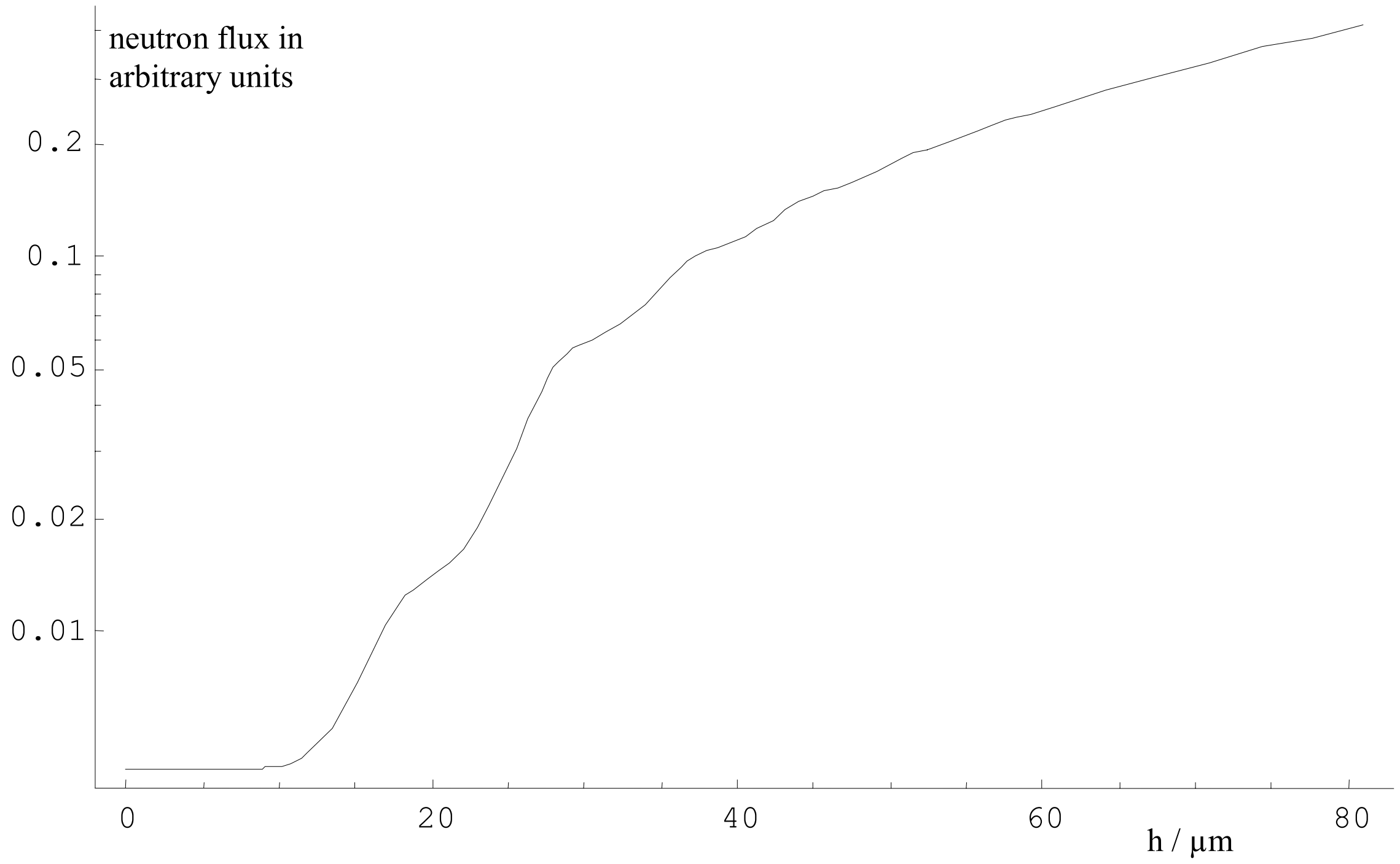

As one can see, the convolution smears out to some amount the “knees” of the curve.

This final prediction can now be fitted to the experimental data. There, one can test two ways of fitting:

a) One can use the background flux measured independently from the data in question here. This leads to a fixed offset parameter, and leaves two unknown quantities, that have to be obtained from the fit:

- the scaling of the total transmitted neutron flux, denoted as *K*,

  *and*

- an offset of the *z*-position, $z_{offset}$, which is constrained to about $\pm 1 \mu m$, since the spatial error of 1 µm (see above) also smears the zero position.

b) The background flux can vary in certain constraints. This leads to three unknown quantities, that have to be obtained from the fit:

- the scaling of the total transmitted neutron flux, denoted as *K*,
- an offset of the *z*-position, $z_{offset}$, which is constrained to about $\pm 1 \mu m$, since the spatial error of 1 µm (see above) also smears the zero position,

  *and*

- the background neutron flux, denoted as $y_0$ .

The results of these fits are shown in Fig.s 3.6a and 3.6b:

Fig. 3.6a

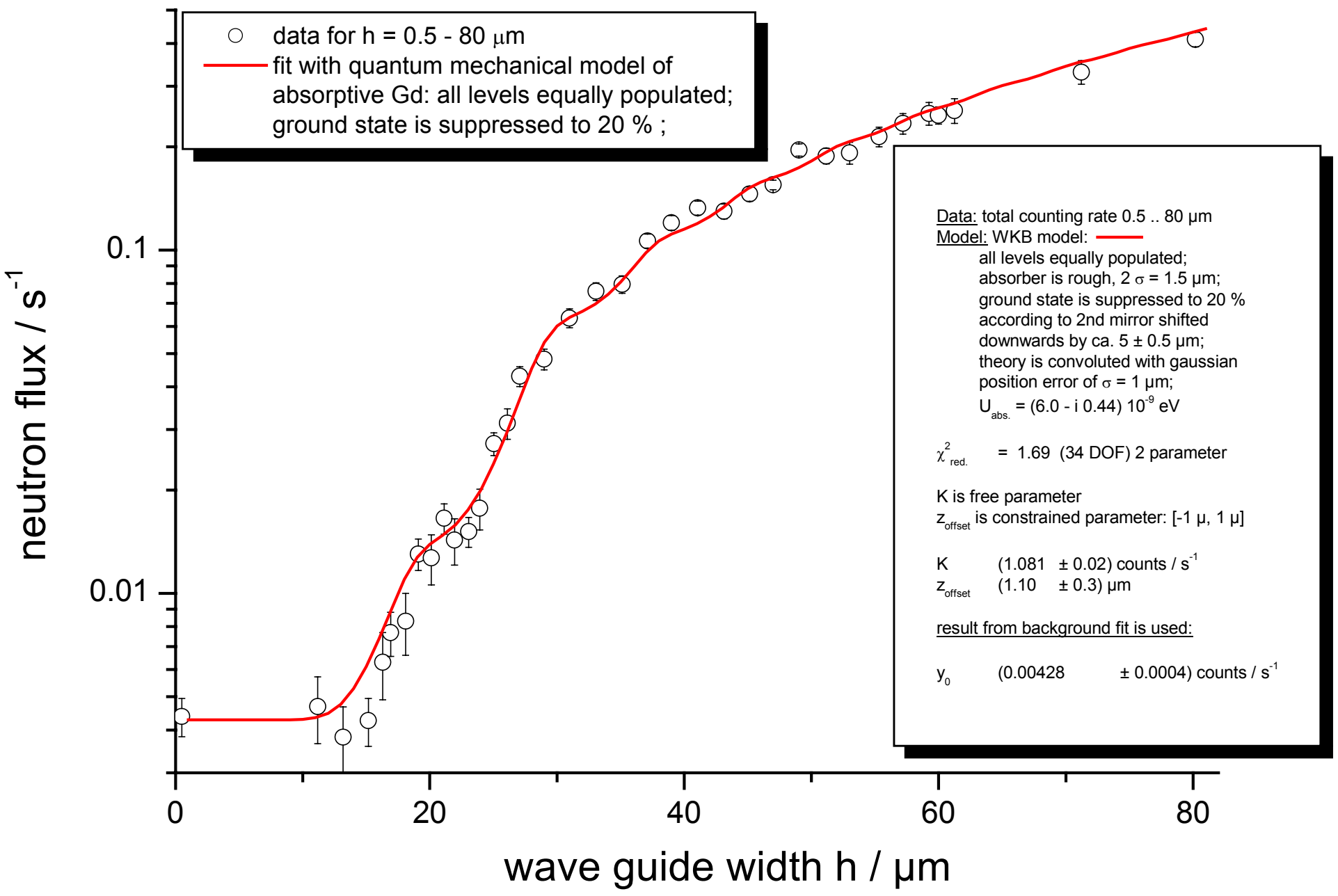

Fig. 3.6a: The theoretical prediction derived from eq.s (3.3.1.31) and (3.3.1.25'), where the eigenvalues of WKB quantization have been used, after convolution with the experimental spatial error of 1 μm is fitted to the data. Total scaling $K$ and position offset $z_{offset}$ are the two parameters of the fit.

Fig. 3.6b

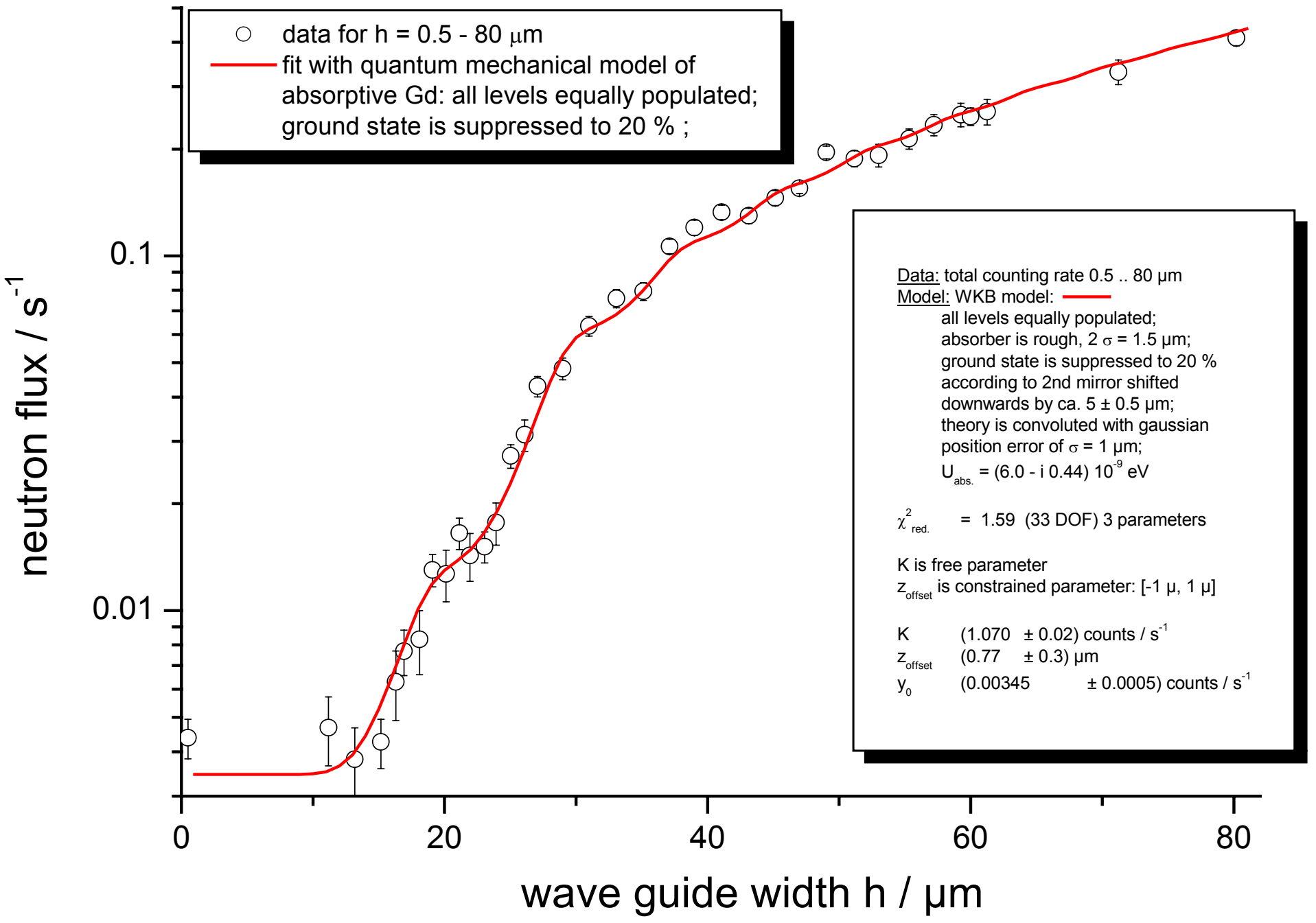

Fig. 3.6b: The theoretical prediction derived from eq.s (3.3.1.31) and (3.3.1.25'), where the eigenvalues of WKB quantization have been used, after convolution with the experimental spatial error of 1 μm is fitted to the data. Total scaling $K$ , position offset $z_{offset}$ , and background $y_0$ are the three parameters of the fit.

As one can see, the prediction fits to the data. All important features of the data, which include the smooth semi-classical behaviour for $h$ > 50 µm, the range of non-penetration for $h < 14\mu m$, and the "knee" at about $h$ = 24 µm, are resembled by the theory. If one bears in mind the low statistics of the data especially at $h$ < 20 µm, a $\chi^2_{red.}$ below 1.6 is satisfactory.

Now one may remember, that despite the fact, that WKB quantization fits extremely well to a linear potential (for an approximative method), its errors are the largest for the lowest states. Therefore, it may be useful to search for a method of constructing exact eigenstates of the vertical motion inside the wave guide for the lowest two states at least, since these states control the emergence of the "knee" at $h$ = 24 µm.

Since the error function shaped potential of the absorber cannot be solved analytically, one may neglect roughness for the moment, and describe the absorber potential as a rectangular one, with a complex magnitude and $d$ = 200 *nm* thickness. This is a potential, that behaves exactly, as it is seen in Fig. 3.3.

In this case, the eigenfunctions of the Schrödinger equation eq. (3.3.1.10'c) of the vertical motion inside the wave guide can be given analytically for each region of the vertical potential $U(z)$ .

Inside the glass substrates there the eigenfunction can be described by:

$$e^{\mp\kappa \cdot z} \quad where \quad \kappa = \sqrt{2m/\hbar^2 \cdot \left(U^{Fermi}_{glass} - E_R \cdot \varepsilon\right)} \ .$$

In the vacuum and inside the absorber the eigenfunctions are given by a linear combination of Airy functions, since there the potential is linear.

Yet, inside the absorber, the eigenfunctions will acquire an additional energy shift, which is equal to the absorber's Fermi pseudopotential magnitude. Therefore, one has to look carefully, whether the typical Airy functions $Ai(z)$ and $Bi(z)$ form a numerically stable set of linearly independent functions inside the absorber with its strongly imaginary potential.

A look into the Appendix A shows the problem. The typical Airy functions used as the function base only behave non-pathologically on the real axis.

Since the imaginary parts of the energy eigenvalues, which enter the Airy functions' argument, are usually small compared to the imaginary part of the absorber's Fermi potential, and compared to one (in units of $E_R$) except for very small wave guide widths, one can use the typical linear combination in the vacuum.
However, inside the absorber *Bi(z)* must be replaced by a complex rotated version of *Ai(z)* given in the Appendix A in eq. (A.3). Thus, the eigenfunctions here must be written as:

$$\psi_{absorber} = A \cdot Ai\left(\eta - \varepsilon + U_{abs.}^{Fermi} / E_R\right) + B \cdot Ai\left(e^{\pm i \cdot 2\pi/3}\eta - \varepsilon + U_{abs.}^{Fermi} / E_R\right).$$

The quantization condition here is then derived from the continuity, that the logarithmic derivative of each eigenstate must have at each boundary between two adiacent regions of the vertical potential. That means, that at the three locations:

- $z = 0$
- $z = h - d$
- $z = h$

the condition

$$\left.\frac{\partial\psi/\partial z}{\psi}\right|_{z_{boundary}}^{left} = \left.\frac{\partial\psi/\partial z}{\psi}\right|_{z_{boundary}}^{right} .$$

must be satisfied.
Applying these conditions to the eigenfunctions given above, and respecting the fact, that the states must vanish inside the glass for $z \to \pm\infty$, respectively, one gets three equations

$$z = 0: \quad \kappa = \left.\frac{Ai'(\eta - \varepsilon) + C_1 \cdot Bi'(\eta - \varepsilon)}{Ai(\eta - \varepsilon) + C_1 \cdot Bi(\eta - \varepsilon)}\right|_{\eta = 0}$$

$$z = h - d: \quad \left.\frac{Ai'(\eta - \varepsilon) + C_1 \cdot Bi'(\eta - \varepsilon)}{Ai(\eta - \varepsilon) + C_1 \cdot Bi(\eta - \varepsilon)}\right|_{\eta = \eta_h - \eta_d} = \left.\frac{Ai'\left(\eta - \varepsilon + U_{abs.}^{Fermi}/E_R\right) + C_2 \cdot e^{\pm i \cdot 2\pi/3} \cdot Ai'\left(e^{\pm i \cdot 2\pi/3}\eta - \varepsilon + U_{abs.}^{Fermi}/E_R\right)}{Ai\left(\eta - \varepsilon + U_{abs.}^{Fermi}/E_R\right) + C_2 \cdot Ai\left(e^{\pm i \cdot 2\pi/3}\eta - \varepsilon + U_{abs.}^{Fermi}/E_R\right)}\right|_{\eta = \eta_h - \eta_d}$$

$$z = h: \quad \left.\frac{Ai'\left(\eta - \varepsilon + U_{abs.}^{Fermi}/E_R\right) + C_2 \cdot e^{\pm i \cdot 2\pi/3} \cdot Ai'\left(e^{\pm i \cdot 2\pi/3}\eta - \varepsilon + U_{abs.}^{Fermi}/E_R\right)}{Ai\left(\eta - \varepsilon + U_{abs.}^{Fermi}/E_R\right) + C_2 \cdot Ai\left(e^{\pm i \cdot 2\pi/3}\eta - \varepsilon + U_{abs.}^{Fermi}/E_R\right)}\right|_{\eta = \eta_h} = -\kappa$$

for the three unknowns $C_1$ , $C_2$ , and $\varepsilon$ . Eliminating $C_1$ and $C_2$ from these equations leads to one remaining equation in $\varepsilon$, which forms the quantization condition.

This equation can be solved for different values of the wave guide width $h$ for the lowest two states, thus producing their energy eigenvalues as a function of $h$. The higher states one may again derive via WKB quantization.

This way, one derives a theoretical prediction for the total transmission of the wave guide, which should be nearly correct for the lowest two states, since their eigenstates have been obtained analytically, whereas WKB quantization naturally forms a good approximation for the higher states.

There remains one question. The roughness must be incorporated into the exact solution of the two lowest states somehow afterwards. Yet, there is a way to achieve this. One may remember, that the neutrons, which enter the wave guide, have very small coherence lengths, especially in the forward direction along the optical axis of the guide.

Therefore, it makes sense to adopt the point of view, that the main effects of the roughness can be incorporated into the analytical results of the two lowest states by a convolution of the resulting transmission coefficients with an area-normalized gaussian, which $\sigma$ equals the roughness of the absorber.

Since this procedure has already to be done to account for the position error of the measurement, this argument leads to the fact, that a prediction, that is made according to eq.s (3.3.1.31) and (3.3.1.25'), and using the exact eigenvalues of the two lowest states and WKB result for the higher states, must be convoluted with a gaussian, that has a $\sigma$ of about 2 µm. That is combined value resulting from the position error and the roughness amplitude.

If one bears these facts in mind, one can recalculate the predictions made above for a pure WKB approach. However, the determination of the eigenvalues for the two lowest states with the exact method given above still is numerical delicate and time consuming.

Therefore, two successful fits of this combined "exact-and-WKB"-prediction to the data are shown, where the absorber's Fermi potential was $U_{abs.}^{Fermi} = (2.0 - i \cdot 0.44) \cdot 10^{-9} eV$, which is at least of the same order of magnitude as the actual value. Thus, these results are close to the ones valid for the actual value of the absorber's potential.

The fits again have been performed in the ways a) and b), that are described above. Their results are seen in Fig.s 3.7a and 3.7b:

Fig. 3.7a

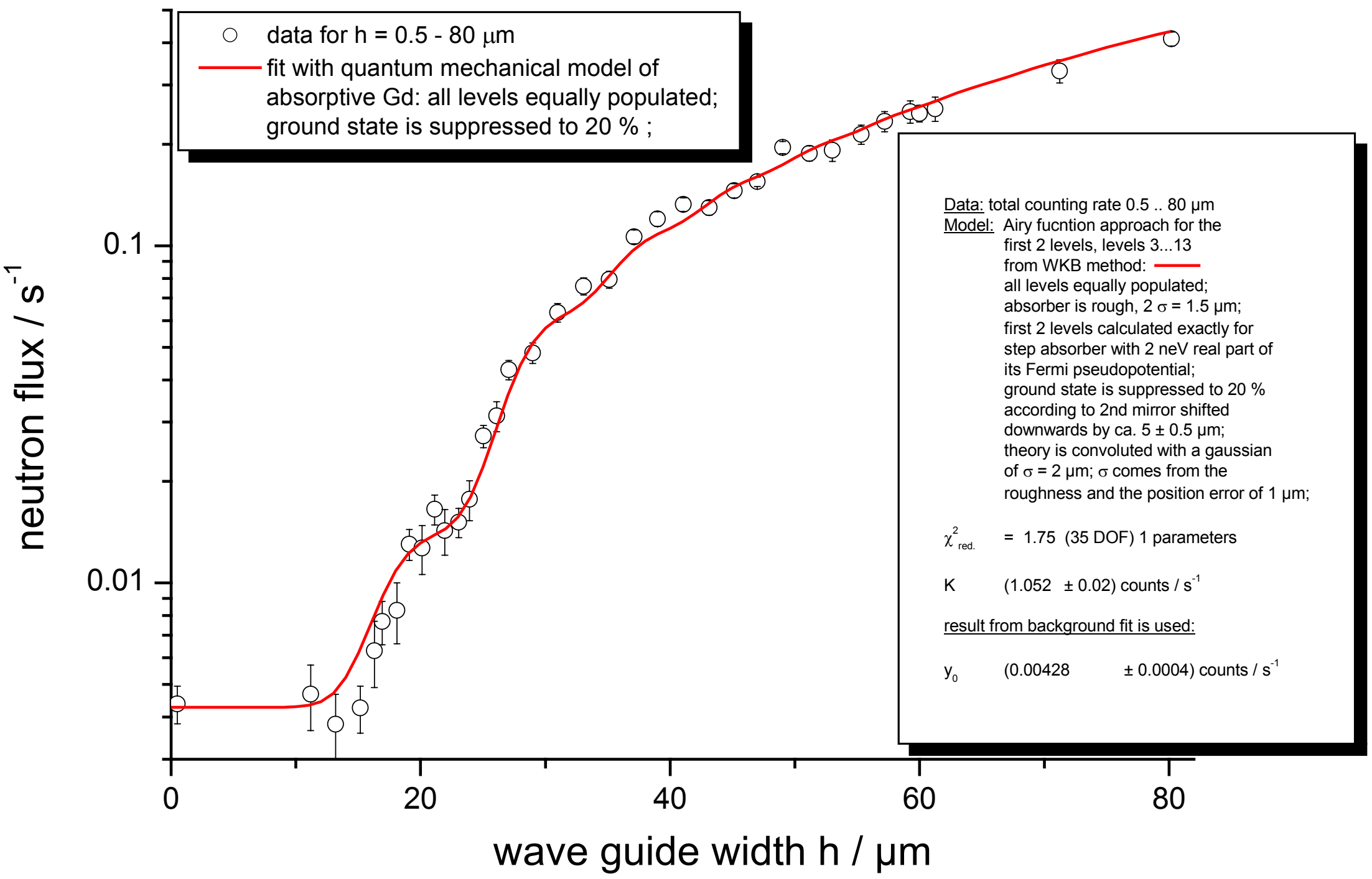

Fig. 3.7a: The theoretical prediction derived from eq.s (3.3.1.31) and (3.3.1.25'), where the eigenvalues of the first two states are derived exactly, whereas the higher states are given by WKB quantization, after convolution with the combined smearing of 2 µm is fitted to the data. Total scaling $K$ is the one parameter of the fit. A position offset is not needed.

Fig. 3.7b

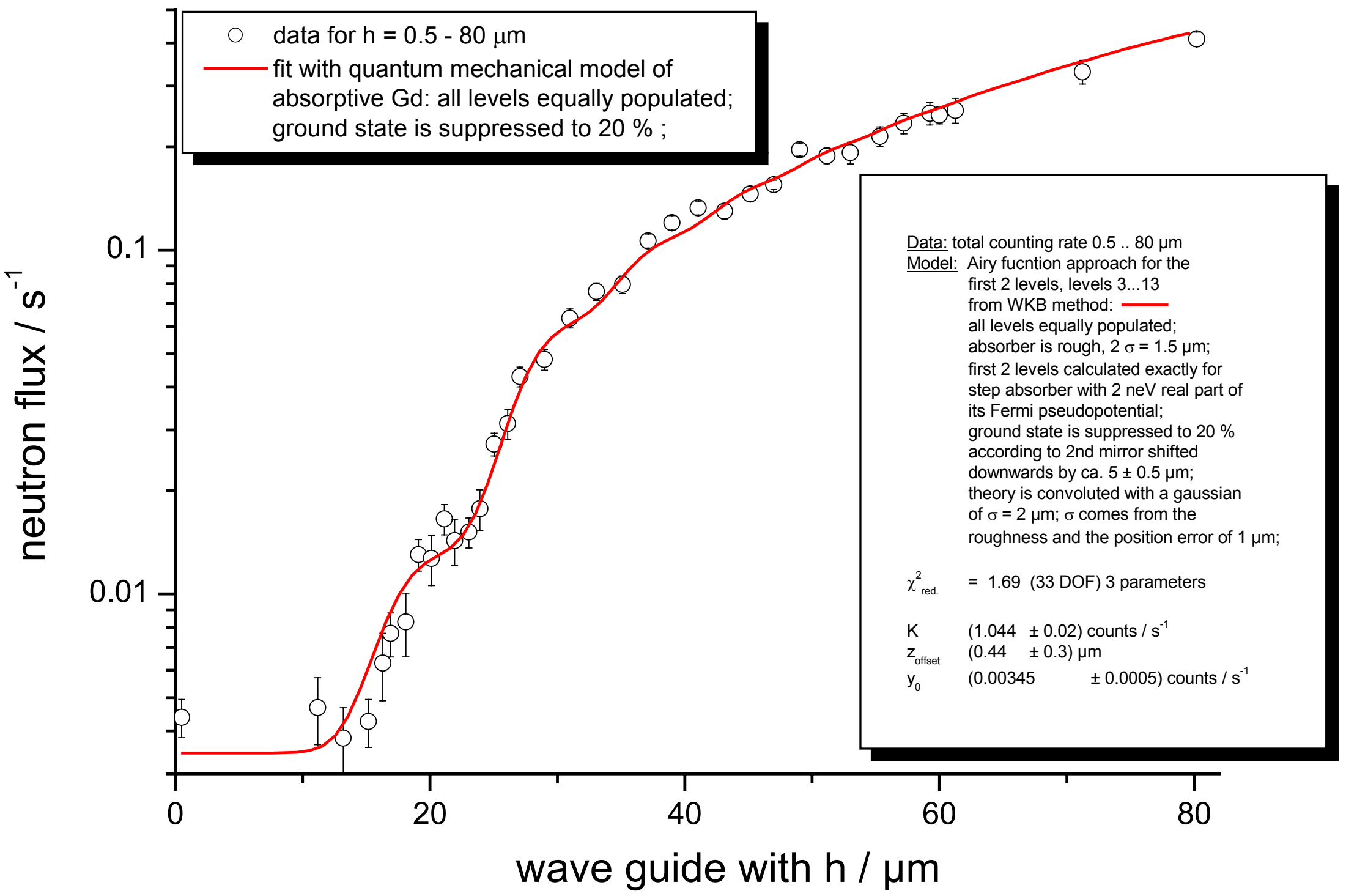

Fig. 3.7b: The theoretical prediction derived from eq.s (3.3.1.31) and (3.3.1.25'), where the eigenvalues of the first two states are derived exactly, whereas the higher states are given by WKB quantization, after convolution with the combined smearing of 2 μm is fitted to the data. Total scaling $K$, position offset $z_{offset}$, and background $y_0$ are the three parameters of the fit.

One sees, that these calculations confirm the results of the (numerically much more stable) WKB quantization model.

If Fig.s 3.6a and 3.6b as well as Fig.s 3.7a and 3.7b are viewed together, then it is justified to say, that this quantum mechanical model is able to describe the experimental results.

### 3.3.3) Gravity – no gravity – an analytical approach to the wave guide and a synopsis of all views

In section 3.2) the classical behaviour of the wave guide was discussed for the case, where no gravitation is present. It was shown, that the transmission of a classical waveguide is governed by a (counterintuitive) quadratic dependency on the wave guide width *h*. Furthermore, the exponent of *h* shifts gradually from two towards one, if the range of the ratio of the vertical velocity to the horizontal velocity becomes comparable to the range of the ratio, that the wave guide width forms with its length:

$$\frac{v_{vert.}}{v_{hor.}} \approx \frac{h}{l} \quad .$$

However, if one plots this quadratical prediction into the data, one has to choose a correct normalization factor with regard to the classical $h^{3/2}$-prediction valid in presence of a gravitational field. When plotting both predictions ($h^2$ for the case without gravity and $h^{3/2}$ with gravity) to the data, one arrives at:

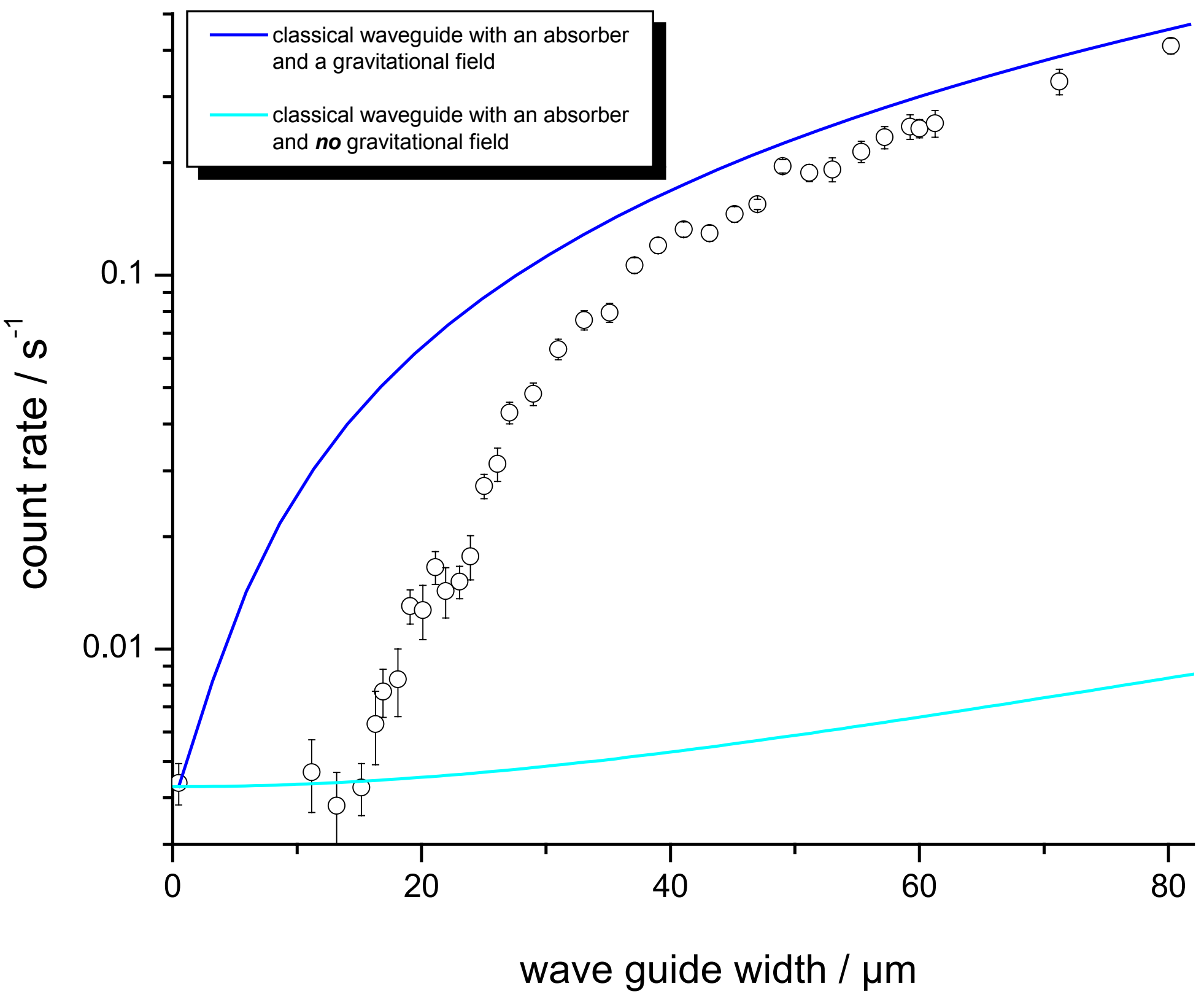

The classically predicted transmission without a gravitational field appears to be very weak! This now is not completely surprising since the neutrons follow linear trajectories inside the wave guide if gravity is turned off, which leads naturally to far higher collision rates with the absorber especially at larger values of $h$. Whereas for larger values of $h$ the transmission in presence of a gravitational field should be enhanced, since a certain part of the neutrons can there pass the wave guide without hitting the absorber due to their curved trajectories.

One should look now, whether it is possible to recover these results in a quantum mechanical treatment of this comparison "gravity – no gravity", since this would underline the results of section 3.3.2), if the quantum mechanical transmission of the field-free wave guide differed in a similarly radical way from the case with a gravitational field as in the classical treatment.

Fortunately, it is possible to reduce the quantum mechanics to an analytical treatment of both cases, if one is willing to accept fitting an effective value of the absorber's Fermi pseudo potential to the data. The line of such an argument goes as follows:

Consider first the case without gravity. The setup is given in Fig. 3.3, where the bottom mirror is now placed at $z = -h$ and the absorber at $z = 0$ for convenience. $h$ here is the wave guide width. Now to zeroth order the potential walls at $z = 0$ and $z = -h$ are infinitely high, thus providing vanishing wave function beyond these points, and giving quantized bound states according to eq.s (3.3.1.16) and (3.3.1.17). Now one has to take into account that (at least) the absorber is described by finite complex potential, the imaginary part of which parametrizes the absorption in the language of potential energy:

$$(3.3.3.1) \qquad U_{absorber} = W - i \cdot V, \qquad W >> V \geq 0$$

Thus, the wave function will enter the absorber exponentially suppressed, as long as $W >> E_{state}$ is valid. The ground state, e.g., outside the absorber, now is shifted somewhat to:

$$|0\rangle_I = A - \sqrt{\frac{2}{h}} \cdot \sin\left(\frac{\pi}{h} \cdot z\right)$$

which gives close to the absorber:

$$|0\rangle_I = A - \frac{\sqrt{2\pi^2}}{h^{3/2}} \cdot z$$

where at $z = 0$ it has to be matched to the wave function inside the absorber, which happens to be of the form:

$$(3.3.3.2)\qquad \left|0\right\rangle_{II} = A\cdot e^{-\kappa\cdot z},\qquad \kappa = \sqrt{\frac{2m}{\hbar^2}W}$$

Matching then yields:

$$(3.3.3.3)\qquad A = \frac{\pi\sqrt{2}}{\kappa\cdot h^{3/2}}\quad .$$

The shift of the energy eigenvalue of the ground state due to the imaginary part of eq. (3.3.3.1) can now be calculated in first order perturbation theory as the overlap of the state $\left|0\right\rangle$ given by eq.s (3.3.3.2) and (3.3.3.3). This is roughly $\sim A^2\cdot iV$ and yields exactly:

$$(3.3.3.4)\qquad E_1^{(1)} = iV\cdot\frac{\pi^2}{\kappa^3 h^3}\quad .$$

The result shows the expected power law dependency of the energy shift on the wave guide width $h$.

The case with gravity now simply deviates from these results by the fact, that the "floor" of the potential well is now rising with a constant slope, as seen in Fig. 3.3.
Yet, this linear potential makes a great difference. If the absorber is sufficiently far away compared to the energy of a given neutron, the linearly rising potential itself bounds the state causing it to decay rapidly exponentially beyond the classical turning point. Thus, a neutron in a given state with gravity will feel the absorber only, if it is close enough. Therefore, one will have to use the bound states of eq. (3.3.1.17) as the starting point. The line of the argument now goes the same way: Close to the absorber the wave function just outside and just inside is to be parametrized by a exponentials. Yet, one has to bear in mind, that this is only valid as long, as the absorber remains outside the classical turning point, since only there the Airy function is guaranteed to decay exponentially fast.
In this way the ground state just outside the absorber is given by:

$$(3.3.3.5)\qquad \left|0,g\right\rangle_I = B\cdot e^{-\kappa'\cdot\Delta z}$$

where $B$ is given by expanding the states eq. (3.3.1.17) around $z = h$:

$$(3.3.3.6)\qquad \begin{aligned} B &= \frac{1}{2\sqrt{\pi R}}\cdot\left(\xi_h - \xi_1^{(0)}\right)^{-1/4} e^{-\frac{2}{3}\left(\xi_h-\xi_1^{(0)}\right)^{3/2}} \\ \xi_h &= \frac{h}{R} \\ \xi_1^{(0)} &= \frac{E_1^{(0)}}{E_R} := \frac{h_0}{R} \\ \kappa' &= \sqrt{\frac{2m}{\hbar^2}\left(m\cdot g\cdot h - E_1^{(0)}\right)}\qquad . \end{aligned}$$

Inside the absorber the state is again given by:

$$(3.3.3.7) \qquad |0\rangle_{II} = A \cdot e^{-\kappa \cdot z}, \qquad \kappa = \sqrt{\frac{2m}{\hbar^2} W}$$

Again, matching at $z = h$ yields $A$, and we have:

$$(3.3.3.8) \qquad A = \sqrt{\frac{E_R}{4\pi W} \frac{\sqrt{\frac{1}{R}}\left(h - h_0\right)}{R}} \cdot e^{-\frac{2}{3}\left(\frac{h-h_0}{R}\right)^{3/2}} \quad .$$

The overlap integral then gives us the correction to the energy eigenvalue, which is now:

$$(3.3.3.9) \qquad E_{1,g}^{(1)} = iV \cdot \frac{E_R}{W} \cdot \frac{\sqrt{h - h_0}}{8\pi\kappa \cdot R^{3/2}} \cdot e^{-\frac{4}{3}\left(\frac{h-h_0}{R}\right)^{3/2}} \quad .$$

Comparison with eq. (3.3.3.4) now clearly shows the difference. The imaginary correction to the energy eigenvalues in presence of a gravitational field decays exponentially fast, whereas without gravity only a power law dependency is present.

In this approach now the roughness has yet to be taken into account. For the purpose of this analytical first order approximation to the dominating effects of the wave guide, it will be sufficient to do this by taking the imaginary part $V$ of the potential eq. (3.3.3.1) as an effective parameter, that has to be fitted to the measured transmission of the wave guide. Thus, it can not be expected, that $V$ will take realistic values – instead one will expected it to be large in order to incorporate effectively the roughness, which enhances absorption efficiency greatly via upscattering of neutrons.

The transmission coefficients of the ground states can be directly calculated from the imaginary corrections to the energy eigenvalues, as follows:

$$(3.3.3.10) \qquad \begin{aligned} T_0 &= e^{-2i\frac{E_1^{(1)}}{\hbar} \cdot T_{flight}} = e^{-2i\frac{E_1^{(1)}}{\hbar} \cdot \frac{L}{v_{\|}}} \\ T_{0,g} &= e^{-2i\frac{E_{1,g}^{(1)}}{\hbar} \cdot T_{flight}} = e^{-2i\frac{E_{1,g}^{(1)}}{\hbar} \cdot \frac{L}{v_{\|}}} \end{aligned}$$

where $L$ is the length of the wave guide and $v_{\|}$ is the horizontal forward velocity of the neutrons. $E_1^{(1)}$ and $E_{1,g}^{(1)}$ herein are given by eq.s (3.3.3.4) and (3.3.3.9), respectively.

The spacing of the gravitational eigenstates now suggests that in a range of wave guide width $h$ between 15 $\mu m$ and 22 $\mu m$ the ground state should give the dominant contribution to the total transmitted flux. Thus, looking at eq. (3.3.3.10), one sees, that one can fit the value of $V$ directly from the measured transmission data of [Ne01] if one fits eq. (3.3.3.9) to the logarithmic derivative of the measured data in a $h$-range between about 15 $\mu m$ and 22 $\mu m$. Fitting to the logarithmic derivation of the data automatically removes the basic exponentials from the transmission coefficients eq. (3.3.3.10). If this is done, one finds numerically values for the imaginary part of the potential eq. (3.3.3.1) of order of magnitude: $V \sim 10^{-6}\, eV$, which is larger than the typical values of the real part $W$ by one to three orders of magnitude, which clearly shows, how the roughness effectively enlargens the imaginary part of the potential.

If one compares the penetration depth $\lambda_{flat}$ of a $peV$-eigenstate entering a rectangular potential step of about 6 $neV$, which is given to be

$$\lambda_{flat} = \kappa^{-1} = \left(\frac{2m}{\hbar^2} W\right)^{-1/2} \approx 5 \cdot 10^{-8} m \quad .$$

to the depth the state can penetrate into the roughness-smeared potential formfactor shown in Fig. 3.4, which is about $\lambda_{rough} \sim 1\, \mu m$, then in first order perturbation theory the formfactor enhances the overlap integral by the square of the ratio of the two penetration depths, which is about

$$\left(\frac{\lambda_{rough}}{\lambda_{flat}}\right)^2 \sim 300 \quad .$$

Now this is exactly ratio, by which the value of $V$ of about 5 $neV$ calculated in sect. 3.1 from the alloy data has to be enlarged in order to fit the effective value found right above. Therefore, the use of a description of the absorber by a complex potential shaped according to Fig. 3.4 seems to be valid, at least on the 10%-level.

If this value range for $V$ is now used to calculate values for the transmission coefficient according to eq. (3.3.3.10), one finds:

$$T_0 \approx 10^{-16}$$
$$T_{0,g} \approx 0.3 \quad .$$

Finally, one may then fit the prediction for the transmission coefficient derived from the eq.s (3.3.3.9) and (3.3.3.10) to the measured data by determining the total normalization of the transmitted flux:

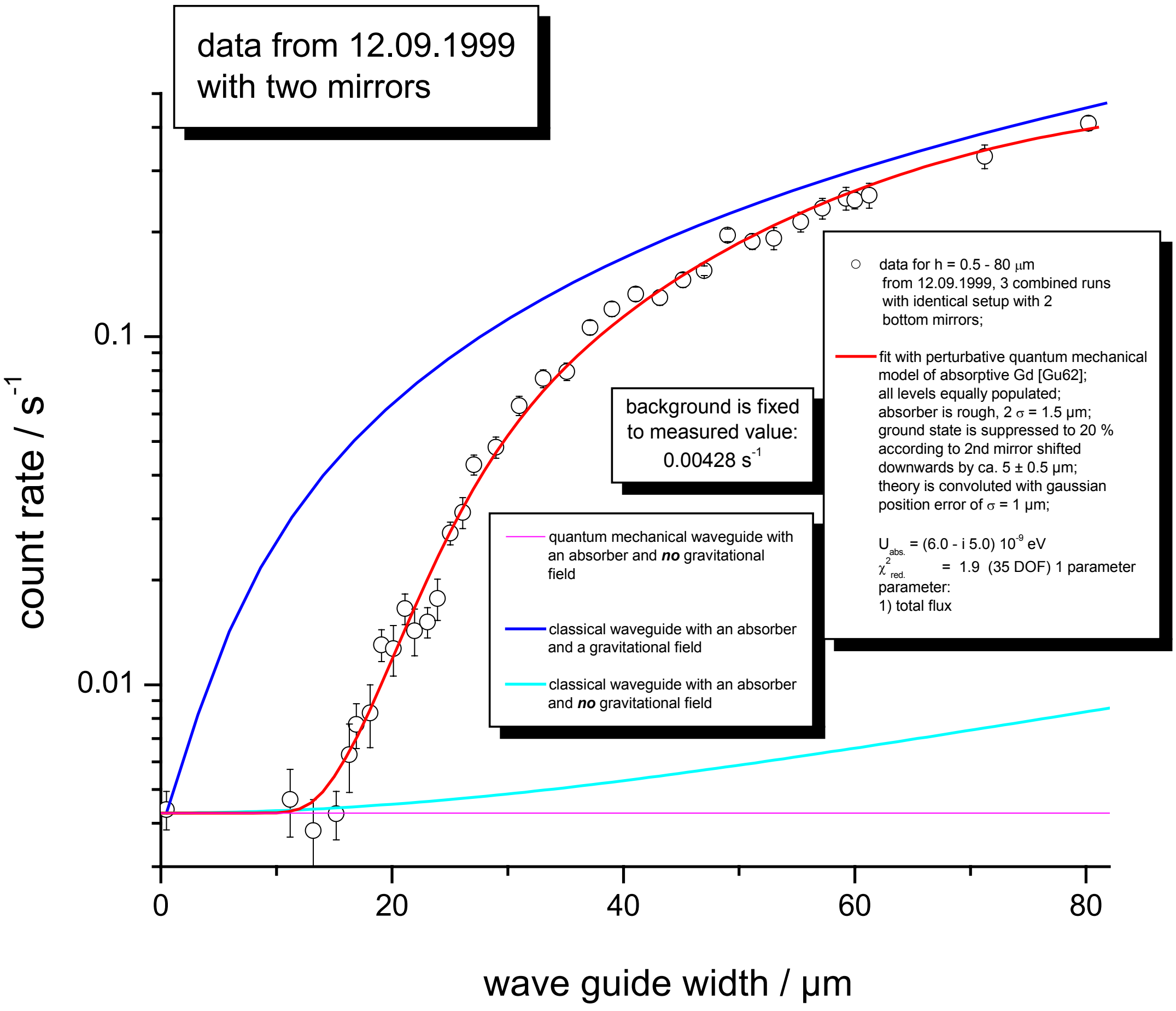

Therefore, it is now clear, that if quantized bound states show up in the experiment, their behaviour in presence of gravity and without are largely different. Without gravity there should be virtually no transmission at all over huge ranges of $h$ up to 100 $\mu m$, since the states eq. (3.3.1.17) constantly feel the absorber, whereas in a gravitational field the system should start to transmit once the absorber crosses the classical turning point height $h_0$ of the ground state. In the light of this analysis, the fact [Ne01], that transmission starts readily beyond $h = 15\ \mu m$, shows with clarity, that gravitational effects substantially influence the bound states found there.

And at last, it is now the time to present a synopsis of all views, which one can use to describe the transmission of the wave guide. There are:

- the classical view without a gravitational field (section 3.2)),
- the classical view with a gravitational field (section 3.2)),
- the quantum mechanical prediction for the case without a gravitational field (shown above),
  *and*
- two quantum mechanical predictions with a gravitational field (section 3.3.2)).

All these results are shown together in Fig. 3.10:

Fig. 3.10

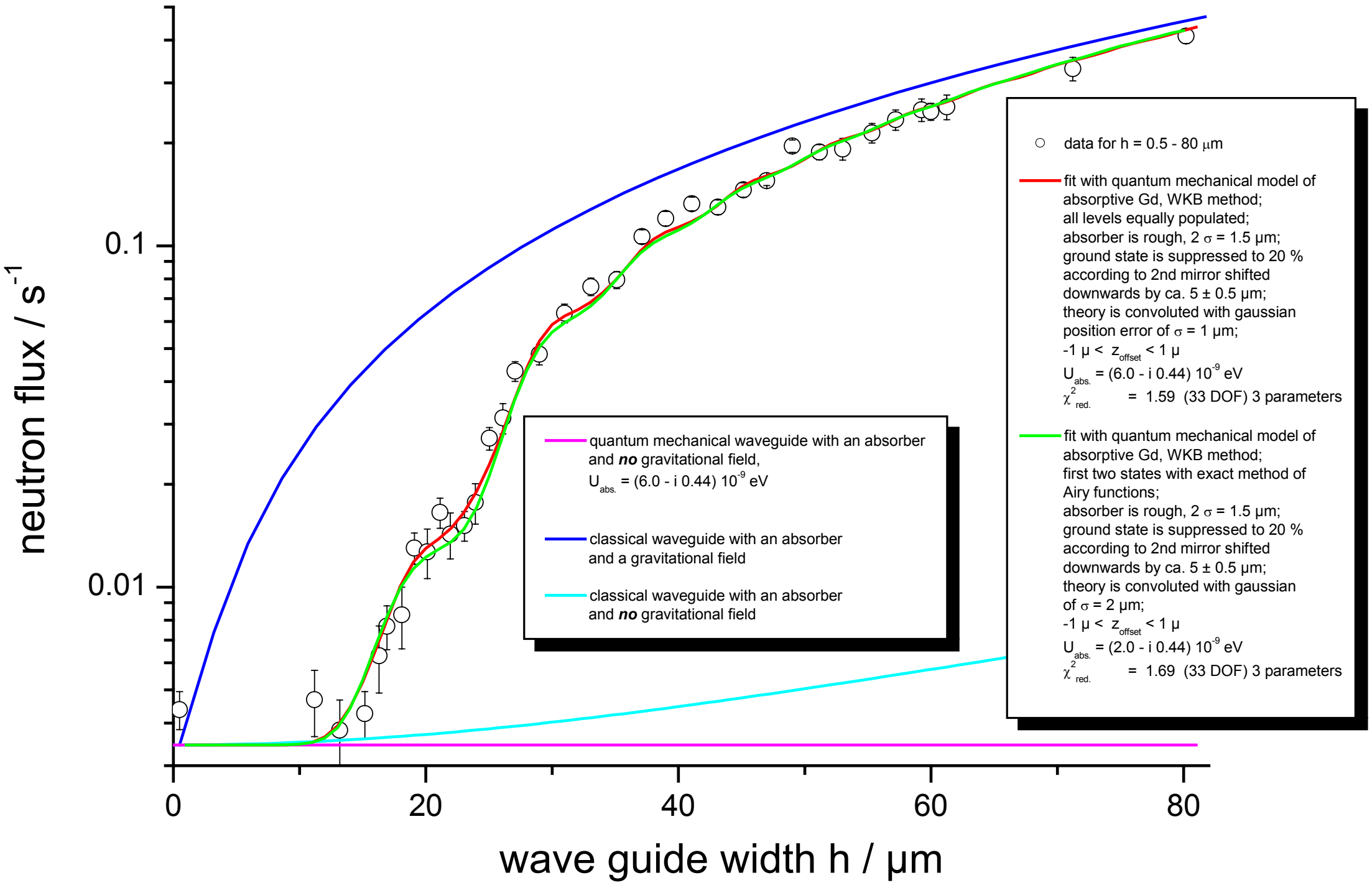

Fig. 3.10: All results of the different views of the wave guide system are plotted together.

If one looks at this synopsis, then it is justified to say, that the present experiment was able to prove the existence of quantum mechanically bound states of massive particles in a gravitational field. Therefore, it is now possible to state, that the formalism of so-called "first quantization" has been established for the classical, non-relativistic gravitational field in its full consequences including the existence of bound states.

### 3.3.4) The spatial distribution of the neutrons after the wave guide

For the moment, these calculations exhaust the quantum mechanical model, that has been present here as a description of a measurement, that is shown in Fig. 3.2. However, a second measurement has been performed. Instead of determining the variation of the transmittivity of the apparatus with the wave guide width *h*, one can try to detect the spatial density distribution of the neutrons, which leave the wave guide. According to Fig. 3.11, a measurement of the neutron density distribution after the wave guide may also show up signs of the gravitationally bound states:

Fig. 3.11

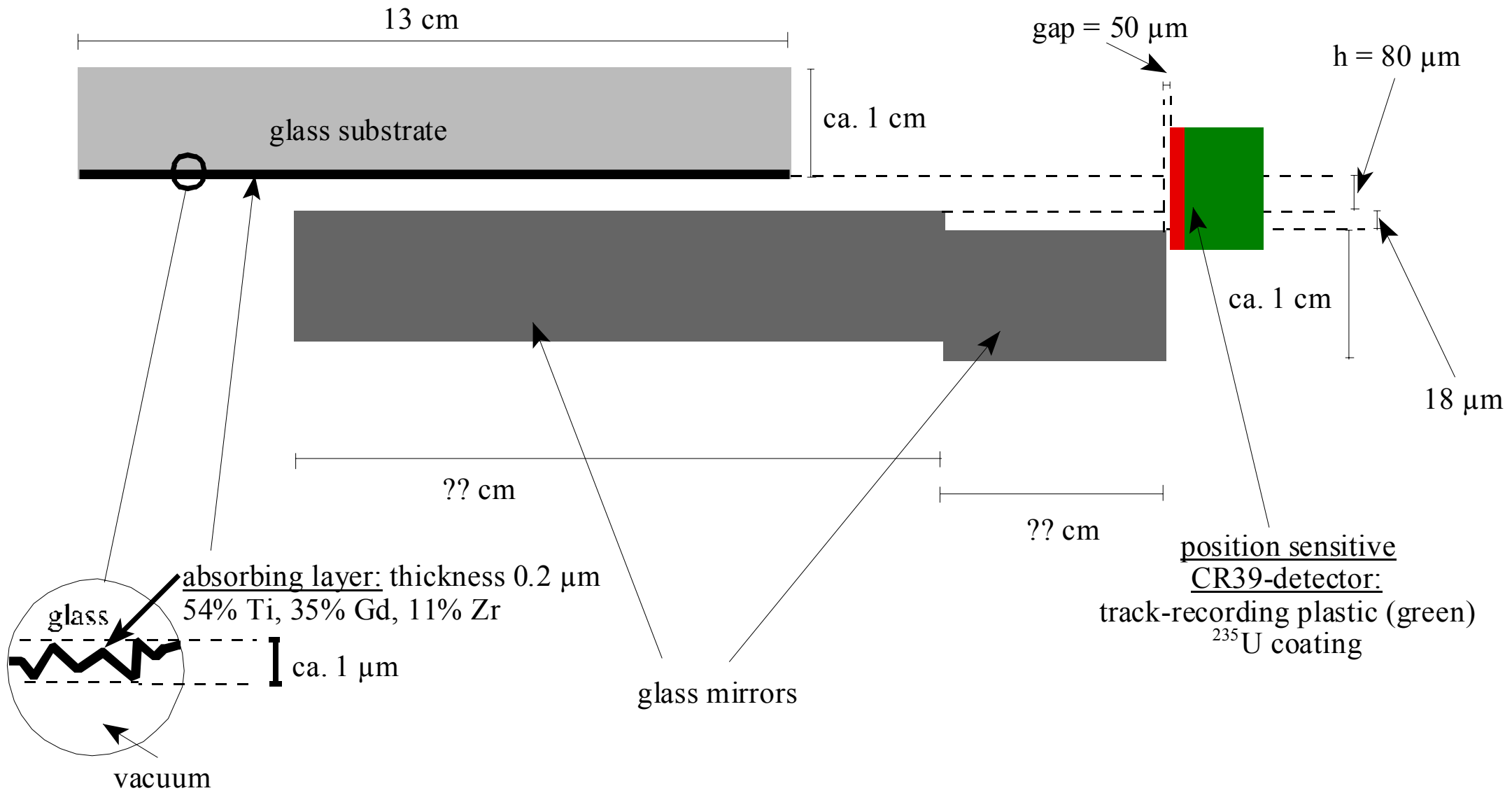

With the formalism developed above, it is now possible to try a description of this other kind of measurement (see chapter one). There the absorber is placed at a constant height *h* and a position-sensitive uranium-coated CR39 track detector records the *z*- (and *y*-) dependency of $j_x^{III}$, given by eq. (3.3.1.27). The setup for this measurement is shown in Fig. 3.11 in chapter one.

A look at eq. (3.3.1.27) shows, that the current density contains off-diagonal interference terms *nn'* with $n \neq n'$, of course. The question is, whether these interferences will be seen from the CR39-detector or not. The measurement of the total flux (see above) does not depend on this detail, since the spatial integration of the current density eq. (3.3.1.27) removes the interference terms due to the orthonormality of the function base. The interference terms of eq.(3.3.1.27) will only show up in the detector, if the whole system can be prepared into a coherent superposition of pure base states. Since we know, that the collimator is a highly incoherent and more or less thermal neutron source (see eq.s (3.3.1.29) and (3.3.1.30) for estimates of the coherence lengths), this coherence condition is not satisfied. Therefore, the probably best knowledge about the system is provided by the density matrix $\rho$ of the final states. Consider a case, where the states $\phi_n(\eta)$ of our system are forming an incoherent mixture with a population $p_n$ for each state. Then $\rho$ is given by:

$$(3.3.4.1)\quad \rho = \sum_n p_n \cdot \left|\varphi_n^{III}\right\rangle\left\langle\varphi_n^{III}\right| .$$

Then, a position-sensitive detection means, that the detection process projects $\rho$ onto the spatial eigenstates $\delta(z\text{-}z_0)$. The trace of these projections of eq. (3.3.4.1) is the analogon to eq. (3.3.1.27). The result for this trace of the projection of $\rho$ onto the spatial eigenstates is:

$$(3.3.4.2)\quad Sp[P_z(\rho)] = Sp(|z\rangle\langle z|\rho) = \sum_n p_n \cdot \left|\varphi_n^{III}(\eta)\right|^2 .$$

Comparison with eq. (3.3.1.27) then shows, that the position-sensitive measurement with the CR39-detector will only show the first interference-free diagonal term of $j_x^{III}$. The prediction for the position-sensitive measurement of Fig. 3.11 must therefore be obtained from:

$$(3.3.4.3)\quad j_x^{III}\Big|_{incoherent} = \frac{\hbar}{m} \cdot \sum_n \Re\left(k_n^{(x)III}\right) \cdot \left|\varphi_n^{III}\right|^2 \cdot \left|T_n\right|^2 .$$

Yet we should realize, that eq.s (3.3.1.27) and (3.3.4.3) describe the two extremes of either a totally lost or a completely preserved phase. Therefore from a phenomenological point of view, it is wise to introduce one parameter, which merges eq.s (3.3.1.27) and (3.3.4.3). We will call that number the "coherence strength" $\gamma$ of the system. $\gamma$ is one for total coherence and zero for the case of a completely incoherent mixture like eq. (3.3.4.2). Therefore, eq.s (3.3.1.27) and (3.3.4.3) merge to:

(3.3.4.3′)

$$j_x^{III} = \frac{\hbar}{m} \cdot \left[ \sum_n \Re\left(k_n^{(x)III}\right) \cdot \left|\varphi_n^{III}\right|^2 \cdot \left|T_n\right|^2 + \gamma \cdot 2 \cdot \Re\left( \sum_{n<n'} k_n^{(x)III} \cdot \left(\varphi_{n'}^{III}\right)^* \varphi_n^{III} \cdot e^{i\cdot\left(k_n^{(x)III} - k_{n'}^{(x)III}\right)\cdot x} \cdot T_{n'}^* T_n \right) \right] .$$

$\gamma$ can be obtained by fitting eq. (3.3.4.3’) to the data. To elucidate the effects of $\gamma$, one may calculate two graphs of eq. (3.3.4.3’) for $\gamma = 0$ and $\gamma = 1$, respectively. These behaviours of the cases of total coherence and total incoherence look like:

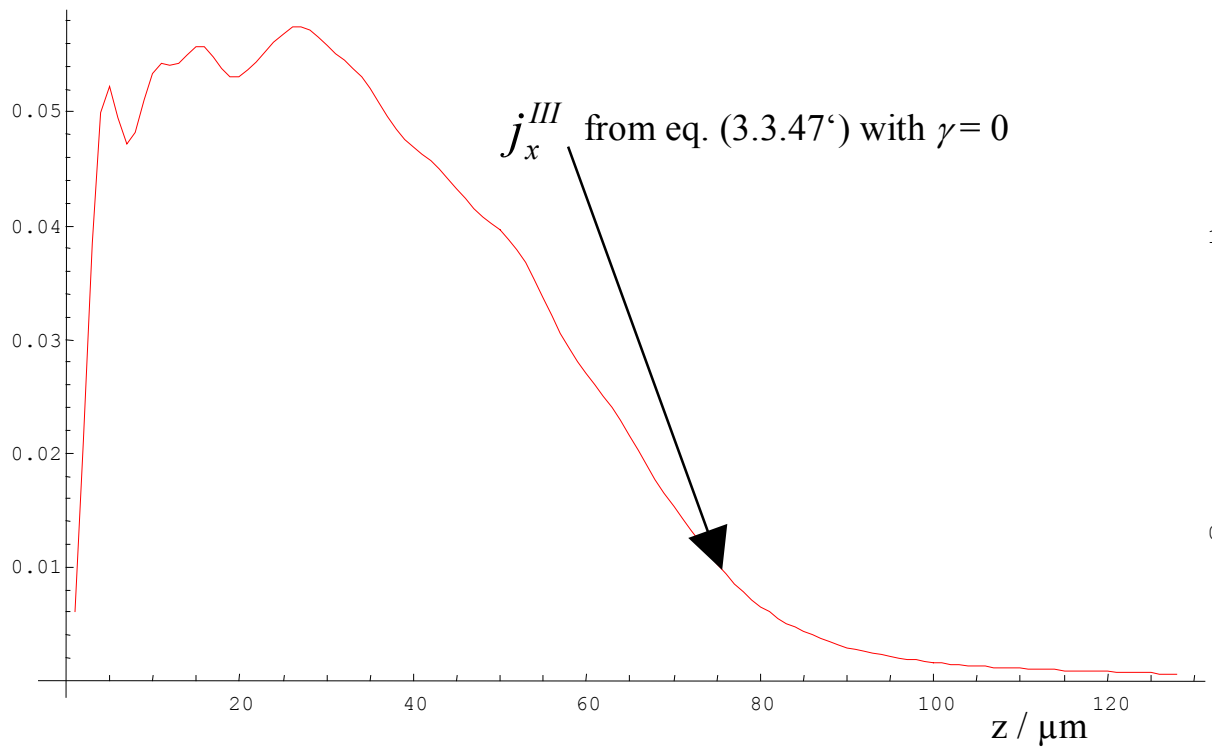

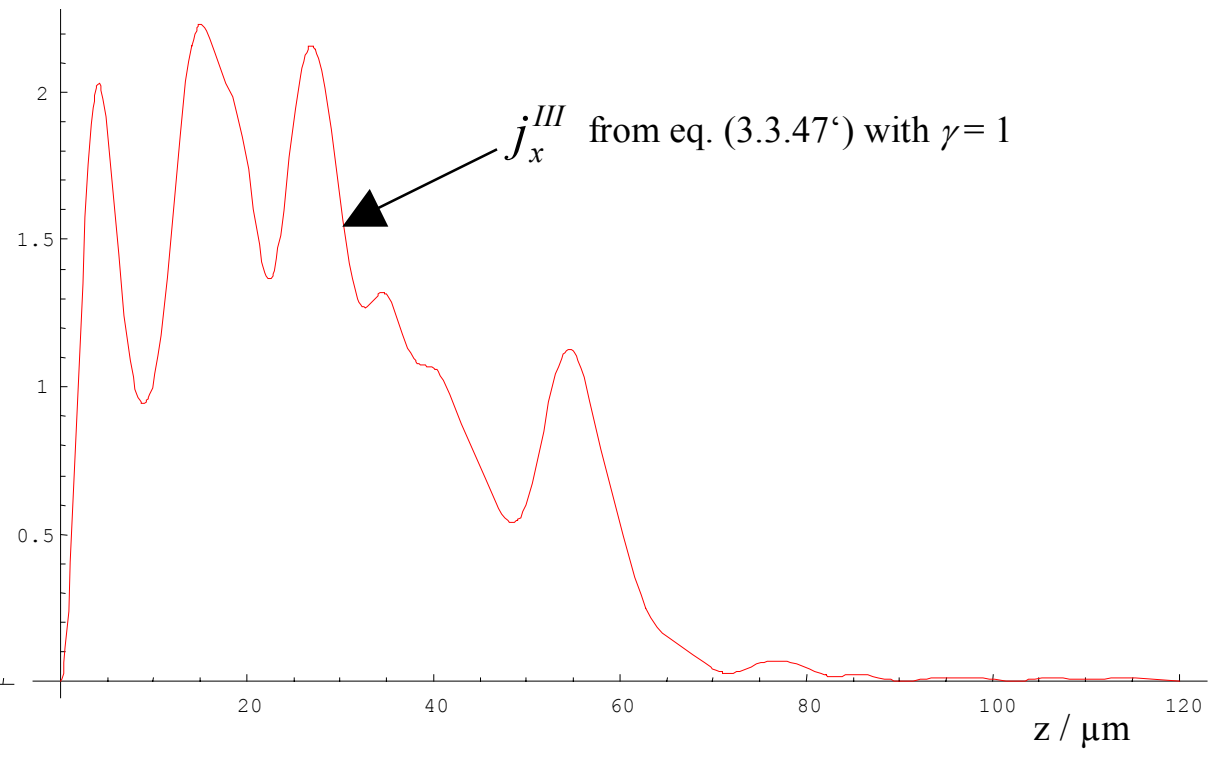

Looking at Fig. 3.11, we realize, that region *III* is split into two parts *IIIa* and *IIIb*. The mirrors are shifted relatively against each other by about $\Delta z$ = 18 µm. Therefore, eq. (3.3.1.25), describing the transmission coefficients $T_n$ , will have to be extended to three boundaries *I/II*, *II/IIIa* and *IIIa/IIIb*. This will lead to an additional summation over an additional matrix of the overlap integrals between states of *IIIa* and *IIIb*, as it has been argued for the case of a mirror step below the absorber (see above). Fortunately, the $T_n$ have only to be calculated for one fixed absorber height. Furthermore, the eigenstates of the regions *IIIa* and *IIIb*, which form the corresponding overlap integral matrix, are given analytically by the first type of the Airy functions, since both regions are after the absorber. Therefore, the step can be included directly this time by calculating the expression eq. (3.3.1.25) for the $T_n$ once more, including the third boundary *IIIa/IIIb*. The result is:

$$(3.3.4.4) \quad T_n = 32 \cdot k_x \cdot \sum_{l,j} \frac{c_{nl}^{IIIb} c_{lj}^{IIIa} c_j^{II}(k_z^I) \cdot k_j^{(x)II} \cdot k_l^{(x)III} \cdot e^{i \cdot \left[ \left( k_j^{(x)II} + k_l^{(x)III} \right) \cdot l + \left( k_l^{(x)III} - k_n^{(x)III} \right) \cdot l' \right]}}{2e^{i \cdot \left( k_j^{(x)II} \cdot l + k_l^{(x)III} \cdot l' \right)} \cdot c_1 - 2e^{i \cdot \left( k_j^{(x)II} + 2 \cdot k_l^{(x)III} \right) \cdot l} \cdot c_2},$$

$$c_1 = \left( k_l^{(x)III} - k_n^{(x)III} \right) \cdot \left[ i \cdot \left( \left[ k_j^{(x)II} \right]^2 - k_x^I k_l^{(x)III} \right) \cdot \sin\left( k_j^{(x)II} l \right) - k_j^{(x)II} \cdot \left( k_x^I - k_l^{(x)III} \right) \cdot \cos\left( k_j^{(x)II} l \right) \right]$$

$$c_2 = \left( k_l^{(x)III} + k_n^{(x)III} \right) \cdot \left[ -i \cdot \left( \left[ k_j^{(x)II} \right]^2 + k_x^I k_l^{(x)III} \right) \cdot \sin\left( k_j^{(x)II} l \right) + k_j^{(x)II} \cdot \left( k_x^I + k_l^{(x)III} \right) \cdot \cos\left( k_j^{(x)II} l \right) \right]$$

where:

$$\begin{cases} c_j^{II}(k_z^I) = \left\langle \varphi_j^{II}\left(\frac{z}{R}\right) \middle| k_z^I, z \right\rangle = \int dz \cdot \left[\varphi_j^{II}\left(\frac{z}{R}\right)\right]^* \cdot e^{i \cdot k_z^I \cdot z} \\ {}_{lj}^{IIIa} = \left\langle \varphi_l^{IIIa}(\eta) \middle| \varphi_j^{II}(\eta) \right\rangle \end{cases} \quad and \quad c_{nl}^{IIIb} = \left\langle \varphi_n^{IIIb}(\eta) \middle| \varphi_l^{IIIa}(\eta) \right\rangle \quad .$$

As $T_n$ is known, eq. (3.3.4.3') can be calculated and fitted to the data of Fig. 1.6. This fit consists of 5 parameters.

Four of them emerge naturally if one looks at Fig. 1.6. Normally, one would need *background*, *z-position offset*, *vertical scaling of the data*, and the coherence $\gamma$ to adjust (3.3.4.3'). Such a result is seen in Fig. 3.12a:

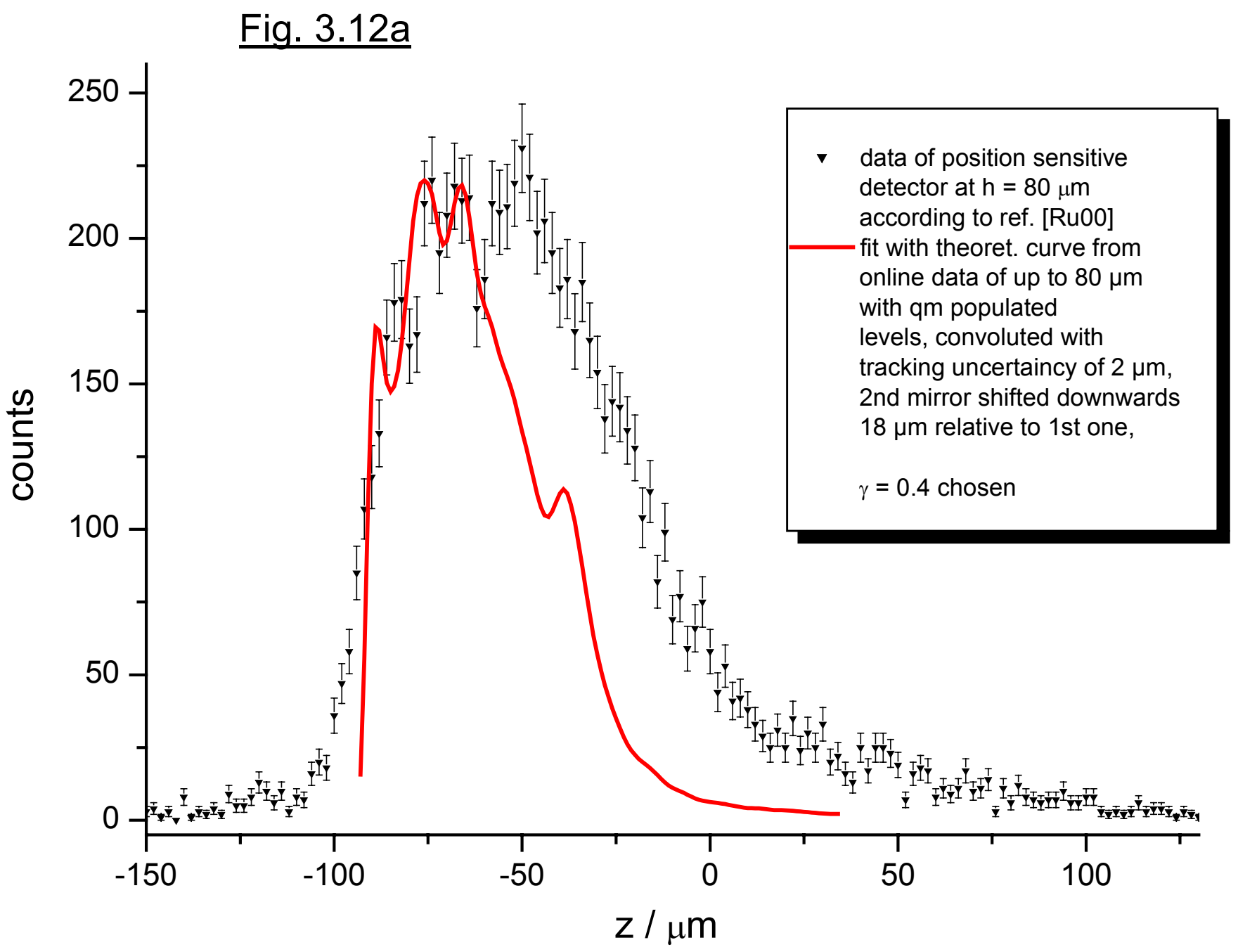

Fig. 3.12a: The prediction of eq. (3.3.4.3') for the actual measurement.Coherence $\gamma$ is set to 0.4 .

This curve does not resemble the data at all. However, a closer look shows, that the prediction shows a certain homothety to the data – it is just to small. One could of possibly stretching the predicted curve along the $z$-axis, until it resembles the width of the data.

Thus, one may introduce a fifth parameter. It expands (or shrinks) the width of eq. (3.3.4.3') in the $z$-direction and is therefore called "*z-scale*". It seems, that the measured curve is some-how scaled differently compared to the prediction eq. (3.3.4.3'), as it can be seen in the fitting result in Fig. 3.12b:

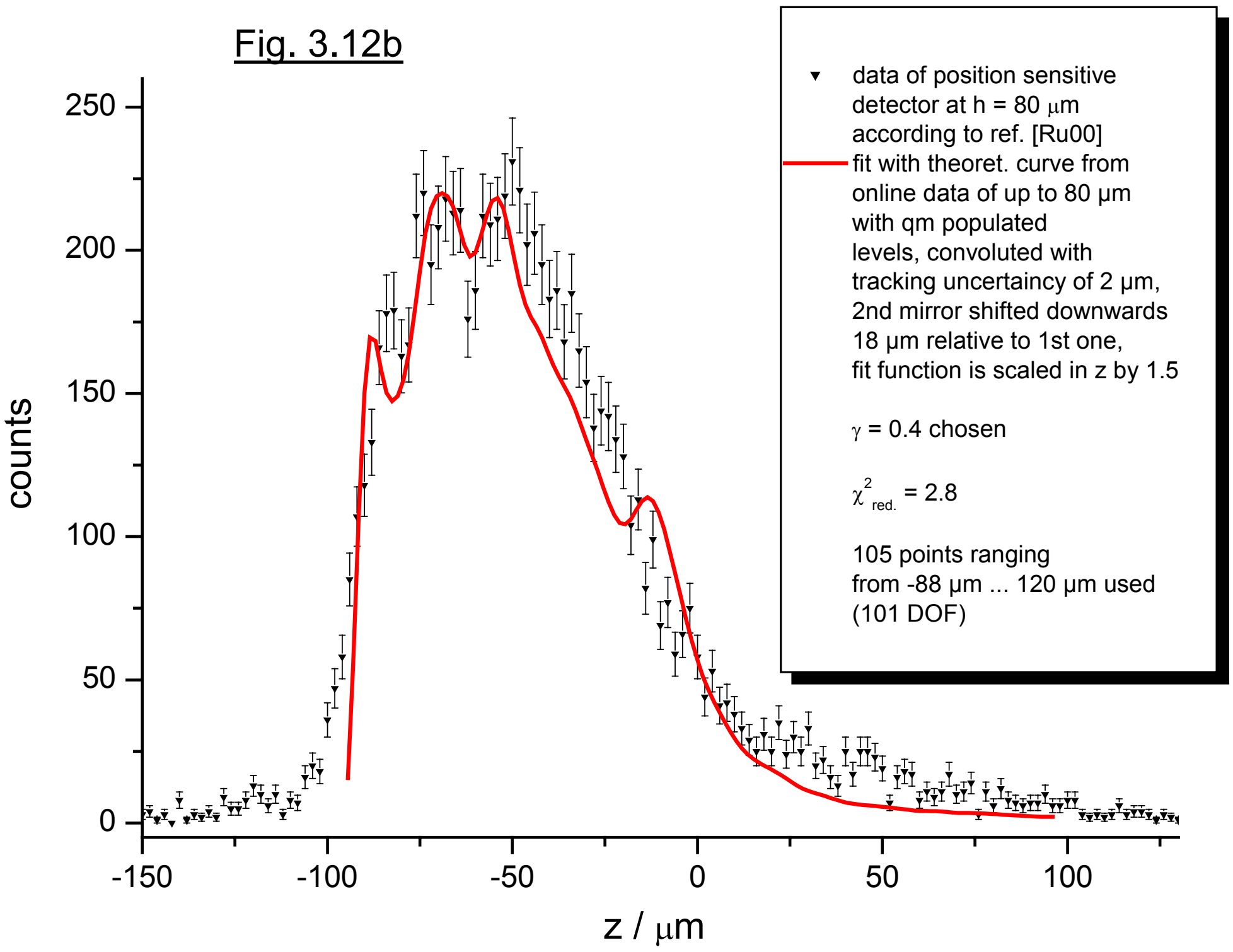

Fig. 3.12b: The prediction of eq. (3.3.4.3') for the actual measurement.Coherence $\gamma$ is set to 0.4 . Here the width of the prediction is scaled with the parameter *z-axis* = 1.5.

Yet, the scaled eq. (3.3.4.3') shows variations around $z$ = -15 µm, which are not seen in the data. On the contrary, if one looks at the error bars, the data appears to be readily smooth besides the two 1.5 sigma and 2 sigma variations around z = -80 µm and z = -60 µm.

Therefore, we should ask for possible causes of a vertical enlargement of the distribution eq. (3.3.4.3'). The most apparent effect producing such a widening is the possibility of diffraction at the exit edge of the absorber and at the 18 µm step between the two bottom mirrors. As we have vertically bound states, we should normally try to calculate the diffraction in terms of the bound states. However, a plane wave diffraction is much easier to calculate and it should provide us with the approximately results as long as we do not look at high orders in the diffraction expansion. The exit edge of the absorber as well as the edge of the step between the mirrors can be approximately regarded as a semi-infinite plane, which then diffracts the incident waves. The distribution of the wave field strength after these edges therefore will be described by properly attached functions of the type of eq. (*Fresnel*). After both the absorber's exit and the mirror step, there will be the product of two such contributions. One then detects a convo-lution of eq. (3.3.4.3') and the two diffraction patterns of the type of eq. (*Fresnel*) in the detector plane:

$$(3.3.4.5)\quad j_x^{diffr}(z)=\frac{1}{2}\left\|\left[F\left(-(z-h_{absorber})\cdot\sqrt{\frac{2}{\lambda\cdot\Delta\rho}}\right)-F(-\infty)\right]\cdot\left[F\left(-z\cdot\sqrt{\frac{2}{\lambda\cdot\Delta\rho}}\right)-F(-\infty)\right]\right\|*j_x^{(133')}(z)$$

where $*$ denotes the convolution with respect to $z$.

Yet, if the diffraction is the dominating part in the convolution eq. (3.3.4.5),then one should directly fit the pure diffraction pattern to the data. The result, that is seen in Fig. 3.13, justifies this assumption:

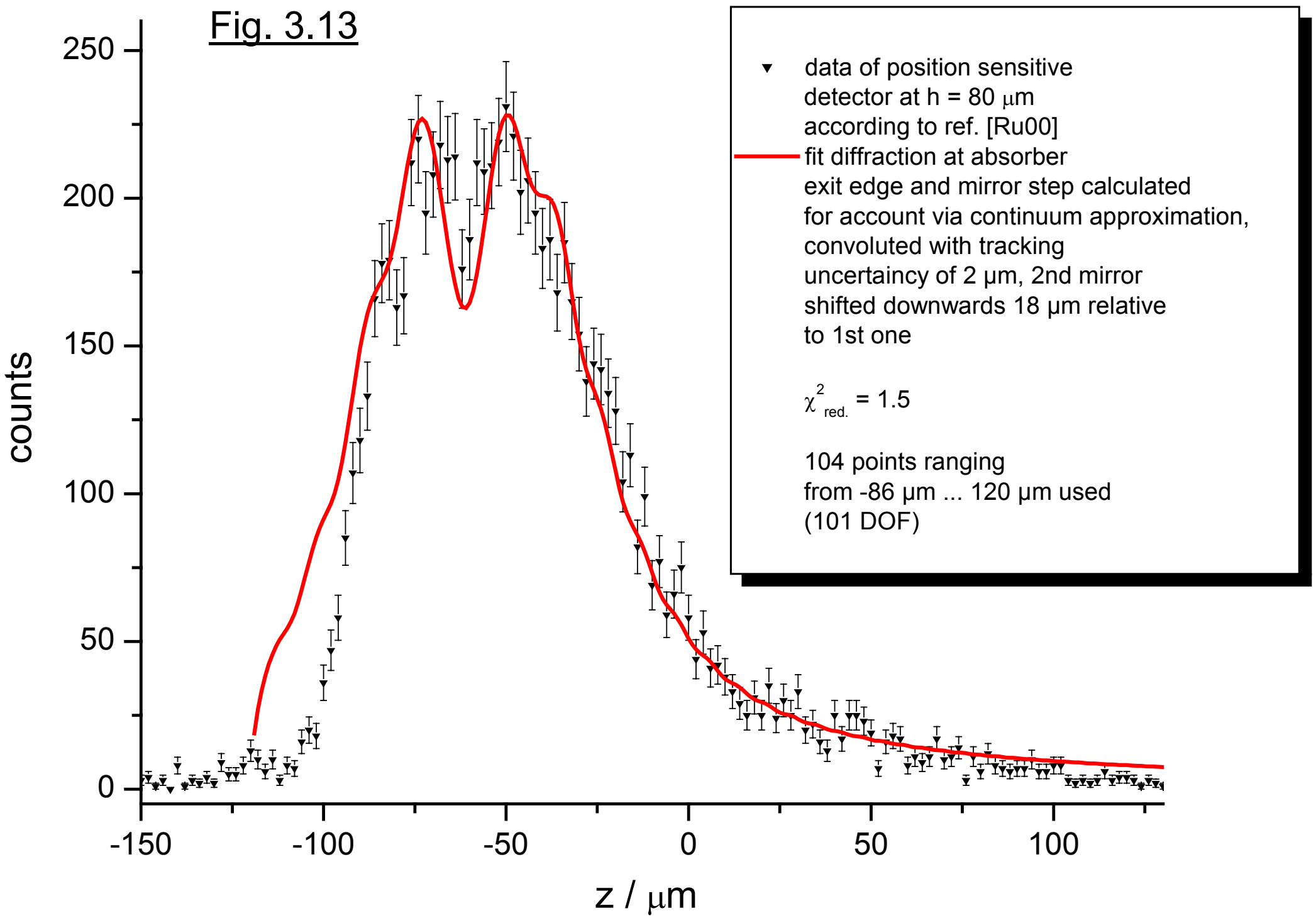

Fig. 3.13: The prediction of the diffraction pattern in eq. (3.3.4.5) for the actual measurement. The total scaling, a position offset, and the tracking uncertainty are the three fit parameters.

$\chi^2$ is quite better and additionally, this provides us with a well working explanation for both the tail in the data, which extends more than twice as wide as the central region of the curve, and the variations in the peak of the data at about $z$ = -50 μm.

Therefore, I regard the main structure and form of the vertical neutron distribution to be generated from a large impact of diffraction onto the final pure distribution eq. (3.3.4.3').

As an extension, we should now discuss the significance of possible variations in the data in the light of the fact, that even at the most pronounced peaks the error takes about 7% of the data value itself. As it was already done for the case of the measurement of the total flux as a function of the wave guide width, one can perform Monte Carlo simulations to obtain the vertical density distribution, which classical neutrons would show in the CR39-detector. If one again assumes, that the mirrors and the absorber (if it is not successfully absorbing the neutron) reflect specularly, then it is easy to incorporate both the facts, that the absorber has a classical absorption probability per collision of below one, and the presence of a step between the bottom mirrors. Fig. 3.14 shows a simulation's result conducted with about $2 \cdot 10^9$ neutrons arriving at the wave guide's entrance. The resulting approximation to the classical density distribution is therefore very smooth:

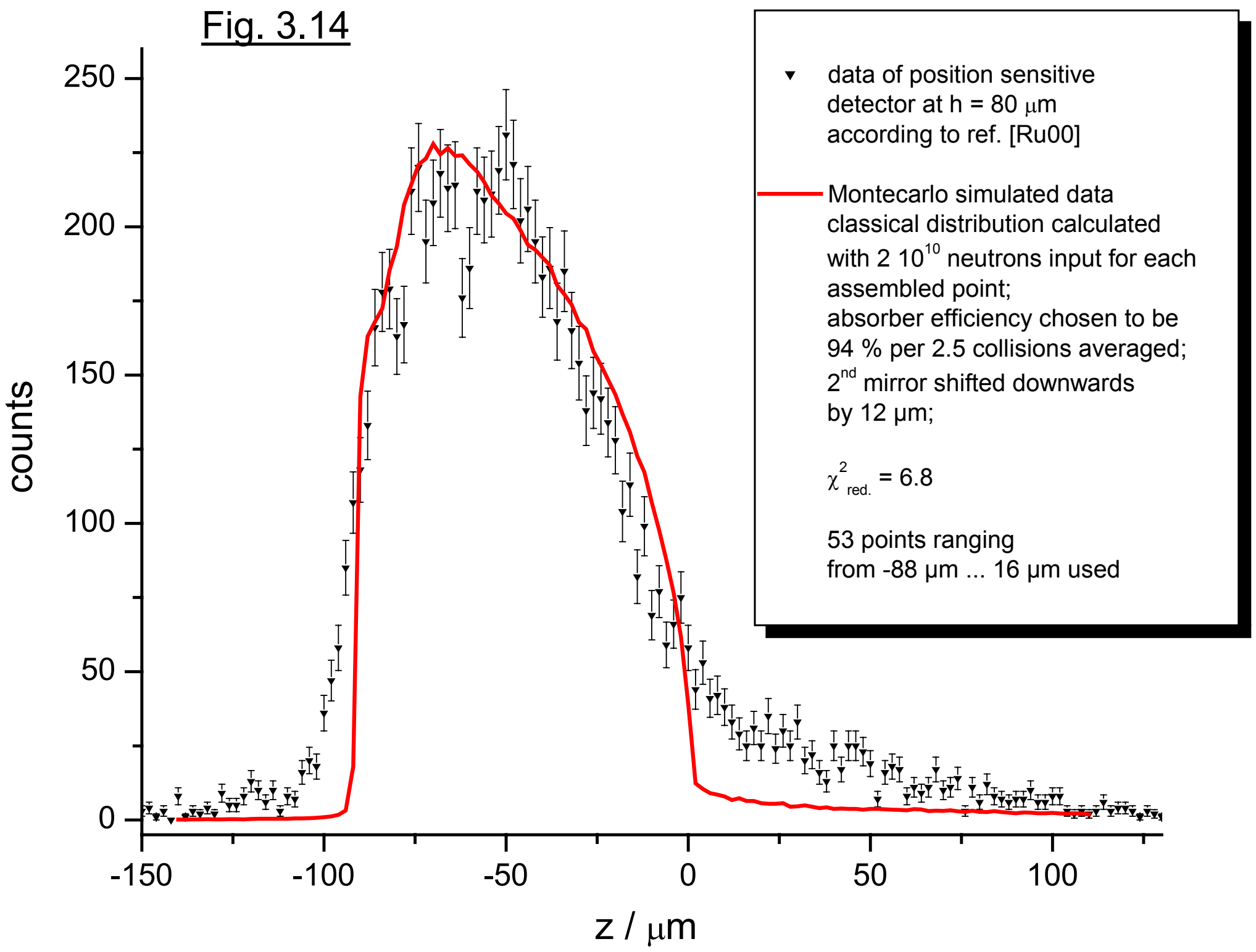

Fig. 3.14: Classical Monte Carlo simulation of the neutron distribution after the wave guide. A very high statistics of about $2 \cdot 10^{10}$ has been used to produce a smooth prediction.

This simulation shows an additional decrease in the intensity at $z < -70$ µm, which is due to the mirror step. But there is small agreement between this result and the data over the remain-ing range of $z$. In particular, the tails of the data cannot be reproduced by giving the absorber an absorption-probability per collision of below one.

However, more insight can be gained, if we perform the Monte Carlo simulation with the same final statistics as achieved in the measurement. This result is visible in Fig. 3.15:

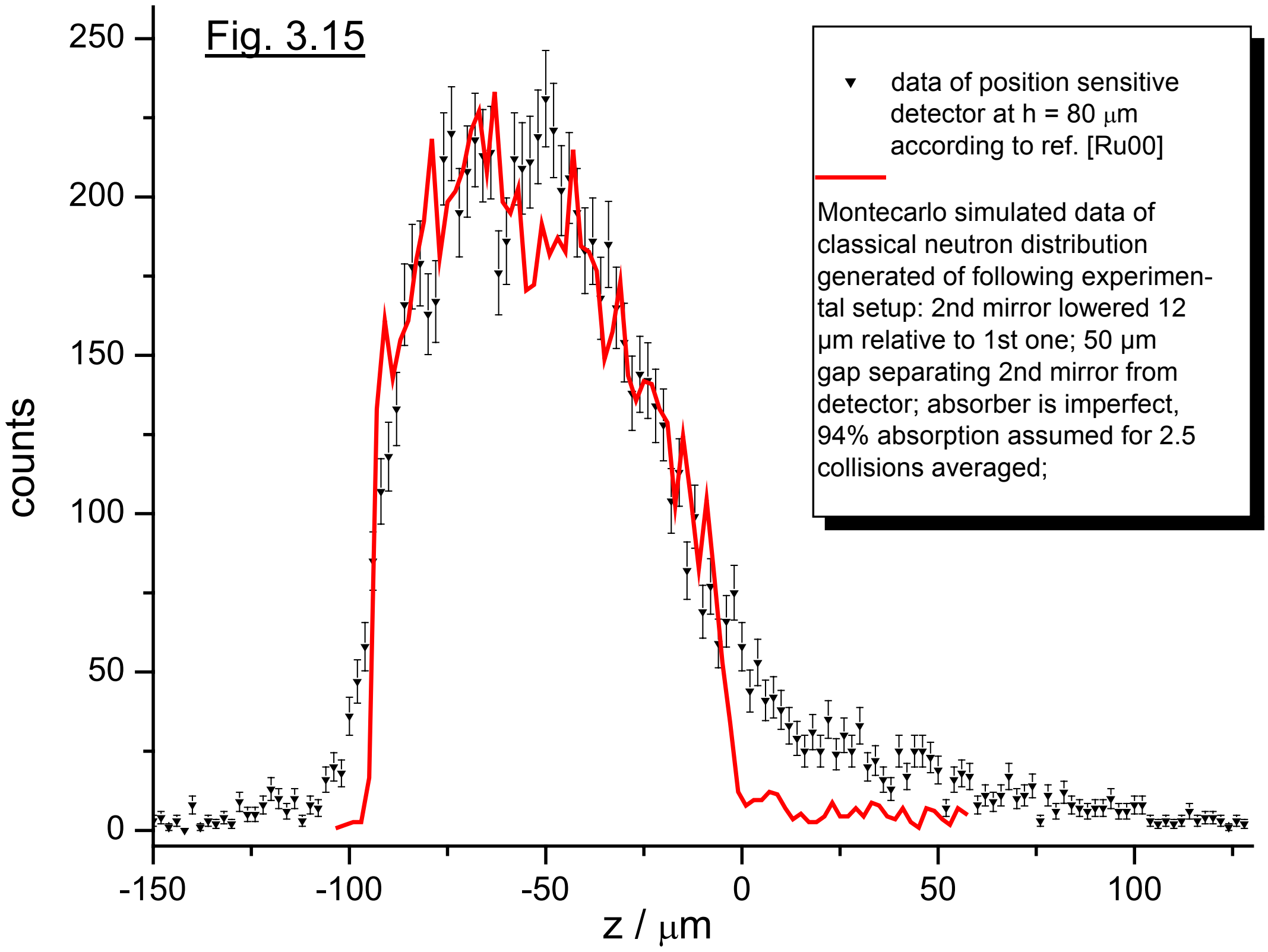

Fig. 3.15: Classical Monte Carlo simulation of the neutron distribution after the wave guide. The actual statistics of the measurement, called “real-time-statistics” has been used to check for the magnitude of statistical fluctuations in the measured data.

Here it is clearly visible, that the classical “real-time statistics” simulation shows exactly the same kind of intensity variations as are seen in the data. A probability estimate showed, that such fluctuations emerge in about 10% of all results. Therefore, here again is a need to perform measurements with an average (not minimal!) relative error of about 1%.

The theoretical analysis of the two type of measurements described in [Ru00], and shown in Fig. 3.2 and Fig. 3.11, respectively, here now comes to an end.

# Chapter 4

# How strange gravitation can be

What is gravitation? – I do not know the answer. Yet, this chapter will deal with some recent theoretical developments, that could show up new aspects of the fourth and weakest of all fundamental forces – provided one of them proves to be correct. These theories predict deviations from the Newtonian law of gravitation at distances below 1 mm. Since our experiment deals with quantized states bound by earth's gravitational field on energy scales of 1 peV corresponding to 10 μm length scale, deviations in the law of gravitation on a scale of 10 μm to 100 μm could lead to predictions like eq.s (3.3.1.31) and (3.3.4.3'), which differs from the case discussed in chapter three. The quantum mechanical model presented there assumed validity of Newtonian gravitation for the whole range of wave guide widths of about 5 μm to 160 μm.

The best description of the "general attraction of the masses" known today is given by Einstein's General Theory of Relativity. Since the energies and gravitational field strengths discussed here are very tiny fractions of the speed of light, Einstein gravity reduces to Newtonian gravity, its non-relativistic weak field limit. Newtonian gravitation is, in the language of local field theory, governed by a Poisson equation, which can be derived from Einstein's field equations for $\mu=\nu=0$ in the limit of weak fields:

$$R_{\mu\nu} - \frac{1}{2} \cdot g_{\mu\nu} R = -\kappa \cdot T_{\mu\nu} \quad , \quad g_{\mu\nu} = \eta_{\mu\nu} + h_{\mu\nu} \; with \left| h_{\mu\nu} \right| << 1 \; \forall \mu, \nu$$

$$\Rightarrow R_{00} = -\kappa \cdot T_{00} \quad , \kappa = \frac{8\pi G_4}{c^4} \qquad ,$$

$$\Rightarrow \quad (4.1) \quad \Delta\phi(\vec{r}) = 4\pi G_4 \cdot \rho(\vec{r})$$

where $\phi$ denotes the gravitational potential field and $\rho$ the mass density distribution. The general solution of this equation can be written using the Green's function of eq. (4.1) in the integral form:

$$(4.2)\quad \phi(\vec{r}) = -G_4 \cdot \int d^3r' \cdot \frac{\rho(\vec{r}\,')}{|\vec{r} - \vec{r}\,'|} \ .$$

For a spherically symmetric mass distribution on a compact carrier this reduces to the known form of Newton's law of gravitation:

$$(4.2')\quad \phi_{sphere}(r) = -G_4 \cdot \frac{M}{r} \ ,$$

where $M$ denotes the total mass of the mass distribution, $r$ the radial distance from its center of mass, and $G_4$ the three-dimensional constant of Newtonian gravity.
The most attractive modifications of eq. (4.2) now arise from two possibilites:

1) Dimensionality of space-time. This changes, for example, the power of distance $r$ in the gravitational law depending on how much spatial dimension a given manifold consists of.

2) There could be an additional "fifth" force coupling to the mass. This force could depend on the chemical composition of matter, thus violating the equivalence principle.

Both cases have been examined theoretically in recent research papers for the following reasons.
For more than sixty years theoreticians have been searching for a completely unified description of all four fundamental forces as well as all elementary particles within one consistent mathematical framework – a so-called "theory of everything". The first promising candidate for such a theory arose only within the last fifteen years. There the fundamental particles and field quanta are replaced by tiny, one-dimensional objects called "strings". They generate their own manifolds, that is space-time, on which their motion takes place. The quantization of their dynamics, which is somehow analogous to the quantized vibrational modes of a mechanical fiber or string, then gives the spectrum of the known particles, their masses and their four-dimensional couplings by means of an effective low-energy field theory. String theory possesses internal mathematical structures, that require a formulation of this theory on a background manifold of either exactly 10 or 26 spatial dimensions. All other values will lead to internal inconsistencies of the theory. Therefore, this is one reason to expect possible additional spatial dimensions in space-time, which is in principle point 1).
Furthermore, since Kaluza and Klein worked on modifications of Einstein gravity, it is known, that formulating the Einstein field equations in a (4+1)-dimensional space-time, and back-projection of these five-dimensional equations onto an effective four-dimensional mani-fold by means of compactification of the fourth spatial dimension onto small compact manifolds, like tori, leads to a system of two effective field equations: Four-dimensional Einstein gravity and Maxwell's equations! Therefore, the presence of additional dimensions seems to provide a (rather old) way to unify forces.

Since about 1997, however, the extra dimensions of modern string theories were thought to be compactified on manifolds with extensions in the order of the Planck length of about $10^{-35}$ m. This was assumed for a long time, since it is known, that the strength of classical Einstein gravity bcomes equal to the other three forces at this scale. Furthermore, if all interactions can use all spatial dimensions to propagate, then much larger compactification scales would have lead to discrepancies in present collider data and other experimental results. Colliders have proved the effective four-dimensionality of the three non-gravitational interactions to a very high precision down to a scale of $10^{-18}$ m.

Yet, recent progress on this subject has been made, since it was realized in [Ar98], that one can possibly localize the three non-gravitational interactions on (3+1)-dimensional sheets, so-called "branes", even in the case, that the compactification scale of the extra dimensions of the string theory is much larger than the Planck length. If the compactification scale of some of the extra dimensions is of order of µm or mm, then the non-gravitational interactions will behave quite normally due to this confinement to (3+1)-dimensional branes, while at distances below about 1 mm, gravitation shows up new behaviour, since it can use all the extra dimensions and is not thought to be tied to a brane. Such frameworks represent detailed realizations of point 1).

During the last ten years, there was an increasing number of circumstantial evidence, that the standard model of elementary particle theory has to be extended towards a theory called "supersymmetry". Furthermore, string theories predict the existence of scalar bosons coupling to the mass of ordinary particles via Higgs scalars. These so-called "moduli fields" then gather mass by breaking of the supersymmetry. If the supersymmetry breaking is not mediated by ordinary gravitation itself at the Planck scale but by gauge couplings, it can take place at energies close to the electro-weak scale. The forces originating from such massive moduli fields (in the case of gauge-mediated supersymmetry breaking) would have macroscopic interaction ranges of up to 1 mm. They provide the most thoroughly studied realization of possibility 2) (see [Di96]).

The first task now is to derive the modifications of eq. (4.2), which will arise from the scenarios described above. Therefore, I will give two short overviews over the realizations of the possibilities 1) and 2), which have been worked out by [Ar98] and [Di96]:

The framework of possibility 1), which I will use here, has been developed by [Ar98], whereas detailed calculations of the non-relativistic force laws within this framework can be studied in [Ke99], for example.

Now consider the case, that space-time is described by an $((n+p)+1)$-dimensional manifold, where $n = 3$ and $p$ is the number of additional spatial dimensions of this manifold. On such a manifold, non-relativistic gravitation would be governed by an $(n+p)$-dimensional Poisson equation:

$$(B.1) \quad \Delta_{3+p}\phi(x) = 4\pi G_{4+p} \cdot \rho(x) \quad , x = \left(\vec{r}, x_1, ..., x_p\right) .$$

Here, $G_{4+p}$ is the $(3+p)$-dimensional gravitational coupling on a $((3+p)+1)$-dimensional manifold. The general solution is again obtained via the $(3+p)$-dimensional Green's function.
Now one can impose compactification on this solution. In general this means, that the $p$ additional dimensions are transformed into $p$ subspaces, which curve back into themselves. Therefore, attached to each point $(r_1, r_2, r_3)$ of our normal configuration space there is a $p$-dimensional closed curved internal space. Because of the compactness of these curled up space coordinates, the $p$ additional dimensions will be periodic after compactification.
This compactified solution can be calculated for extremal cases of very small and very large distances as well as for intermediate distances, where the solution must be formulated in terms of the harmonical functions of eq. (B.1).
These expressions then show indeed, that the hierarchy problem of the interactions can be solved, if two of the $p$ compactified extra dimensions habe radii of several ten μm, whereas the lasting $p - 2$ extra dimension are compactified to the Planck scale. In this case, the $(4+p)$-dimensional strength of gravitation becomes unified with the other interactions at the weak scale – the extra dimensions scale the effective four-dimensional strength of gravitation.
Furthermore, the $r^{-1}$-dependency of Newtonian gravitation is modified by the extra dimensions, since the two large of them cause an additional Yukawa-like potential term, that has a range, which is equal to the compactification radius of the two large extra dimensions.
These results, that have been first obtained by [Ar98], as well as a summary of the detailed calculations, that can be found in [Ke99], one may resume in Appendix B. It is shown there, that the experimentally relevant result of this extra dimensions framework is given by a modified non-relativistic gravitational law eq. (B.12):

$$(B.12) \quad \phi(r) = -\frac{G_4 \cdot M}{r}\left(1+\alpha \cdot e^{-\frac{r}{\lambda}}\right) \quad where: \begin{cases} \alpha = 2 \cdot s \\ \lambda = \dfrac{1}{R} \end{cases} .$$

Here we see the emergence of an additional Yukawa-interaction generated by macroscopically compactified extra dimensions.

In the Appendix B, eq. (B.12) together with the values from eq. (B.7) show, that even present experimental data (see [Ad00]) cannot exclude the existence of even two macroscopically compactified additional dimensions in space-time, if the 6-dimensional unification scale is larger than 5 TeV, since this experiment excludes $\alpha > 1$ only for distances > 218 µm.

Regarding possibility 2), it should be noted here, that the moduli fields, which realize point 2) (the fifth force), and emerge in string theory, generate additional gravitation-like forces after the supersymmetry is broken, that have exactly the same form of eq. (B.12), as it is the case for additional dimensions. Such results have been obtained by [Di96] and are there discussed in detail. In particular, moduli forces can violate the equivalence principle, since they generally depend on the kind of matter they interact with. However, the moduli predict values of $\alpha$ in the range of $10^2$ to $10^4$ [Di96], where experimental constraints [Ad00] are more severe.

**4.1) "Online" measurement and the law of gravitation**

The experimental data discussed here obviously shows the formation of quantum mechanically bound states of ultacold neutrons in the earth's gravitational field. As these states form between the gravitational potential and the Fermi pseudopotential of a mirror and have spatial extension of 15 μm to 150 μm, they should react sensitively to the presence of modifications in the gravitational law like eq. (B.12).

The mirror contributes to the total gravitational potential felt by neutrons. Whereas the earth's potential is dominated by distances, which are orders of magnitude above the critical distance R < 1 mm, and is therefore genuinely Newtonian, the gravitational potential generated by the mirror glass plate will be felt by neutrons in distances as close as 15 μm to the glass' surface. If then $\lambda$ of eq. (B.12) is in the range of 15 μm to 150 μm, and $\alpha$ is strong enough, dynamics of neutrons above the mirror should change, since they feel the non-Newtonian part of the gravitational potential generated by the mirror itself.

Therefore, a very precise performance of the two types of measurements discussed in the first three chapters might be able to check the bound states formed above the glass mirror for possible deviations of the total gravitational potential seen by neutrons from the Newtonian form.

The statistics of present measurements are weak. This is particularly true for the regime of wave guide widths smaller than 50 μm, where the behaviour of the transmitted flux as a function of that width shows the strongest "quantum" behaviour.

However, according to [ILL01] the phase space density of UCN will be increased by five times due to use of enhanced neutron guides at the high flux research facility of the ILL, Grenoble. Together with a successor version of the apparatus, which is partially re-developed to provide much better long-term stability and determinability of parameters such as the wave guide width, this will provide a near-future prospect to repeat themeasurements with about 100 times the present statistics.

Furthermore, there are proposals [ILL01], which promise to give a new generation of UCN sources providing a UCN phase space density of more than $10^3$ times of the presently achievable densities. Once such sources are available, measurements can be performed with about $10^4$ times the present statistics.

Bearing this in mind, one should do the following:

First, the calculations of chapter three for a gravitational potential eq. (B.12) should be repeated as a function of the modification's strength $\alpha$ and its range $\lambda$.

Next, a Monte Carlo simulation from the quantum mechanical prediction eq. (3.3.1.31) for the total flux as a function of the wave guide width calculated for Newtonian gravity should be generated for the increased level of statistics discussed above.

Third, the results of eq. (3.3.1.31) for a gravitation eq. (B.12) will then be fitted to this Monte Carlo simulation.

The obtained $\chi^2$ – values form a function of $\alpha$ and $\lambda$, since the results of eq. (3.3.1.31) using eq. (B.12) are obtained for pairs ($\alpha$, $\lambda$). These $\chi^2$ – dependency will then give predictions of confidence interval regions in an $\alpha$-$\lambda$-plane, that can be reached by future measurements with increased statistics – provided, these measurements will not find deviations from Newtonian gravity.

Therefore, let us first compute the gravitational potential of a homogenous glass plate of thickness $D$. Since the considered distances above the surface are very small (<150 µm) compared to the extensions of the plate ( 10cm in length, D @ 1 cm), the rectangular form of the plate will not contribute to the result as long as we are not close to the boundaries of the glass plate. Therefore, it is convenient to represent the plate by a circular disk of radius $R$, thickness $D$ and mass density $\rho_0$.

The gravitational potential of a general mass distribution eq. (4.2) becomes, according to eq. (B.12):

$$(4.3)\quad \phi(\vec{r}) = -G_4 \cdot \int d^3r' \cdot \frac{\rho(\vec{r}\,')}{|\vec{r}-\vec{r}\,'|} \cdot \left(1+\alpha \cdot e^{-\frac{|\vec{r}-\vec{r}\,'|}{\lambda}}\right) .$$

We Now calculate $\phi$ for positions $z$ along the rotational symmetry axis of the circular disk, which is chosen to be the $z$-axis. Then it is convenient to use cylindrical coordinates $(\xi,\phi,z)$ , where $\xi$ and $\phi$ denote the radial distance and the angle in the plane of the disk, respectively.

Thus eq. (4.3) becomes:

$$(4.4)\quad \phi(z) = -G_4 \cdot \rho_0 \cdot \int_{-D}^{0} dz' \int_{0}^{2\pi} d\varphi \int_{0}^{R} d\xi \cdot \xi \cdot \frac{\left(1+\alpha \cdot e^{-\frac{\sqrt{\xi^2+(z-z')^2}}{\lambda}}\right)}{\sqrt{\xi^2+(z-z')^2}} \quad .$$

This integral can be split into the Newtonian and the non-Newtonian part and transformed further to give:

$$\begin{aligned}
(4.5)\quad \phi(z) &= -2\pi \cdot G_4 \cdot \rho_0 \cdot \int_{-D}^{0} dz' \cdot \left(\sqrt{R^2+(z-z')^2} - (z-z')\right) \\
&\quad - 2\pi \cdot \alpha \cdot G_4 \cdot \rho_0 \cdot \int_{-D}^{0} dz' \cdot \int_{z-z'}^{\sqrt{R^2+(z-z')^2}} du \cdot e^{-\frac{u}{\lambda}} \\
&= +2\pi \cdot G_4 \cdot \rho_0 \cdot \int_{z+D}^{z} d\eta \cdot \left(R - \eta + \frac{\eta^2}{2R} + O(\eta^3)\right) \\
&\quad - 2\pi \cdot \alpha \cdot \lambda \cdot G_4 \cdot \rho_0 \cdot \int_{z+D}^{z} d\eta \cdot \left(e^{-\frac{R+O(|\eta|^2)}{\lambda}} - e^{-\frac{\eta}{\lambda}}\right) \quad ,|z| << D << R \\
&\cong const. + 2\pi \cdot G_4 \cdot \rho_0 \cdot D \cdot z \cdot \left(1 - \frac{D}{2R}\right) \\
&\quad + 2\pi \cdot \alpha \cdot \lambda \cdot G_4 \cdot \rho_0 \cdot \int_{z+D}^{z} d\eta \cdot e^{-\frac{\eta}{\lambda}} \\
&= \phi_0 + 2\pi \cdot G_4 \cdot \rho_0 \cdot D \cdot z \cdot \left(1 - \frac{D}{2R}\right) - 2\pi \cdot \alpha \cdot \lambda^2 \cdot G_4 \cdot \rho_0 \cdot e^{-\frac{z}{\lambda}} \cdot \left(1 - e^{-\frac{D}{\lambda}}\right) \quad .
\end{aligned}$$

This potential contains two parts. The first one is linear in $z$ – reminiscence of the Newtonian part of eq. (B.12). The second one is an exponential, which describes the Yukawa interaction. If we now combine two parallel glass plates of the same thickness $D$, extension $R$ and density $\rho_0$ , then the linear Newtonian parts of the plates cancel each other. Thus, the earth's dominating linear potential, and the two Yukawa terms of the plates survive in the potential of two identical and parallel glass plates. Therefore, we have:

$$(4.6)\quad \phi_{tot}(z) = g \cdot z - 2\pi \cdot \alpha \cdot \lambda^2 \cdot G_4 \cdot \rho_0 \cdot \left(e^{-\frac{z}{\lambda}} + e^{\frac{z-h}{\lambda}}\right) \quad .$$

The classical prediction of the total flux as a function of the wave guide width $h$, which corresponds to eq. (4.6), is proportional to the available phase space volume of the system. This is given according to eq.s (3.2.3) and (3.2.2”) as:

$$(4.7) \quad \dot{N}_{tot} \propto \cdot \int_{o}^{h} dh' \cdot \sqrt{2g \cdot \left(\phi(h) - \phi(h')\right)}$$

as long as the potential eq. (4.6) is a monotonically increasing function of z. This is guaranteed, if:

$$\phi'_{tot}(z) \geq 0 \quad \forall z \quad \Leftrightarrow \quad g \geq \left|2\pi \cdot \alpha \cdot \lambda \cdot G_4 \cdot \rho_0\right| \ .$$

Since glass has a density of about 3 $g/cm^3$ and we are interested in ranges of the Yukawa terms of $\lambda$ ° 50 µm, this condition corresponds to $\alpha < 10^{11}$. Present experimental limits (see [Ad00]) for $\lambda < 50$ µm, however, are below $\alpha$ ° $10^7$, therefore this constraint on $\alpha$ will be no problem. A look at both the eq.s (4.7) and (4.6) together now shows, that eq. (4.7) will essentially be proportional to $h^{1.5}$ , the known classical behaviour, if $h$ t 3 $\lambda$ .
Therefore, it will be possible to predict confidence levels on $\alpha$ and $\lambda$ from the classical formulas fitted to Monte Carlo Simulations for $h$ > 50 µm, since the measured curves will only show semi-classical behaviour in this regime.

Monte Carlo simulations of the classical Newtonian behaviour, which is $h^{1.5}$, can be performed the same way, as described in section two of chapter three. The result of such simulations for a statistics 100 times and $10^4$ times the present measurement’s statistics is given in the two plots of Fig.s 4.1a and 4.1b :

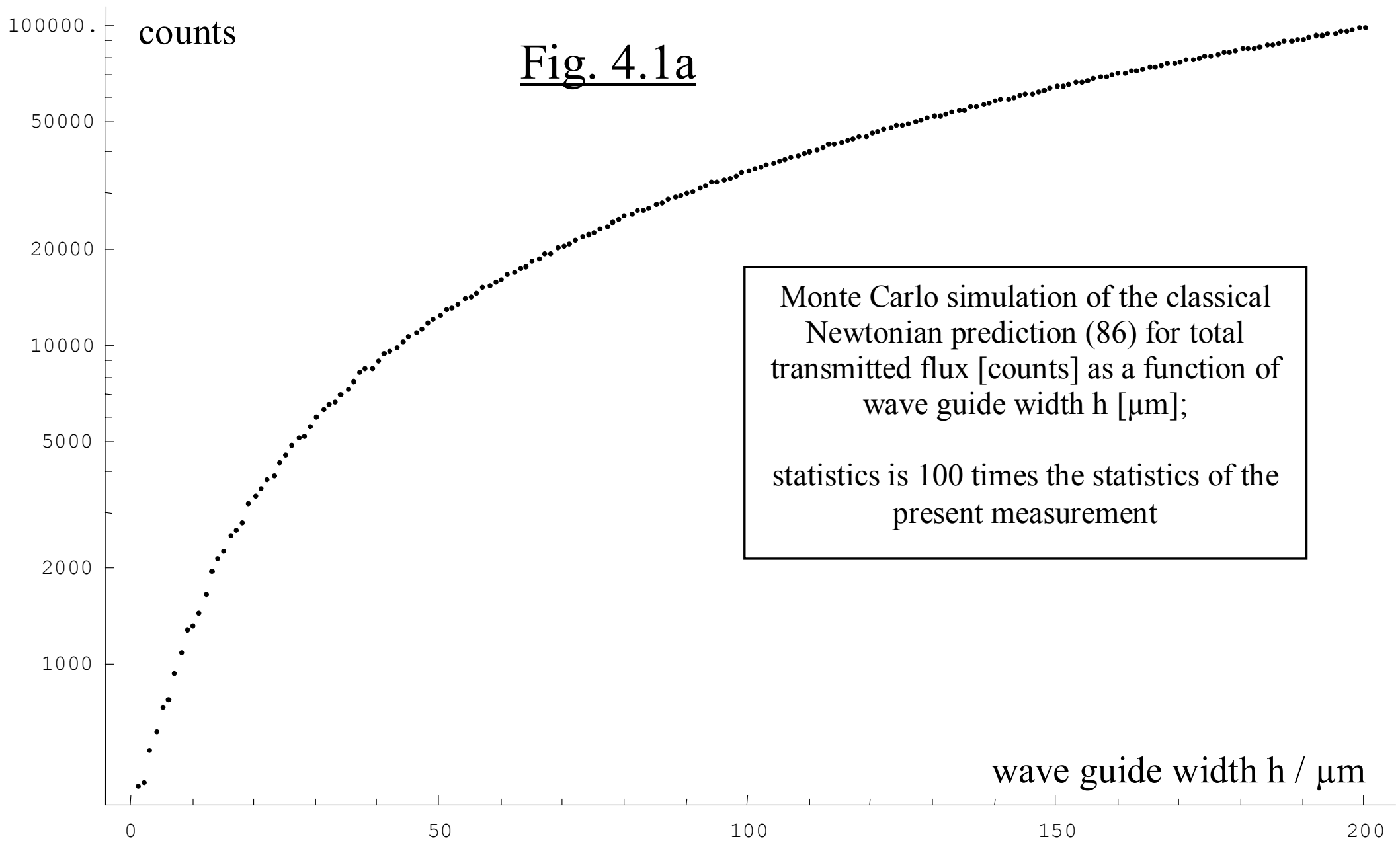

Fig. 4.1a

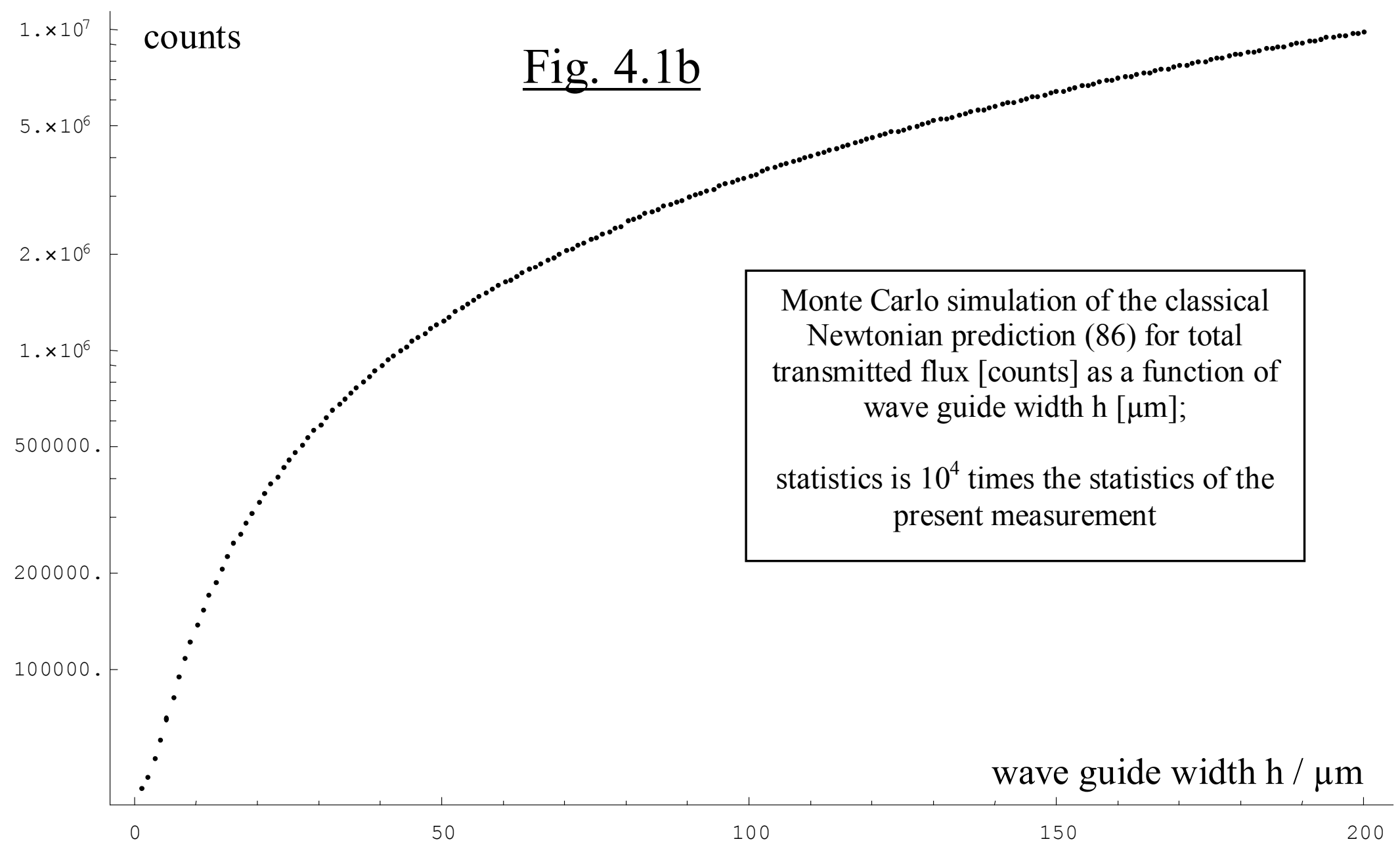

This simulated data was fitted with the eq.s (4.7) and (4.6) for $h$ > 50 µm, while a certain "grid" of values of α and λ was used subsequently in eq. (4.6). Since the $\chi^2$ of the Newtonian prediction for $\alpha = 0$ is close to unity, the other values of $\alpha$ and $\lambda$ will usually lead to worse $\chi^2$. This allows one to derive confidence limits on both $\alpha$ and $\lambda$. Such confidence limits have been calculated for both kinds of simulated statistics using $\Delta\chi^2$ for two independent parameters at a confidence level (C.L.) of 90 % and 67 %, respectively, as an exclusion criterion for the simulation with $10^4$ times the present statistics, and one parameter at a confidence level of 67 % as an exclusion criterion for the simulation with 100 times the present statistics.

The limits are seen in Fig.s 4.2a and 4.2b:

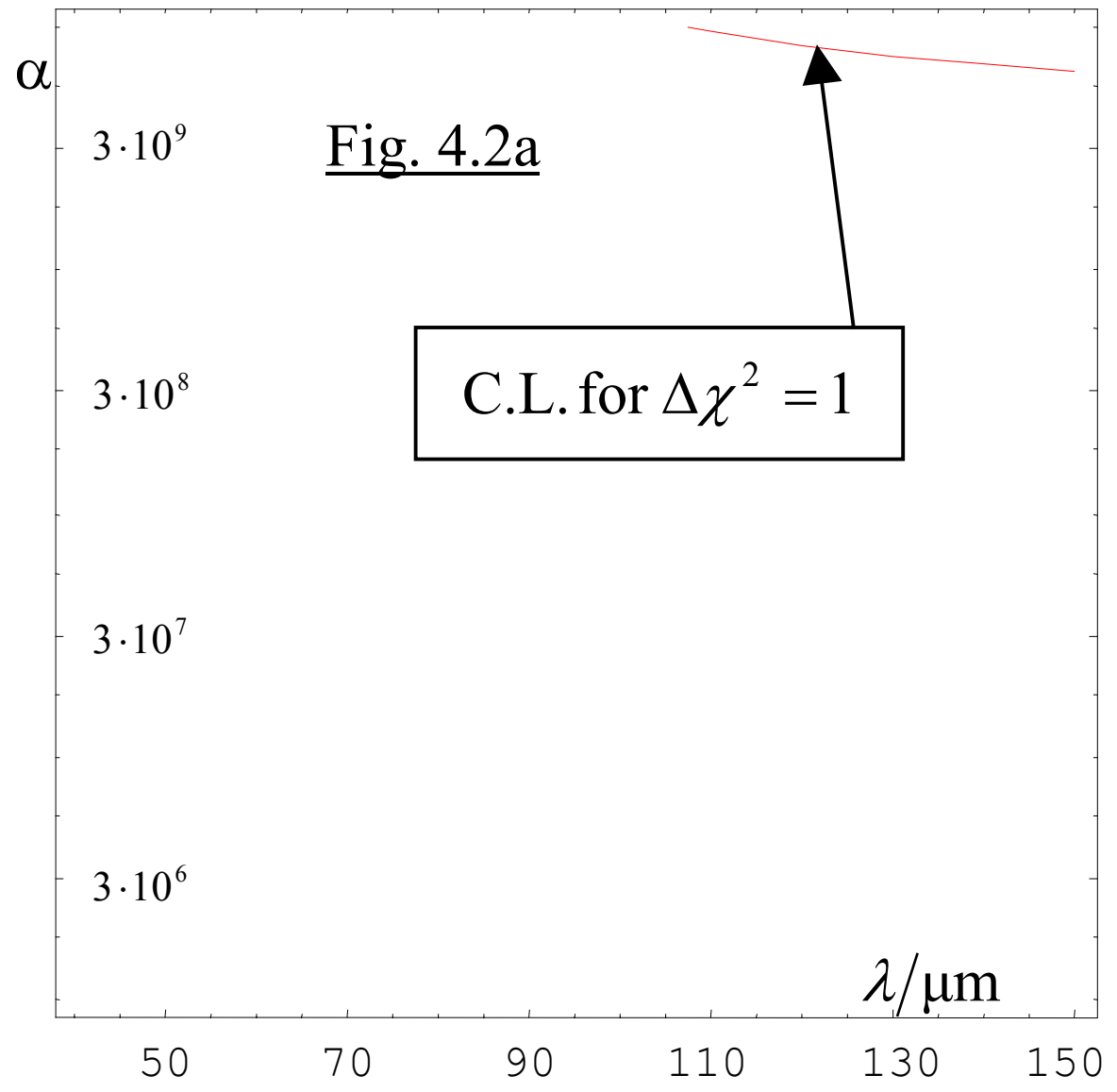

Fig. 4.2a: Prediction of confidence limits (C.L.) on $\alpha$ and $\lambda$ for 100 times the present statistics

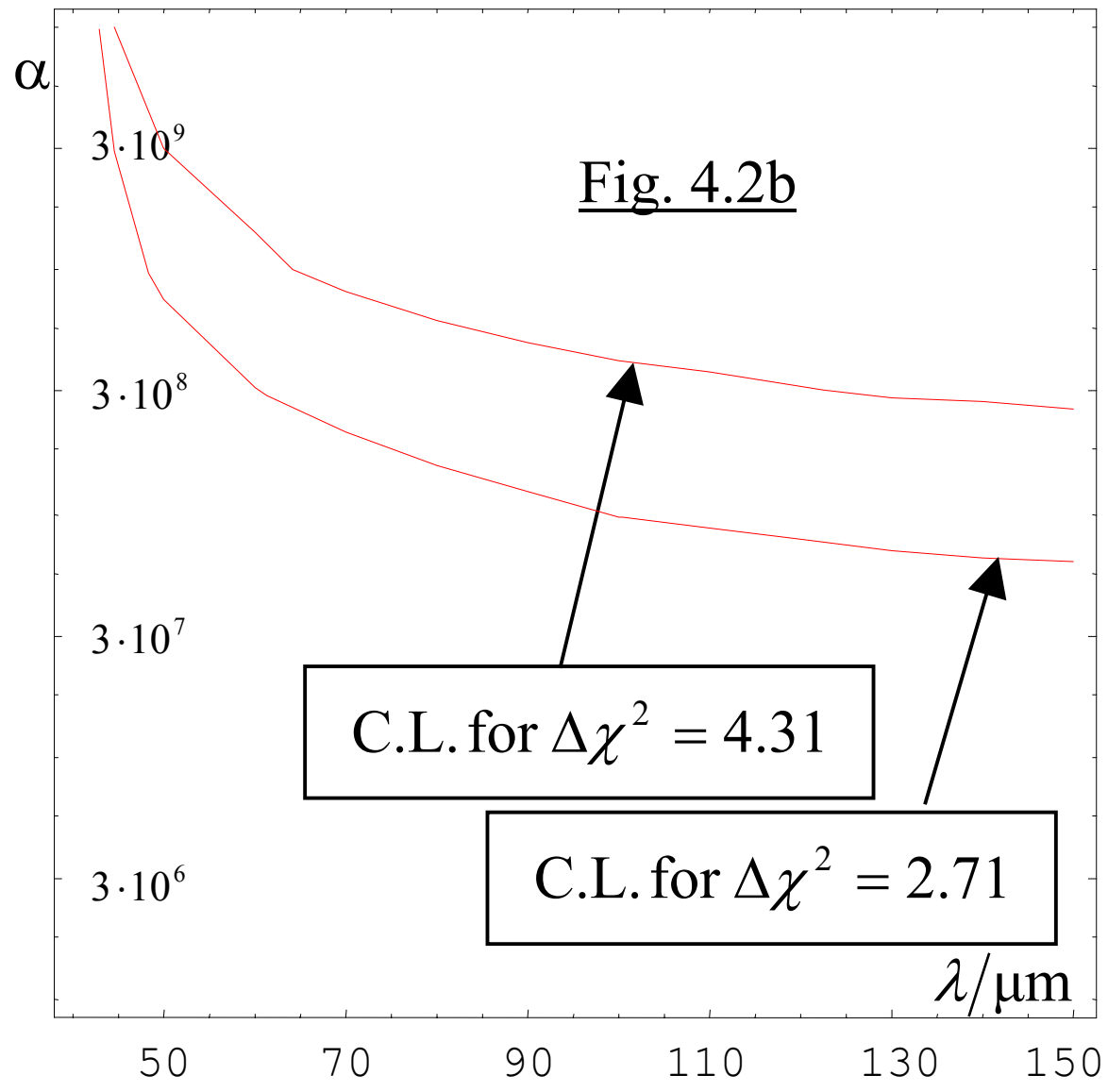

Fig. 4.2b: Prediction of confidence limits (C.L.) on $\alpha$ and $\lambda$ for $10^4$ times the present statistics

The same procedure can now be applied to the quantum mechanical prediction as well. The prediction for the dependency of the fuls on the wave guide width, eq. (3.3.1.31), for Newtonian gravity is used to generate a Monte Carlo simulation for 100 times and $10^4$ times the present statistics.

Then the prediction eq. (3.3.1.31) must be calculated for the Yukawa terms in eq. (4.6) as a modification to the Newtonian potential. The eigenvalues have to be calculated independently for each single pair $(\alpha, \lambda)$ for this purpose. Here one can again use the WKB approach presented in section 3.3.2 to treat eq. (4.6) as long as this potential is monotonically increasing over the whole range of $z$-values. Since this is the case for $\alpha < 10^{11}$, even if $\lambda$ is chosen as large as 50 µm, the Yukawa terms of eq. (4.6) can be treated with the WKB method. $\alpha$-values of about $10^{10}$ are exluded for all $\lambda > 20$ µm by present experiment in [Ad00].

Fig. 4.3 now displays the resulting confidence limits on that follow from the fits of eq. (3.3.1.31) using the potential eq. (4.6) as a function of $\alpha$ and $\lambda$ to data, that has been Monte Carlo simulated from the Newtonian prediction eq. (3.3.1.31) with $10^4$ times the present experimental statistics.

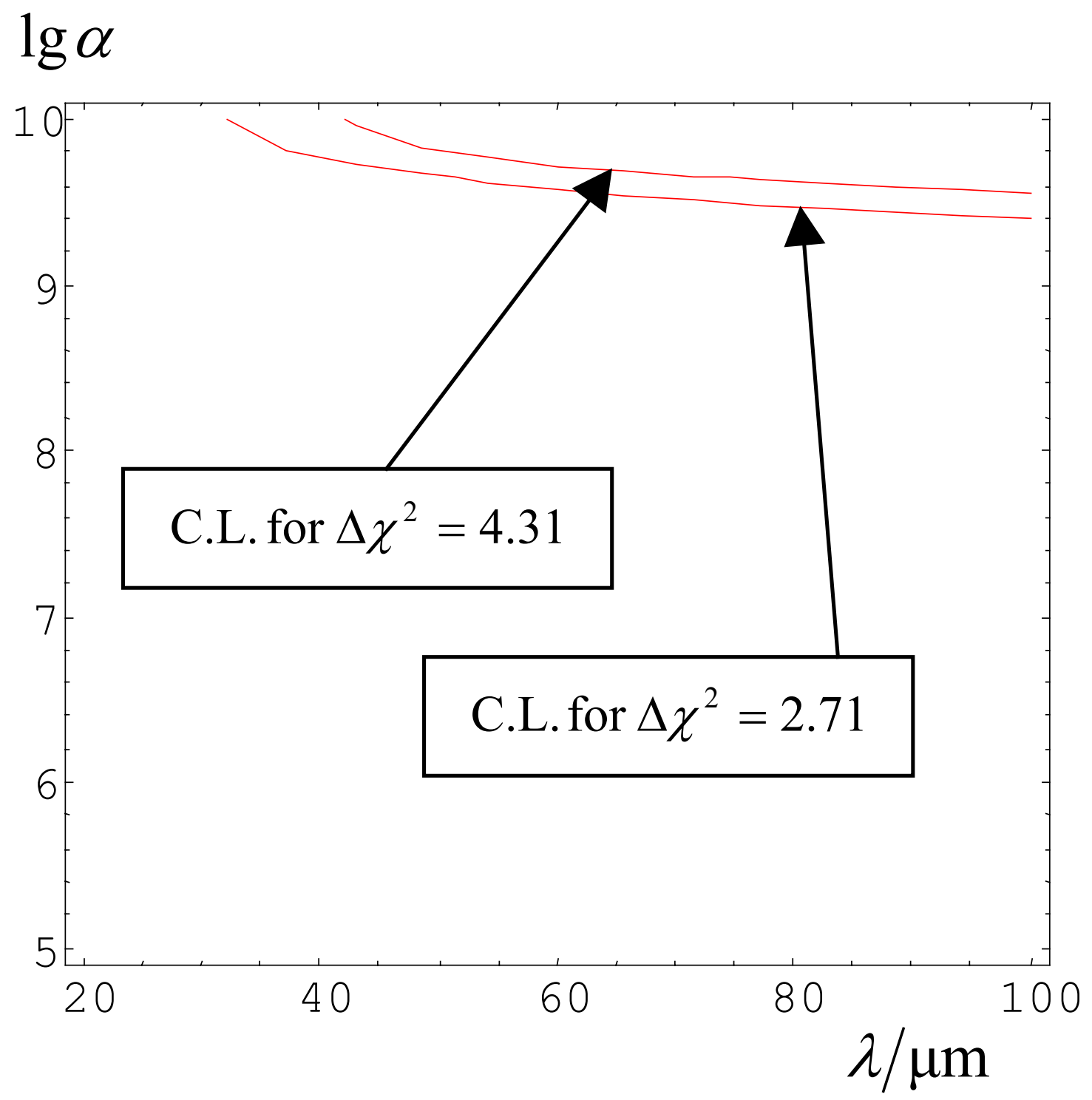

Fig. 4.3: Prediction of confidence limits (C.L.) on $\alpha$ and $\lambda$ for $10^4$ times the present statistics derived from the quantum mechanical theory given by eq.s (3.3.1.31) and (4.6)

The fits of the Yukawa dependent predictions eq. (3.3.1.31) to the Monte Carlo simulated quantum mechanical curves, which are valid for Newtonian gravity, have been performed nearly the same way as done for the Newtonian prediction in section 3.3.2).
However, there was one difference: There is no offset parameter for the *z*-position any more. The offset has been determined for the Newtonian prediction in section 3.3.2), and it has been used as a constant offset for the non-Newtonian fits. This was done, because eventually arising additional offsets in *z* will be due to the effects of the Yukawa potential terms in eq. (4.6) and should therefore not be fitted "away". On the contrary, they will contribute to possible increases of the $\chi^2$-values of the non-Newtonian fits.

Finally, we have to face the possibility, that this indirect measurement of gravitationally bound quantum states (it accesses the vertical motion by determining the transmission factors of the horizontal motion) may be not sensitive enough to detect deviations from Newtonian gravity within the interesting ranges of parameters. Despite the fascinating primary energy resolution of this measurement of around $10^{-12}$ eV, determining the eigenvalues of vertically bound motions by connecting their imaginary parts to the transmission factors of horizontal motion using an absorber might "smeare" out more precise information, since we do not directly measure the canonical observables of the Hamiltonian of quantized vertical motion.

# Summary

In the summer of 1999 an experiment was carried out at the Institute of Max Laue and Paul Langevin, Grenoble, that for the first time tried to prove the existence of quantized bound states of massive particles in a gravitational field ([Ru00], [Ne01], [Ne00]). If quantum mechanics and gravitation work together in the regime of the non-relativistic, classical Newtonian law, one does expect the formation of bound states of particles, that move within a gravitational cavity. Such a device can be formed by a bottom mirror of glass and the earth's gravitational field, if neutrons of very low velocities, so-called "ultracold neutrons" (UCN) move above the glass mirror. UCN reflect from ordinary solid state materials under all angles of incidence.

If an absorber is added and placed several ten μm above the bottom mirror, then this system can be regarded as a one-dimensional wave guide filled by a homogenous gravitational field. Inside such a wave guide bound states should form in the vertical motion of the neutrons. If the neutrons flow through wave guide, then neutrons in different bound states of the vertical motion will be differently absorbed by the absorber. Thus, the transmittivity of this device could show up a quantized "stepwise" behaviour as a function of the width of the wave guide.

This transmittivity was measured in the experiment. In this work now, a description of this kind of measurement was derived from the first principles of quantum mechanics and the Newtonian law of gravitation. The joint application of these principles give rise to a strict formalism, that finally leads to a prediction for the transmittivity of the wave guide as a function of its width.

Whereas classical estimates of the transmittivity as well as descriptions that neglect the gravitational field fail to describe the measured data, the prediction, that is derived from both quantum mechanics and gravitation, is able to reproduce the result of the transmittivity measurement in all its important features with a three-parametric non-linear least squares fit, that yields $\chi^2 / DOF = 1.6$ .

This result, that has been checked by using two different methods to obtain the energy eigenvalues of the quantized vertical motion, finally allows one to fully establish the formalism of "first quantization" for non-relativistic classical gravitation. The formation of quantum states of massive particles in a gravitational was indeed shown by the experiment in question.
The experiment performed a second measurement, that recorded the vertical spatial distribution of the neutrons, that left the wave guide. It is shown within the model, that is developed here, that this distribution can be properly described by the effects of diffraction, that takes place at the backward egdes of the absorber and the mirrors. A fit to the data of the neutron distribution with this diffraction approach yields a complete description with a $\chi^2/DOF = 1.5$. Three parameters have been adjusted here.

Finally it should be realized, that this experiment detects the influence of gravitation on the motion of massive particles on a distance scale of 10 μm to 100 μm. Since recent theoretical developments within the framework of string theory suggest the existence of macroscopically compactified additional dimensions of spacetime ([Ar98]), there is some (new) motivation to search for deviations from the Newtonian law of gravitation on distance scales between 1 μm and 1 mm. Therefore, the analysis of the transmittivity of the "gravitational" wave is refined in this work to yield predictions as a function of the parameter space ($\alpha$, $\lambda$), that controls the non-Newtonian modifications of the gravitational law, that arise from the extra dimensions. These results are then used to search for confidence limits on the parameter space of the modifications.
It comes out, that the present experimental data is statistically too weak to allow for the declaration of reasonable confidence limits on $\alpha$ and $\lambda$. Therefore, predictions are carried out on confidence limits, that could be reached on (near) future measurements with increased statistics.

# Appendix A

# Airy functions

**(A.1)** Defining differential equation:

$$\varphi''(\eta) - \eta \cdot \varphi(\eta) = 0 \quad .$$

**(A.2)** Airy functions can be expressed via Bessel functions:

$$\begin{cases} \phi_A(\eta) = \frac{1}{3}\sqrt{\eta} \cdot \left[ I_{-1/3}\left(\frac{2}{3}\eta^{3/2}\right) - I_{+1/3}\left(\frac{2}{3}\eta^{3/2}\right) \right] \\ \phi_B(\eta) = \frac{\sqrt{3}}{3}\sqrt{\eta} \cdot \left[ I_{-1/3}\left(\frac{2}{3}\eta^{3/2}\right) + I_{+1/3}\left(\frac{2}{3}\eta^{3/2}\right) \right] \end{cases} \quad .$$

The basic types of Airy functions are:

$$Ai(\eta) = \phi_A(\eta) \quad ,$$
$$Bi(\eta) = \phi_B(\eta) \quad .$$

**(A.3)** There exist three pairs of linearly independent solutions of eq. (A.1), which are called “Airy functions”. These are:

$$Ai(\eta) \quad , \quad Bi(\eta)$$
$$Ai(\eta) \quad , \quad Ai(e^{i \cdot 2\pi/3}\eta)$$
$$Ai(\eta) \quad , \quad Ai(e^{-i \cdot 2\pi/3}\eta) \quad .$$

**(A.4)** The relations between $Bi(\eta)$ and the complex rotated $Ai(\eta)$ of (A.3) are given by:

$$Bi(\eta) = e^{i\cdot\pi/6} \cdot Ai(e^{i\cdot 2\pi/3}\eta) + e^{-i\cdot\pi/6} \cdot Ai(e^{-i\cdot 2\pi/3}\eta)$$
$$Ai(e^{\pm i\cdot 2\pi/3}\eta) = \frac{1}{2} e^{\pm i\cdot\pi/3} \cdot \left[Ai(\eta) \mp i \cdot Bi(\eta)\right] \quad .$$

**(A.5)** Behaviour of the linearly indepent solutions (A.3):

$\phi_A(z)$:

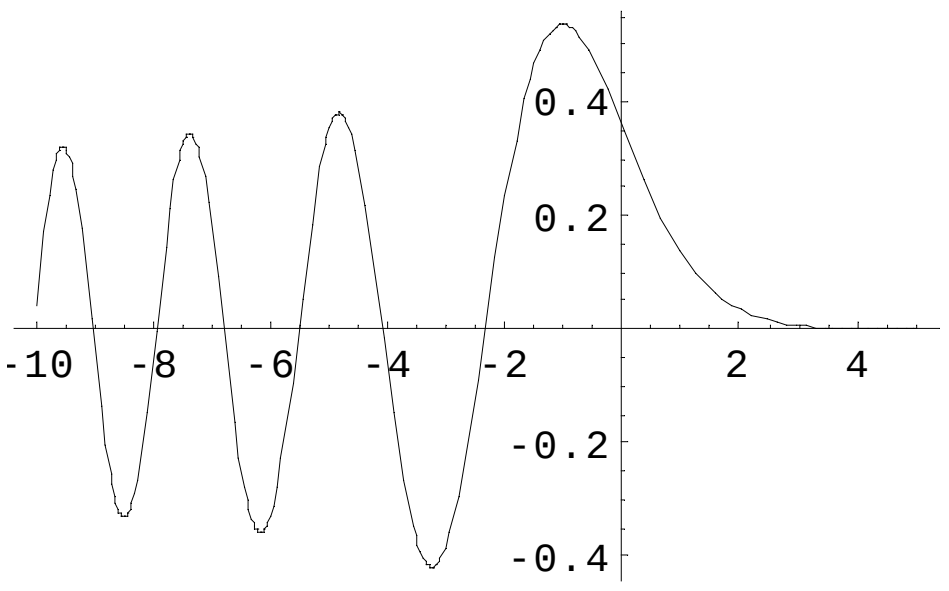

$\phi_B(z)$:

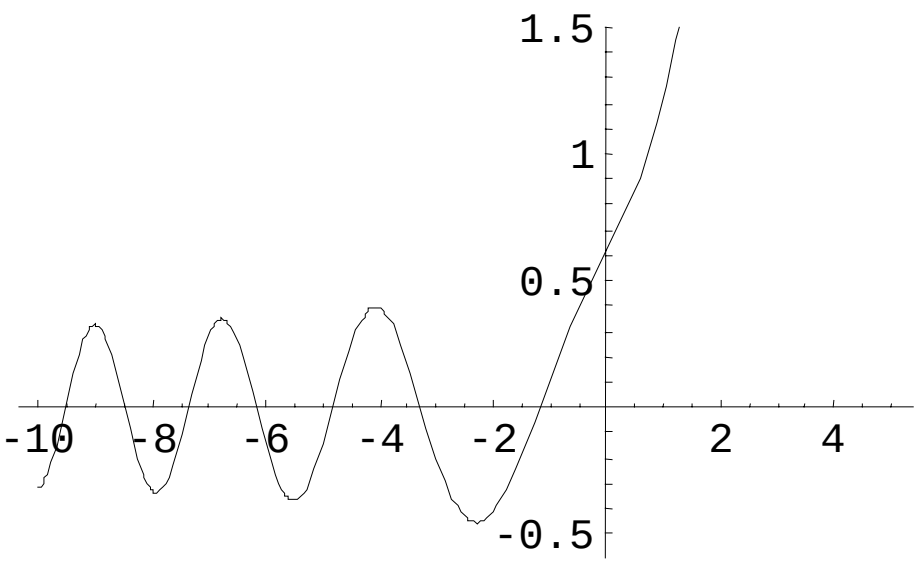

<u>$Ai(e^{i\cdot 2\pi/3}\eta)$:</u>

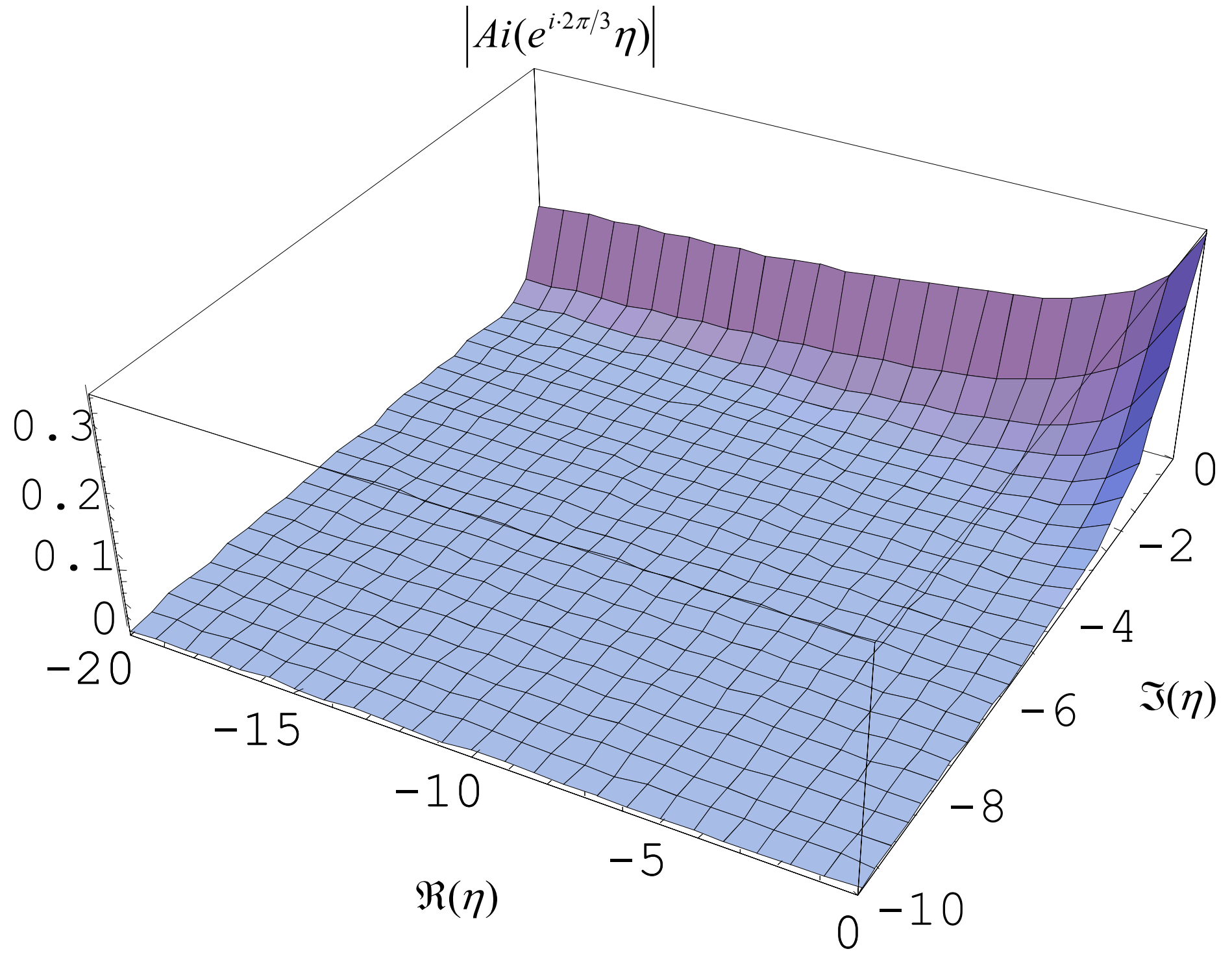

$\underline{Ai(e^{-i\cdot 2\pi/3}\eta):}$

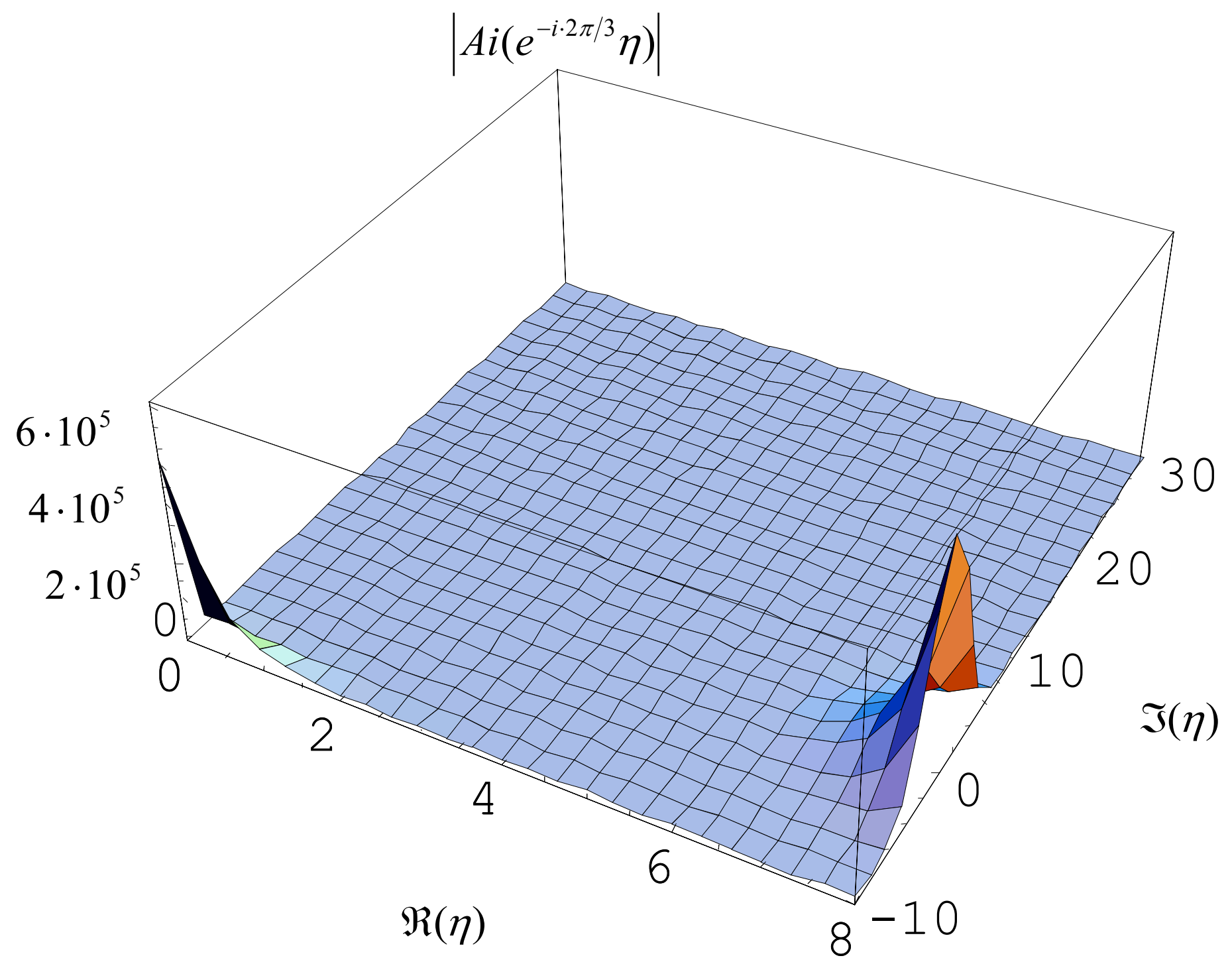

We see, that Ai and Bi behave non-pathologically only on the real axis, whereas the complex rotated Ai show controllable behaviour in certain quadrants of the complex plane. The behaviour outside the argument regions, for which (A.3) is plotted here, is marked by exponential explosion.

**(A.6)** Generalization of eq. (A.1):

$$\varphi''(\eta)+\left(\varepsilon-(1+\kappa)\cdot\eta\right)\cdot\varphi(\eta)=0$$

$$\Rightarrow \quad \varphi(\eta)=\begin{cases} Ai\left(\sqrt[3]{1+\kappa}\cdot\eta-\varepsilon/(1+\kappa)^{2/3}\right)\\ Bi\left(\sqrt[3]{1+\kappa}\cdot\eta-\varepsilon/(1+\kappa)^{2/3}\right)\\ \textit{or linear combinations according to } \text{(A.3)} \end{cases} .$$

# Appendix B

# Additional dimensions of space-time – the "fifth" force

The framework presented here has been developed by [Ar98], whereas detailed calculations of the non-relativistic force laws within this framework can be studied in [Ke99], for example.

Now consider the case, that space-time is described by an (($n$+$p$)+1)-dimensional manifold, where $n$ = 3 and $p$ is the number of additional spatial dimensions of this manifold. On such a manifold, non-relativistic gravitation would be governed by an ($n$+$p$)-dimensional Poisson equation:

$$(B.1)\quad \Delta_{3+p}\phi(x) = 4\pi G_{4+p}\cdot\rho(x) \quad , x = \left(\vec{r}, x_1,\ldots,x_p\right) .$$

Here, $G_{4+p}$ is the (3+$p$)-dimensional gravitational coupling on a ((3+$p$)+1)-dimensional manifold. The general solution is again obtained via the (3+$p$)-dimensional Green's function, resulting in:

$$(B.2)\quad \phi(x) = -G_{4+p}\cdot\int d^{3+p}x'\cdot\frac{\rho(x')}{\left[\left(\vec{r}-\vec{r}'\right)^2+\sum_{j=1}^{p}\left(x_j-x_j'\right)^2\right]^{\frac{p+1}{2}}} .$$

Now we will impose compactification on this solution. In general this means, that the $p$ additional dimensions are transformed into $p$ subspaces, which curve back into themselves. Therefore, attached to each point ($r_1$, $r_2$, $r_3$) of our normal configuration space there is a $p$-dimensional closed curved internal space. Because of the compactness of these curled up space coordinates, the $p$ additional dimensions will be periodic after compactification.

Consider now the special case where all $p$ additional dimension are compactified on tori with radii $R_j$ . Then the coordinates $x_j$ will be (2π $R_j$)-periodic. We will now calculate from eq. (B.2) the gravitational potential of a spherically symmetric and compact mass distribution of total mass $M$ for the case of toroidal compactified $p$ additional dimensions. In transforming eq. (B.2) to this case, the appropiate boundary conditions have to be applied.

$\phi$ must vanish for transfinite three-dimensional distances $r$ and it must be periodic in the $p$ additional coordinates on the $p$ tori:

$$(B.3)\quad \begin{cases} \lim\limits_{r\to\infty}\phi(x) = 0 \\ \phi(\vec{r},x_1,\ldots,x_j,\ldots,x_p) = \phi(\vec{r},x_1,\ldots,x_j + 2\pi\cdot R_j,\ldots,x_p) \quad \forall j \end{cases}$$

$$where: \quad x = (\vec{r},\vec{x}) = (\vec{r},x_1,\ldots,x_j,\ldots,x_p)$$

With these boundary conditions, the analogon to eq. (4.2') can be written as:

$$(B.4)\quad \phi(x) = -\sum_{\vec{m}\in Z^p} \frac{G_{4+p}\cdot M}{\left[r^2 + \sum\limits_{j=1}^{p}\left(x_j - 2\pi\cdot R_j\cdot m_j\right)^2\right]^{\frac{p+1}{2}}} \quad ,\ \vec{m} = (m_1,\ldots,m_p) \quad .$$

As denoted, $\vec{m}$ is a $p$-dimensional vector of the $p$-dimensional integer lattice Ù$^{p}$ . This is the general solution for a spherically symmetric body living in the (3+1)-dimensional space-time, where the $p$ additional spatial dimensions are compactified on tori with radii $R_j$ .

Two limiting cases can be constructed now:

a) We are deep inside the $p$ additional curved spaces, that is, $r \ll R_j$ for all $j$. Then the dominating term in the sum over $\vec{m}$ in eq. (B.4) is the one with $\vec{m}$ = 0. All other terms just give tiny contributions due to $R_j \gg r$ for all $j$. Then we have:

$$(B.4a)\quad \phi(x) \cong -\frac{G_{4+p}\cdot M}{r_{3+p}^{p+1}} \quad ,\ r_{3+p} = \sqrt{r^2 + \sum_{j=1}^{p} x_j^2} \quad .$$

This is a result already expected from Gauß's law in (3+$p$) flat and infinite spatial dimensions.

b) Far away from the gravitating mass, i.e. $r >> R_j$ for all $j$, the sum over $\vec{m}$ can be replaced by an integral over the $p$ tori. Then eq. (B.4) becomes:

$$(B.4b)\quad \phi(x) \cong -\frac{G_{4+p}\cdot M}{\int\limits_{p\ tori} d^p x}\cdot \int\limits_{p\ tori} \frac{d^p x}{\left[r^2+\vec{x}^2\right]^{\frac{p+1}{2}}} \quad , \begin{cases} \int\limits_{p\ tori} d^p x = \Sigma_p = (2\pi)^p \prod\limits_{j=1}^{p} R_j \\ \int\limits_{p\ unit\ spheres} d^p \rho = \Omega_p = \dfrac{(2\pi)^{\frac{p+1}{2}}}{\Gamma\left(\frac{p+1}{2}\right)} \end{cases}$$

$$= -\frac{G_{4+p}\cdot M}{\Sigma_p}\cdot\frac{1}{r^{p+1}}\cdot r^p\cdot \int\limits_{p\ scaled\ tori} \frac{d^p\rho}{\left[1+\vec{\rho}^2\right]^{\frac{p+1}{2}}} \quad , |\vec{\rho}| << 1$$

$$\cong -\frac{G_{4+p}\cdot M}{\Sigma_p}\cdot\frac{1}{r^{p+1}}\cdot r^p\cdot \int\limits_{p\ unit\ spheres} d^p\rho$$

$$= -\frac{\Omega_p G_{4+p}\cdot M}{\Sigma_p}\cdot\frac{1}{r} = -G_4\cdot\frac{M}{r} \quad .$$

Therefore, far outside the compactification we recover Newtonian gravity. However, we realize, that in this limit the original ((3+$p$)+1)-dimensional gravitational coupling $G_{4+p}$ has been replaced by an effective four-dimensional coupling $G_4$, which is given by:

$$(B.5)\quad G_4 = \frac{\Omega_p}{\Sigma_p}\cdot G_{4+p} \quad .$$

From these extremes one can proceed to the intermediate case. There we have $R_j >> r$ for $0 \le j \le n$ and $R_j << r$ for $n < j \le p$. $s$ of the $p$ additional dimensions shall now have radii much larger than those of the other $p$-$s$ additional dimensions. Then the result is a mixture of eq.s (B.4a) and (B.4b) resulting in:

$$(B.4c)\quad \phi(r) \cong -\frac{\Omega_{p-s} G_{4+p}\cdot M}{\Sigma_{p-s}}\cdot\frac{1}{r^{s+1}} = -G_{4+s}\cdot\frac{M}{r} \quad ,$$

where the effective gravitational coupling $G_{4+s}$ is now defined as:

$$(B.6)\quad G_{4+s} = \frac{\Omega_{p-s}}{\Sigma_{p-s}}\cdot G_{4+p} \quad \Rightarrow \quad G_4 = \frac{\Omega_p}{\Sigma_p}\cdot\frac{\Sigma_{p-s}}{\Omega_{p-s}}\cdot G_{4+s} \quad .$$

As shown in [Ar98], this expression now provides a unique solution to the hierarchy problem. Consider the case of string theories like the $E8 \times E8$ heterotic superstring or like eleven-dimensional supergravity. Then *p* = 7. Now consider further a situation where the 7-*s* small additional dimensions are compactified to the Planck length, whereas the *s* larger additional dimensions shall have a radius *R*. A solution of the hierarchy problem, the problem of the enormous discrepancy between the four-dimensional gravitational coupling strength and the gauge couplings of the standard model, would exist, if the gravitational coupling for a particle at the electro-weak scale of about 1-2 TeV at distances of the corresponding Compton wave length, which is about $10^{-19}$ m, would be of the same order of magnitude. From this condition the hypothetical value of $G_{4+s}$ can be determined by use of eq. (B.4c), if it is further assumed that $R >> ƛ_{ew}^{C}$ where $ƛ_{ew}^{C}$ is the electro-weak Compton wave length. Let us here express all quantities in Planck units. Then we have, using (B.4c):

$$(B.7) \quad G_{4+s} \approx \frac{\left(ƛ_{ew}^{C}\right)^{s+1}}{M_{ew}} = \frac{\left(2\pi^2\right)^{\frac{s+1}{2}}}{M_{ew}^{s+2}} \quad .$$

Since $G_4$ equals unity in Planck units, we can achieve $G_4 = 1$ by eq. (B.6) if we choose the radius *R* of the *s* large additional dimensions to be:

$$G_4 = \frac{\Omega_p}{\Sigma_p} \cdot \frac{\Sigma_{p-s}}{\Omega_{p-s}} \cdot G_{4+s} = \frac{(2\pi)^{\frac{1-s}{2}} 2^s \cdot (p-s)!}{p! \cdot R^s} \cdot G_{4+s} \quad , G_4 = 1$$

$$\Leftrightarrow \quad (B.7') \quad R = \sqrt[s]{\frac{(2\pi)^{\frac{1-s}{2}} 2^s \cdot (p-s)!}{p!} \cdot G_{4+s}} \approx \sqrt[s]{\frac{\pi^{\frac{3+s}{2}} 2^{s+1} \cdot (p-s)!}{p!} \cdot \frac{1}{M_{ew}^{s+2}}} \quad .$$

For $s = 1, 2, 3$ and $M_{ew} = 2$ TeV, this leads to the following values fo *R*:

$$s = 1 \quad \Rightarrow R \approx 2 \cdot 10^{13} \cdot m \approx 140 \cdot Astronomical\ Units$$
$$s = 2 \quad \Rightarrow R \approx 1 \cdot mm$$
$$s = 3 \quad \Rightarrow R \approx 4 \cdot 10^{-9} \cdot m \quad .$$

These values change for $M_{ew} = 10$ TeV to:

$$s = 1 \quad \Rightarrow R \approx 2 \cdot 10^{11} \cdot m \approx 1 \cdot Astronomical\ Unit$$
$$s = 2 \quad \Rightarrow R \approx 40 \cdot \mu m$$
$$s = 3 \quad \Rightarrow R \approx 3 \cdot 10^{-10} \cdot m \quad .$$

Therefore we see, that the case $s$ = 1 is excluded by the data of the planetary motions. Yet, already the case $s$ = 2 might escape present experimental data [Ad00], if the (4+$s$)-dimensional gravitational scale is at 5 – 10 TeV and the strength of gravitation at distances of about ten times $R$ is already close to the Newtonian limit. Therefore, these higher dimensional superstring theories show together with the presence of $s$ = 2 or $s$ = 3 rather macroscopically compactified additional dimension (the remaining (7-$s$) dimensions are compactified at the four-dimensional Planck scale) a possibility to resolve the hierarchy problem, as this is presented in [Ar98]!

To evaluate the strength of the gravitation at intermediate ranges between infinity and $R$, one has to solve the Poisson equation eq. (B.1) in terms of harmonical functions, resuming eq.s (B.4) and (B.4a). We then have:

$$(B.8)\quad \phi(x) = -G_{4+p}\cdot M\sum_{\tilde{m}\in Z^p}\int d^p x\cdot\frac{e^{i\cdot\tilde{m}\cdot\tilde{x}}}{\left[r^2+\sum_{j=1}^{p}\left(x_j-2\pi\cdot R_j\cdot m_j\right)^2\right]^{\frac{p+1}{2}}}\quad,\begin{cases}\tilde{m}=(\frac{m_1}{R_1},\ldots,\frac{m_p}{R_p})\\ \tilde{x}=(x_1,\ldots,x_p)\end{cases}.$$

As shown in [Ke99], this integral can be transformed into:

$$(B.9)\quad \phi(x) = -G_{4+p}\cdot M\sum_{\tilde{m}\in Z^p}\frac{e^{-i\cdot\tilde{m}\cdot\tilde{x}}}{\left|\tilde{m}\right|^{n/2-1}}\int d\rho\cdot\frac{\rho^{n/2}\cdot J_{\frac{n}{2}-1}\left(\left|\tilde{m}\right|\cdot\rho\right)}{\left[r^2+\rho^2\right]^{\frac{p+1}{2}}}$$
$$= -\frac{G_4\cdot M}{r}\sum_{\tilde{m}\in Z^p}e^{-r\cdot\left|\tilde{m}\right|}\cdot e^{-i\cdot\tilde{m}\cdot\tilde{x}}\quad.$$

Here $\left|\tilde{m}\right| = \sqrt{\frac{m_1^2}{R_1^2}+\ldots+\frac{m_p^2}{R_p^2}}$ can be interpreted as the masses of Kaluza-Klein modes (KK-modes), the quantized eigenmode vibrations of strings in the directions of the compactified dimensions. Since we look at the non-relativistic limit, concerning distances $r$ being large compared to the electro-weak scale, it is possible to omit the detailed structure of the $p$-dimensional internal space thus setting $\tilde{x}$ = 0 in eq. (B.9). Then we get:

$$(B.10)\quad \phi(r) = -\frac{G_4\cdot M}{r}\sum_{\tilde{m}\in Z^p}e^{-r\cdot\left|\tilde{m}\right|}\quad.$$

To obtain the more precise behaviour of eq. (B.10) in the case, that is given by the intermediate case discussed above, except, that the distance $r$ shall be larger than all of the compactifi-cation radii, we note, that only terms up to first order have to be taken into account in eq. (B.10), since higher orders will be less dominant due to the exponential.

Zeroth order is given for $|\tilde{m}| = 0$:

$$(B.11) \quad \phi^{(0)}(r) = -\frac{G_4 \cdot M}{r} \ .$$

This is Newtonian gravity. The next order arise from all terms, where one $m_j$ is 1 or -1. $s$ of $p$ compactified dimensions have equal radii $R$, where $R$ is much larger than the radii of the ($n$-$s$) remaining dimensions, which are at the Planck scale. The Planck scale compactified $p$-$s$ dimensions will therefore contribute very massive KK-modes even for $m_j$ = 1 or $m_j$ = -1. Their fraction in eq. (B.10) is thus negligible. Therefore only the KK-modes of the macroscopically compactified dimensions survive in the sum of eq. (B.10). Since there are $s$ macroscopically compactified dimensions with a radius $R$, eq. (B.10) in first order of $|\tilde{m}|$, if one takes only the dominant contributions elaborated above, becomes:

$$(B.12) \quad \phi(r) = -\frac{G_4 \cdot M}{r}\left(1 + \alpha \cdot e^{-\frac{r}{\lambda}}\right) \quad where: \begin{cases} \alpha = 2 \cdot s \\ \lambda = \dfrac{1}{R} \end{cases} \ .$$

Here we see the emergence of an additional Yukawa-interaction generated by macroscopically compactified extra dimensions. Further we see, that $\alpha < 14$ for all eleven-dimensional theories, which will be compactified on tori.

In [Ke99] it is further shown, that for the cases, where $n$-spheres or Calabi-Yau manifolds are used as manifolds to compactify on, it can be shown, too, that $\alpha < 20$.

# Appendix C

# Scattering from rough surfaces

According to [Si88], the Born approximation for elastic scattering of a plane wave from a rough surface with perfect local reflectivity independent of the wavelength can be formulated as given by the following differential scattering cross section:

$$(C.1)\quad \left.\frac{d\sigma}{d\Omega}\right|_{one\ nucleus\ at\ \vec{r}} = C\cdot\left|\int\limits_{V\{z'|z'\geq z_{surface}\}} d^3r\cdot e^{i\cdot\vec{q}\cdot\vec{r}}\right|^2 \quad , \quad \vec{q}=\vec{k}-\vec{k}'\ , \quad \left|\vec{k}\right|=\left|\vec{k}'\right| \quad ,$$

where C is a proportionality constant. Introducing the spatial displacement $\vec{r}-\vec{r}'$, this can be rewritten as:

$$(C.2)\quad \frac{d\sigma}{d\Omega}=C\cdot\int\limits_{V\{z|z\geq z_{surface}\}} d^3r \int\limits_{V\{z'|z'\geq z_{surface}\}} d^3r'\cdot e^{i\cdot\vec{q}\cdot(\vec{r}-\vec{r}')} \ .$$

The integral (C.2) now can be reformulated in terms of surface integrals using Gauß' theorem. Suppose we have two vector fields $\vec{v}$ , $\vec{v}'$, chosen to be:

$$(C.3)\quad \begin{cases} \vec{v}=\dfrac{\vec{A}}{\vec{q}\vec{A}}\cdot e^{i\cdot\vec{q}\cdot\vec{r}} & \Leftrightarrow\quad \vec{\nabla}_{\vec{r}}\cdot\vec{v}=i\cdot e^{i\cdot\vec{q}\cdot\vec{r}} \\ \vec{v}'=\dfrac{\vec{A}}{\vec{q}\vec{A}}\cdot e^{-i\cdot\vec{q}\cdot\vec{r}'} & \Leftrightarrow\quad \vec{\nabla}_{\vec{r}'}\cdot\vec{v}'=-i\cdot e^{-i\cdot\vec{q}\cdot\vec{r}'} \end{cases}$$

$$\Rightarrow\quad (C.4)\quad \left(\vec{\nabla}_{\vec{r}}\cdot\vec{v}\right)\left(\vec{\nabla}_{\vec{r}'}\cdot\vec{v}'\right)=e^{i\cdot\vec{q}\cdot(\vec{r}-\vec{r}')}$$

Here $\vec{A}$ is an arbitrary unit vector. Then applying Gauß's theorem:

$$(C.5)\quad \int\limits_{\partial V} d\vec{f}\cdot\vec{v}=\int\limits_{V} d^3r\cdot\vec{\nabla}\vec{v}\ , \quad d\vec{f}\perp\partial V$$

yields together with eq.s (C.3) and (C.4) the following expression for eq. (C.2):

$$(C.6)\quad \frac{d\sigma}{d\Omega}=\frac{C}{\left(\vec{q}\vec{A}\right)^2}\cdot\int\limits_{\partial V}\left(d\vec{f}\cdot\vec{A}\right)\int\limits_{\partial V}\left(d\vec{f}'\cdot\vec{A}\right)\cdot e^{i\cdot\vec{q}\cdot(\vec{r}-\vec{r}')}\ , \quad \partial V \text{ matter surface}\,.$$

Now let $\vec{A}$ be the unit vector in *z*-direction, where the *z*-direction is the direction perpendicular to the average zero plane of the surface. Then eq. (C.6) simplifies to:

$$(C.7)\quad \frac{d\sigma}{d\Omega}=\frac{C}{q_z^2}\cdot\iint dxdy\cdot dx'dy'\cdot e^{i\cdot q_z\cdot[z(x,y)-z(x',y')]}\cdot e^{i\cdot\vec{q}_\rho\cdot\vec{\rho}}\ .$$

The next step is to use the notion that the height variation *z* of the surface is described by a gaussian random variable. That means that the functional dependence of *z(x,y)* in eq. (C.7) can be rewritten as an integral over the *z* variation directly, but weighted with a gaussian distribution eq. (2.2):

$$\frac{d\sigma}{d\Omega}=\frac{C}{q_z^2}\cdot\iint dXdY\cdot dxdy\cdot e^{i\cdot\vec{q}_\rho\cdot\vec{\rho}}\cdot\int\Delta z\cdot e^{i\cdot q_z\cdot\Delta z}\cdot\frac{1}{\sqrt{2\pi\cdot G(\vec{\rho})}}\cdot e^{-\frac{\Delta z^2(\vec{\rho})}{2\cdot G(\vec{\rho})}}$$

$$\text{where:}\quad X=x-x',\ Y=y-y',\ \vec{\rho}=(X,Y)$$

$$\Rightarrow\quad (C.8)\quad \frac{d\sigma}{d\Omega}=\frac{C}{q_z^2}\cdot L_xL_y\cdot\int dX\,dY\cdot e^{-\frac{1}{2}\cdot q_z^2\cdot G(X,Y)}\cdot e^{i\cdot\vec{q}_\rho\cdot\vec{\rho}}$$

for a given rough surface of extensions $L_x$ and $L_y$. The latter integral can be rewritten in terms of bessel functions using polar coordinates:

$$(C.8)\quad\Rightarrow\quad \frac{d\sigma}{d\Omega}=\frac{C}{q_z^2}\cdot L_xL_y\cdot\int d\rho\cdot\rho\cdot e^{-\frac{1}{2}\cdot q_z^2\cdot G(\rho)}\int_0^{2\pi}d\varphi\cdot e^{i\cdot q_\rho\cdot\rho\cdot\cos(\varphi)}\ .$$

$$\Rightarrow\quad (C.9)\quad \frac{d\sigma}{d\Omega}=2\pi\cdot\frac{C}{q_z^2}\cdot L_xL_y\cdot\int d\rho\cdot\rho\cdot e^{-\frac{1}{2}\cdot q_z^2\cdot G(\rho)}\cdot J_o(q_\rho\cdot\rho)$$

To proceed further, one now has to enter the specific form of the height-height correlation function, which is for our case the gaussian roughness given by eq. (2.1). It is convenient to split eq. (2.1) into the main height variance and the height autocorrelation function *C(X,Y)* as follows:

$$(2.1)\quad\Rightarrow$$

$$G(X,Y)=\left\langle[z(X,Y)-z(0,0)]^2\right\rangle=\left\langle z^2(X,Y)+z^2(0,0)-2\cdot z(X,Y)\cdot z(0,0)\right\rangle$$

$$=\left\langle z^2(X,Y)\right\rangle+\left\langle z^2(0,0)\right\rangle-2\cdot\left\langle z(X,Y)\cdot z(0,0)\right\rangle$$

$$\Rightarrow\quad (C.10)\quad \begin{cases}G(X,Y)=2\cdot\sigma^2-2\cdot C(X,Y)\\ C(X,Y)=\sigma^2\cdot e^{-\frac{\rho}{\xi}}\end{cases}\ .$$

Entering this result into eq. (C.9) and considering the case of scattering dominated by the spatial extension of the hills on the surface, that is $\rho \leq \xi$, one gets with the expansion of C(X,Y) to first order in $\rho / \xi$:

$$(C.11) \quad \frac{d\sigma}{d\Omega} = 2\pi \cdot \frac{C}{q_z^2} \cdot L_x L_y \cdot e^{-q_z^2 \cdot \sigma^2} \int d\rho \cdot \rho \cdot e^{-q_z^2 \cdot \sigma^2 \cdot \left(1 - \frac{\rho}{\xi}\right)} \cdot J_o(q_\rho \cdot \rho) \quad .$$

This integral can be evaluated analytically for the two limiting cases of arbitrary lateral *k*-vector transfer $q_\rho$ and of vanishing lateral *k*-vector transfer $q_\rho = 0$ :

$$(C.12) \quad \lim_{q_\rho \to 0} \frac{d\sigma}{d\Omega} \propto 2\pi \cdot \frac{C}{q_z^2} \cdot L_x L_y \cdot e^{-q_z^2 \cdot \sigma^2}$$

$$(C.13) \quad \left.\frac{d\sigma}{d\Omega}\right|_{q_\rho \neq 0} \propto 2\pi \cdot C \cdot L_x L_y \cdot \frac{\frac{\sigma^2}{\xi}}{\left(q_z^4 \cdot \frac{\sigma^4}{\xi^2} + q_\rho^2\right)^{3/2}} \quad .$$

The limiting case eq. (C.12) describes specular scattering from a rough surface, whereas eq. (C.13) accounts for the diffuse, i.e. non-specular scattering. Because there is no momentum vector transfer in specular scattering, eq. (C.12) depends solely on the mean height roughness $\sigma$. This way the results eq.s (C.12) and (C.13) already suggest a kind of measurement to determine the roughness parameters $\sigma$ and $\xi$.

First, one needs to choose a kind of radiation with a wavelength within one order of magnitude of the length scale, to which the roughness shall be known. Then, one has to perform a scan of the specularly scattered intensity as a function of the angle of incidence. The analysis of such a scan will then provide information about $\sigma$ according to (C.12). Second, one has to scan the diffusely scattered intensity. With $\sigma$ known from the analysis of the specular scan, non-specular scans will provide information about the height-height correlation length $\xi$ as shown in eq. (C.13).

For an analysis of the mirrors' roughness a suitable wavelength has to be chosen. These mirrors are polished to give reflections as purely specular as possible for UCN, which have wavelengths of about 50 Å. Mirrors appearing to be flat in this range of wavelengths are expected to have mean height roughnesses of about 10 Å or even less. Additionally, their correlation lengths should be orders of magnitude above the mean height roughness to provide sufficient flatness at this scale. Thus x-rays with wavelengths of 1 Å ... 10 Å should provide an efficient tool to determine the roughness by scattering at length scales, where roughness should show up.

Before one proceeds now towards such measurements, one should take another look at the result eq. (C.12). It is clearly visible, that the cross section diverges for a vanishing vertical wave vector transfer $q_z$ . Furthermore, eq. (C.12) does not describe the phenomenon of total external reflection, which one observes for grazing incidence using for instance x-rays. Yet, total reflection usually occurs at very small $q_z$ , which often leads to a situation of $q_z \cdot \sigma \leq 1$ in the regime, where total reflection occurs. There, however, the validity of first order BA approach presented above for very small angles of incidence is questionable at least.
Since the scattering will be done with x-rays with wavelengths 1 Å ... 10 Å under very small angles of incidence, scattering will naturally approach the regime of total external reflection. If the mean height roughness $\sigma$ shall be extracted from data correctly, one therefore will have to account for total external reflection.
According to ref. [Si88,Bo94] one may solve this problem, that first order BA does not resemble Fresnel reflectivity for very small angles of incidence, by replacing the plane waves as initial wave functions with the so-called Fresnel eigenstates, which provide exact solutions for the reflectivity and the transmittivity of matter with perfectly flat surfaces. This method is therefore called "distorted wave BA (DWBA)". To take into account, that both BA and DWBA of first order are likely to be in a $q_z \cdot \sigma \leq 1$ - regime if applied to the range of total reflection, one has systematically to proceed beyond first order DWBA to include at least second order DWBA, as it has been done in [Si88].
It is shown in these references that first order DWBA changes the expression of the specular reflectivity eq. (C.12) to:

$$(C.14) \quad \frac{d\sigma}{d\Omega} \propto \left| \frac{q_z - q_z^+}{q_z + q_z^+} \right|^2 \cdot \left| e^{-q_z \cdot q_z^+ \cdot \sigma^2} \right|, \quad q_z^+ = q_z \sqrt{1-\delta} \ , \delta = 1 - n^2 \ .$$

The first factor describes the Fresnel reflectivity with respect to the average zero plane of the surface of a material with a refractive index *n*. The second one is a modified form of eq. (C.12) called "Nevot-Croce factor" after Nevot and Croce [Né76]. The emergence of $q_z^+$ in the Nevot-Croce factor compared to (C.12) leads to a constant reflectivity of unity for angles of incidence below a certain value given by $\delta$ if $\delta > 0 \wedge \delta \in$ Ñ. In the case of complex $\delta$, the Nevot-Croce factor shows a small suppression of the reflectivity below unity even inside the region of total external reflection. This is due to the fact, that a complex $\delta$ corresponds to absorptive media. A typical plot of eq. (C.14) for a mean height roughness of $\sigma$ = 20 Å and an x-ray wavelength of $\lambda$ = 1.5 Å is shown in Fig. 2.2:

Fig. 2.2

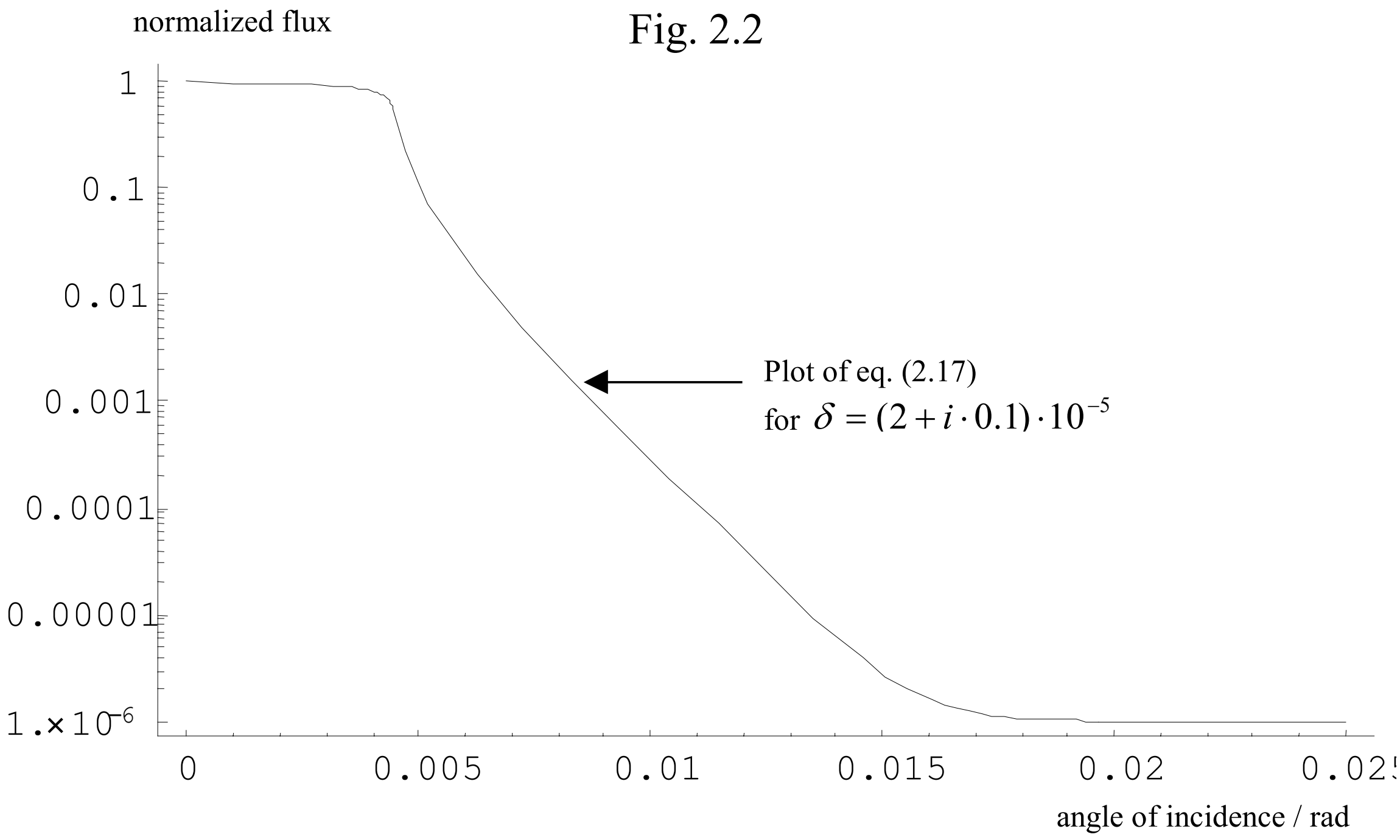
normalized flux
1
0.1
0.01
0.001
0.0001
0.00001
1.×10^-6
Plot of eq. (2.17)
for $\delta = (2 + i \cdot 0.1) \cdot 10^{-5}$
0
0.005
0.01
0.015
0.02
angle of incidence / rad

# Acknowledgments

Here now, at this point where the circle closes in, it is the time to thank everybody, who contributed to this work, namely:

Priv. Doz. Dr. Hartmut Abele for his guidance towards this subject, his fortifying support throughout the months (and years), his spirited enthusiasm, and – yeah! – his enduring patience.

Prof. Dr. Dirk Dubbers for handing me the chance to carry out this fascinating piece of work, and his interest in this subject.

Prof. Dr. Otto Nachtmann who took over the part of the second referee, for his highly elucidating discussions of the theoretical backgrounds of the subject, and his thorough interest in this work.

Dr. Valery Nesvizhevsky for the deep discussions on the subject throughout all the years, that passed, for his friendliness, and his unrestricted support. Dr. Irina Snigireva and her crew of the Microimaging Laboratory at the ESRF, Grenoble, for their incredible fast support concerning the high-resolution scans of the absorber's surface.

All the members of the research group. In particular, there should be named Dr. Martin Klein, for the talks with him, as well as his support concerning the depths of Mathematica, and Dr. Ulrich Schmidt, for his clarifying talks on the techniques of dealing with Airy functions.

Michaela Tscherneck for her incredible effort in proof-reading and many discussions throughout the time.

The few people, I really call my friends, for their support. They were my backbone.